\documentclass[a4paper,11pt]{article}
\usepackage{jinstpub} % for details on the use of the package, please see the JINST-author-manual
\usepackage{float}
\usepackage{siunitx}
\usepackage{multirow}
\usepackage{subcaption}
\usepackage{makecell}
\usepackage{appendix}
\title{\boldmath Short-Baseline Near Detector (SBND): Design and Initial Performance}
\collaboration{SBND Collaboration}
\author[aq]{P.~Abratenko,}
\author[p]{R.~Acciarri,}
\author[c]{C.~Adams,}
\author[p]{T.~‌Alafaro,}
\author[at]{L.~Aliaga-Soplin,}
\author[j]{R.~Alvarez-Garrote,}
\author[ac]{J.~‌Ameel,}
\author[u]{D.~Andrade Aldana,}
\author[z]{C.~Andreopoulos,}
\author[as]{A.~Antonakis,}

% ...
\affiliation[c]{Argonne National Laboratory, Lemont, IL 60439, USA}
% ...
\affiliation[j]{Centro de Investigaciones Energ{\'e}ticas, Medioambientales y Tecnol{\'o}gicas, Madrid E-28040, Spain}
% ...
\affiliation[p]{Fermi National Accelerator Laboratory, Batavia, IL 60510, USA}
% ...
\affiliation[u]{Illinois Institute of Technology, Chicago, IL 60616, USA}
% ...
\affiliation[z]{University of Liverpool, Liverpool L69 7ZE, United Kingdom}
% ...
\affiliation[ac]{University of Michigan, Ann Arbor, MI 48109, USA}
% ...
\affiliation[aq]{Tufts University, Medford, MA 02155, USA}
% ...
\affiliation[as]{University of California Santa Barbara, Santa Barbara, CA 93106, USA}
\affiliation[at]{University of Texas at Arlington, Arlington, TX 76019, USA}
% ...

\emailAdd{sbnd\_info@fnal.gov}

\abstract{

The Short-Baseline Near Detector (SBND) is a 112\,tonne active mass liquid argon time projection chamber (LArTPC) situated in the Booster Neutrino Beam at Fermilab, forming part of the Short-Baseline Neutrino (SBN) Program. SBND began operation in 2024 and has since accumulated the world's largest sample of neutrino-argon interactions. This paper describes the as-built detector and presents performance results from the first year of beam operation. All detector systems performed at or above design requirements. The liquid argon purity achieved an electron lifetime exceeding 30\,ms, ten times the design requirement, for more than 95\% of Run 1. The TPC wire plane readout achieved noise levels at the theoretical minimum for the given wire geometry (an average of 388~$e^-$ for the 4\,m long collection plane wires) with 99.3\% of readout channels live. Together, the liquid argon purity and TPC performance are enabling extremely clean event reconstruction in the SBND TPC. The photon detection system, consisting of passive reflectors and multiple active components (PMTs and X-ARAPUCAs), has achieved a light yield exceeding 10 photoelectrons per MeV throughout the detector volume and a timing resolution of a few nanoseconds, sufficient to resolve the 19\,ns bunch structure of the Booster Neutrino Beam. A nearly 4$\pi$ cosmic ray tagger system surrounds the TPC to mitigate cosmic backgrounds in neutrino and BSM physics analyses. The data acquisition system has demonstrated stable performance up to an event rate of 8\,Hz, and a data collection efficiency of 98.6\% was achieved throughout SBND’s first neutrino run in 2025. These results demonstrate that SBND represents the state of the art in multi-tonne-scale LArTPC detector performance.

}


\arxivnumber{}

\begin{document}
\maketitle
\flushbottom

\section{Introduction}
\label{sec:intro}
The Short-Baseline Near Detector experiment (SBND), located in the Booster Neutrino Beam (BNB) \cite{BNB} at Fermi National Accelerator Laboratory (Fermilab) in Batavia, Illinois (USA), is the near detector of the Short-Baseline Neutrino Program (SBN)~\cite{SBNproposal}. The mission of the SBN Program is a high-statistics exploration of short-baseline neutrino oscillations, $\nu-$Ar interactions, and Beyond the Standard Model (BSM) physics. %The BNB uses 8\,GeV protons from Fermilab's Booster synchrotron. 
Three large liquid argon time projection chamber (LArTPC) detectors are located along the BNB at increasing distance from the neutrino source: SBND (112\,tonnes active mass) at 110\,m, MicroBooNE~\cite{Microboone} (85\,tonnes) at 470\,m, and ICARUS~\cite{Kochanek:2011chl,ICARUS:2023gpo} (476\,tonnes) at 600\,m. 

The primary physics goal of the SBN Program is to definitively explore short-baseline neutrino anomalies such as seen by the LSND~\cite{LSND} and MiniBooNE~\cite{MiniBooNE:2008yuf,MiniBooNE:2013uba} experiments. The location of the detectors is optimized for detection of oscillations induced by a fourth (sterile) neutrino with $\Delta m^2 \sim 1\,\rm{eV}^2$. The ICARUS detector is located near the anticipated oscillation maximum while the SBND detector is as close to the $\nu-$\-source as practical to minimize any oscillation effect in the neutrino beam. The SBN Program is designed to provide $5\sigma$\ exclusion potential over the full LSND 99\% CL allowed region (for 3+1 sterile neutrino) with $6.6 \times 10^{20}$\ protons on target (POT)  combined running at the BNB.

The SBND experiment, which started operation in 2024, is accumulating the largest sample of $\nu-$Ar interactions ever recorded. In the first two years of operation SBND collected $7.0 \times 10^{20}$\ POT and is expected to collect data at least for one more year of beam operation. 
The first year of data, which this paper reports on, SBND collected $3.48 \times 10^{20}$\ POT   
corresponding to about a factor of 2 more $\nu-$Ar interactions than all previous or currently operating experiments combined.

The SBND design builds on the experience of other large LArTPC detectors built over the past twenty five years including ICARUS~\cite{Kochanek:2011chl,ICARUS:2023gpo}, ArgoNeuT~\cite{Anderson:2012vc}, and MicroBooNE~\cite{Microboone}. From inception, the design of SBND was planned as a prototyping opportunity for the Deep Underground Neutrino Experiment (DUNE) detectors in parallel with the ProtoDUNE single phase~\cite{ProtoDUNE} program at CERN. Multiple parts of SBND share designs with DUNE or explore alternate implementations.   

In this paper, we describe the as-built design, construction, and initial performance of the SBND detector. After a general overview of the detector design in section~\ref{sec:overview}, we start a description of supporting infrastructure: the cryostat in section~\ref{sec:cryostat} and the LAr purification and circulation system in section~\ref{sec:cryogenics}. This is followed by descriptions of the active detector components: the TPC cathode and drift high voltage (HV) in section~\ref{sec:TPCdrift}, the TPC readout planes, wire biasing and readout electronics in section~\ref{sec:TPCreadout}, the photon detection system (PDS) in section~\ref{sec:PDS}, and the cosmic ray tagger (CRT) system in section~\ref{sec:CRT}. Section~\ref{sec:DAQnTrigger} describes the data acquisition (DAQ), timing, and trigger systems. Each section includes both design descriptions and performance measurements from the first year beam operation in 2024-25. An overall summary of performance, demonstrating that all systems performed at or above design requirements is given in section~\ref{sec:summary}.

\section{Detector overview}
\label{sec:overview}

\begin{figure}[hbt]
  \centering
 % \subfloat{\includegraphics[width=0.60\textwidth]{figures/CryostatMarkIIILayout.png}}
  \subfloat{\includegraphics[width=0.95\textwidth]{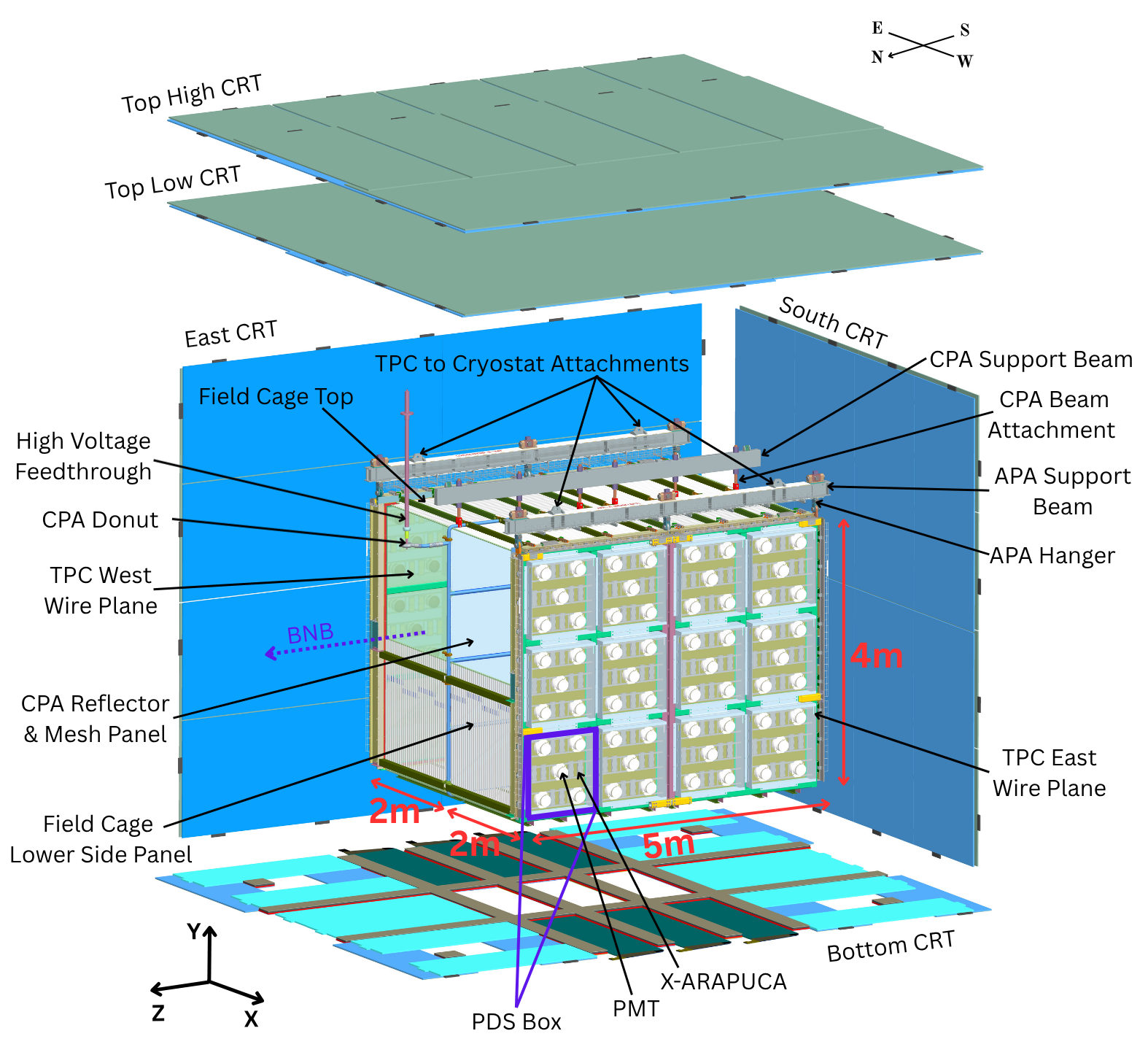}}
\caption{3D model of the SBND time projection chamber (TPC) detector viewed from the downstream (north) end. The cryostat, cosmic ray tagger (CRT) north and west walls are omitted for clarity. Also omitted is the upper half of the north field cage to reveal views of the cathode (CPA) and east anode plane (APA) and photon detection system (PDS).} 
\label{fig:TPC3Dmodel}
\end{figure}

 The central detector of SBND consists of the TPC and PDS which are completely immersed in a bath of LAr contained in a rectangular cryostat. Surrounding the cryostat of the central detector is the cosmic ray tagger (CRT) system used to identify charged particles entering from outside the cryostat. A 3D model of the SBND detector is shown in figure~\ref{fig:TPC3Dmodel} with the cryostat omitted for clarity. Table~\ref{tab:SBNDPar} shows the key dimensional specifications of the SBND detector.

\begin{table}[htb!]
\centering
\caption{Dimensional specifications and channel counts of the SBND detector.}
\label{tab:SBNDPar}
\begin{tabular}{c c}
\hline \hline{}
LAr total mass & 270\,tonnes \\ \hline
LAr active mass & 112\,tonnes \\ \hline
%TPC dimensions (H $\times$\ L $\times$\ W) & 4\,m $\times$\ 5\,m $\times$\ 4\,m \\ \hline
TPC H $\times$\ L $\times$\ W & 4\,m $\times$\ 5\,m $\times$\ 4\,m \\ \hline
TPC drift distance & 2\,m \\ \hline
TPC drift field & 500\,V/cm \\ \hline
TPC readout channels & 11264 \\ \hline
Photomultipliers & 120 \\ \hline
X-ARAPUCA modules & 192 \\ \hline
Central detector total weight &  7.7$\,$tonnes \\ \hline
%LAr Purity (e$^-$\ lifetime) & >\,3\,ms \\ \hline 
Cosmic ray tagger channels & 4540 \\ \hline
\hline
\end{tabular}
\end{table}

\begin{figure}[hbt]
  \centering
 % \subfloat{\includegraphics[width=0.60\textwidth]{figures/CryostatMarkIIILayout.png}}
  \subfloat{\includegraphics[width=0.98\textwidth]{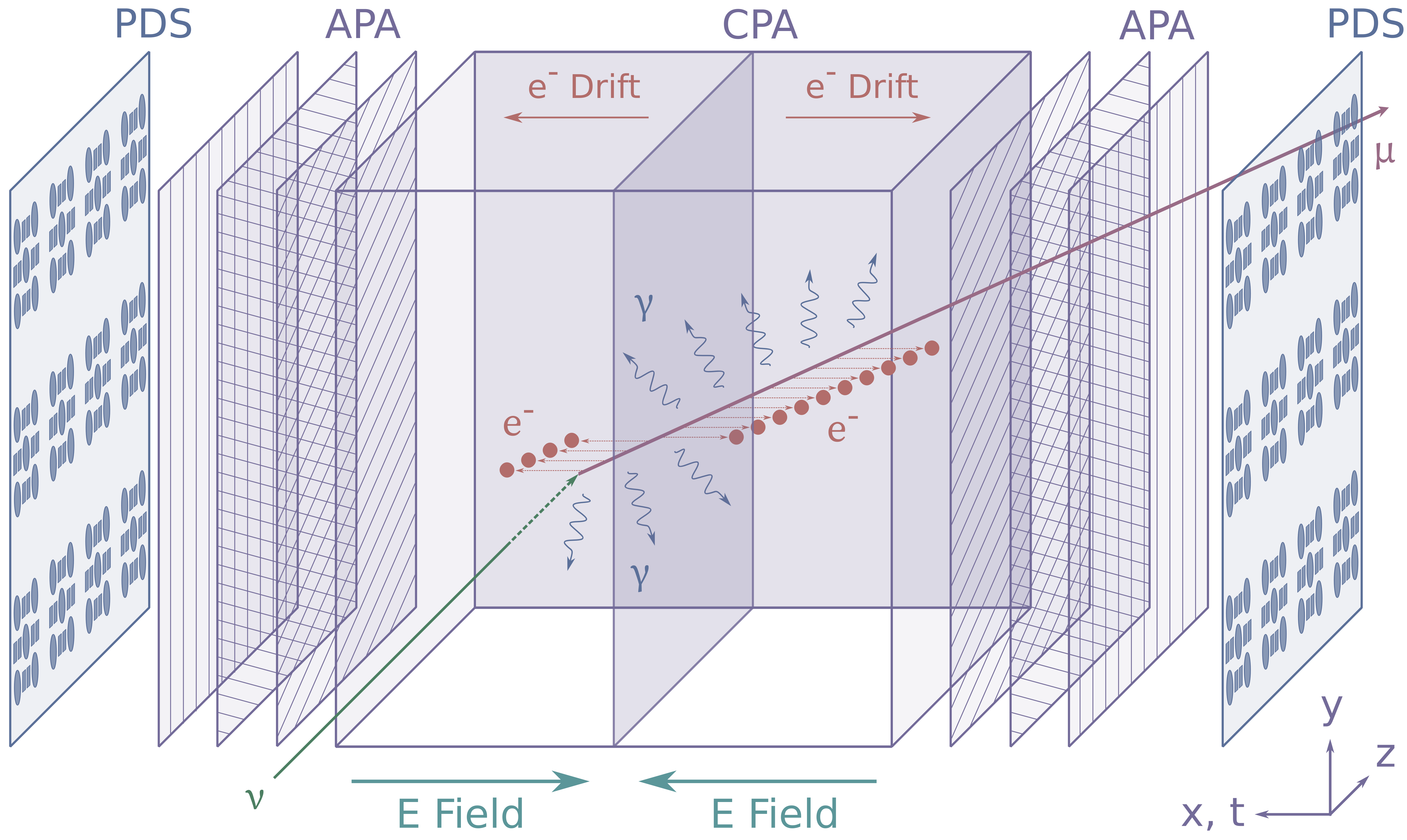}}
\caption{Conceptual diagram of how the SBND central detector functions. The SBND detector has two TPC drift volumes separated by a central cathode plane (CPA), ionization electrons are drifted horizontally towards wire readout planes known as anode plane assemblies (APAs), scintillation photons are collected by a PDS located behind each APA.} 
\label{fig:TPCconcept}
\end{figure}

LAr acts both as a nuclear target for the neutrino beam and the medium for detecting particles produced in those interactions.  A $\nu_\mu$\ interacting with an argon nucleus produces various particles, mainly a charged muon, pions and protons.  These charged particles traverse the LAr, ionizing argon atoms along their paths, losing energy but leaving a trail of ionization electrons (released from the outer atom shell) and scintillation light (from excited atoms and ionization recombination).  The ionization electrons drift in the electric field to the anode planes where their charge produces a proportional signal in the anode readout devices.  
The omni-directional vacuum ultraviolet scintillation light is detected by PDS sensors mounted behind the anode planes; the portion of emitted light which hits the cathode plane is wavelength shifted and reflected towards the anode planes.  Figure~\ref{fig:TPCconcept} shows a conceptual view of these basic TPC detection functions for a muon from a neutrino interaction.  

The SBND TPC consists of two equal-sized drift volumes sharing a common central cathode plane. With respect to the beam, the detector is positioned with the cathode plane parallel and with a horizontal offset of 74\,cm, and centered vertically, as shown in figure~\ref{fig:TPC3Dmodel}. Each TPC volume is read out by anodes forming the east (beam right) and west (beam left) sides of the central detector. Each anode consists of two planes of induction wires and one plane of collection wires. Wire signals from all three planes are amplified and digitized using electronics mounted directly to the detector and immersed in LAr to minimize electronic noise.

To ensure efficient collection of ionization electrons from the whole TPC volume, electronegative impurities must be removed from the argon using a filter. A minimum electron lifetime in LAr of 3\,ms was specified as the design goal, which corresponds to more than two times the drift time for an electron originating at the cathode.

For a ton-scale LArTPC, SBND's PDS is notable in both high photocathode coverage and granularity, with a total of 312 photo-sensitive devices.
The PDS employs two different sensor technologies: photomultiplier tubes (PMTs) and silicon photomultipliers (SiPMs). In addition, reflective plates with a wavelength-shifting coating are mounted on the TPC cathode plane to enhance light yield especially for interactions near the cathode. The SiPMs are mounted within a light trap system known as X-ARAPUCA~\cite{Machado:2018rfb}. Wavelength shifting material is employed for many of the sensors to shift the VUV scintillation light emitted by argon to the sensitive range of the sensors.
One fifth of the PMTs and half of the X-ARAPUCAs are deployed without wavelength shifter to detect exclusively the reflected (and wavelength-shifted) light from the cathode. SBND is the only large LArTPC with both PMTs and X-ARAPUCAs, offering the possibility of a direct performance comparison with a large sample of neutrinos.

Since the SBND detector is located on the surface, the TPC is exposed to a significant cosmic ray flux with ${\sim}2.9$\ cosmic muon tracks traversing the TPC within the time it takes for an electron to drift from the anode to the cathode (1.28\,ms). %window required to record a neutrino interaction. 
To help separate cosmic-ray activity from neutrino interactions, an array of scintillator walls 
forming the CRT system surrounds the cryostat, as shown in figure~\ref{fig:TPC3Dmodel}, with ${\sim}95\%$\ geometric coverage for tracks entering from above or the sides. The geometric coverage (${\sim} 60\%$) below the detector, used for tagging through-going cosmics, is limited by the necessity of fitting the CRT around the cryostat support frame.  These scintillator modules are located on the floor underneath the cryostat. 
%(see Figure~~\ref{fig:crttintoutprinciple}) 
Cosmic rays are tagged by comparing TPC track trajectories with hit space-points formed from coincident hits in orthogonal CRT modules and nanosecond-level timing matches between CRT and TPC track-associated PDS signals. 
%muons from cosmics in each 1.7\,ms long readout window of the TPC. 

Data from the TPC, PDS, CRT detectors are collected by the DAQ system in coincidence with delivery of neutrinos by the BNB. The neutrino beam timing is driven by the structure of the 8\,GeV proton beam provided by the Fermilab Booster. Protons are delivered to the BNB at a rate up to 5\,Hz in 1.6\,$\mu$s long bunches each consisting of 81 narrow buckets each separated by 19\,ns. The trigger system is used to synchronize detector readout with the period of active neutrino beam thus rejecting most cosmic rays. The finer beam structure can also be used with the fast PDS and CRT signals for further cosmic ray rejection offline.

\subsection{Assembly and installation strategy}
\label{sec:assembly}

\begin{figure}
    \centering
    \includegraphics[width=0.75\linewidth]{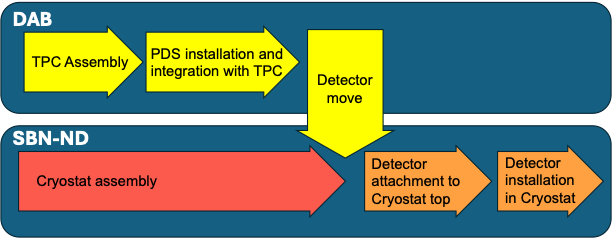}
    \caption{Construction and assembly of the central detector and cryostat happened in parallel, in two buildings at Fermilab, D0-Assembly building (DAB) and the near detector hall (SBN-ND).}
    \label{fig:detassembly}
\end{figure}

Components of the SBND central detector were built in Brazil, the United Kingdom, and the United States then assembled at Fermilab.  The detector assembly strategy is summarized in figure~\ref{fig:detassembly}.

The central detector (TPC with PDS) was assembled in a clean environment within a dedicated facility in the D0 Assembly Building (DAB) at Fermilab.  In parallel, the large-scale construction and installation of the cryostat took place at the SBN Near Detector building (SBN-ND). The central detector was then transported approximately three miles between the two buildings on a flat-bed truck.

\begin{figure}
    \centering
    \includegraphics[height=0.28\linewidth]{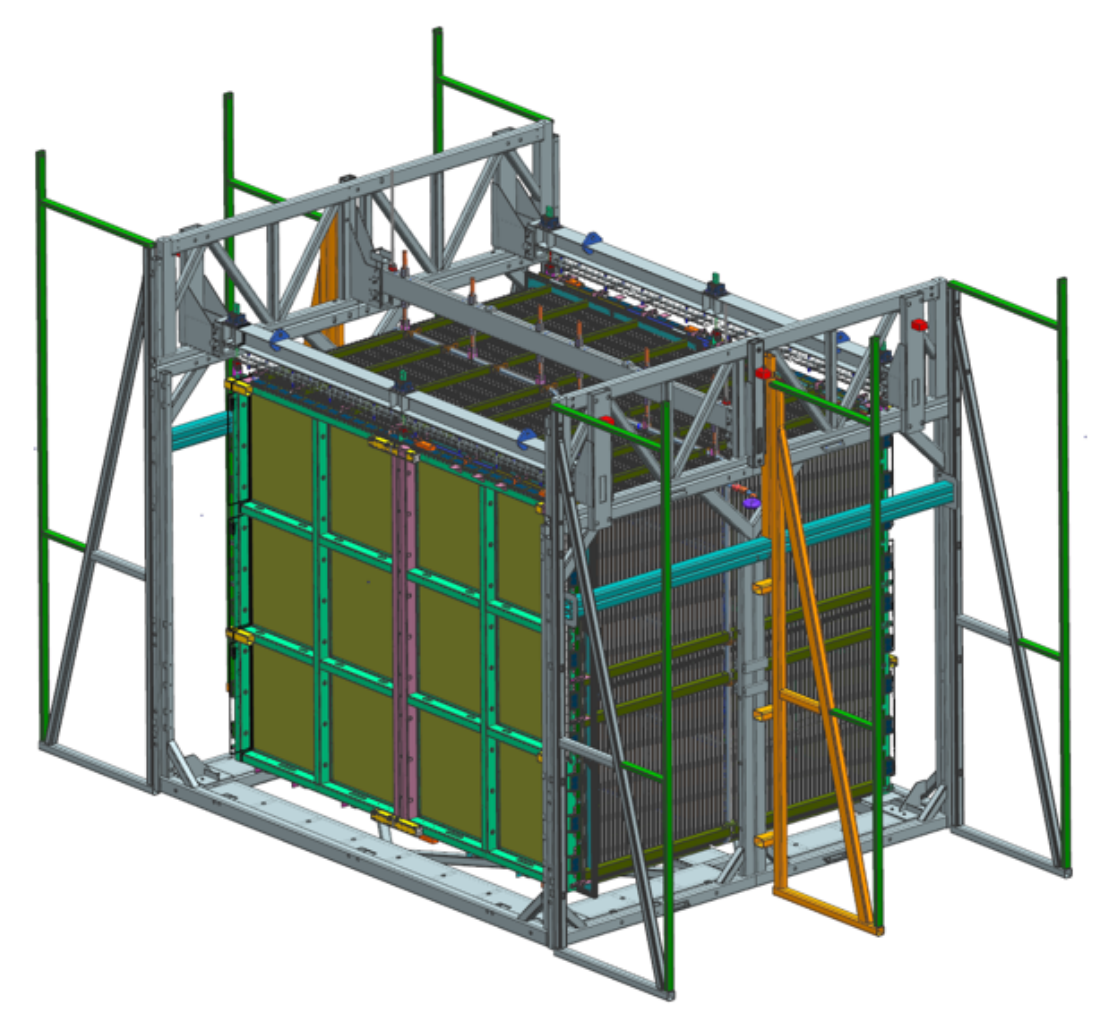}
    \includegraphics[height=0.28\linewidth]{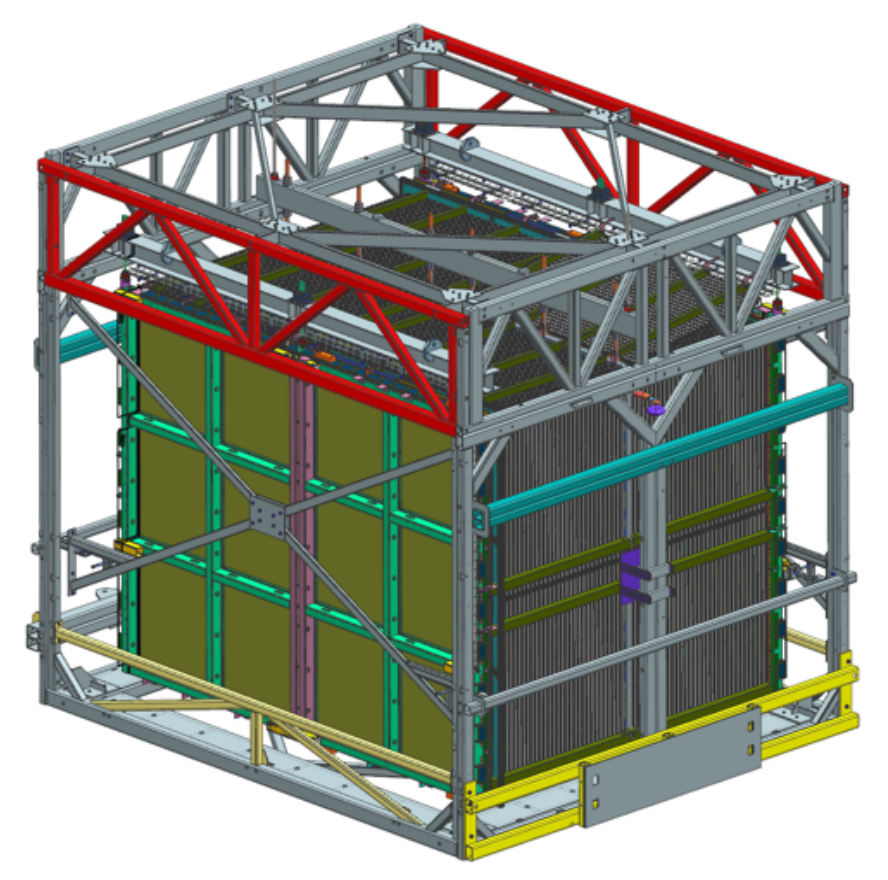}
    \includegraphics[height=0.28\linewidth]{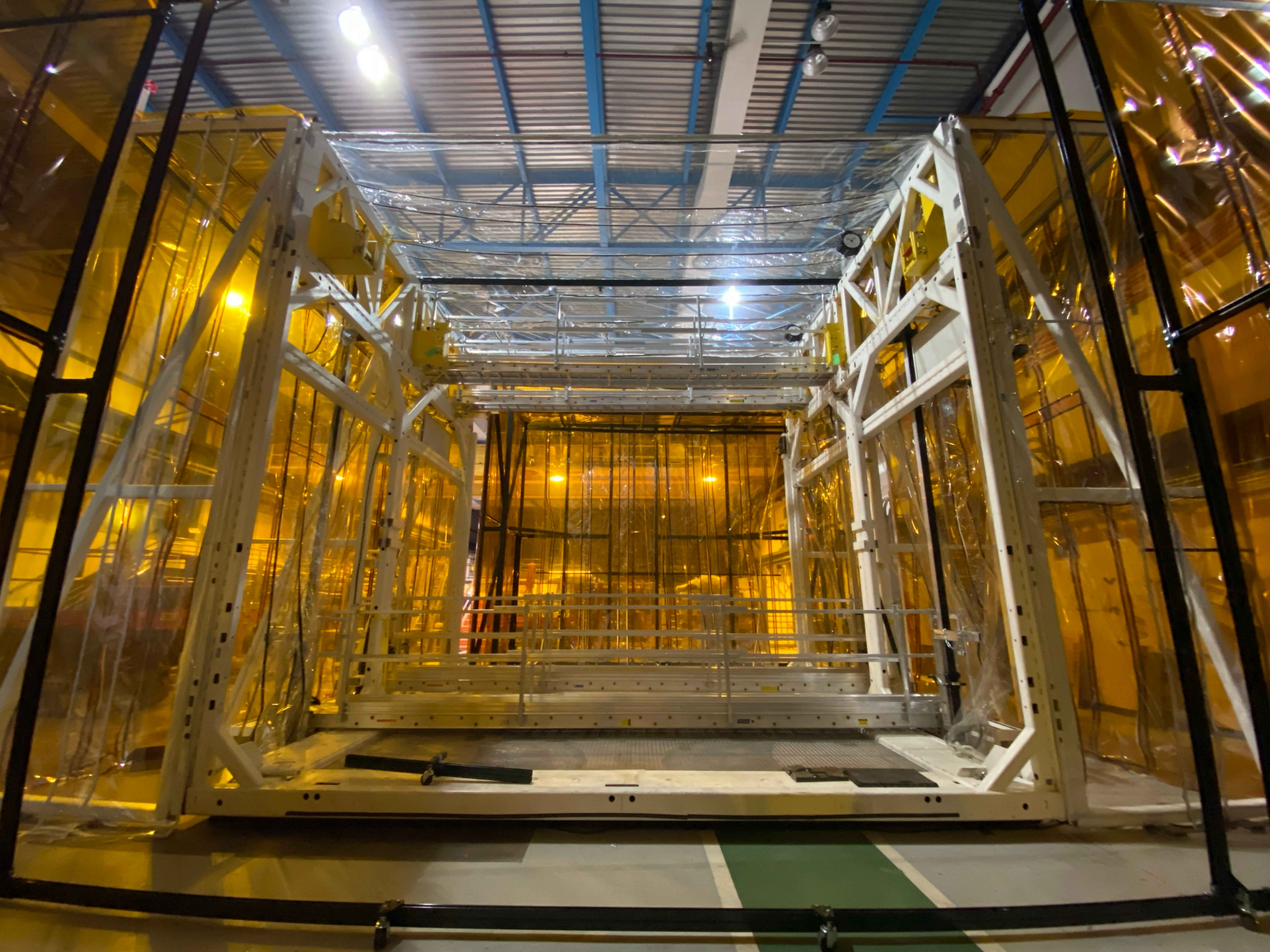}
    \caption{Left: the ATF supporting the SBND TPC in assembly configuration, with outriggers. Center: the ATF supporting the TPC in transportation configuration with cross-braces, stops, push-plate and tow-bar. Right: the central assembly space of the ATF with external light filtering clean tent structures, and scaffolding.}
    \label{fig:atf}
\end{figure}

The central detector was supported during both assembly and transport within a custom-engineered structure, the Assembly and Transportation Frame (ATF), as shown in figure \ref{fig:atf}. This structure had the capability to support the central detector during component assembly and attachment to the cryostat top, with features enabling precision alignment and adjustment in situ. It also formed the basic structure for a clean working environment with plastic sheeting selected to block out UV and blue light to protect the photo-sensitive materials of the PDS. 

During detector assembly, three of the primary mechanical components of the TPC -- namely the wire readout planes and the cathode plane -- were independently installed into the ATF. 
Each plane was supported at two points, and the ATF gave flexibility to adjust the position and angle of each plane in three dimensions while the detector was hanging. This facilitated assembling a TPC which was aligned to be within 5mm of square in all dimensions. Upon completion of the TPC assembly, the PDS was installed. A laser tracker system was used for a comprehensive measurement scheme to verify and maintain the relative positions of the TPC components during assembly, installation of PDS, transportation, and installation into the cryostat.  

Bracing and attachment points for the detector were connected to the ATF to ensure the safe transportation of the TPC.
Transportation between DAB and SBN-ND involved five main operations, and preparation to ensure that the detector was protected from exposure to light and moisture, while maintaining contact points for lifting and attachment: (1) the detector was moved out of DAB on Hillman rollers, using fork trucks attached to the ATF; (2) the ATF was lifted onto a flat-bed truck, using a 100\,ton capacity mobile crane, and fastened; (3) the flat-bed truck moved from DAB to SBN-ND at a speed of around 2\,mph, with constant monitoring of acceleration, angular tilt and light; (4) the ATF was craned from the truck onto steel tracks outside of SBN-ND; (5) the ATF was pushed into the SBN-ND building on Hillman rollers using a fork truck.  The one-day transport operation is documented in this video~\cite{movevideo}.

\begin{figure}[hbt]
  \centering
 % \subfloat{\includegraphics[width=0.60\textwidth]{figures/CryostatMarkIIILayout.png}}
  \subfloat{\includegraphics[width=0.49\textwidth]{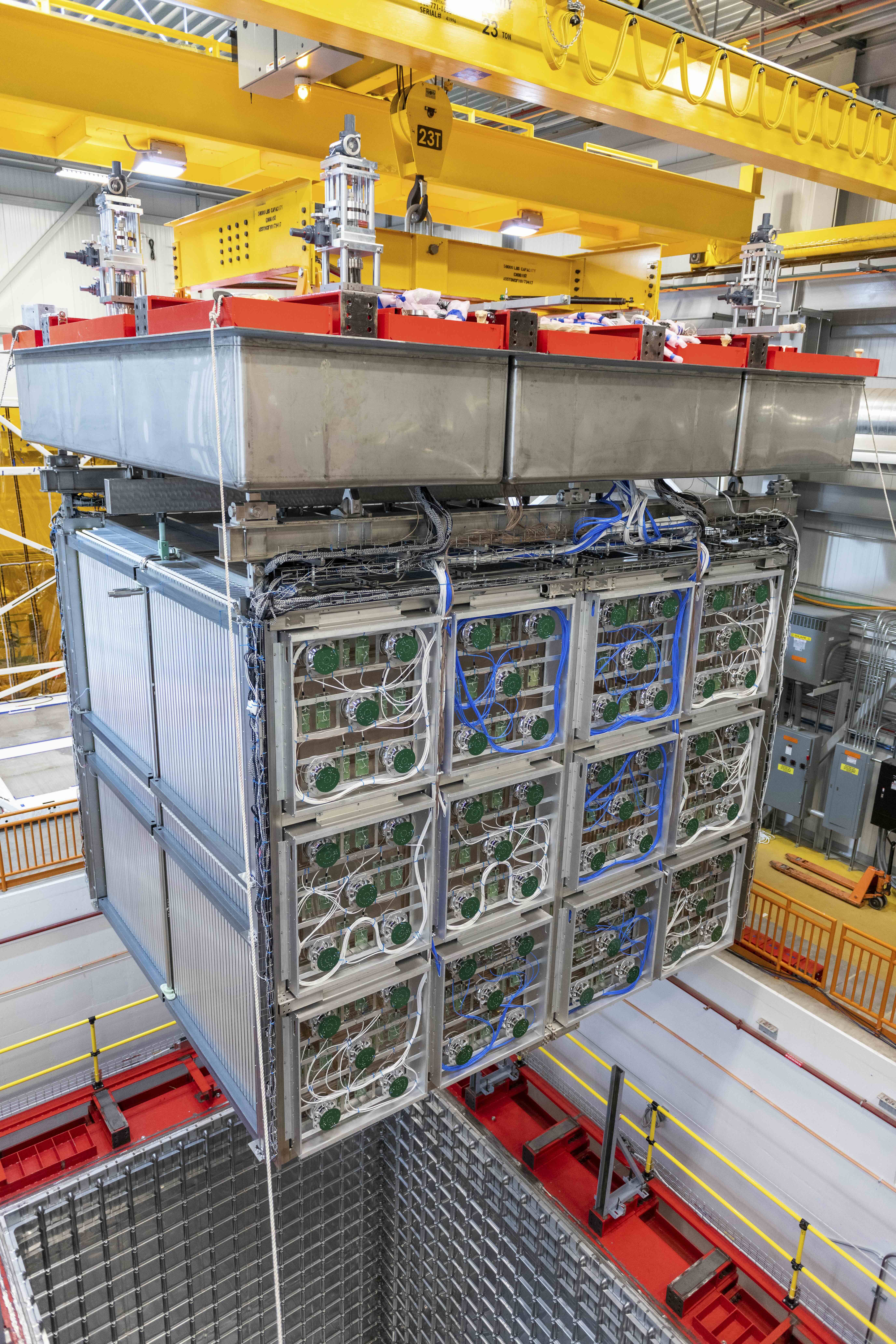}}
\caption{SBND central detector attached to the cryostat top cap being rotated before lowering into the cryostat.} 
\label{fig:rigging}
\end{figure}

The central detector is supported by suspension from appropriately engineered structures.  During assembly, the engineered structure was the ATF.  During operation, the engineered structure is the detector top cap of the cryostat.  The ATF was designed to support the combined weight of the detector with its top cap (estimated at around 22 tons) and perform the function of transferring the load.  Once the ATF with the detector was within the SBN-ND building, the top cap was placed onto the ATF and hangers attached, which shifted the suspended detector load from the ATF to the cryostat top cap; these hangers could be adjusted to enable the alignment of the TPC hanging position to a tolerance of 1\,cm.  The ATF has a hinged opening section to allow removal of the detector and top cap for installation of the detector into the cryostat.
Figure~\ref{fig:rigging} shows the completed  SBND central detector being lowered into the cryostat in the SBN-ND building.

After the central detector was installed in the cryostat, the electronics crates and cabling to the DAQ system were installed on top of the cryostat.  In parallel, cryogenics piping connections were made between the cryostat and  the cryogenic subsystems that had already been installed at three levels inside the building and on the surface outside the building.  

The modules of the CRT detector were constructed and tested in Switzerland then stored at Fermilab prior to installation. Installation followed a sequence driven by the physical constraints of other systems. The bottom CRT modules, located underneath the cryostat, were installed prior to assembly of the cryostat. The north side wall was installed prior to the final cryogenics piping which crosses over to the top of the north wall. The east, south, and west walls were installed after filling the cryostat with LAr to provide access to the cryostat walls for final qualification of the cryostat. 
The top CRT layers were installed after full TPC operation was confirmed with cosmic rays and beam neutrinos.

Assembly of the central detector in the ATF began in the summer of 2021 and completed in the fall of 2022 with installation of the PDS systems. Assembly of the cryostat in the SBN-ND building was completed at the same time. The detector was moved to the SBN-ND building in December 2022 and installed in the cryostat in April 2023. Installation of the detector readout electronics and cryogenic systems was completed in December 2023 with cryogenics commissioning commencing in January 2024.

\section{Cryostat}
\label{sec:cryostat}

The cryostat, designed to maintain a stable cryogenic environment for the SBND detector, consists of a free-standing warm carbon steel outer structure (the tertiary membrane) with a removable, independently insulated roof, two insulation layers sandwiching a secondary membrane, and a cold inner stainless steel liner (the primary membrane).
The outer structure provides the mechanical support for the insulation, the liquid argon containment system, and the detector. It is engineered to withstand the hydrostatic pressure of the liquid argon and the pressure of the gas volumes.
The cryostat roof, designed to support the suspended detector, is constructed in four sections that provide penetrations for both detector and cryogenic systems. The main cryostat parameters are shown in table~\ref{tab:CryostatPar}. The cryostat inner dimensions are designed to provide sufficient separation of the TPC high voltage components from the cryostat floor, walls, and roof. Additional interior space is provided downstream of the detector at the north end of the cryostat for cryogenics piping and personnel access. Section~\ref{sec:cryostatbox} describes the mechanical construction and assembly process for the main cryostat structure, section~\ref{sec:cryostatroof} describes the construction of the roof, and section~\ref{sec:cryostatperf} describes the performance of the cryostat after filling with LAr. 

\begin{table}[htb!]
\centering
\caption{SBND cryostat main parameters.}
\label{tab:CryostatPar}
\begin{tabular}{c c}
\hline \hline{}
%Cryostat external dimensions (H $\times$\ L $\times$\ W) & 7646\,mm $\times$\ 9414\,mm $\times$\ 7658\,mm \\ \hline 
%Cryostat external dimensions (H $\times$\ L $\times$\ W) & 7646 $\times$\ 9414\ $\times$\ 7658\,mm$^3$ \\ \hline 
Cryostat external dimensions (H $\times$\ L $\times$\ W) & 7650 $\times$\ 9410\ $\times$\ 7660\,mm$^3$ \\ \hline
%Cryostat internal dimensions (H $\times$\ L $\times$\ W) & 5423\,mm $\times$\ 7027\,mm $\times$\ 5202\,mm \\ \hline
%Cryostat internal dimensions (H $\times$\ L $\times$\ W) & 5423 $\times$\ 7027 $\times$\ 5202\,mm$^3$ \\ \hline
Cryostat internal dimensions (H $\times$\ L $\times$\ W) & 5420 $\times$\ 7030 $\times$\ 5200\,mm$^3$ \\ \hline
Gaseous Argon (GAr) ullage & ${\sim}5$\% of volume \\ \hline
Tertiary membrane thickness & 10\,mm \\ \hline
Insulation thickness & 790\,mm \\ \hline
Primary membrane thickness & 1.2\,mm \\ \hline
Roof thickness (insulation plus tertiary membrane) & 810\,mm \\ \hline
Walls and floor insulation density & 90\,kg/m$^3$ \\ \hline
Roof insulation density & 75\,kg/m$^3$ \\ \hline
%Total LAr volume & 189\,m$^3$ \\ \hline 
Design cryostat heat leak & 10\,W/m$^2$ \\ \hline
Actual cryostat heat leak (includes detector power) & 13\,W/m$^2$  \\ \hline
Maximum allowable operating pressure & 250\,mbarg \\ \hline
Standard operating gauge pressure & 50-100\,mbarg \\
\hline \hline
\end{tabular}

\end{table}

\subsection{Main cryostat structure}
\label{sec:cryostatbox}

The cryostat design is based on Mark-III technology~\cite{MarkIII} of Gaztransport \& Technigaz (GTT), developed through their collaboration with the CERN Neutrino Platform during the construction of the ProtoDUNE cryostats. GTT's contributions focused on the design of the insulation and containment (primary and secondary membrane) systems, while CERN's encompassed the design of the outer structure, component procurement, and cryostat installation. This technology is an evolution of liquefied natural gas (LNG) storage and transport techniques.  

\begin{figure}[hbt]
  \centering
%  \subfloat{\includegraphics[width=0.49\textwidth]{figures/CryostatMarkIIILayout.png}}
   \subfloat{\includegraphics[width=0.49\textwidth]{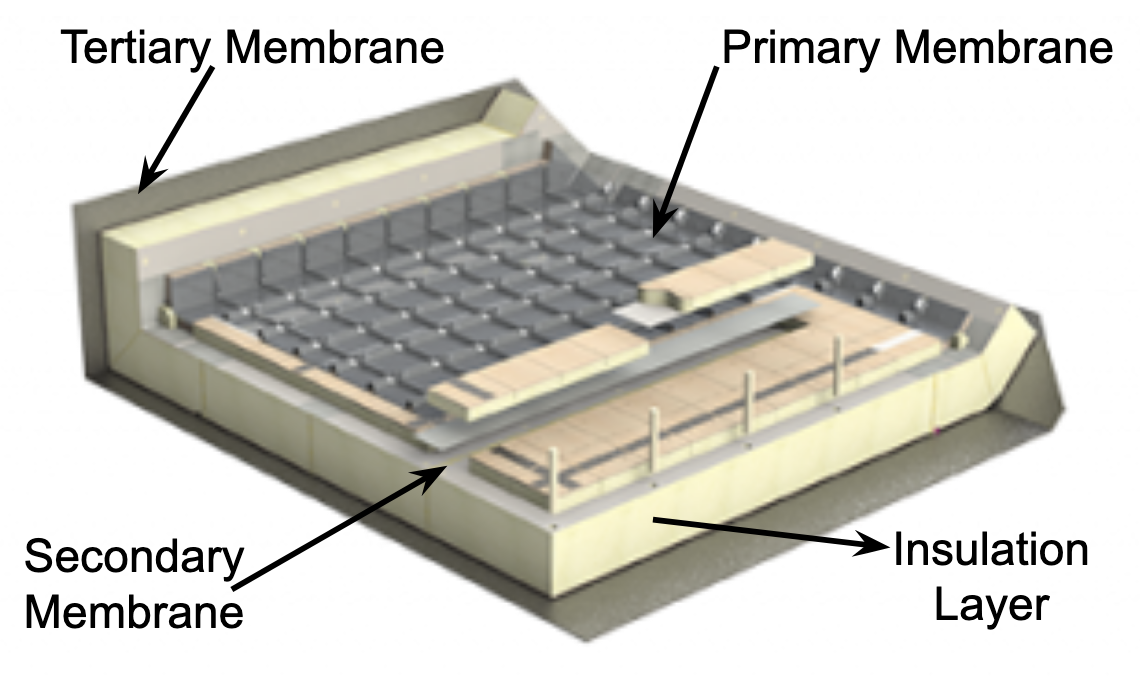}}
  \subfloat{\includegraphics[width=0.49\textwidth]{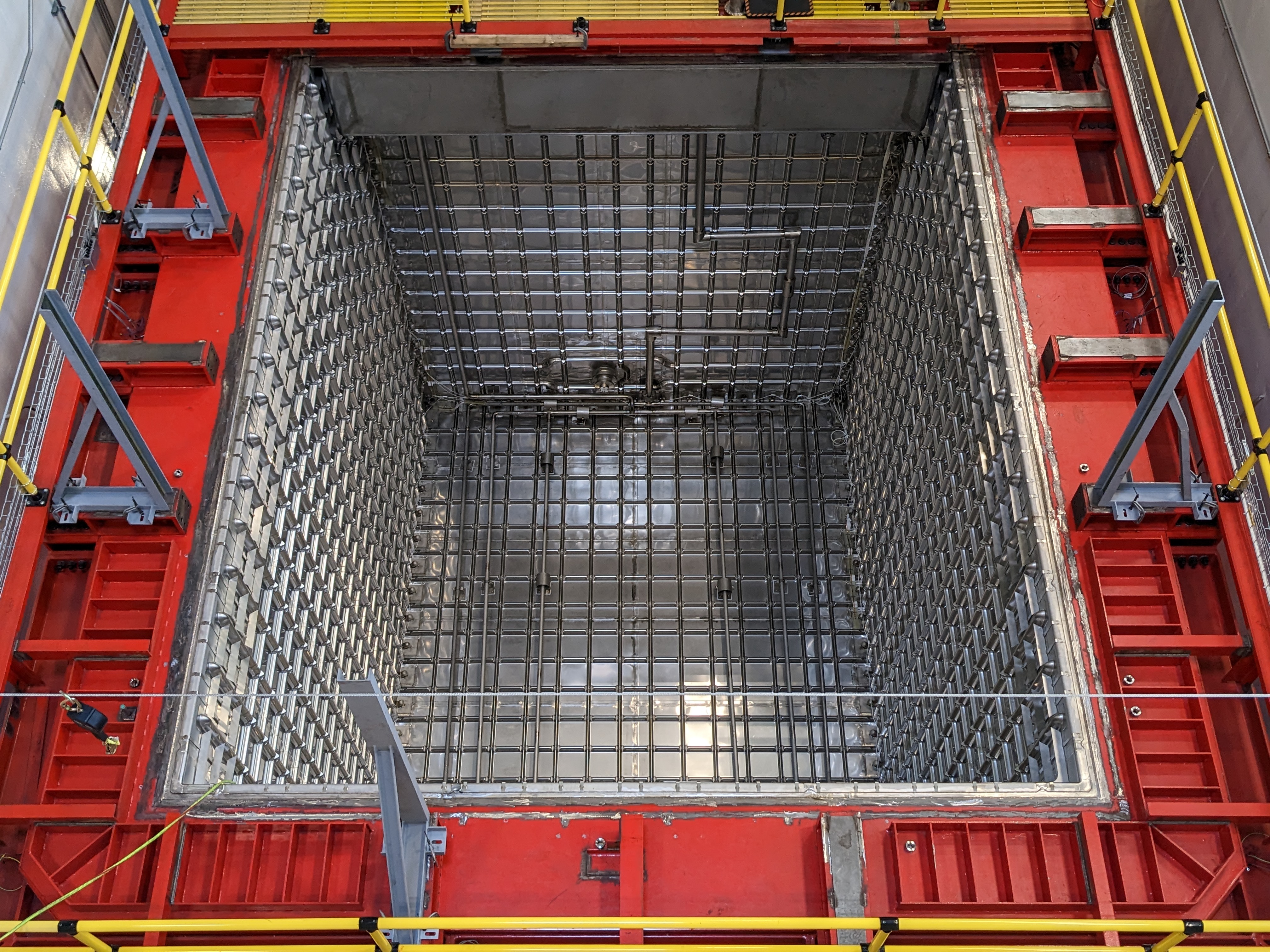}}
\caption{Left: Cross-section view of the layers of the SBND cryostat (figure courtesy of GTT). Right: View of the interior of the cryostat before detector insertion.} 
\label{fig:cryostat}
\end{figure}

We describe the components of the cryostat walls and floor starting from the outside moving in. The outermost layer is composed of 10\,mm thick steel plates welded together, forming the tertiary membrane, which are enclosed in a web of large 300\,mm wide vertical and horizontal I-beams bolted together. The plates feature ribs welded on the outside for additional strength and are secured to the I-beams by custom-designed clamps. The tertiary membrane provides an effective gas enclosure and allows for the maintenance of a positive nitrogen atmosphere of 4-10\,mbarg
inside the insulation volume, preventing condensation in the insulation layer. The full assembled tertiary membrane plus I-beams is known as the warm structure.
Moving inward the insulation layer is found, composed of two thick layers of reinforced polyurethane foam sandwiched between plywood plates.
The outer insulation layer is attached to the tertiary membrane by a set of rods and mastic adhesive. 
Embedded in this insulation layer is the secondary LAr containment membrane.
In contact with the innermost side of the foam insulation is the 304L stainless steel 1.2\,mm primary membrane that holds the LAr. The primary membrane can expand or shrink in two dimensions as a function of the temperature, thanks to a special corrugation configuration that allows it to respond like a bi-dimensional spring. 
Figure~\ref{fig:cryostat} shows a cross-sectional view of the insulation and membrane layers (left), as well as the interior of the cryostat before detector installation (right).

\begin{figure}[hbt]
    \centering
    \includegraphics[width=0.98\textwidth]{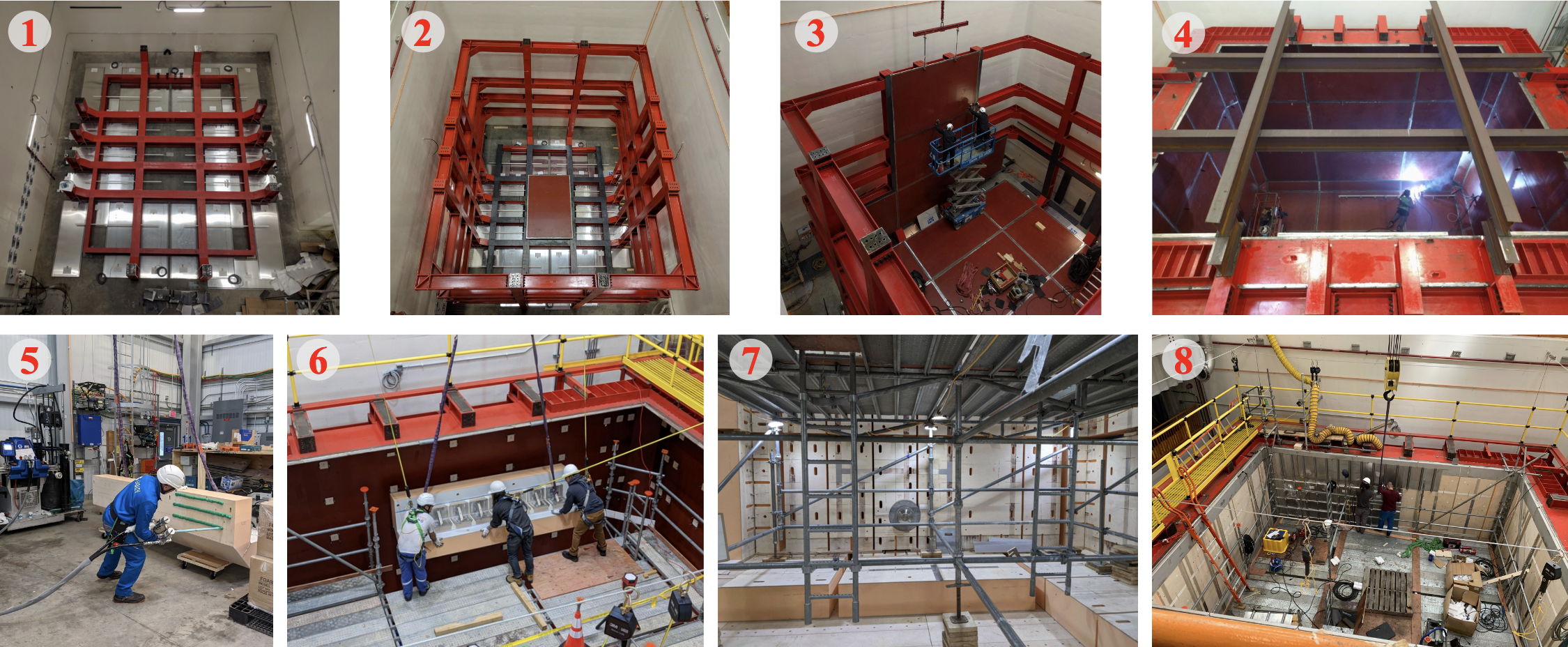}
    \caption{Cryostat assembly sequence: installation of the outer I-beams (1,2); positioning of the tertiary membrane plates (3); welding of the plates together (4); mastic applied to the outer insulation layer sections (5); outer insulation layer installation inside the vessel (6); second insulation layer installation (7); positioning and welding of the primary membrane sections (8).}
    \label{fig:CryostatAssembly}
\end{figure}

Figure~\ref{fig:CryostatAssembly} shows the sequence of steps for the cryostat assembly. The installation procedure started with a planarity survey of the building floor and the positioning of elastomer pads and G10 strips, for seismic protection and electrical isolation, on which the cryostat is placed.
The I-beam outer structure was assembled wall by wall and put into place one wall at the time.
Tertiary membrane plates were installed and seam welded together once they were all in place. A global three-dimensional laser scan was performed to verify the final internal dimensions.
Once the warm structure was in place, the pre-cut sections of insulation layers were installed, starting from the corners and moving towards the center of each wall. 
Triplex, a GTT-proprietary composite material consisting of a thin sheet of aluminum laminated on both sides by a layer of glass cloth and resin, was glued across the edges of the secondary membrane of adjacent insulation sections to guarantee the overall membrane tightness.
After each insulation layer was completed, all voids were filled with glass-wool or expanding spray-foam to reduce the gas convection as much as possible.
The installation proceeded by layers, with the inner (primary and corrugated) membrane installed last.
The integrity of each of the three membranes was verified through leak testing upon its completion: a bubble leak test for the tertiary membrane, a vacuum test for the secondary membrane (the outer insulation layer was evacuated and pressure rise monitored over several days to assess membrane tightness), and a helium leak test for the primary membrane.

\subsection{Cryostat roof}
\label{sec:cryostatroof}

\begin{figure}[hbt]
  \centering
    \includegraphics[width=0.95\textwidth]{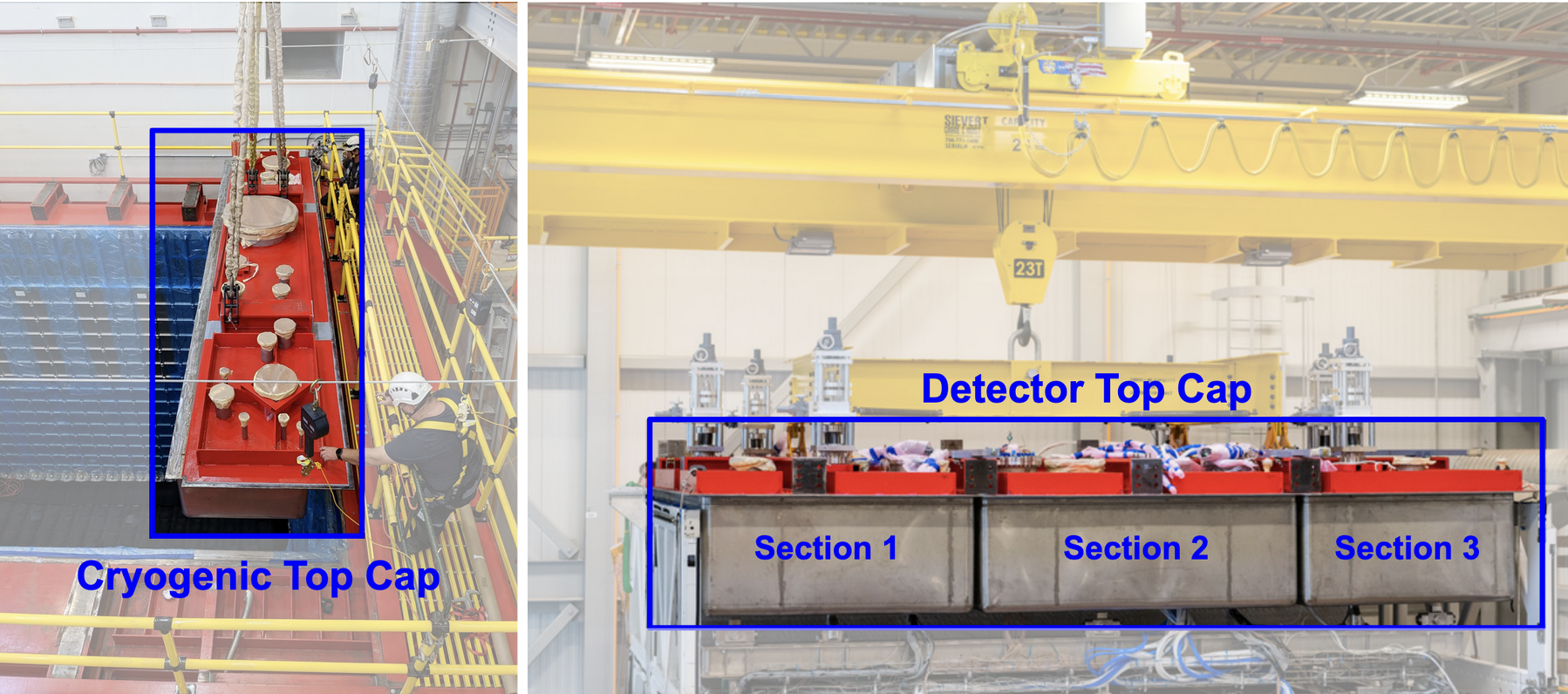}
\caption{Left: SBND cryostat cryogenic top cap as it was lowered into place on the cryostat. Right: SBND cryostat detector top cap, composed of three roof sections.} 
\label{fig:cryostattopcap}
\end{figure}

The cryostat roof is constructed in four sections: the cryogenic top cap, which accommodates penetrations for cryogenic systems and select detector instrumentation (purity monitors, high-voltage feedthrough, temperature sensors, and cameras), and three sections together forming the detector top cap, that mechanically supports the detector and provides penetrations for all detector power and signal cables.
The sections were sized to allow for transportation in standard 40\,ft shipping containers.
The outer layer of each roof section closely resembles that of the cryostat walls and floor: 10\,mm thick steel plates featuring ribs for additional strength, welded together and covered, once the roof is fully assembled, by a web of large 300\,mm wide I-beams that are bolted together. However, the insulation layer is simpler consisting of a single 800\,mm thick layer of polyurethane foam without a secondary membrane layer. The primary membrane of the top cap sections is formed by a 1.2\,mm thick 304L stainless steel flat sheet that encapsulates the insulation foam on all sides except the top (which is covered by the outer layer). This primary membrane can be flat because the top of the cryostat is exposed only to gaseous argon temperatures with thermal contraction low enough that a bi-dimensional spring effect is not required. Furthermore, the insulation and primary membrane layers do not extend all the way to the edges of the tertiary layer, leaving a border around the perimeter of each section. This creates, upon assembly, a 30\,mm gap between adjacent top cap sections, as well as between each top cap section and the main cryostat body to allow for welding during final installation.
Figure~\ref{fig:cryostattopcap} illustrates the cryogenic top cap (left) during installation, as well as the three sections of the detector top cap (right).

\begin{figure}
    \centering
    \includegraphics[width=0.9\linewidth]{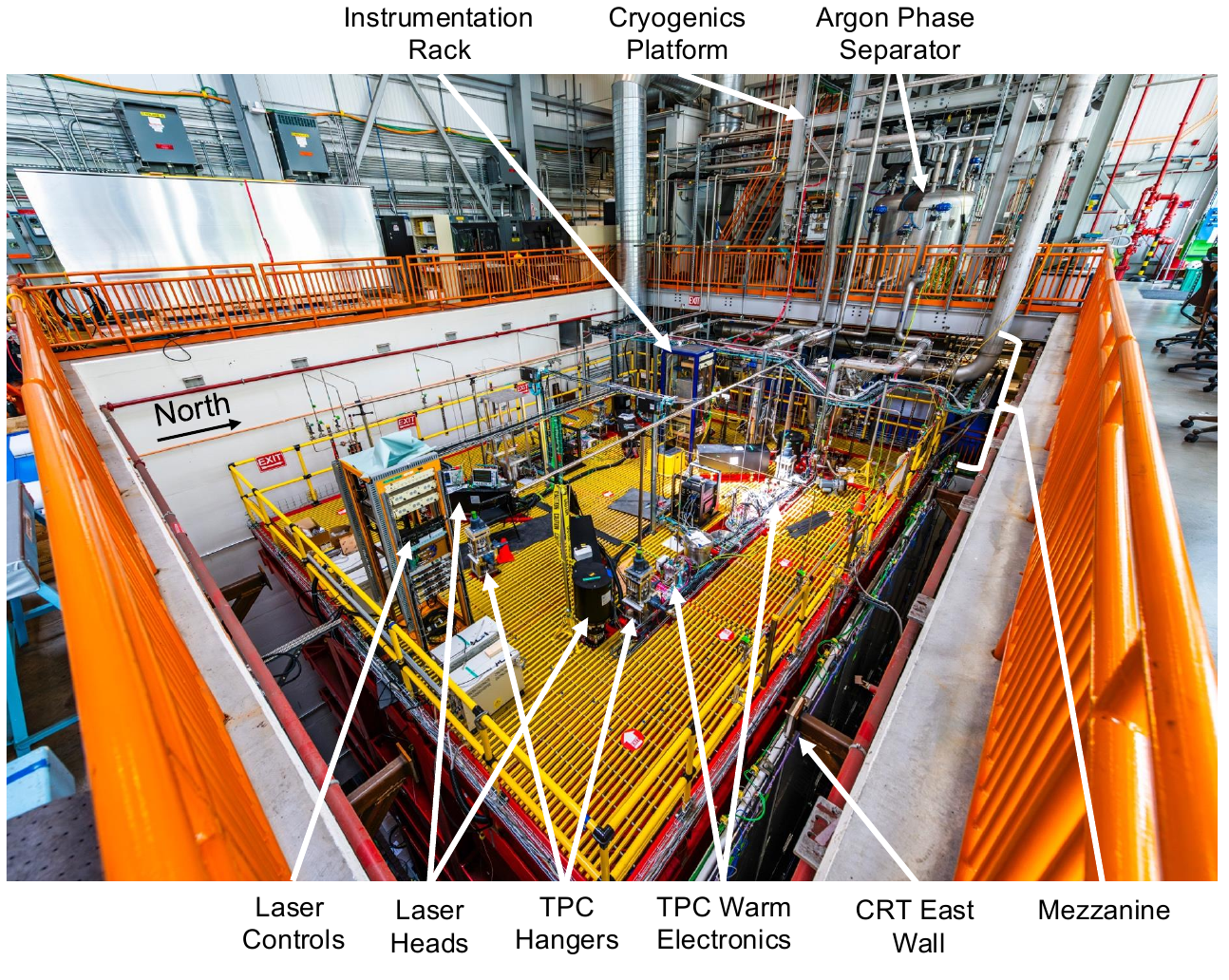}
    \caption{SBND Cryostat and detector during detector commissioning and prior to completion of CRT installation. In this view looking to the northwest, the cryogenics platform which holds the LAr condenser is located at the top rear of the photo.  Behind the cryostat just below the cryostat top is the  mezzanine which contains relay racks for detector power supplies, detector readout, data acquisition servers, and detector slow controls.}
    \label{fig:cryoincommissioningphase}
\end{figure}

Prior to installation, all sections were aligned by positioning them on top of the cryostat. The cryogenic top cap was welded to the cryostat to enable cryogenic system installation. The three sections of the detector top cap were welded together, but not to the rest of the cryostat, forming a single assembly. This assembly was then removed from the cryostat and placed atop the ATF, facilitating connection to the detector and routing of power and signal cables through the top cap penetrations. Following cable routing, the detector top cap, with the assembled detector hanging underneath, was installed onto the cryostat and welded in place. All welds, both between top sections and between top and cryostat, underwent dye penetrant inspection to ensure the overall cryostat integrity. The fully assembled cryostat with installed central detector is shown in figure~\ref{fig:cryoincommissioningphase}. This photograph was taken during the detector commissioning after the cryostat was filled with LAr.

\subsection{Performance}
\label{sec:cryostatperf}

\begin{figure}[htb!]
  \centering
    \includegraphics[width=0.95\textwidth]{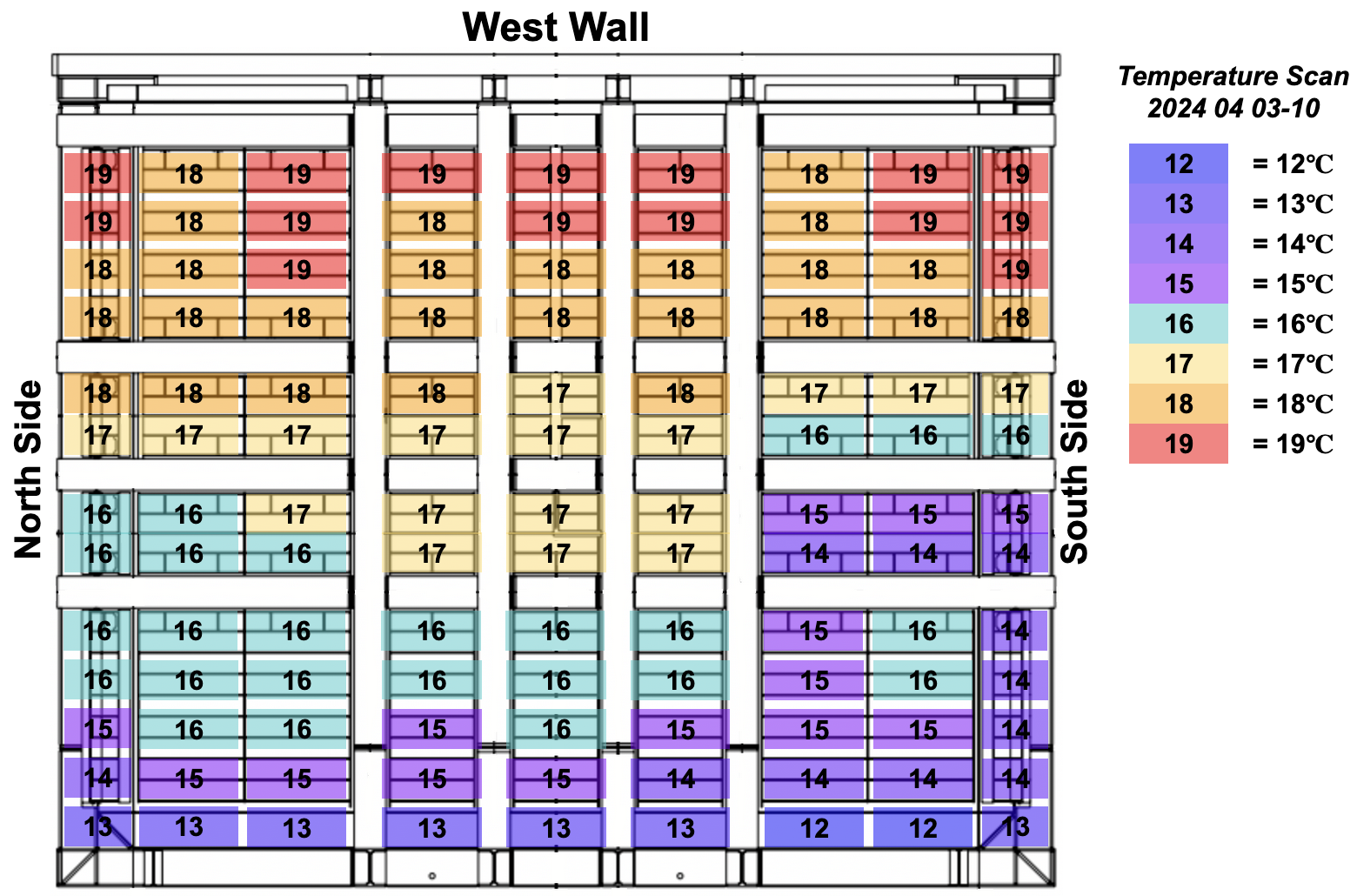}
\caption{Cryostat West wall temperature distribution, measured via infrared imaging under stable liquid argon conditions, demonstrating thermal homogeneity and close correlation with the adjacent pit wall temperature.} 
\label{fig:cryoTscan}
\end{figure}

During and following the filling of the cryostat with LAr (described in section~\ref{sec:cryocooldown}), a comprehensive set of temperature scans of the outer walls of the cryostat was conducted to identify potential cold spots. The scans were performed with a smartphone-connected infrared camera and corroborated with a handheld infrared scanning thermometer. The uncertainty in these measurements is estimated to be less than $\pm \rm{1^oC}$. Due to access limitations, only the upstream (south) and side (east and west) walls were scanned. During the filling process, the scans focused on the easily accessible bottom 2\,m of the walls to monitor changes as the liquid level increased. No significant cold spots were identified during the filling. Three weeks after the fill was completed, a scan was performed over the full height of the cryostat. This scan revealed a remarkably homogeneous temperature distribution across each cryostat wall, with a gradual temperature decrease from the top to the bottom, consistently remaining above the dew point maintained by the building HVAC system which is typically below $\rm{10^oC}$. All three walls exhibited similar overall temperatures and trends. Scanned temperatures were  also consistent with readout of RTDs  permanently mounted on the outer wall. The vertical gradient in cryostat surface temperature matches the gradient measured on the adjacent pit walls, located approximately 90\,cm from the cryostat walls. A representative plot of the temperature data recorded on the west wall of the cryostat is shown in figure~\ref{fig:cryoTscan}.

%\clearpage
\section{Argon purification and recirculation system}
\label{sec:cryogenics}
The SBND detector requires complete submersion in a stable liquid argon environment with a level of impurities that provides a minimum electron lifetime of three milliseconds, roughly equivalent to an oxygen concentration of 100 parts per trillion (ppt)~\cite{Geynisman_2017, Schwartz_2025}. Commercially available argon, such as was used to fill the SBND cryostat, typically contains impurities of up to five parts per million (ppm). Through a joint effort between FNAL and CERN, a fully automated cryogenic system was constructed to achieve the necessary argon level and purity in the detector cryostat, as well as to provide alarm notification and operate life-safety systems. Consisting of the major equipment listed in table\,\ref{tab:CryoMajorEquipment}, the system maintains pressure and level control of the cryostat argon volume and, through continuous circulation, improves its purity. 

\begin{table}[htb!]
\centering
  \caption{SBND Major Cryogenic Equipment.}
\label{tab:CryoMajorEquipment}
\begin{tabular}{c c}
\hline \hline
Liquid Argon Storage Dewar & 30000\,L \\ \hline 
Liquid Nitrogen Storage Dewar & 36000\,L \\ \hline
Liquid Nitrogen Supplied Argon Condenser & 14\,kW, 1100\,L \\ \hline
Argon Phase Separator & 800\,L \\ \hline
Liquid Argon Circulation Filter & 114\,L 4\AA \ mol. Sieve, 770\,L Cu \\ \hline
Argon Gas Filter & 10\,L 4\AA \ mol. Sieve, 40\,L Cu \\ \hline
Liquid Argon Circulation Pump (x2) & 1\,kg/s, 2.4\,bar diff. \\ \hline
Condenser Pump & 1\,kg/s, 2.4\,bar diff. \\ \hline
\hline
\end{tabular}

\end{table}

\begin{figure} [hbt]
    \centering
    \includegraphics[width=1\textwidth]{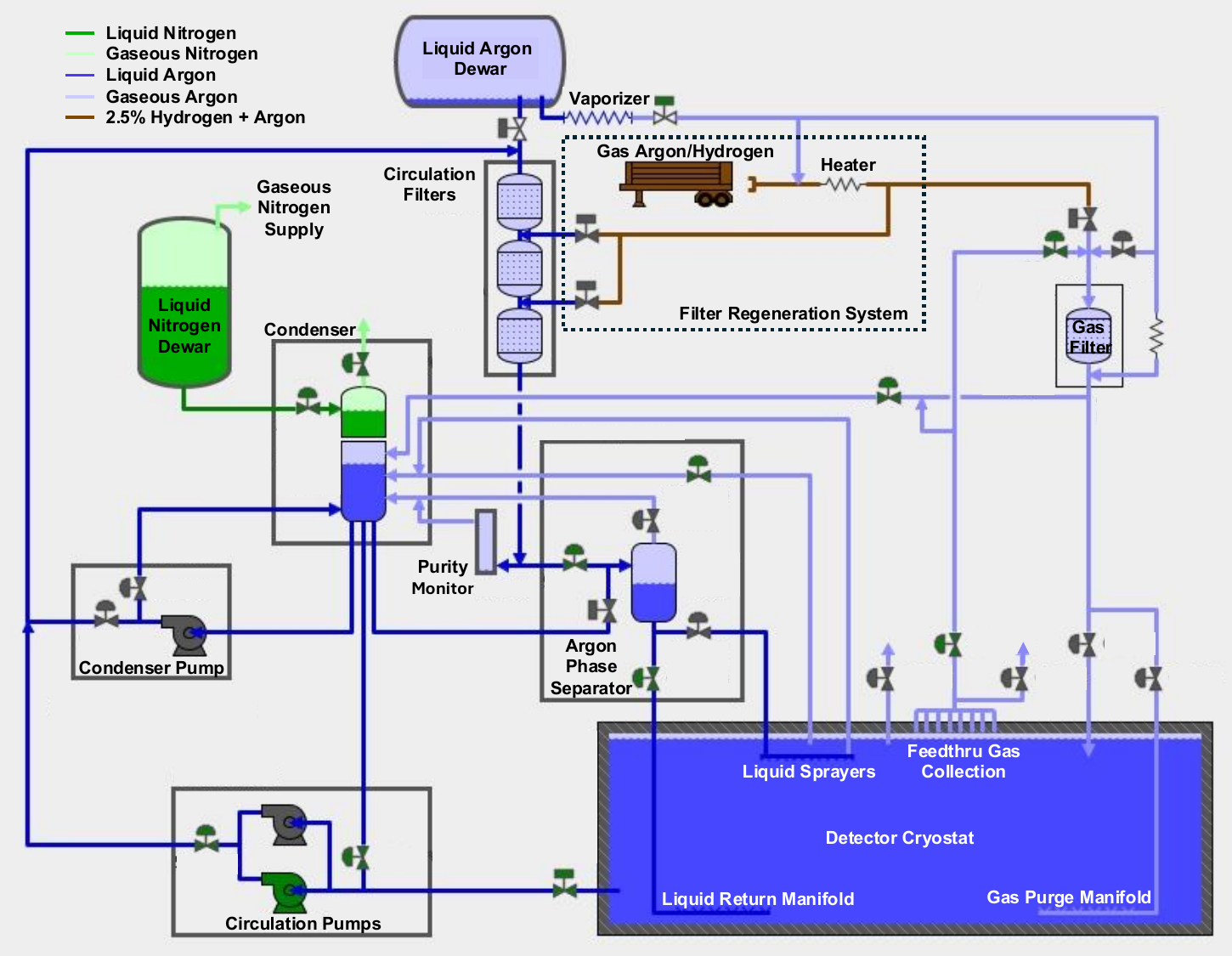}
    \caption{Process flow diagram from control system. Open valves are shown in green and indicate the normal operating configuration of the system in steady-state. The line colors represent different fluids as indicated by the key at the upper left. }
    \label{fig:cryogenicsPFD}
\end{figure}

\begin{figure} [hbt]
    \centering
    \includegraphics[width=1\textwidth]{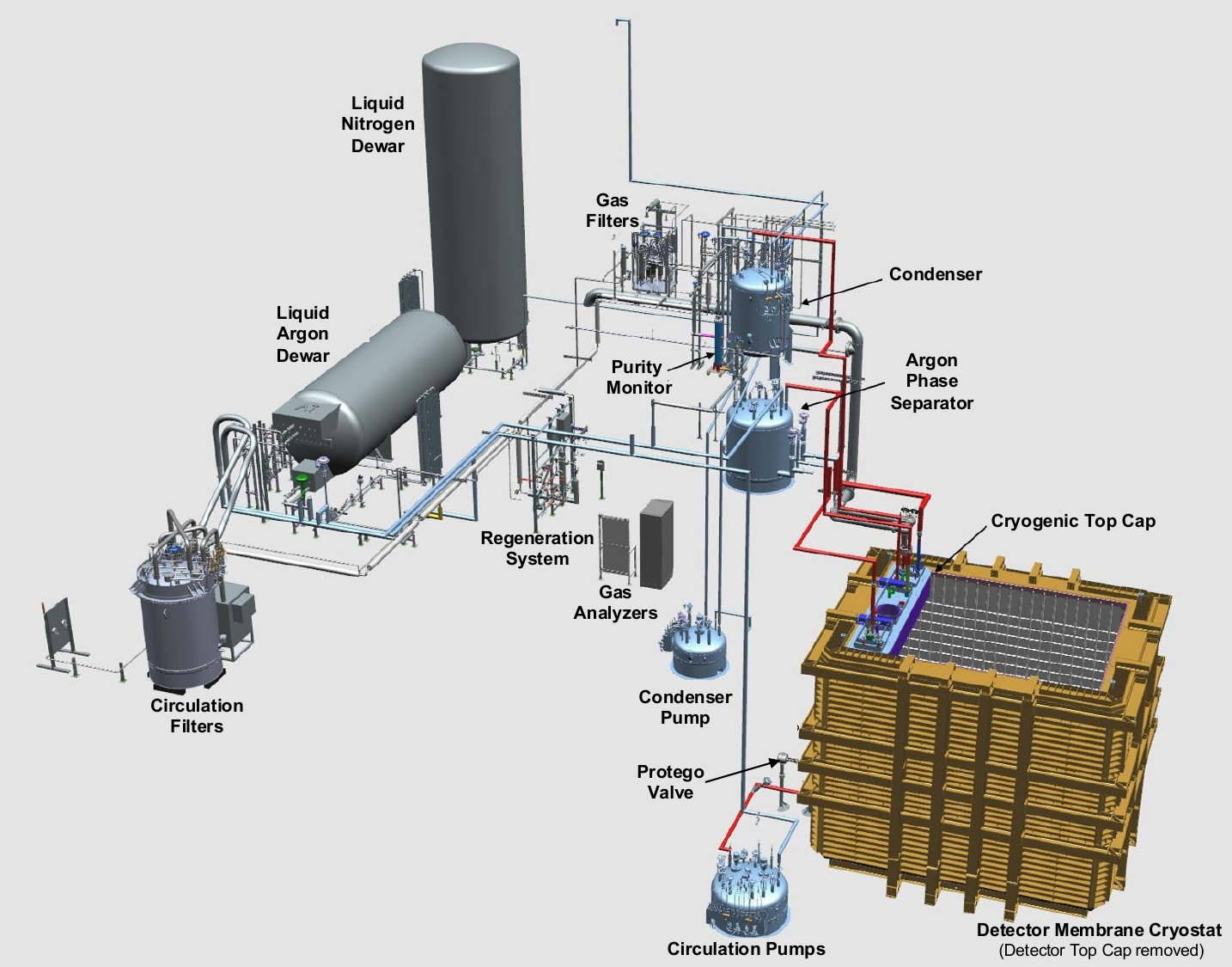}
    \caption{Cryogenic system 3D model with building and related structures removed to improve visualization. External to the building on ground level sit the liquid argon and nitrogen storage dewars, along with the circulation filters. The remainder of the system equipment resides within the building. The detector cryostat and circulation pumps are situated in the pit. The phase separator, regeneration system, and gas analysis equipment are installed on the ground level. A mezzanine level, between ground and pit bottom, is where the condenser pump resides. At the highest point is the condenser, as well as the warm gas filter and in-line purity monitor, installed on the cryogenics platform.}
    \label{fig:CryoModel}
\end{figure}

A simplified flow diagram and an annotated 3D model of this system are shown in figures \ref{fig:cryogenicsPFD} and \ref{fig:CryoModel}, respectively. Although the focus of this section will be on the primary process equipment, it should be noted that there are many ancillary systems that provide the necessary utility functions to that equipment, such as filter regeneration and argon purity monitoring.

\subsection{System description}

In steady-state operation, liquid argon is withdrawn from the cryostat near the bottom, circulated through a series of filters, then re-condensed back into the cryostat. The liquid is withdrawn from the cryostat through an integrated isolation valve manufactured by Protego~\cite{Protego}. The body of the valve extends horizontally through the tertiary membrane to the inner membrane, welded to each membrane to create a leak-tight seal. This specialized valve, considered a critical safety component of the system, is pneumatically operated and designed to fail closed inside the cryostat in the event an oxygen deficiency hazard (ODH) emergency is detected due to an unintended release of cryogens in the building. 

Immediately downstream of the Protego valve on the same level as the cryostat floor are the two Barber-Nichols liquid argon pumps~\cite{Barber-Nichols} that pump the liquid argon to generate the necessary circulation. These are housed in a vacuum vessel, more commonly referred to as a valve box, as it also contains the necessary control and isolation valves, as well as interconnecting piping. The pumps are installed in a parallel configuration, as a single pump is sufficient for full system operation, allowing for minimal time without circulation should one pump reach its service interval or otherwise need to be taken offline. Each pump provides a stable argon circulation rate of nominally 1\,\si[per-mode=symbol]{\kilo\gram\per\second} and a discharge pressure of 4\,bar absolute, effectively completing a volume exchange of the argon within the cryostat every three days.

The operating circulation pump directs flow through vacuum-jacketed piping to another valve box for purification. Installed at ground level approximately 10\,m above the pumps and outside the building, this valve box contains three vessels connected in series and filled with purification media: one with 114\,L of 4\,\AA \  molecular sieve to remove water and two with 385\,L each of activated copper (BASF Cu-0226 S) on an alumina substrate to remove oxygen. Should saturation occur, both types of media are able to be regenerated using an integrated regeneration system. 

Following purification, the argon flow enters the phase separator valve box, where the vapor produced by heat ingress and pressure losses is separated from the liquid in a large vessel. The purified liquid in this phase separator vessel returns to the bottom of the cryostat through a custom designed distribution manifold inside the cryostat manufactured by Kriosystem~\cite{Kriosystem}. The vapor is directed to a separate valve box containing a condenser directly above it, where a tube heat exchanger returns the vapor to a liquid state using nitrogen from the external storage dewar. In normal operation, the liquid argon in the condenser gravity drains to the circulation pump valve box and is reintroduced at the pump suction and back through the process. Conversely, when the circulation pump is inactive, the argon can also flow back to the cryostat volume through the open Protego valve, thereby preserving operation of the condenser in such instances. An alternate path through the condenser pump valve box directly to the circulation filters is provided but not used. 

The cryostat itself is well insulated but does generate boil-off argon vapor as expected due to heat ingress through the insulation and top caps, as well as internal electronics. This vapor is also directed to the condenser to maintain a stable pressure within the cryostat. While the majority flows directly through a vacuum-insulated pipe, a small amount is collected through feedthroughs on the detector top caps and is routed through a gas filter prior to the condenser. This significantly reduces the amount of impurities outgassed from cables and electronics in those feedthroughs from reaching the liquid argon in the cryostat.

System automation is provided by hardware in the form of programmable logic controllers (PLC) made by Beckhoff~\cite{beckhoff} paired with SCADA and HMI software called Ignition~\cite{Ignition}. There are six PLCs distributed around the system that communicate with a coordinator PLC to control the cryogenic equipment, a design that simplifies the installation. An additional PLC, which also communicates with the main coordinator PLC,  manages the life safety system monitoring ODH and alerts personnel, as well as configures the system to a safe state if such an event should occur.

\subsection{Monitoring instrumentation}\label{sec:interlocks}

The status of the detector, the cryostat and the liquid argon is monitored by temperature sensors, purity monitors, cold cameras and level meters inside of the cryostat and cryogenic system. Some were used primarily during the cooldown and filling phases, and some remain essential throughout detector operation, enabling real-time monitoring of key parameters for safe and effective operation. \\

\noindent\textbf{Temperature sensors} \\
 SBND uses 111 resistance temperature detectors (RTDs) to monitor the temperature of the cryostat at key locations in order to validate the cryostat performance. 
 These sensors are distributed throughout the cryostat, on inner walls (42), within the insulation space (33), and on the outer walls (24), as well as mounted to key parts of the TPC (12). Four sensors are mounted on the cryostat floor and twenty eight sensors are mounted in a vertical array on the inner wall spaced 15-30\,cm apart. An additional ten sensors are mounted on the detector and roof of the cryostat. 
 The temperature sensors played a crucial role during the cryostat cooling and filling processes, as shown in figure~\ref{fig:cooldownplot}, monitoring the liquid level and measuring the temperature gradient across the detector. 
 Section~\ref{sec:cryocooldown} describes the detector requirements and cooldown process. \\

\noindent\textbf{Purity monitors} \\
%\label{sec:PurityMonitor}
SBND uses purity monitors~\cite{Adamowski_2014} to monitor the electron lifetime in LAr primarily during commissioning. Purity monitors operate by generating electrons from a photocathode through the photoelectric effect~\cite{BONETTI1990135} using a Xe flash lamp. These electrons drift toward an anode under an applied electric field. By measuring the charge difference between the cathode and anode, the fraction of charge lost to electronegative impurities over this distance is measured and hence the electron lifetime in the liquid argon can be determined, providing a measurement of its purity. SBND employs three purity monitors from previous experiments at Fermilab which have been refurbished to ensure reliable measurements.

\begin{figure}[hbt]
    \centering
    \includegraphics[width=0.45\textwidth]{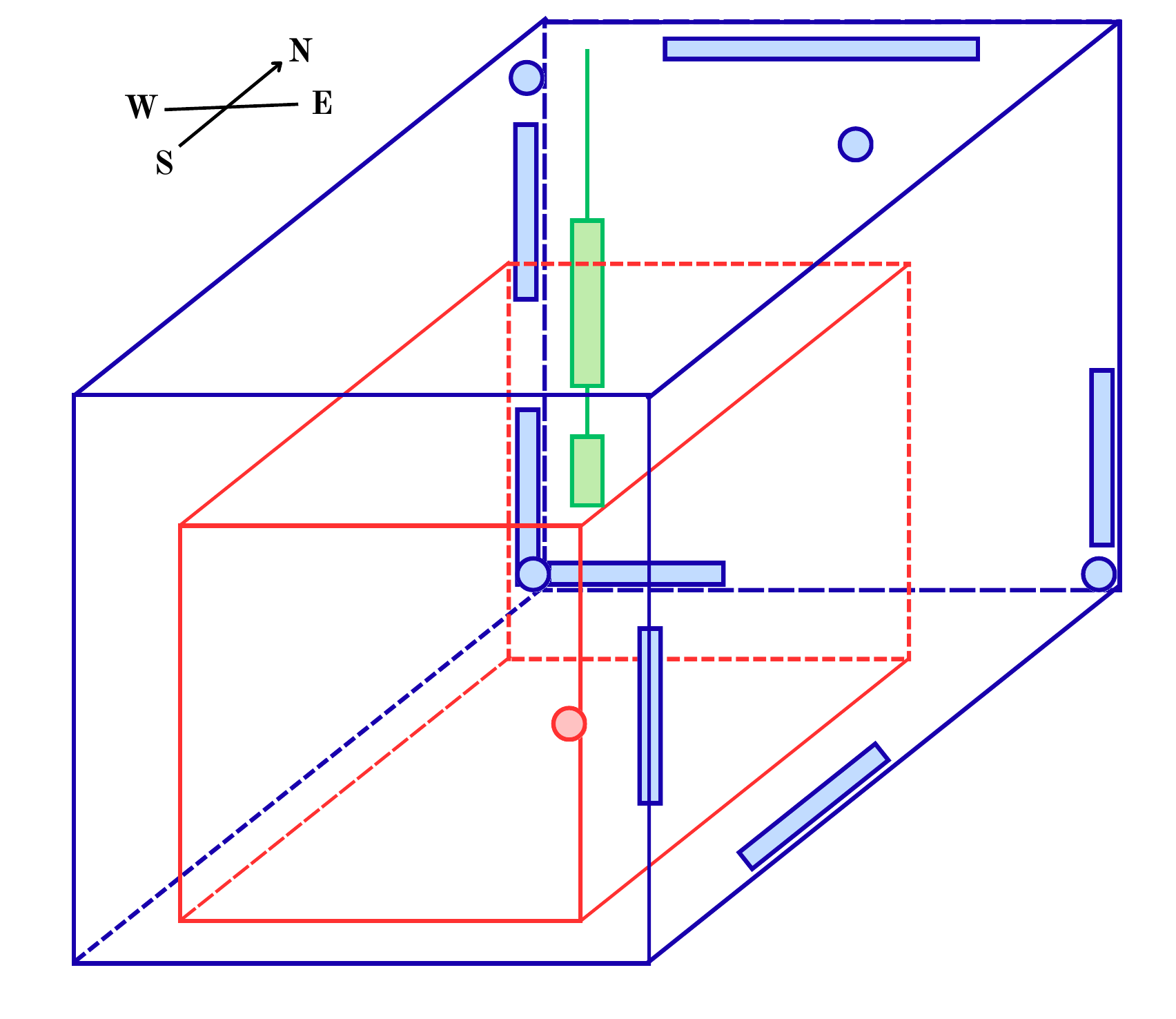}
    \includegraphics[width=0.35\textwidth]{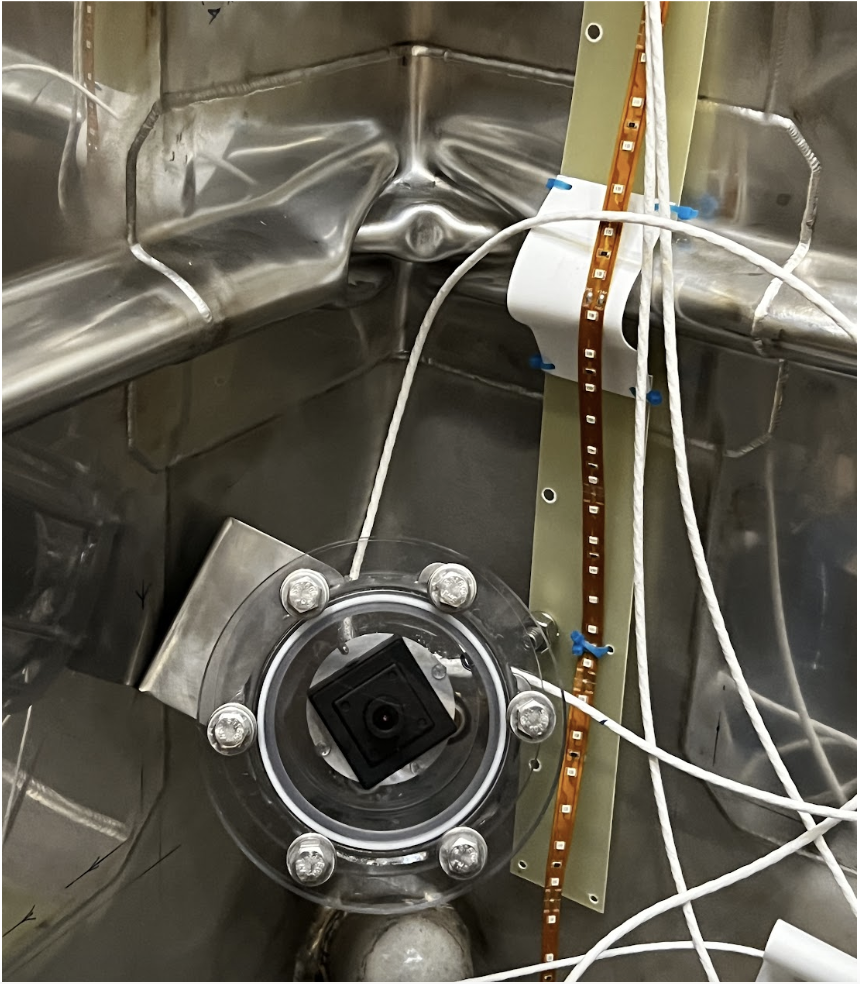}
    \caption{Left: Locations of cameras (circles) and LEDs (blue rectangles) and purity monitors (green rectangles) inside the cryostat. The blue enclosure represents the cryostat and the red one represents the TPC. Cameras and LEDs indicated in blue are mounted on the cryostat walls and the camera indicated in red is mounted on the TPC frame.  Right: A camera and LED strip mounted on the cryostat wall.}
    \label{fig:camera1}
\end{figure}

The Inline Purity Monitor is positioned outside the cryostat as shown in figure~\ref{fig:CryoModel}. It is located just downstream of the LAr filters and provides a measurement of LAr purity before entering the cryostat.
The Short Purity Monitor is located in the northwest corner of the cryostat, as shown in figure~\ref{fig:camera1}, this monitor is optimized to detect short electron lifetimes (up to 1\,ms) and is positioned closer to the bottom of the cryostat. The Long Purity Monitor is also positioned in the northwest corner; this monitor is designed to measure longer electron lifetimes (up to 5\,ms). It is placed around the mid-height of the cryostat.
Since the internal purity monitors produce light inside the cryostat, they are disabled during normal data taking and integrated into the PDS interlock to ensure that no light source is turned on while the PDS is powered. \\

\noindent\textbf{Level meters} \\
SBND is equipped with two capacitance-based level meters to track the level of liquid argon within the cryostat. These meters were calibrated during the filling process using data from the RTDs. During normal operations, the level meters are integrated into the interlock system ensuring that the TPC high voltage (HV) is safely shut down if the liquid level falls below a critical threshold. \\

\noindent\textbf{Cold cameras} \\
Five cold cameras are installed within the cryostat to visually monitor critical aspects of the detector, including mechanical connections, the high-voltage feedthrough, and the wire planes. Four cameras are mounted on the inner walls of the cryostat, while one is positioned on the TPC frame, with a field of view covering the wire planes on the southeast side of the detector. Standard commercially available cameras that were not designed for cryogenic environments are isolated from the LAr using specially designed polycarbonate enclosures, based on a successful design for ProtoDUNE, with integrated heating systems. An aluminum plate supports an electronic circuit distributing power to the camera and heating circuit. The enclosures were filled with  gaseous argon (GAr) at 200\,mbar to minimize contamination of the liquid argon (LAr) in case of leaks from the enclosure into the cryostat. 

Since these cameras require visible light to operate, 
 a set of LEDs was installed at strategic locations to illuminate the detector. Yellow LED tape was cut into 5-foot-long strips and secured to G10 boards using Teflon cable ties. 
Yellow LEDs were chosen to minimize degradation of light-sensitive components of the PDS system. At cryogenic temperatures, the LEDs emit green light, which is still within an appropriate wavelength range. Like the internal purity monitors, the LED power supply is integrated into the PDS interlock to  ensure that no light source is on while the PDS is powered.
Figure~\ref{fig:camera1} illustrates the camera and LED setup and their positions within the cryostat.

\subsection{System performance and cool-down}
\label{sec:cryocooldown}
The cryogenic system was purged, cooled, and filled over a period of approximately three months with initial purging of the system piping and cryostat volume beginning in January 2024. Liquid argon was supplied by a commercial vendor in trucks carrying an average of 14000\,L each. All deliveries were sampled for nitrogen, oxygen, and water using a suite of gas analyzers prior to being transferred to the external storage dewar. To purge the cryostat, purified argon gas vaporized from the liquid argon storage dewar was injected at the bottom of the cryostat using a technique referred to as a piston-purge. This process allowed the denser argon to effectively form a gas piston that displaced the air to the top and out to atmosphere through various vent paths. Non-flowing piping segments connected to the cryostat top plate, such as electronics feedthroughs, were vented to further evacuate impurities. The cryostat purge was maintained for 14 days with 14\,g/s of GAr, which reduced impurities to less than 1\,ppm of nitrogen and oxygen and 6\,ppm water, as planned. 

\begin{figure} [hbt!]
  \centering
  \includegraphics[width=1\textwidth]{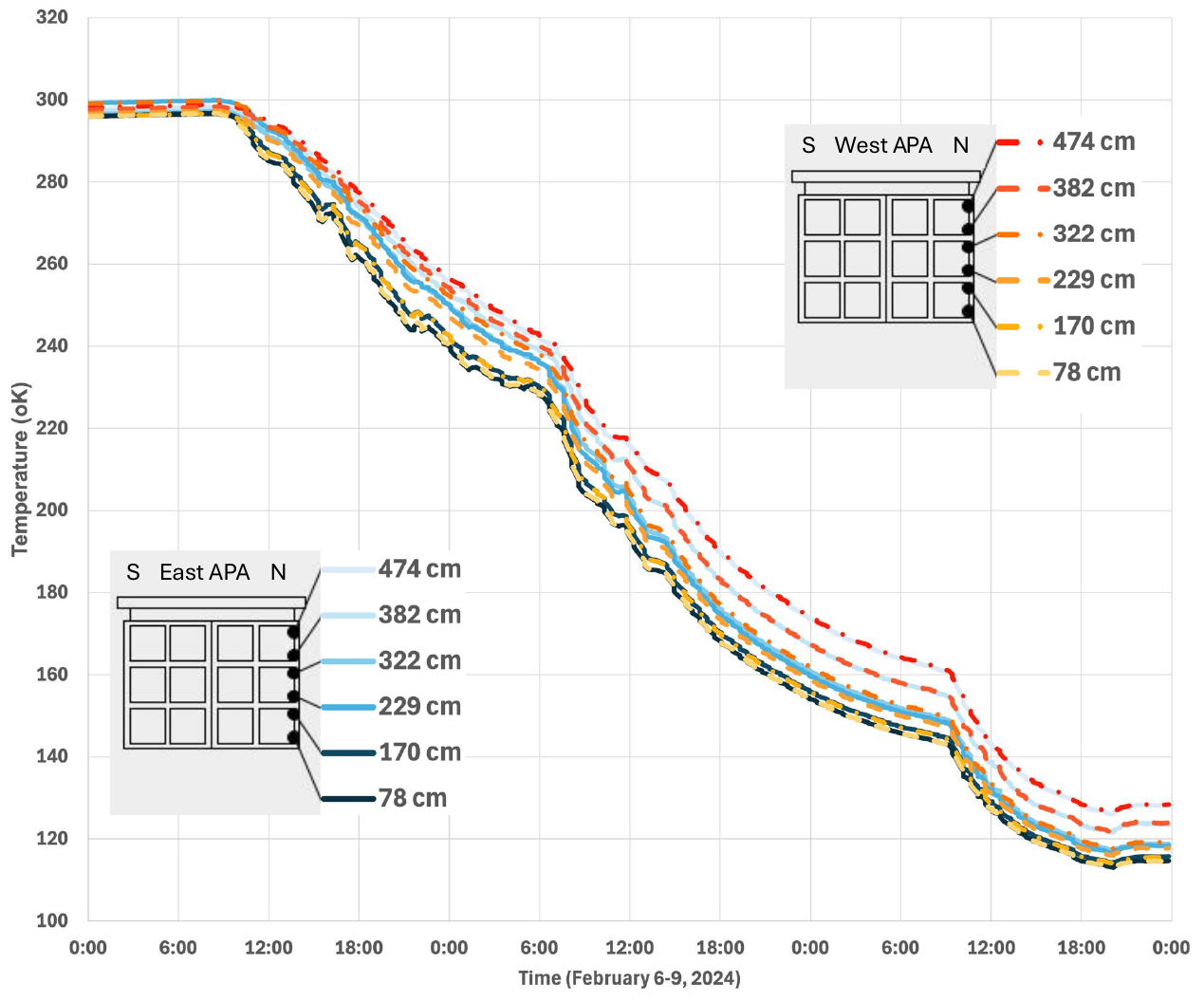}
\caption{Detector temperature distribution during the 72 hour cooling process using the spray manifold. The temperature probe locations are shown in the lower left inset for the East APA and in the upper right inset for the West APA. Locations shown are elevation from the cryostat floor.} 
\label{fig:cooldownplot}
\end{figure}

The cooling process was initiated immediately after the purge. Using a manifold of spiral spray nozzles  
beneath the cryogenic top cap which is supplied with liquid argon from the phase separator vessel where pressure and level are automatically controlled, the nozzles provided a consistent conical mist of cold argon for the duration of the process. The TPC detector required maintaining a spatial temperature differential no greater than 10\,K per meter and an overall cooling rate not exceeding 40\,K per hour. The detector and cryostat reached temperatures below 130\,K in approximately 60 hours, as shown in figure~\ref{fig:cooldownplot}, at which point testing of TPC electronics internal to the cryostat was performed to ensure detector integrity prior to filling.  This testing was concluded within two days, after which point filling commenced. At the start of cooling all probes were within about 1$^{\circ}$K of each other while at the end of this cooling phase there was a 14$^{\circ}$K differential from top to bottom.

It is important to note that the cooling process continued concurrently with the fill process to maintain the thermal condition of the detector and cryostat. To complete the filling, the liquid argon was transferred at rates between 200 and 500\,L per hour through the circulation filters to remove impurities before reaching the phase separator vessel and ultimately the cryostat. In roughly three weeks, the cryostat level reached 522\,cm which submerges the TPC and enough of the support beams to mitigate the risk of high voltage breakdown in gaseous argon. The filling process required 16 argon tanker deliveries on top of 4 initial deliveries required to commission the dewar then purge and cool the cryostat.

\begin{figure} [hbt!]
  \centering
  \includegraphics[width=0.80\textwidth]{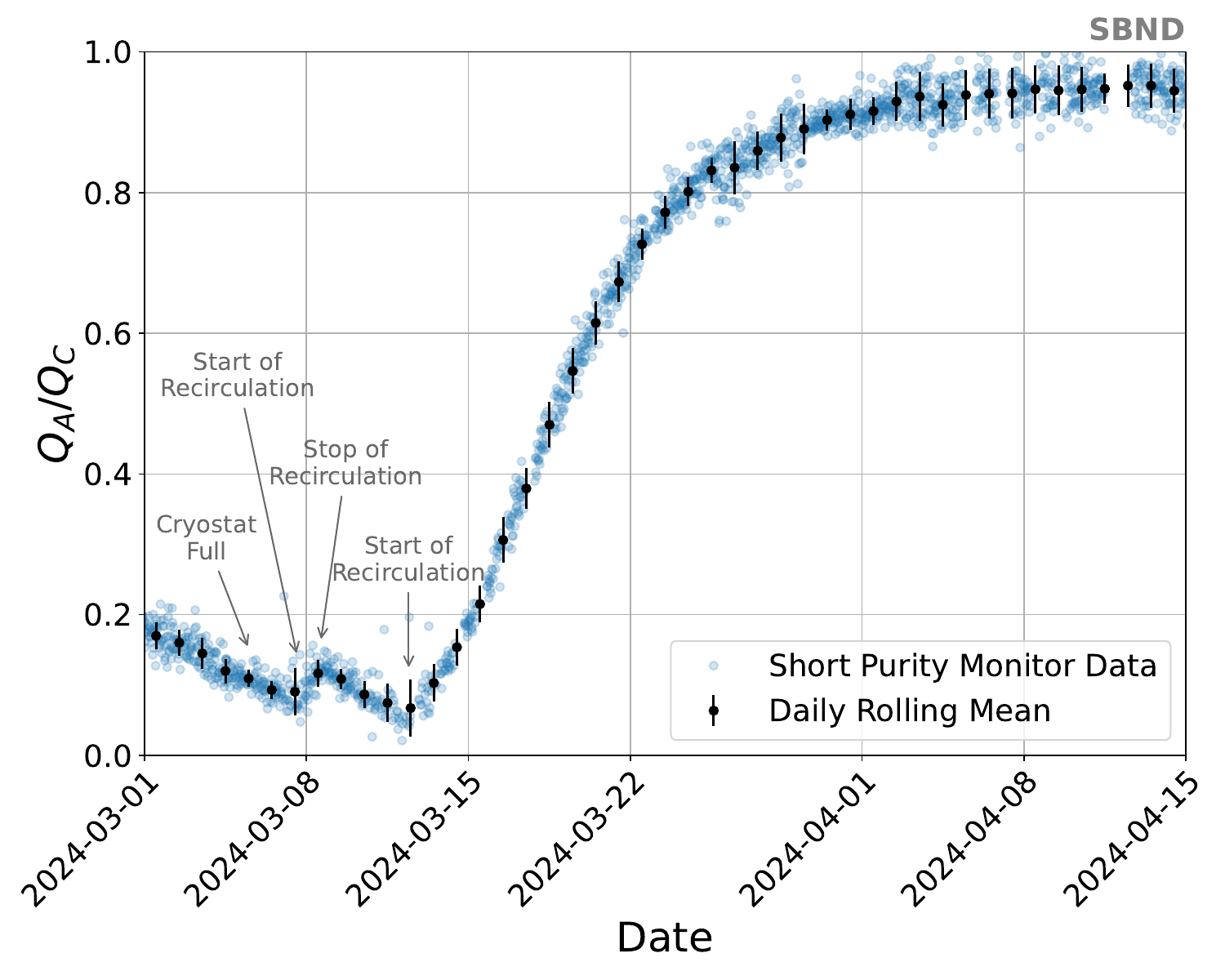}
\caption{Liquid argon purity measured with the short internal purity monitor as a function of time during the cryogenics commissioning. $Q_A$\ is the charge measured on the anode of the purity monitor and $Q_C$\ is the charge measured at the cathode of the purity monitor. During the period March 8 (labeled Stop of Recirculation) and March 12 (labeled Start of Recirculation) commissioning of LAr circulation, GAr condenser, and phase separator were being completed. There is a gap in data on April 12 when the purity monitor was turned off while the PMT system was turned on. Flashes from the purity monitors  could damage the PMTs if they are powered on.   } 
\label{fig:relativeLifetime}
\end{figure}

After completion of filling, the argon was circulated for several weeks to improve the LAr purity prior to turning on detector high voltages. During this commissioning phase, the bulk argon purity was measured with the two purity monitors installed inside the cryostat. The purity monitors indicated an increase of greater than 800\% in free electron lifetime during initial cleanup of liquid argon being circulated and filtered in the liquid argon filter in March and April 2024. Figure~\ref{fig:relativeLifetime} shows the purity measured in the short monitor from the completion of filling the cryostat to the point that cryogenics commissioning was complete. The plot shows the ratio of charge measured at the anode of the purity monitor ($Q_A$) to the charge measured at the cathode of the purity monitor ($Q_C$). Due to the short drift distance in the short monitor, the sensitivity is limited for electron lifetimes in excess of 1\,ms. While the purity monitors are not absolutely calibrated, the absolute value of the free electron lifetime was later measured with the TPC using cosmic tracks. This initial measurement was taken at a single point in time, July 24, 2024. The obtained value was in excess of 10\,ms at that time, which far exceeds the design requirement of 3\,ms. 

In transitioning to steady-state operation of the cryogenic system, it became apparent that mitigating vapor locking was a significant operating issue, which is often the situation in these systems. The SBND system has three inherently problematic piping segments prone to vapor locking. One occurs between the phase separator and the cryostat, where vapor accumulates in a high point of the piping when flow is interrupted or insufficient. This has been mitigated by connecting a supply of pressurized argon gas from the argon storage dewar, isolated by an automated valve that can be opened manually to rapidly increase the pressure in the phase separator and force the vapor through when a vapor lock occurs. Another can occur between the circulation pump discharge and the filter valve box when establishing flow during pump starts. Fortunately, the filter valve box contains vents at several locations to allow vapor to be vented. The final and most consequential vapor lock has a tendency to develop between the condenser and the circulation pump suction line. Using the connection between the condenser and phase separator, this vapor lock can be cleared by the same method as that of the phase separator, but with a different valve arrangement.

In late-May 2024, the circulation filter media was regenerated, in parallel with detector commissioning, to restore full capacity after filling and initial circulation. The process involved operating the filter bypass control valve, which allowed the circulation filter to be isolated to warm up while the rest of the cryogenic system remained at operating conditions. As this is the only circulation filter in the system, minimizing the time required to regenerate is important. Using the integrated regeneration system, the filter media was regenerated in two days after which the filter was slowly cooled and the circulation flow through it was restored. The filter has not required another regeneration and none is expected through the end of 2027. The detailed sequence of the regeneration process is shown in figures~\ref{fig:CuRegen} and \ref{fig:MSRegen} in appendix~\ref{appendixCryo}. The filters and recirculation system have maintained electron lifetimes in the argon above 30\,ms, which is well above the design requirement.

When the piping and components are adequately cooled and circulation is well established, the argon cryogenic system operates very reliably and requires nearly no intervention beyond routine maintenance or power interruptions. The control valve on the pump discharge regulates the differential pressure across the pump, keeping it on its operating curve automatically. Similarly, pressure and level in the detector cryostat are also precisely maintained by the control system operation of the condenser, with fluctuations less than 5\,mbar. The nominal heat load of the entire cryogenic system is 7\,kW, as determined empirically by nitrogen consumption, which is well below the maximum capacity of the condenser. Taking into account only the portion attributed to the cryostat by eliminating parasitic heat load contributions from items such as the circulation pump and detector electronics power, losses through the cryogenic equipment and transfer lines, and heat leak through cables and internal piping, this is determined to be approximately 10.5\,\si[per-mode=symbol]{\watt\per\meter\squared} through the surfaces in contact with the liquid argon. This is slightly higher than the expected heat leak of 10\,\si[per-mode=symbol]{\watt\per\meter\squared} for the design of the membrane cryostat.

\section{TPC cathode and drift field system}
\label{sec:TPCdrift}

The uniform electric field within the SBND TPC is critical for optimal detector performance. The TPC drift field system is composed of 16 Field Cage (FC) elements, a Cathode Plane Assembly (CPA), a High Voltage Feedthrough (HVFT), an external High Voltage Power Supply (HVPS), a high-voltage cable, and 2 high-voltage filters. 
The FC system provides a constant electric field across the TPC volume, transitioning from -100\,kV at the cathode to 0\,V at the anode readout, through a series of resistor-populated printed circuit boards (PCBs). The HVPS, located external to the cryostat, supplies the required voltage. The HVFT, mounted to a cryostat penetration, provides electrical continuity into the cryostat. 
Subsequent sections will detail the individual components of this system, while a list of the main system parameters is shown in table~\ref{tab:HVDriftPar}, with full system parameters in appendix~\ref{appendixTPC}.  The measured as-built location of the TPC relative to the cryostat is detailed in table~\ref{tab:tpcposition}.

\begin{table}[htb!]
\centering
 \caption{SBND TPC Drift System main parameters.}
\label{tab:HVDriftPar}
\begin{tabular}{c c}
\hline \hline{}
%CPA dimensions (H $\times$\ L $\times$\ W) & 4194\,mm $\times$\ 5213\,mm $\times$\ 54\,mm \\ \hline
CPA dimensions (H $\times$\ L $\times$\ W) & 4194 $\times$\ 5213 $\times$\ 54\,mm$^3$ \\ \hline
%CPA tubing OD & 54\,mm \\ \hline
%CPA tubing wall thickness & 2\,mm \\ \hline
%CPA mesh aperture & 5\,mm \\ \hline
%CPA mesh wire thickness & 1\,mm  \\ \hline
CPA mesh transparency & 70\% \\ \hline
%Field Cage resistor model & SM104FE-1000M \\ \hline
%Field Cage resistor value & 1\,G$\Omega$ \\ \hline
%Field Cage MOV model & ERZ-V14D182 \\ \hline
%Field Cage MOV clamping voltage & 1.7\,kV \\ \hline
%N. Field Cage modules per drift volume & 8 \c \hline
Field cage module total resistance & 16.75\,G$\Omega$ \\ \hline
HV Power Supply model & Heinzinger PNChp 150000-1 neg \\ \hline
%HV Power Supply input & 1/N/PE 110V \\ \hline
HV Power Supply output & DC 0–150\,kV, 0–1\,mA\\ \hline
Nominal HV Operating Voltage & -100\,kV \\ \hline
%HV Power Supply readback sampling frequency & 20\,Hz\\ \hline
%HV cable model & Dielectric Sciences 2121 \\ \hline
%HV cable gauge & 12\,AWG \\ \hline
%HV cable voltage rating & 150\,kV DC \\ \hline
%HV Filter resistor model & Vishay Dale DZZ1 \\ \hline
%HV Filter resistor value & 1\,M$\Omega$ \\
%Ground break resistor value & 10\,$k\Omega$ \\ \hline
%Ground break diode & 1N4005 \\ \hline
\hline 
\end{tabular}
 
\end{table}

\subsection{Cathode}
\label{sec:cathode}

\begin{figure}[hbt]
    \centering
    \includegraphics[width=0.95\textwidth]{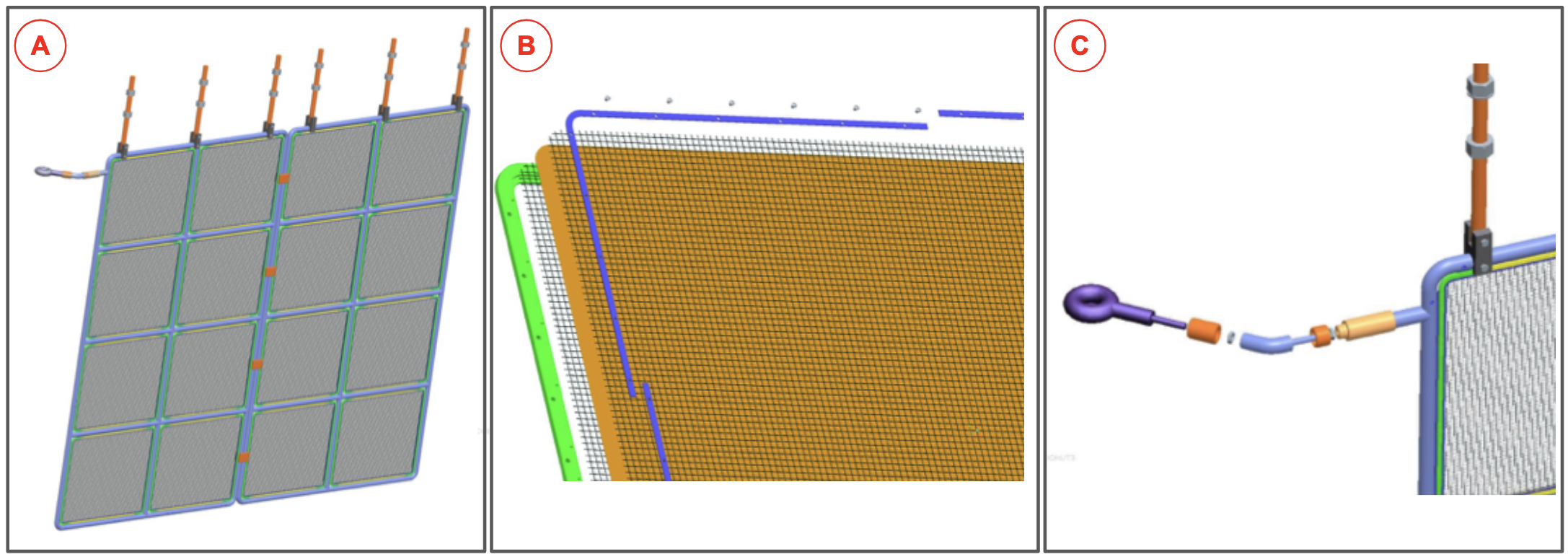}
    \caption{(a) CPA drawing. (b) Detail of the mesh assembly. (c) Detail of the HV donut connection to the CPA main frame.}
    \label{fig:CPADrawing}
\end{figure}

The CPA, positioned centrally within the detector, is biased to -100\,kV, establishing two drift volumes with a 0.5\,kV/cm electric field (see figure~\ref{fig:TPC3Dmodel}). The CPA is an electro-polished stainless-steel structure, consisting of a tubular frame of 54\,mm outer diameter and 2\,mm wall thickness, with 16 square apertures arranged in a 4x4 array, 16 mesh panel assemblies, and a high-voltage feedthrough interface (see figure~\ref{fig:CPADrawing}a).
Holes in the tubular frame facilitate liquid argon ingress into the hollow interior to prevent trapping of impurities. Stainless-steel bars, welded to the interior perimeter of each aperture, provide mounting points for the mesh assemblies via M6 button-head screws.
Each mesh assembly, as shown in figure~\ref{fig:CPADrawing}b, comprises a stainless-steel frame securing two wire meshes, tensioned using a cryofitting installation technique.
Four TPB-coated dielectric reflective plates (see section~\ref{sec:PSDFoilSystem} for details) 
are sandwiched between the mesh layers.
The high-voltage feedthrough interface (figure~\ref{fig:CPADrawing}c) consists of a standard stainless-steel toroidal receptacle (donut), welded to a steel tube and an M24 threaded connection.
A corner joint and an M48 threaded sleeve are employed to provide connection to the CPA main body, adjustable alignment of the feedthrough, and a cover for the threads with a smooth surface to prevent high E fields. The CPA is suspended from a stainless-steel beam using six G10 threaded rods, which provide electrical isolation between the CPA and support elements. These rods pass through apertures in the beam and are secured by four G10 nuts per rod (two above and two below the beam). The rods are then connected to the upper section of an H-shaped G10 bracket, which is in turn connected to the upper section of the CPA tubular frame via a through-bolt and nut.

\begin{figure}[hbt]
    \centering
    \includegraphics[width=0.95\textwidth]{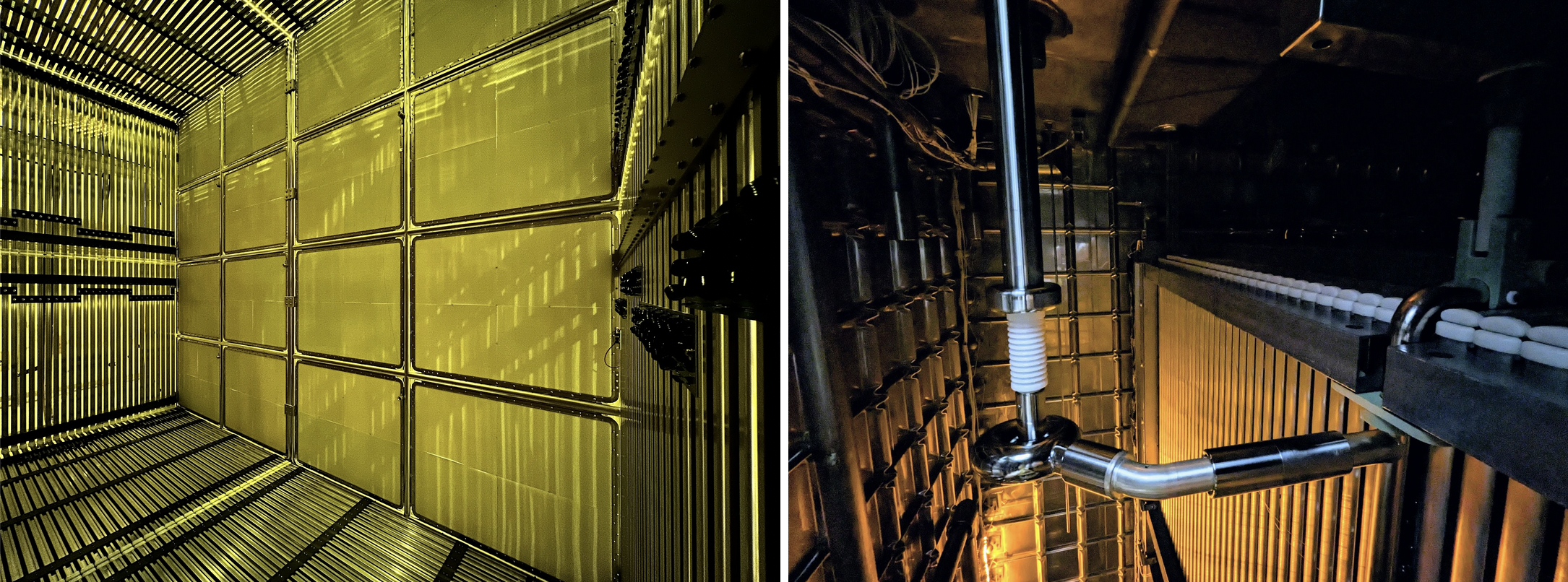}
    \caption{Left: CPA with installed field cages. Resistor boards and TPB-coated reflector plates are visible in the photo. Right: HVFT installation on the HV donut.}
    \label{fig:CPADPhoto}
\end{figure}

The CPA frame was the initial TPC component installed within the ATF (see section~\ref{sec:intro}). Following frame placement, the mesh panels and TPB-coated reflector plates were installed, followed by the bottom FC modules and the anode planes. Once the anode planes were positioned, the top and side FC modules were installed. 
The last components of the HV system to be installed on the detector were the HV donut and the HVFT. The HV donut was mounted to the detector and the HVFT inserted into the donut after the TPC was inserted into the cryostat.
The CPA with assembled field cages, including visible resistor boards and TPB-coated reflector plates, is shown in figure~\ref{fig:CPADPhoto} (left), while figure~\ref{fig:CPADPhoto} (right) details the HVFT connection to the HV donut.

\subsection{Field cage}

\begin{figure}[hbt]
    \centering
    \includegraphics[width=0.99\textwidth]{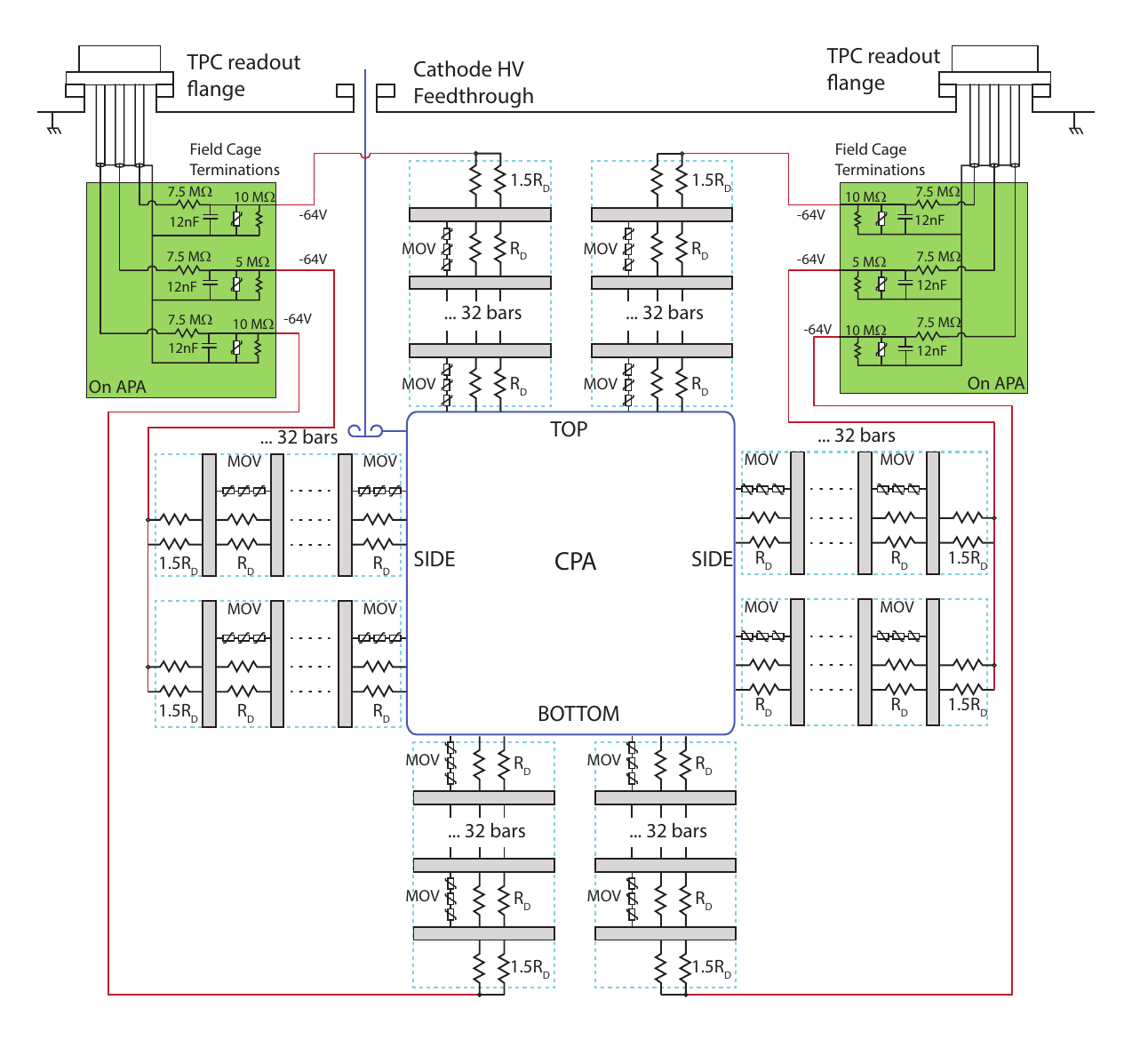}
    \caption{SBND field cage electric schematic for one of the two TPC volumes. The voltage divider resistors have a value $R_D = 1\,\rm{G}\Omega$, and the MOV has a clamping voltage of 1.7\,kV.}
    \label{fig:FCSchematic}
\end{figure}

The field cage (FC) functions as a voltage divider between the cathode and the wire readout plane anodes and defines the electron drift volumes: the active volume of the TPC. The FC comprises 16 modules, eight per drift volume, each of them acting as an individual circuit as shown in figure~\ref{fig:FCSchematic}. Each module is mechanically  connected to the APA and CPA, but electrically isolated from both, excepting the resistive divider chain connection to the CPA. Once installed, the FC modules enclose the entire detector volume. 

The connection at the CPA can act as a hinge, enabling the FC panels to open and close at the APA side for access inside the TPC during installation, maintenance or decommissioning. Each module consists of 32 extruded aluminum profiles running perpendicular to the drift field. The aluminum profiles use a design identical to the ProtoDUNE single phase detector~\cite{ProtoDUNE}. The top and bottom modules, 2.5\,m in length, are supported by four fiber-reinforced polymer (FRP) I-beams.  The top FC modules incorporate an aperture at one corner to accommodate the laser calibration system detailed in section~\ref{sec:LaserSystem}. The side modules, 2\,m in length, are  supported by two FRP rectangular hollow section beams.
Electrically, each module serves as a 32-step voltage divider with a total equivalent resistance of 16.75\,G$\Omega$.  Each voltage division step is realized by two parallel 1\,G$\Omega$ resistors, in parallel with a series of three metal oxide varistors (MOVs) designed to mitigate against any electrical discharge in this HV circuit (see table~\ref{tab:HVDriftPar-appendix} in appendix~\ref{appendixTPC} for the electrical components specifications). Applying -100\,kV at the cathode results in a potential of -4.58\,kV at the last (closest to the APA) profile of the field cages. All electrical components were tested and characterized in liquid nitrogen. While a direct electrical connection is made to the APA frame, the anode connection is established through a field cage termination board (FCTB). This board provides a default termination resistance, and a Safe High Voltage (SHV) cable connection to the outside of the cryostat, via the TPC readout signal feedthrough flange, through which it is possible to either supply a different termination voltage to the FC or monitor the current flowing through the divider chain, or both.
The FC electric schematic is shown in figure~\ref{fig:FCSchematic}. Under normal operating conditions, each FCTB has its return connected to the flange, with the center conductors connected to their respective power supply channels.

To verify FC module connections and drift high-voltage circuit continuity, the cathode was biased to -3\,kV and the total current through the installed FC modules was measured following each FC module installation and major installation milestone (TPC completion, TPC move to the SBND building, TPC installation in the cryostat).
During cooldown and filling, a -1\,kV bias was maintained on the cathode, and continuous current measurements were taken for the entire circuit and each FCTB to monitor the system's electrical continuity.

\subsection{Drift HV supply system}

\begin{figure}[htb]
    \centering
    \includegraphics[width=0.9\textwidth]{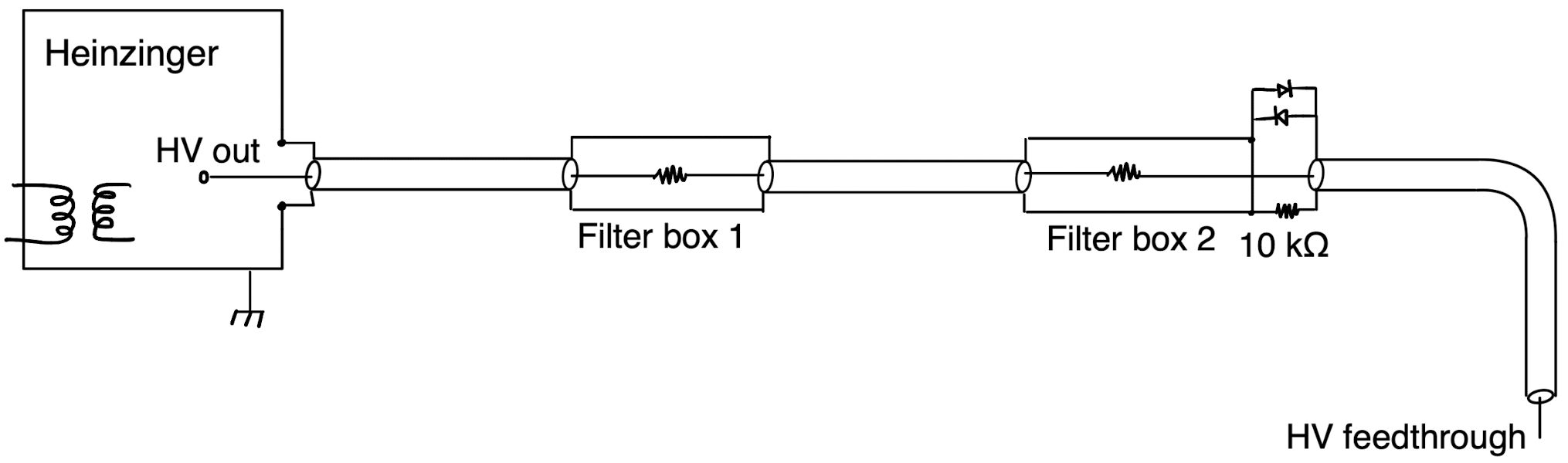}
    \includegraphics[width=0.9\textwidth]{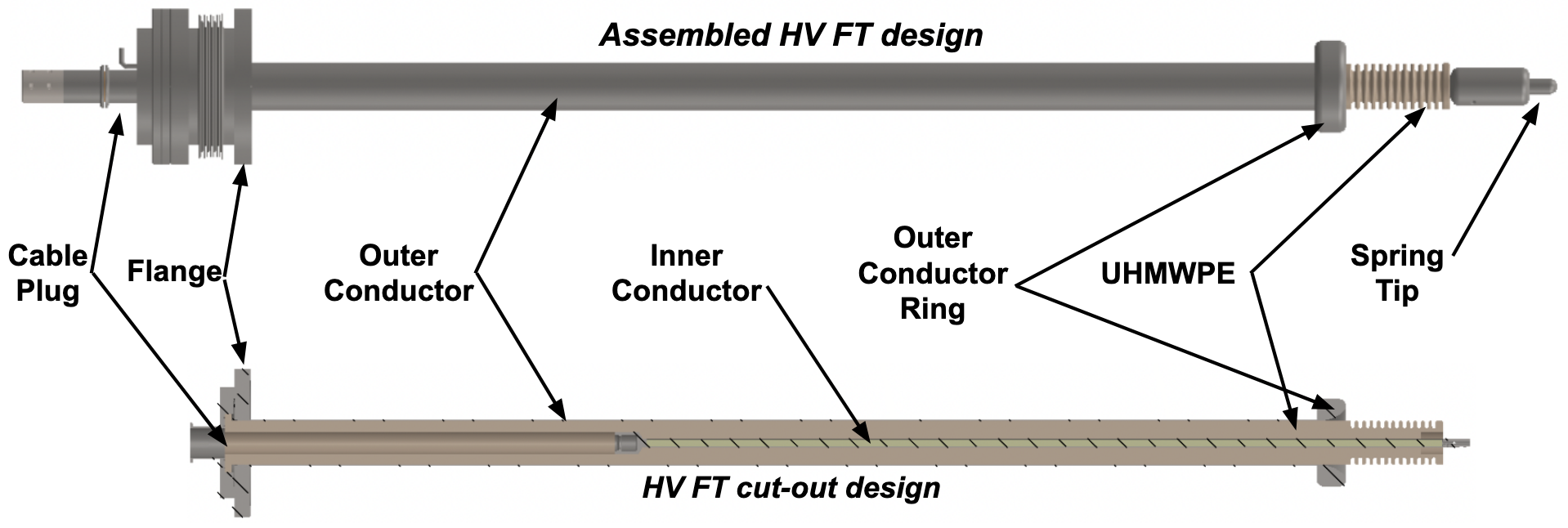}    
    \caption{Top: Schematic diagram of the warm side of the drift high voltage delivery system. Cables: \textit{Dielectric Sciences 2121}, rated to 150kV DC. Filter box resistors: 1M$\Omega$ \textit{Vishay Dale DZZ1}, 100W, 125kV rating. Bottom: High voltage feedthrough design.}
    \label{fig:HVFTDesign}
\end{figure}

The system supplying the drift high voltage to the cathode of the TPC consists of a HV power supply, cables, HV filters, and HV feedthrough, as shown in figure~\ref{fig:HVFTDesign} (top). A low ripple Heinzinger power supply, capable of delivering up to -150\,kV, provides the -100\,kV potential required to generate the TPC electric field.
Remote control and monitoring of the Heinzinger parameters (voltage, current, current thresholds) is performed using its analog controls interface with a 20\,Hz sampling frequency on its read-back. The HVFT design is based on the established construction methodology employed by the ICARUS collaboration~\cite{icarus_HVFT}. The feedthrough features a coaxial geometry, consisting of an inner conductor (HV) and an outer conductor (ground) separated by an ultra-high-molecular-weight polyethylene (UHMWPE) insulator, as depicted in figure~\ref{fig:HVFTDesign} (bottom). The outer conductor, a stainless-steel tube, encloses the insulator and extends into the liquid argon (LAr) within the cryostat. It features a donut-like ring at the bottom, which helps reduce the strength of the electric field within the LAr near the end of the FT. This configuration confines the electric field within regions filled with high-dielectric-strength media (UHMWPE and LAr). The inner conductor, a thin-walled stainless-steel tube, minimizes heat input and prevents the formation of argon gas bubbles near the feedthrough's lower end. A cup-style cable plug connection at the upper end for the HV cable and a spring-loaded tip contact at the lower end for the cathode interface complete the inner electrode.
Two HV filters are interposed between the power supply and the feedthrough to mitigate, in conjunction with the cables, potential discharge effects by distributing the stored energy within the system. Collectively, the resistors in the filters and cables act as a low-pass filter, attenuating the 30\,kHz voltage ripple present at the power supply output. The characteristics of the filters are detailed in table~\ref{tab:HVDriftPar-appendix} in appendix~\ref{appendixTPC}, and the cable in figure~\ref{fig:HVFTDesign}. The HV filters are cylindrical, each encapsulating a 1\,M$\Omega$ resistor immersed in R-Temp fluid, an insulating oil. Each resistor terminal is electrically connected to a cable receptacle filled with Fluorinert FC-40, another insulating oil. Both oil systems are sealed. Each filter is housed within a splash pan capable of containing the full volume of both R-Temp and Fluorinert in the event of a leak. An electrical connection from the ground of the HVFT (which is in turn electrically connected to the cryostat) to the return of the drift HV power supply ensures the entire system is properly grounded.

The HVFT underwent high-voltage breakdown testing by connecting it to an HV power supply, submerging its tip in LAr within a test cryostat, ramping the voltage to -105\,kV, and monitoring for several days. The power supply, cables, and filters were tested independently by disconnecting the cable between the second HV filter and the HVFT, filling the cable receptacle in the HV filter with Fluorinert, and ramping the system to -100\,kV. No significant high-voltage instabilities were observed during these tests.

\subsection{Performance and stability}

The SBND cathode has maintained stable operation at the nominal -100\,kV since the start of the 2024 neutrino run. Occasional, low-amplitude current spikes are observed, of the order of 30 to 50\,$\mu$A above the baseline value of 88\,$\mu$A (a value consistent with the resistive load). When the current exceeds a predefined threshold, the Heinzinger power supply enters current limiting mode, resulting in a transient voltage spike on the voltage readback. These low-amplitude spikes, typically lasting less than one second, generally do not affect the TPC performance. Infrequently (at a rate of approximately once per month or less), higher-amplitude current spikes (order of 150\,$\mu$A above the baseline) occur, causing a significant voltage drop (approximately 60\,kV) at the cathode. These voltage drops, though brief, can induce low-amplitude current spikes in other subsystems, such as the FCTB, located at the opposite end of the field cage circuit, and the wire bias system. These events are not harmful to the HV or other subsystems of the detector.

Concerning the uniformity of the electric field established inside the TPC, the detector non-uniformity amounts to less than $5\%$ in the drift direction, and less than $10\%$ across the fiducial volume of the detector. These values are derived from the observed charge deposition per unit length of minimum-ionizing muons. A comprehensive analysis of the electric field non-uniformity is currently in progress, which will be published in a forthcoming publication~\cite{calibration}.

\subsection{Laser calibration system}
\label{sec:LaserSystem}

SBND employs a UV Laser Calibration System to measure the electric field distortions within the LArTPC. Electric field distortions are predominantly caused by the build-up of slow-moving positive argon ions from cosmic rays, referred to as space charge effect (SCE). This effect needs to be taken into account for precise particle tracking and energy measurements. 

\begin{figure} [htb]
    \centering
   \includegraphics[width=.99\linewidth]{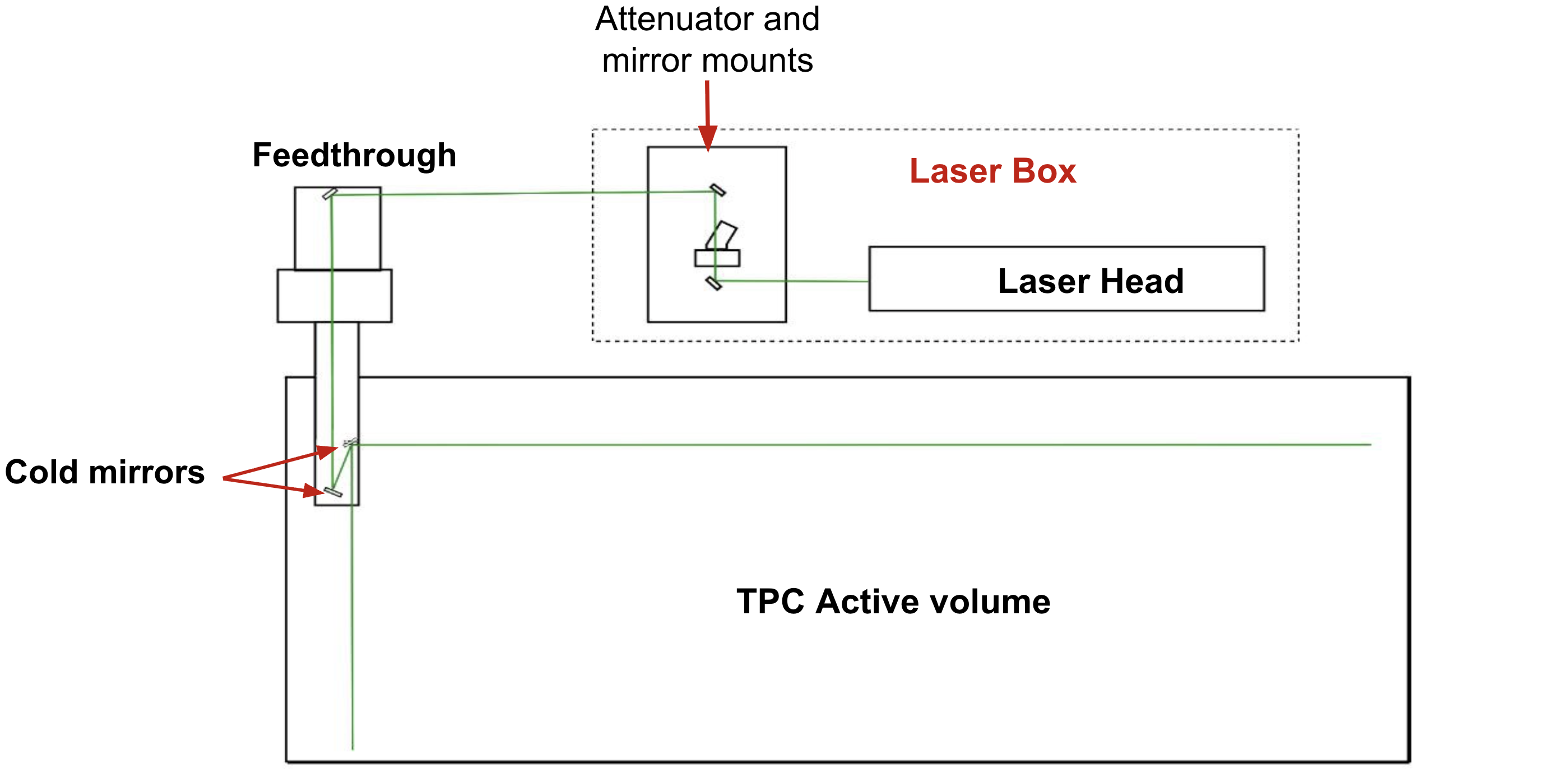}
    \caption{Schematic drawing of the SBND laser system~\cite{Mulleria_Babu_2025}. }
    \label{fig:lcs_schema}
\end{figure}

The calibration system, shown schematically in figure~\ref{fig:lcs_schema}, uses a Class-4 Nd:YAG laser operating at \SI{266}{nm} to produce ionization tracks in LAr via multiphoton absorption~\cite{multiphoton}. 
The calibration laser beam is emitted from the laser head and subsequently passes through an attenuator and a series of dichroic mirrors to reach the TPC. 
The feedthrough is equipped with a rotary motor and a high-resolution ($0.001^\circ$) encoder enabling precise angular control of the laser path. Similarly, a linear motor with a $\mu$m resolution adjusts the vertical position of the cold mirrors. A \SI{2}{m} long vacuum-sealed quartz tube passing through the center of the feedthrough ensures clean laser penetration through the argon interface. 

The feedthrough parts inside the cryostat are made from polyamide-imide (PAI) to maintain stability under extremely low temperatures.
A more detailed description of the system can be retrieved from a similar system used in MicroBooNE~\cite{uboone_laser}. 

\begin{figure}[htbp]
\centering
\includegraphics[width=.5\textwidth]{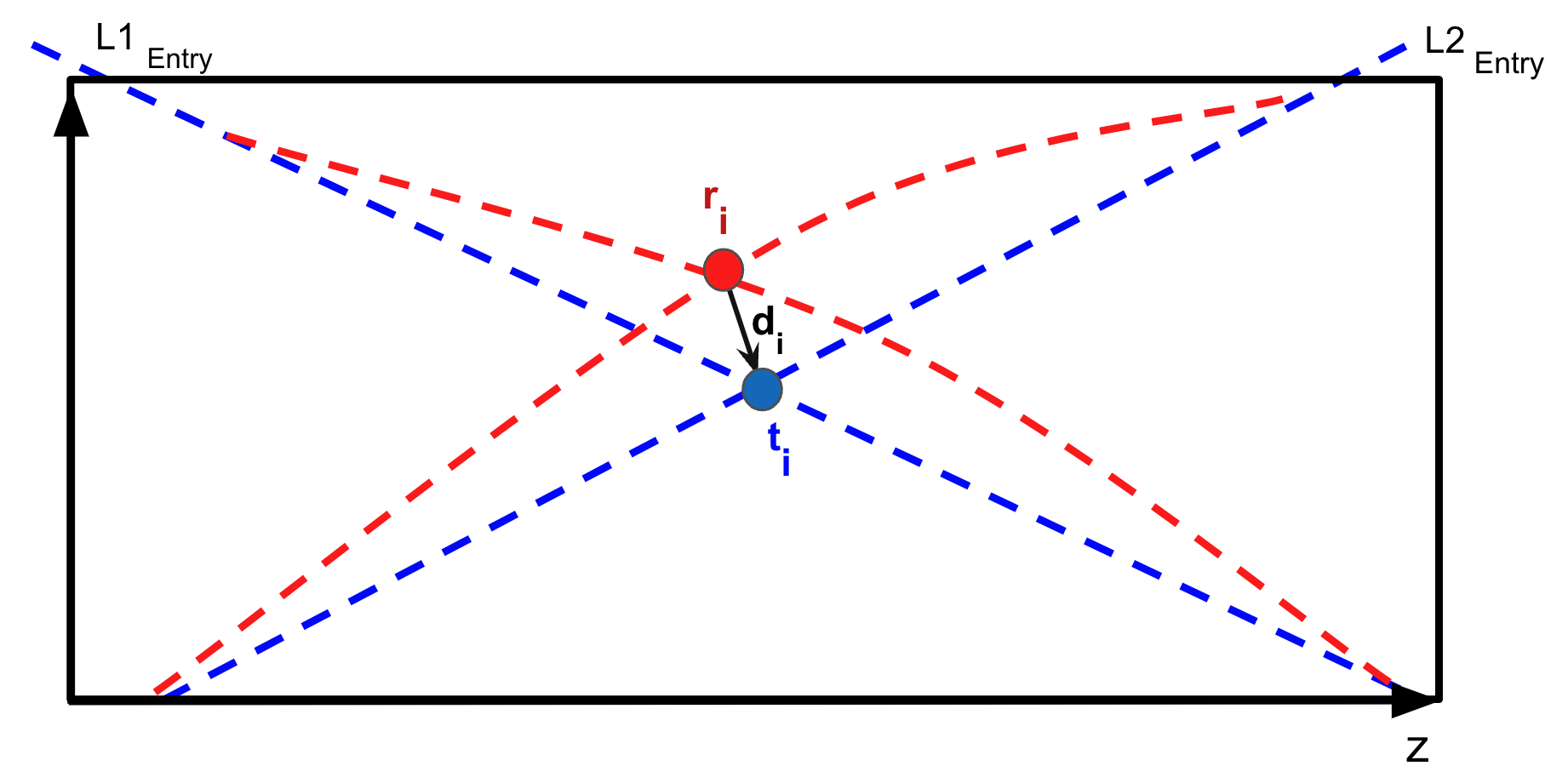}
\caption{Track reconstruction from crossed beams in SBND, where blue is the true track, and red is the reconstructed track. The displacement vector $d_i$ starting from $r_i$ to $t_i$ is the correction vector. \label{fig:cross}}
\end{figure}

SBND has a total of four identical laser systems, which are located at the four top corners of the TPC.
Multiple lasers are used simultaneously to create intersecting ionization tracks, as shown in figure~\ref{fig:cross}. The spacial displacement vectors of the laser tracks and their intersection can be used to project reconstructed track points onto the true track in 3D space. The use of crossing laser beams will be first demonstrated in SBND. This approach creates unique crossing points that can be easily flagged and used as reference points for comparison. The crossing track points are less susceptible to angular dependencies than a single laser beam and offer a clear way to correlate reconstructed points with true track locations. 

%\clearpage
\section{TPC readout: wire planes and electronics}
\label{sec:TPCreadout}

The SBND detector has two TPC volumes, each of which is read out by a wire-readout plane, or Anode Plane Assembly (APA) which monolithically spans the TPC. APA design and assembly is outlined in section~\ref{sec:APAdesign}. APAs detect charge signals induced by ionization electrons, created by charged particles as they pass through the liquid argon volume. The electron drift time for each TPC is around 1.28\,ms. The analog pulses induced on the APA wires by the ionization electrons are around 10\,$\mu$s in width. The digitization and amplification of these signal pulses takes place in electronics mounted directly on the APA frames. These electronics, detailed in section~\ref{sec:coldelec} are submerged in liquid argon and the low temperature environment enables the detector to achieve low levels of electronic noise.

\subsection{Wire readout plane design and assembly}
\label{sec:APAdesign}

\begin{figure}
    \centering
    \subfloat{\includegraphics[width=0.6\linewidth]{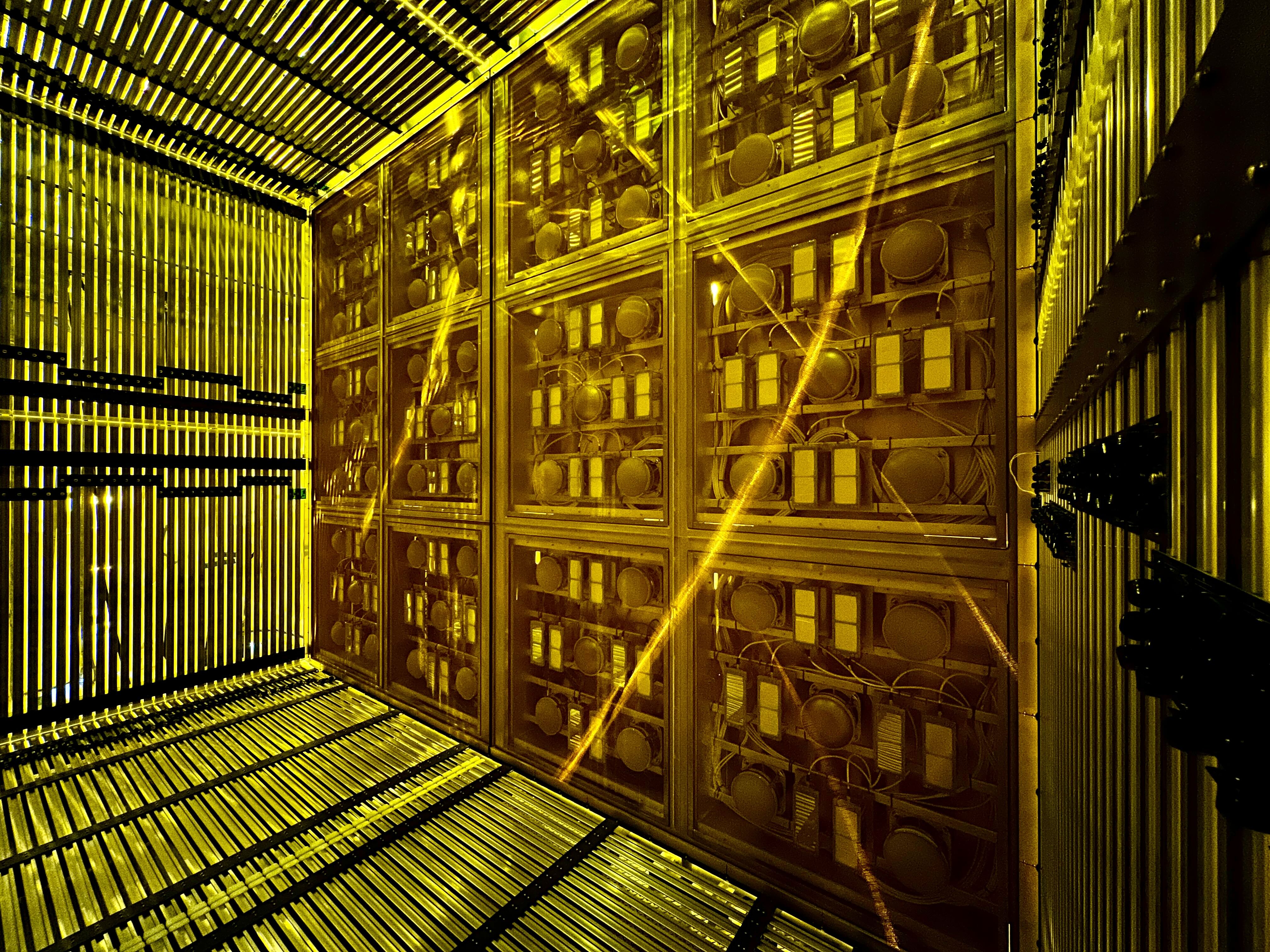}}
        \hspace{0.01\textwidth}
    \subfloat{\includegraphics[width=0.3\linewidth]{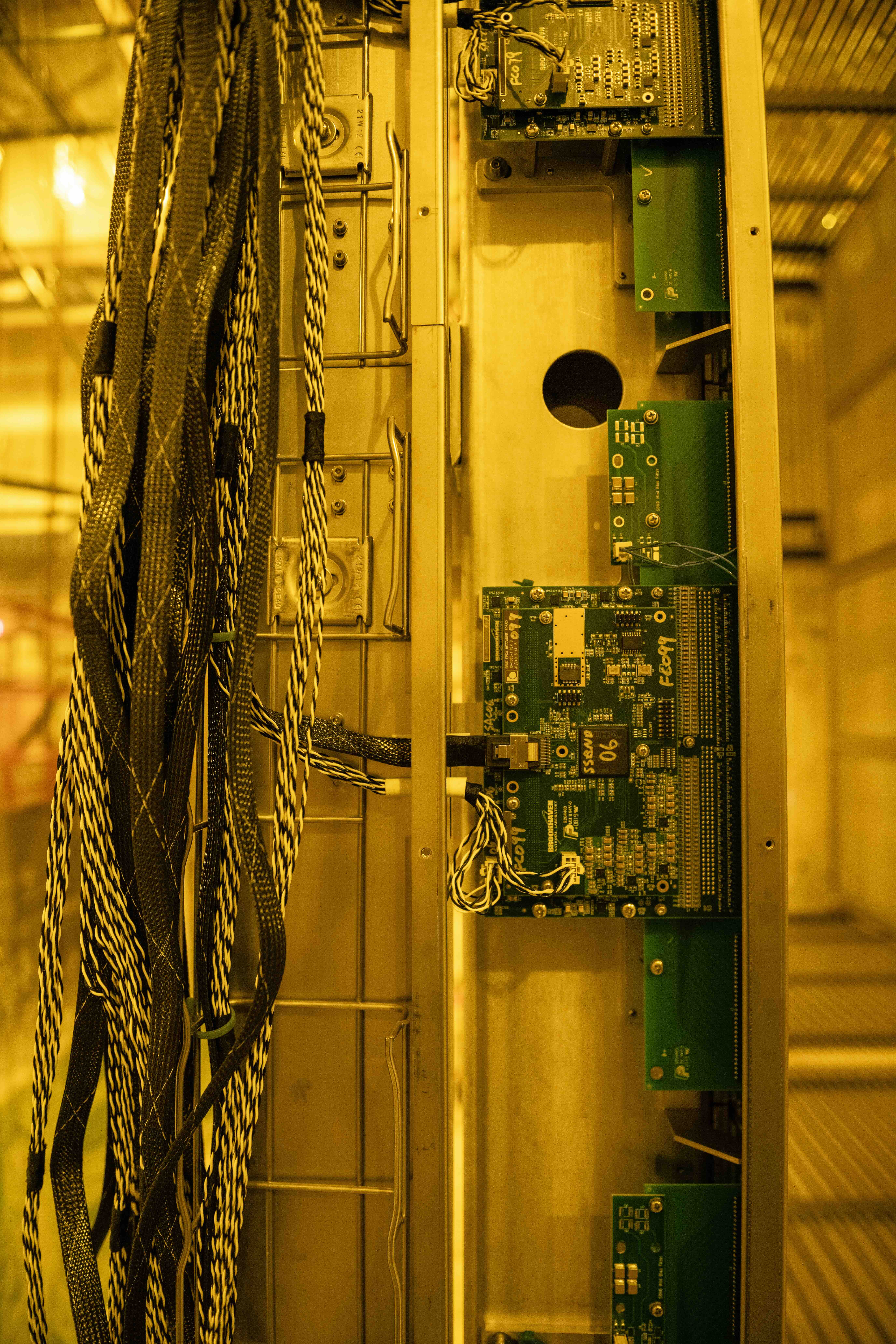}}
    \caption{Left: SBND west TPC; APA-frames coupled together, PDS mounted behind the wires, and field cages connected on all sides. Right: SBND TPC readout electronics.}
    \label{fig:APAinTPC}
\end{figure}

Each APA consists of three wire planes facing towards the cathode: two induction planes, U and V at ±60$^{\circ}$ to vertical and a collection plane, Y, with vertical wires.  Both the nominal wire plane spacing and the nominal wire spacing in each plane are 3\,mm with ±0.5\,mm tolerance.  There are 11,264 wires in SBND, which make up APA East and APA West. Figure~\ref{fig:APAinTPC} shows SBND's West APA as installed in the TPC.

Although each APA spans the width of the TPC, each plane was constructed from two independent half-planes (called APA-frames), due to mechanical construction and transportation constraints. Full details of the design, fabrication and quality assurance of the APA-frames are given in~\cite{SBNDWirePlanes}.  The key parameters of the APAs are given in table~\ref{tab:apaparameters} and are outlined here. 

\begin{table}[]
    \centering
   \caption{Table outlining the key parameters and tolerances for the SBND wire planes}
    \label{tab:apaparameters}
    \begin{tabular}{c c}
    \hline \hline
        \textbf{Parameter} & \textbf{Specification} \\ \hline \hline
        Number of readout channels per APA & 5632 \\ \hline
        Number of wire planes per APA & 3 (U,V,Y) \\ \hline
        Wire spacing & 3 ±0.5\,mm \\ \hline
        Wire plane spacing & 3 ±0.5\,mm\\ \hline
        Wire tension & 7 ±1\,N \\ \hline
        APA frame dimensions & 2566 $\times$ 4160\,mm \\ \hline
        APA plane dimensions & 5162 $\times$ 4160\,mm\\ 
         \hline \hline
    \end{tabular}
 
\end{table}

Each APA-frame consists of a welded stainless steel frame with a stack of FR4 geometry boards to which wires are epoxied and soldered. Each plane of wires is supported by wire support combs to ensure that the longest unsupported wire length is less than or equal to 1.6\,m. %$\leq 1.6$\,m.

The geometry boards are planar to the wires, which means that the wire plane spacing and planarity is directly linked to the flatness of the front face of the stainless steel frame. Strict flatness tolerances of ±0.5\,mm across the APA-plane were maintained by leveling the attachment points of the geometry boards. One side of each APA-frame forms the central join of the APA. This central edge has geometry boards perpendicular to the wire planes, the U and V planes are wrapped over teeth and attached to boards which facilitate electrical connections between two APA-frames.

\begin{figure}[htb]
    \centering
        \subfloat{\includegraphics[width=0.485\textwidth]{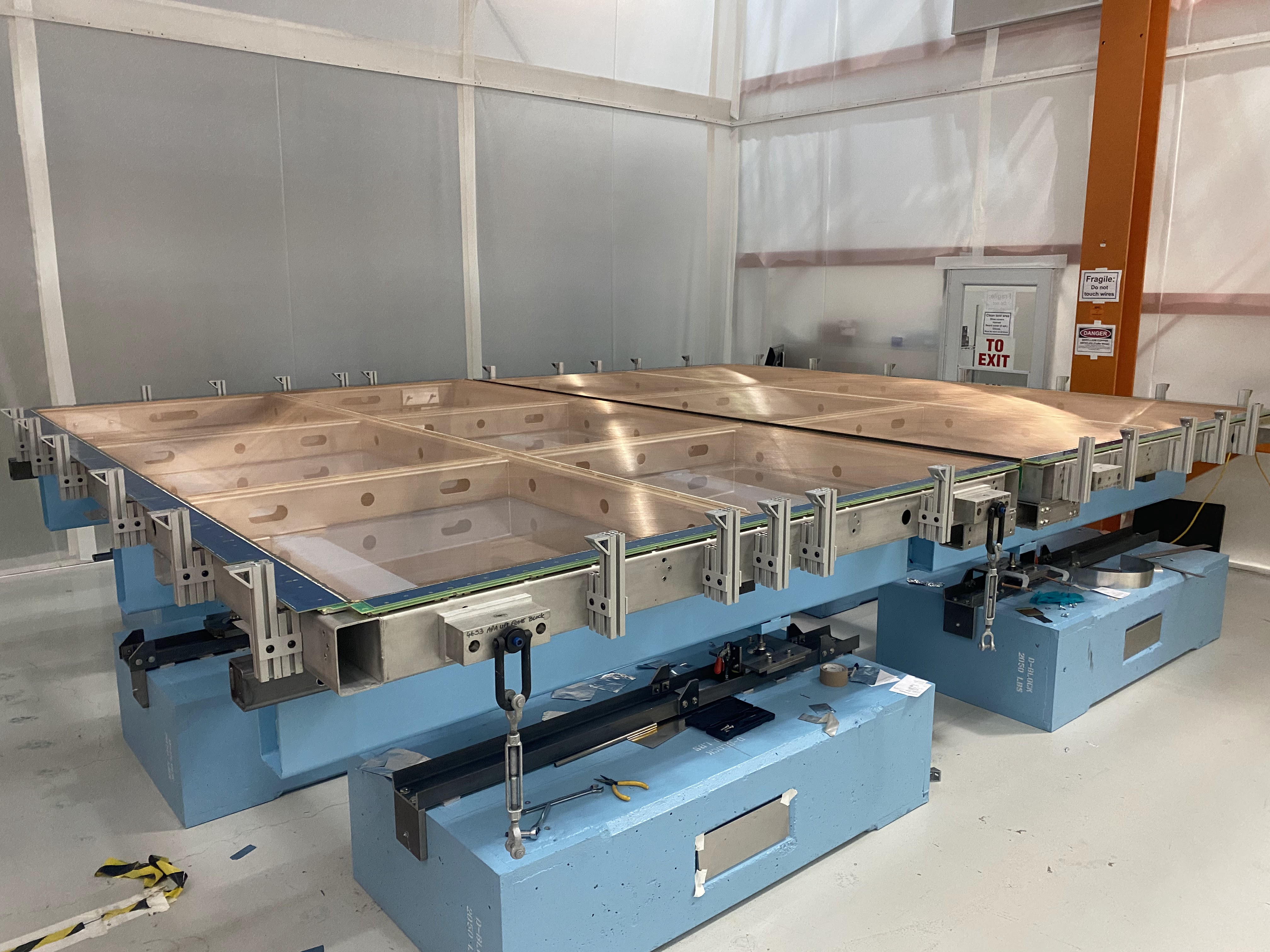}}
        \hspace{0.01\textwidth}
        \subfloat{\includegraphics[width=0.485\textwidth]{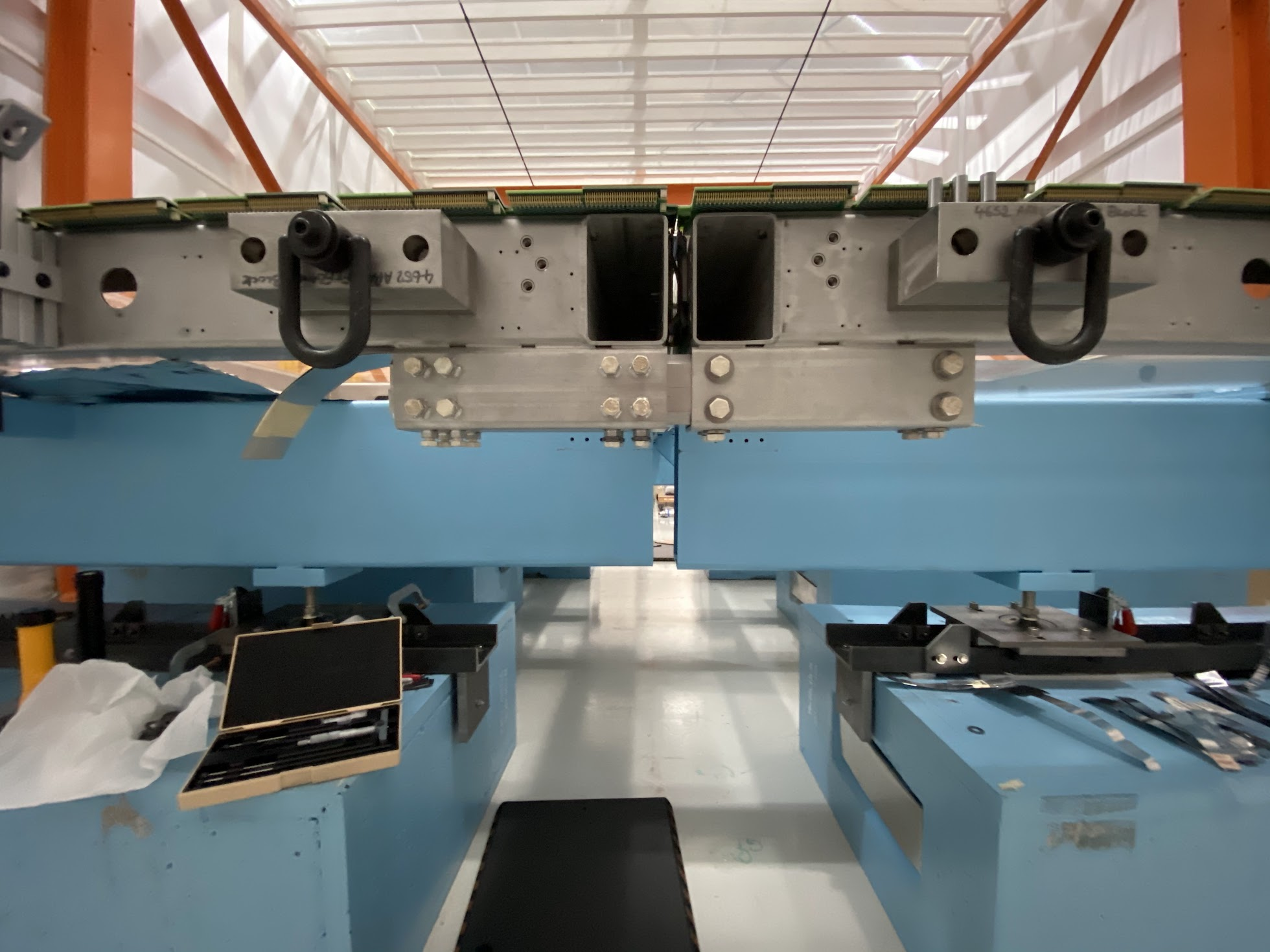}}
\vspace{1mm}
        \subfloat{\includegraphics[width=0.32\textwidth]{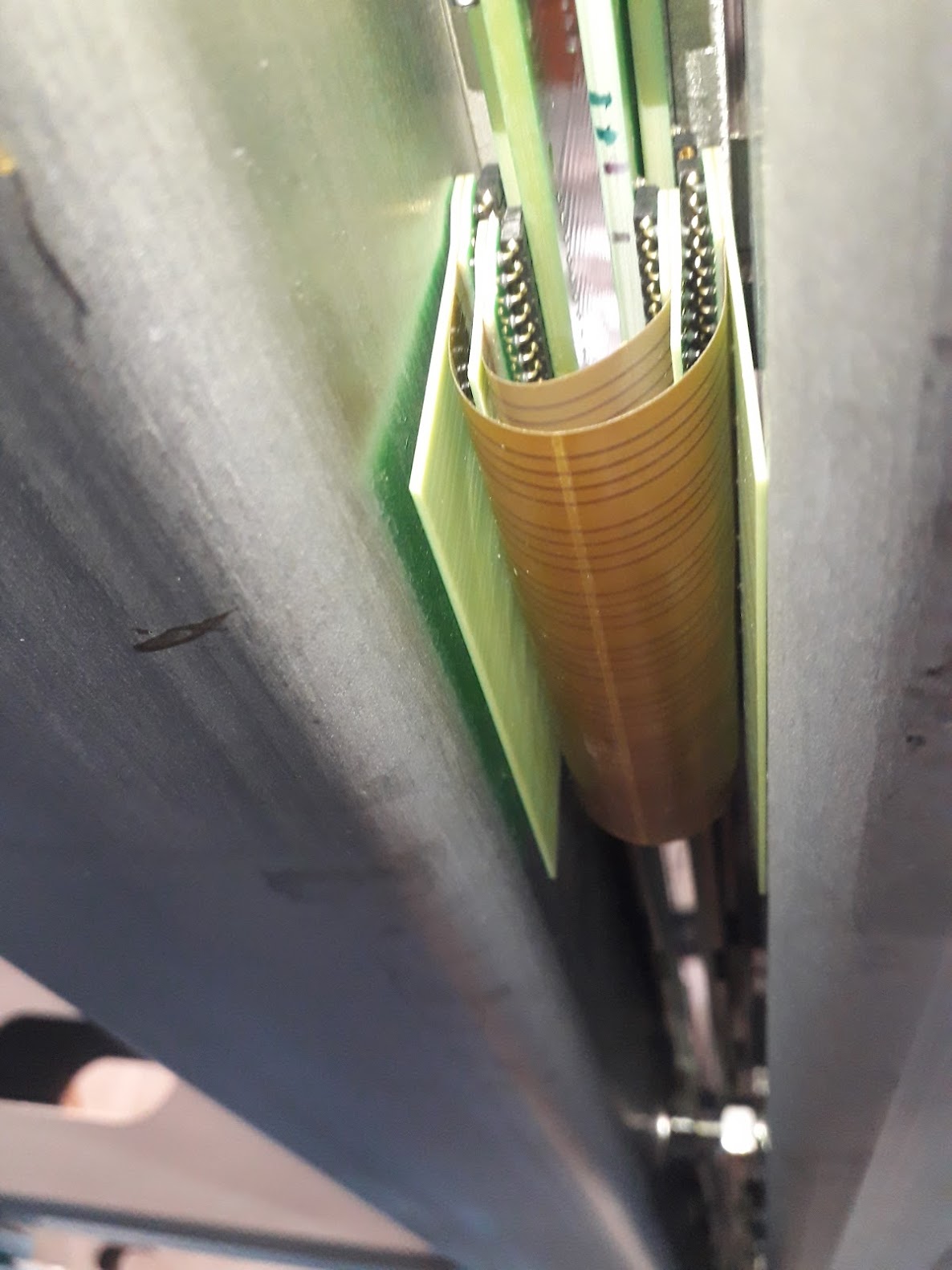}}
        \hspace{0.01\textwidth}
        \subfloat{\includegraphics[width=0.32\textwidth]{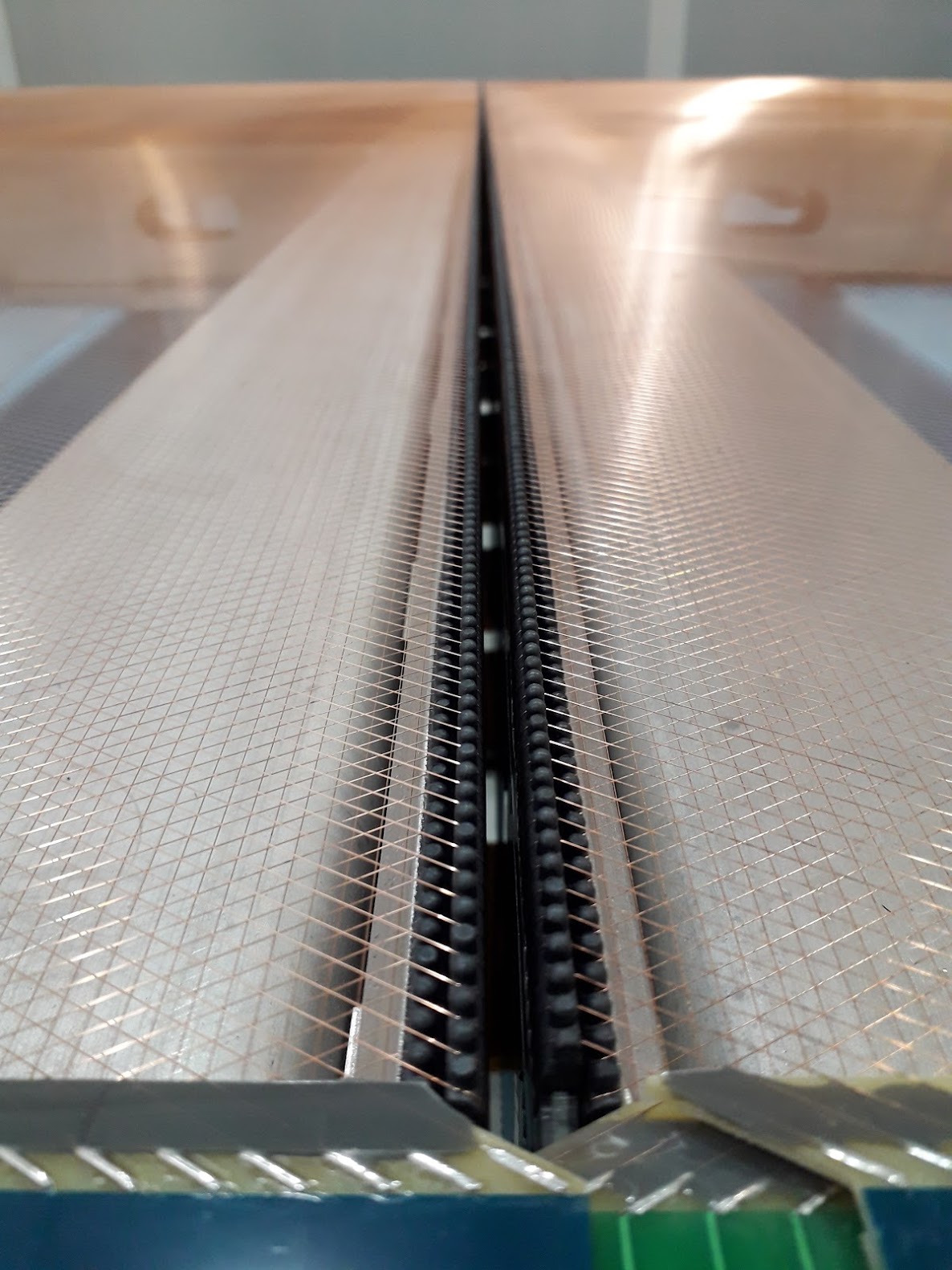}}
        \hspace{0.01\textwidth}
         \subfloat{\includegraphics[width=0.32\textwidth]{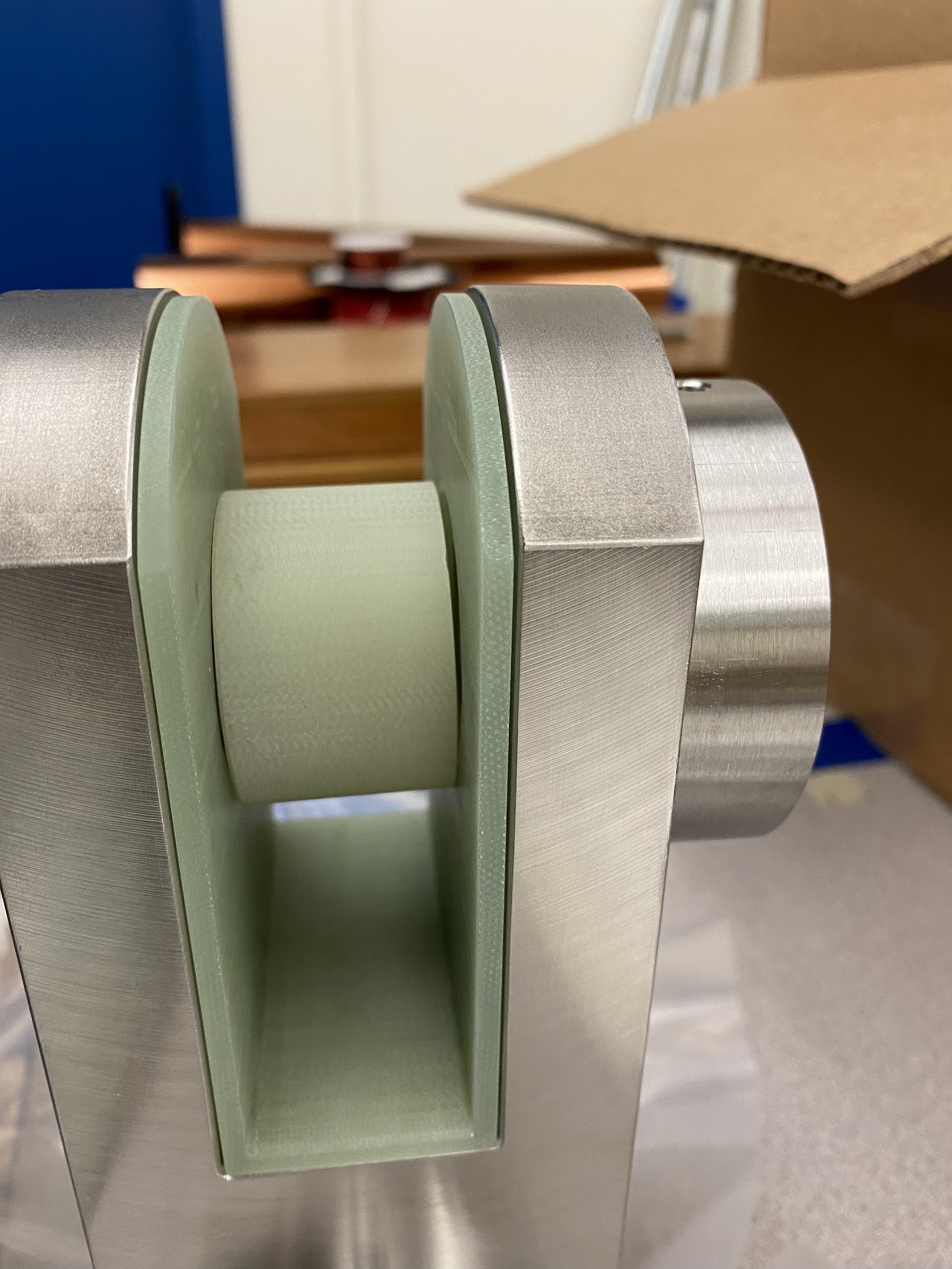}}
    \caption{Components of APA plane assembly: (top left to bottom right) Two coupled APA frames supported on the alignment tables, mechanical coupling of the two APA planes, kapton jumper connections between U and V planes, the aligned center of the APA plane, APA hanger lined with G10 insulator.}
    \label{fig:APAphotos}
\end{figure}

The APA-frames are electrically and mechanically connected to create one APA, whose readout spans the top and sides of the plane and reads out signal from all three planes of wires. 
The precision assembly of these components into a full wire-readout plane is fundamental to the detector operations.  Each pair of APA frames was aligned on custom made alignment tables capable of controlled movement in 3 dimensions. Using a laser tracker, the APA frames were positioned to a tolerance of ±0.5\,mm with an overall flatness of ±2\,mm across the APA plane.  The pairs of APA-frames are mechanically coupled using stainless steel bars. In order to electrically couple the induction planes of the two APA frames, flexible kapton-based jumper cables were installed on the back of the wire plane.  Quality control measurements such as continuity, bias distribution and isolation between planes were carried out to ensure the functionality of each channel.  Figure~\ref{fig:APAphotos} shows photographs of these key elements during APA plane assembly.

Each APA hangs mechanically supported by a support beam, as shown in figure~\ref{fig:TPC3Dmodel}. There is G10 insulation at the connection points, as shown in figure~\ref{fig:APAphotos}, which ensures that the APA is electrically isolated from the cryostat. Grounding is instead routed through the cold electronics return path, as described in section~\ref{sec:grounding}.

In order to maintain a uniform ground plane behind the stainless steel frame structure of the APA, grounding meshes were installed to provide a uniform electric field between the collection plane and the reference ground of the steel frame. The optical transparency of the meshes is 85\% and they are mounted flush to the steel frame behind the wire planes, as detailed in~\cite{SBNDWirePlanes}.
Copper coated G10 cover boards were mounted on the front face of the APA above the wire-mounting geometry boards, to prevent charge accumulation.  The copper face of all cover boards in an APA frame are electrically connected, and are biased to match the local electric field. 

\begin{figure}
    \centering
    \includegraphics[width=\linewidth]{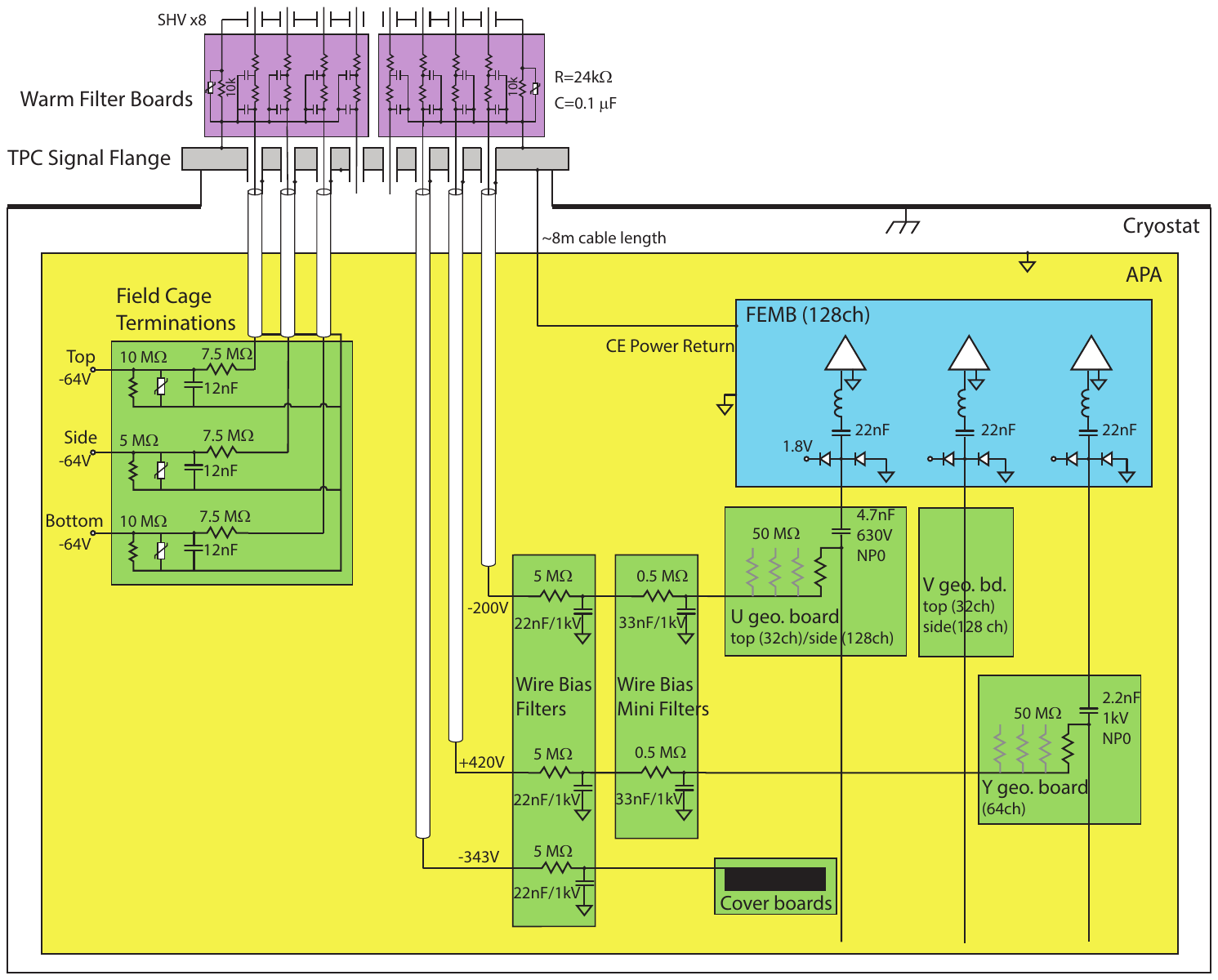}
    \caption{Circuit showing all electrical connections from the warm filter board through to the cryostat for the wire bias and readout, cover board biases and FCTBs, with original design values for wire biases. }
    \label{fig:wirebias}
\end{figure}

Figure~\ref{fig:wirebias} shows the schematic for the wire bias and cover board bias circuits.
The bias connections are isolated from the wire readout electronics, and of particular note are the crossed diodes at the connection to the readout which provide surge protection.
The three wire planes U, V and Y are biased at -200\,V, 0\,V, and +600\,V respectively. The optimization of these bias voltages is discussed in section~\ref{sec:TPCPerformance}.  
These biases ensure uniform electron transparency across the planes, as discussed in section~\ref{transparency}.  Biases for the wire planes and the cover boards are supplied by high precision ISEG~\cite{iseg} power supply modules installed in MPOD crates~\cite{MPOD}, allowing for remote control and monitoring.    During commissioning, the collection planes' biases were increased from the proposed 420V to match the transparency requirements, as discussed in section~\ref{sec:TPCPerformance}.

\subsubsection{APA wire bias operational status}

Table~\ref{tab:tpcbias} shows a summary of the high voltage values currently used at SBND. 

\begin{table}[h]
    \centering
    \caption{Operating voltages of the various TPC system component, including voltage drops.}
    \label{tab:tpcbias}
    \begin{tabular}{cc}
    \hline \hline
    \textbf{System}  & \textbf{Bias Voltage} \\ \hline \hline
    Cathode & -100\,kV \\ \hline
    Cover Boards & -343.35\,V \\ \hline
    U planes & -212.32\,V \\ \hline
    Y planes & 600.606\,V \\ \hline
    \hline   
    \end{tabular}

\end{table}

All TPC bias voltages and currents are continuously monitored by the SBND detector controls system to track stability. The current draw for each of the wire planes and cover boards is expected to be below 10\,nA in normal operating conditions.  In APA West there is a short between one U-plane wire and one V-plane wire, which means the current draw for U-west is around 4.1\,$\unit{\uA}$.  The location and nature of this short has been confirmed by both event display images and raw waveform analysis. While the short impacts signal behavior on the affected channels, it does not compromise the overall safety, stability or functionality of the detector. The diode protection (as shown in figure~\ref{fig:wirebias}) allows the excess current to flow to ground without damaging the readout electronics. 

\subsection{Grounding and ground monitoring}
\label{sec:grounding}

Minimization of electrical noise affecting the TPC APA wire signals is further enhanced by a rigorous grounding and electrical power distribution system.  The electro-magnetic (EM) environment surrounding the SBND  detector reflects the base frequency (60\,Hz in the U.S.) and related harmonics of the local AC power distribution.  Supporting electrical and mechanical systems (power supplies, fans, pumps, et.) contribute additional frequencies at various amplitudes.  Any conductive metals within the environment (structure, wiring, piping) have the capacity to act as EM broadcast antennas, extending the reach of this EM noise.  Meanwhile, the SBND TPC anode wires and readout electronics, designed to detect small ionization signals, are also excellent pickup antennas.  The grounding and AC electrical distribution system is designed to ensure proper power distribution for the facility and supporting systems while protecting sensitive detector electronics from the EM noise environment.
 
The system begins with the building structure itself.  An Ufer ground~\cite{ufer} was implemented during construction of the SBN-ND building, wherein all conductive materials in the structure (concrete reinforcement, framing) were electrically tied and connected at a single point to earth ground.  The quality of the connections was verified at multiple stages of the construction process, providing a robust ground reference for the electrical system.  The building's electrical power distribution was split into two parts, one providing AC power to the building and supporting systems including the cryogenics (\textit{Building Ground}), and the other providing AC power to the detector systems (\textit{Detector Ground}).  Each part, or side, has its own appropriately sized transformer, and both transformers connected to the same ground reference, but with two important modifications for the detector side.  The transformer for the detector side was double-shielded, on both the primary and secondary.  The same transformer's connection to ground went through a saturable inductor, designed to provide a low-impedance path for DC currents (ensuring safety) and a high-impedance path for EM noise.  The scheme effectively isolates the detector's electrical systems from external electrical interference.
 
The scheme also results in two possible paths to the earth ground, and preventing a short between these can become challenging.   For example, the SBND cryostat structure is on the Detector Ground, while the cryogenics system is on Building Ground; the cryogenics piping was designed to include ceramic dielectric breaks to isolate it from the cryostat.  Non-metallic wiring and cable supports were extensively used as these can physically bridge between the building and the detector at many points.  To ensure and maintain the separation of the ground paths, a Ground Current Impedance Monitor (GIZMO)~\cite{gizmo} was utilized.  Any current detected by the device implied a mis-connection between the Building Ground and Detector Ground electrical systems, triggering an alarm.  The device was particularly useful during the construction, installation and commissioning phases, when the risk of unintentional connections was highest.

The low-voltage power distribution for detector components was carefully engineered and reviewed, as it is critical for maintaining signal integrity. Each system was evaluated to ensure proper source and return paths. The ISEG power supplies feature floating returns on each channel and are only referenced at the cryostat. Proper cable shielding terminations were also a key consideration across all subsystems.

Inside the cryostat, specific strategies were employed to preserve the integrity of the detector ground. The TPC APA frames serve as the local reference for the cold electronics. These frames are electrically isolated from the cryostat, and the East and West APA frames are isolated from each other. The only connection between the APA frames and the cryostat occurs through the cold electronics: their DC power and readout cable returns are connected to their associated flanges on the Cryostat top cap.

\subsection{Readout electronics}

\begin{figure} [htb]
    \centering
    \includegraphics[width=0.325\linewidth]{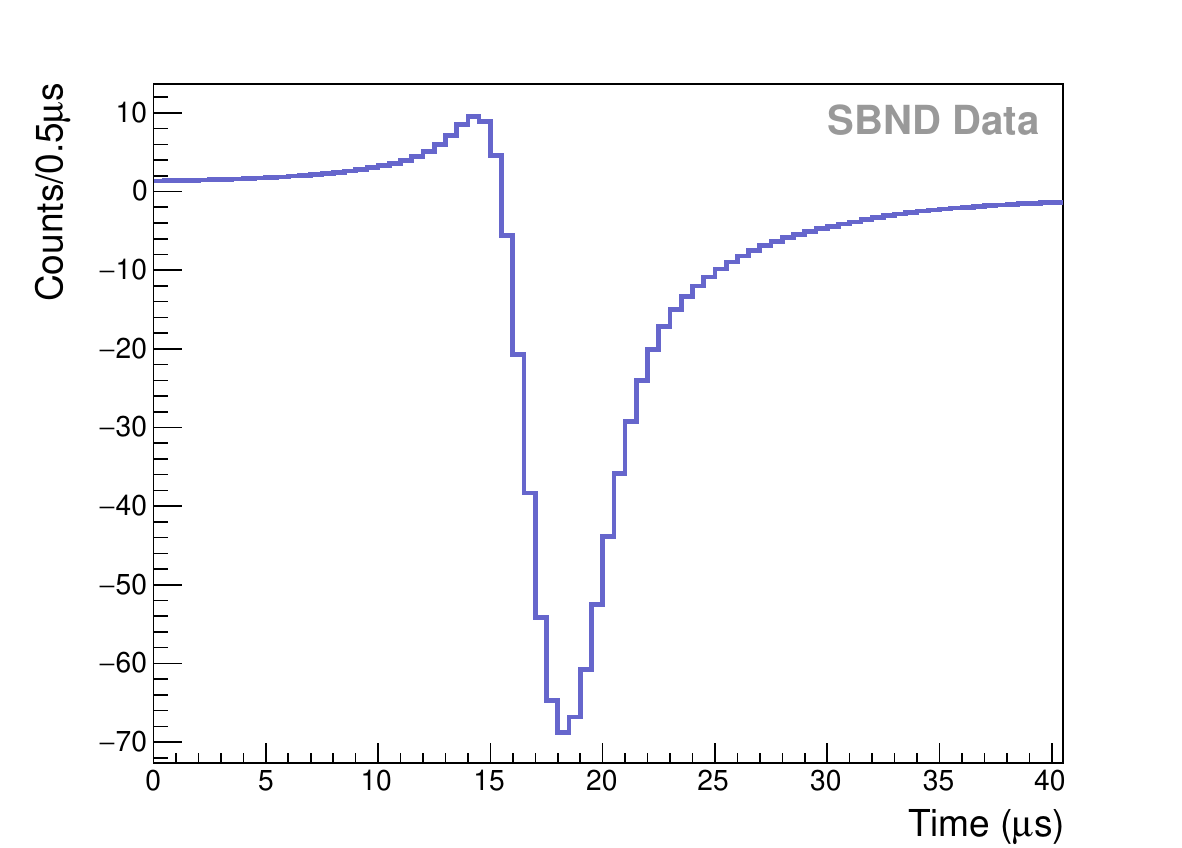}
    \includegraphics[width=0.325\linewidth]{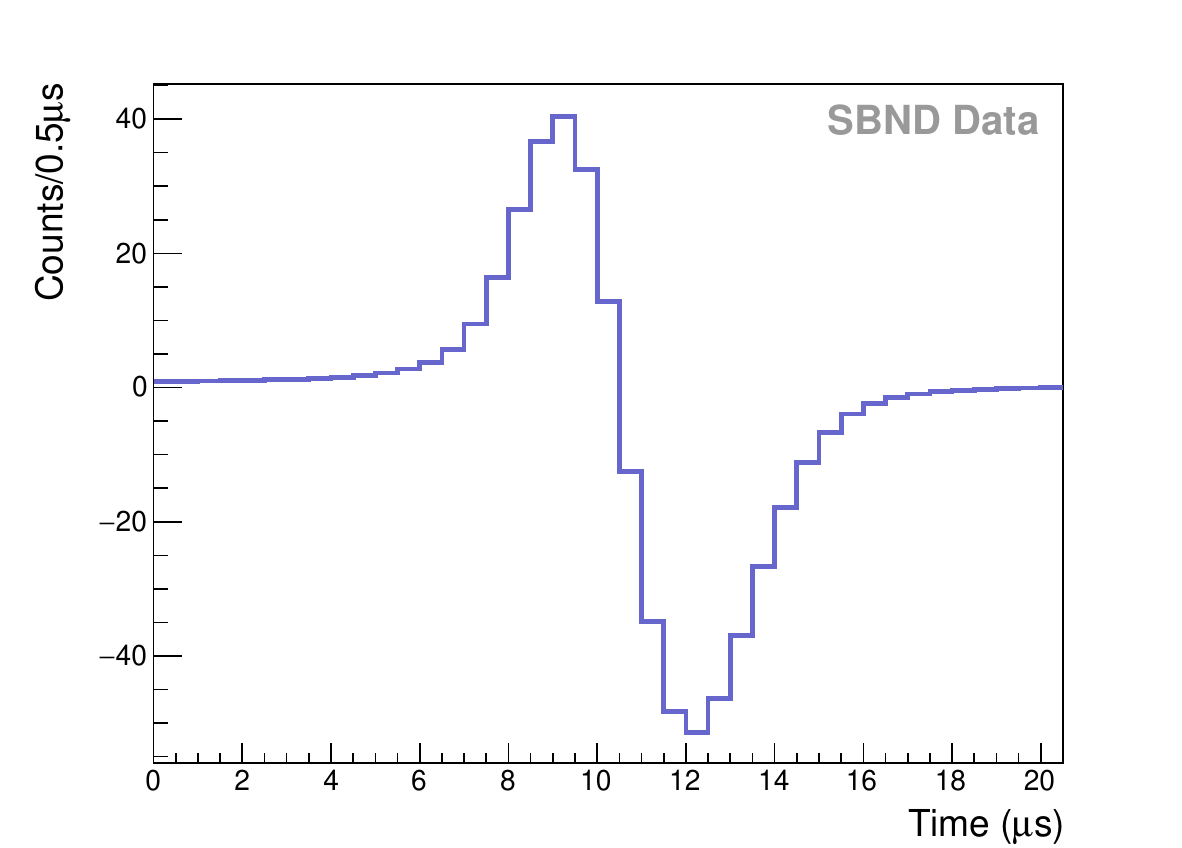}
    \includegraphics[width=0.325\linewidth]{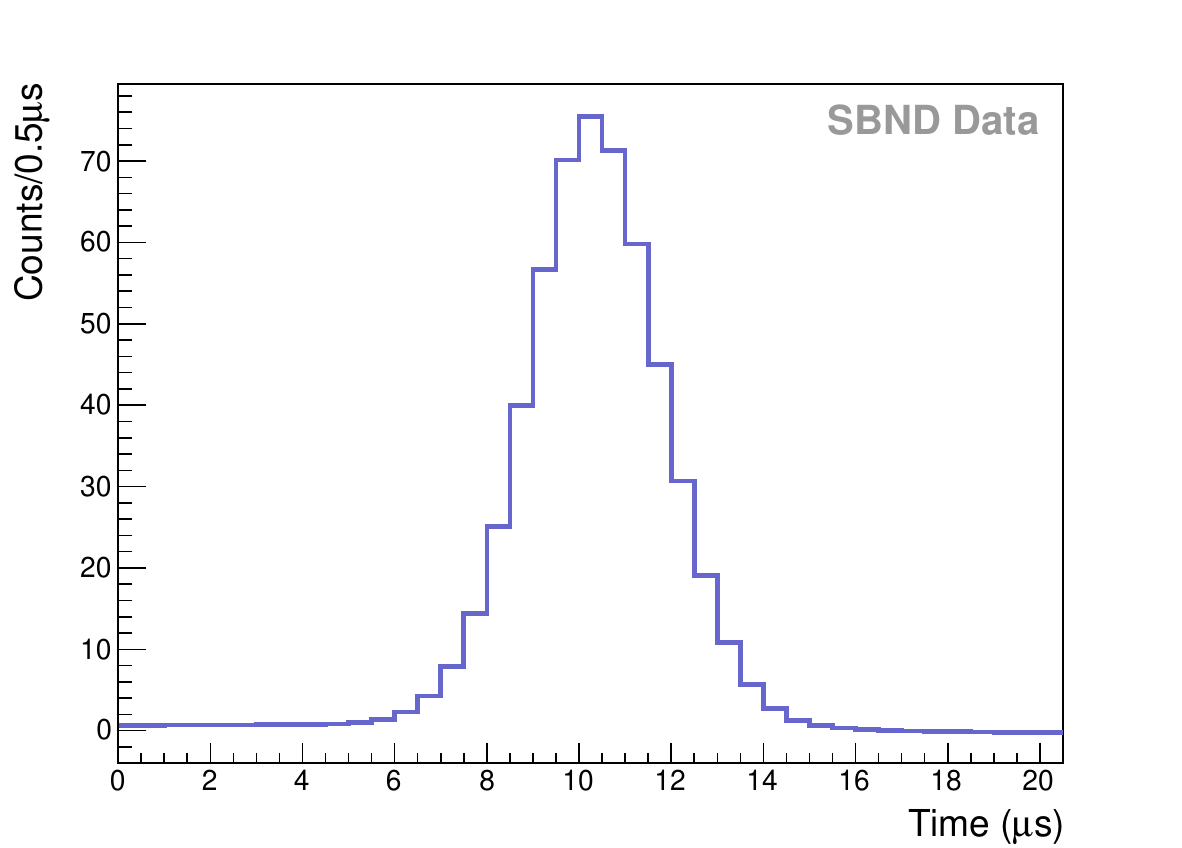}
    \caption{Median of ~75000 raw waveforms obtained from CRT-tagged cosmic-ray muon tracks traveling parallel to the APA in SBND, left: 1st induction, center: 2nd induction and right: collection plane. For averaging, the induction plane waveforms are aligned at zero crossing, collection plane waveforms are aligned at their maxima. The bin width represents the digitization rate of 2\,MHz.}
    \label{fig:avgpulses}
\end{figure}

The goal of the TPC readout electronics is to read in charge and convert it to digital form while minimizing inherent system noise at the input of the front-end pre-amplifiers and providing sufficient precision and dynamic range to achieve the SBND physics goals~\cite{SBNproposal}. 
A typical pulse has a length of length of ${\sim} 10\mu$s. Figure~\ref{fig:avgpulses} shows "raw" signal pulses from each wire plane, prior to offline signal processing. 

\begin{figure} [htb]
    \centering
    \includegraphics[width=0.8\linewidth]{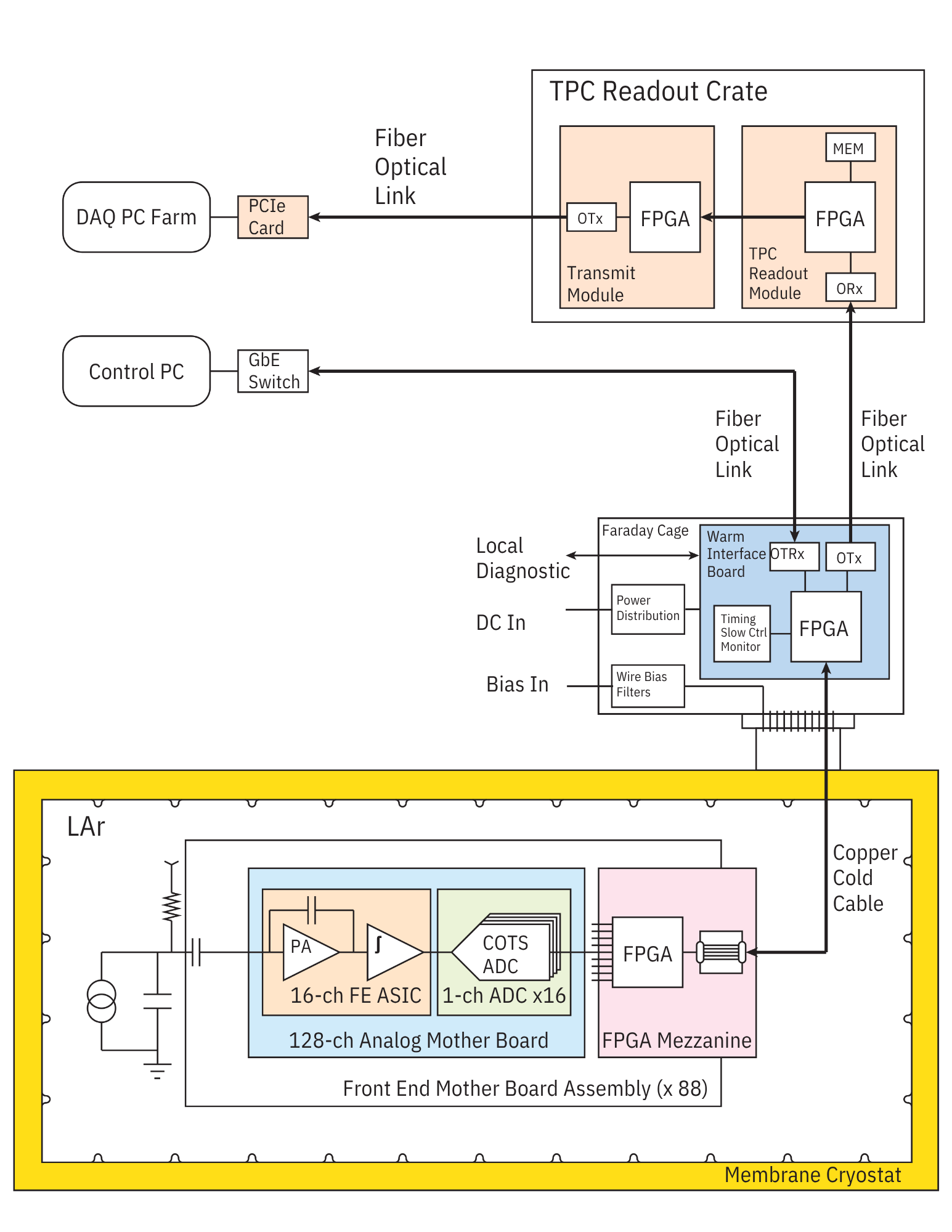}
    \caption{Overview of the TPC readout electronics system, from bottom to top:  Cold Electronics (CE) which provide signal amplification, digitization, and multiplexing at cryogenic temperature immersed within the LAr inside the cryostat (yellow box) as described in section~\ref{sec:coldelec}; warm interface electronics attached to the signal feedthroughs on top of the cryostat, which pass clock and control signals to the CE and data from the CE to the TPC readout crates as described in section~\ref{sec:warmintelec}; and TPC readout crates described in section~\ref{sec:TPCDAQ}.}
    \label{fig:tpc_elec}
\end{figure}

Each wire is read out by an individual channel. The analog input signal is digitized, amplified and multiplexed in the liquid argon, referred to here as the cold electronics (CE), before being sent to filtering and further amplification stages outside of the cryostat, as shown in figure~\ref{fig:tpc_elec}. This front-end (FE) readout electronics system is described in this section. The subsequent handling of the digitized data, the back-end readout electronics, are described in section~\ref{sec:TPCDAQ} as part of the data acquisition system. This schema is similar to that used in ProtoDUNE-SP~\cite{protoDUNEce}, but with some important differences, the foremost being that commercial analog to digital converters are used. 

The FE system~\cite{SBNDce} minimizes intrinsic noise with the following principles:
\begin{itemize}
    \item electrically isolating each APA from the rest of the TPC except through a return to the detector ground at each signal feedthrough as described in section~\ref{sec:grounding} to minimize ground loops;
    \item placement of the FE pre-amplifier ASICs described in section~\ref{sec:coldelec} inside the TPC directly on the APAs to minimize capacitance between the wires and the FE inputs and allow for cryogenic operation of the ASIC CMOS transistors.
\end{itemize}
To place the FE pre-amplifier ASICs on the APAs, the CE system uses front-end motherboards (FEMBs) described in section~\ref{sec:coldelec} which provide signal amplification, digitization, and data multiplexing and transmission to outside the LAr cryostat. The CE is supported by warm interface electronics described in section~\ref{sec:warmintelec} which are attached directly to the signal feedthroughs on top of the cryostat and which pass clock and control signals to the CE and wire data from the cryostat to the TPC readout crates.

\subsubsection{Cold electronics}
\label{sec:coldelec}

The FEMBs attached to the APAs use a "three-chip" architecture with eight custom 16-channel analog FE ASICs, 128 commercial-off-the-shelf (COTS) single-channel 12-bit Analog to Digital Converters (ADCs), and a commercial Field-Programmable Gate Array (FPGA) for clock and control signals in and digital multiplexing and data transmission out, as shown in figure\,\ref{fig:tpc_elec}. Each FEMB receives low-voltage ($<$5V) power from the signal feedthrough via eight twisted-pair cables and transmits 4\,$\times$1\,Gbps link signal data streaming out of the TPC on coaxial cables for 32 channels per link; all cables are 8\,meters long.

Each SBND FEMB contains eight of the same FE ASICs (aka LArASIC) as the ProtoDUNE detector FEMBs described in detail in~\cite{ProtoDUNE}. The FE ASICs are designed to operate with a dynamic range of approximately 1.4\,V at LAr temperatures and provide a fifth-order anti-aliasing filter with a programmable gain and time constant. The commercial FPGA (Altera Cyclone IV) is also identical to the ProtoDUNE detector FEMBs. The commercial single-channel ADCs, used in the SBND FEMBs, are successive approximation ADCs digitizing at 2\,MHz powered at 2.5\,V manufactured by Analog Devices~\cite{CEadc}. The features of the FEMBs are summarized in table~\ref{tab:ceprop}.

\begin{table}[htb!]
\centering
  \caption{SBND FEMB settings (bold denotes value used in SBND) and parameters. }
\label{tab:ceprop}
\begin{tabular}{c c}
\hline \hline{}
FE gain & 4.7, 7.8, \textbf{14}, and 25\,mV/fC \\ \hline
FE shaping time & 0.5, 1, \textbf{2}, and 3\,$\unit{\us}$ \\ \hline
FE baseline mode & \textbf{200 (collection)} and \textbf{900 (induction)} mV \\ \hline
FE leakage current & 100, \textbf{500}, 1000, 5000\,pA \\ \hline
FE dynamic range & 0.2-1.6\,V \\ \hline 
FE calibration charge & 3.41\,fC per step (63 steps)  \\ \hline
ADC bit range & 12-bit \\ \hline
ADC digitization speed & 2\,MHz \\ \hline
ADC dynamic non-linearity & 0.25\,least significant bit (at 1.8\,V ref.)\\ 
\hline \hline
\end{tabular}

\end{table}

Prior to installing on the SBND APAs, the FEMBs were tested both at LN2 and room temperatures, following a similar program described in~\cite{protoDUNEce}. The commercial ADCs on the SBND FEMBs were found to be much more stable than the custom ProtoDUNE ADCs, with lifetime $>10^6$\,years at cryogenic temperatures~\cite{SBNDce}.

\subsubsection{Warm interface electronics}
\label{sec:warmintelec}

\begin{figure}
    \centering
    \includegraphics[width=0.7\linewidth]{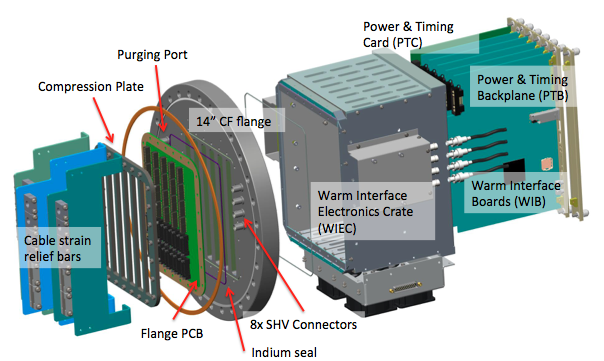}
    \caption{Overview of the warm interface electronics system. One WIEC powers and reads out half of each APA for a total of 4 WIECs on the SBND detector.}
    \label{fig:tpc_warm_elec}
\end{figure}

The warm interface electronics provide low voltage power and clock and control signals to the CE and send data signals out from the CE to the TPC readout crates via custom electronics boards in a custom warm interface electronics crate (WIEC) mounted directly to the TPC signal feedthroughs. Each WIEC powers and reads out the FEMBs on half of each APA, for 4 total WIECs on the SBND detector~\cite{SBNDce}, the main components of which are shown in figure~\ref{fig:tpc_warm_elec}. The two types of boards in each WIEC are:
\begin{itemize}
    \item One Power and Timing Card (PTC). The PTC receives 48V power from a Wiener PL507 module and steps it down to 12\,V via commercial DC-DC converters, filters it, and delivers the 12\,V to each warm interface board (WIB) via a passive Power and Timing Backplane. The PTC also receives the synchronization clock and control signals from the detector control system and distributes it to the WIBs via the backplane;
    \item Six warm interface boards (WIBs). The WIBs further filter the 12\,V power from the PTC and step it down to the $<$5\,V lines to the FEMBs. The WIBs use the clock and control signals from the PTC to synchronize all the FEMBs and configure them. Finally, the WIBs receive the TPC data from the FEMBs, provide error checking and further digital multiplexing, and transmit the data to the TPC readout crates via optical fibers.
\end{itemize}
The warm interface electronics are designed to minimize external noise sources entering the TPC both by filtering the low voltage power on the PTC and WIBs and using optical connections to the TPC readout crates. The WIEC also provides the return of the cold electronics ground to the detector ground at the signal feedthrough as described in section~\ref{sec:grounding}. Prior to final production, testing, and installation of the SBND FE readout electronics, integration tests using the full CE to WIEC system demonstrated equivalent noise charge (ENC)\,$<500$\,electrons~\cite{SBNDce}.

\subsection{SBND TPC performance}
\label{sec:TPCPerformance}
The remainder of this section characterizes the performance of the TPC readout system. In particular, we provide information about the measured electronics noise, the uniformity of the cold electronics channels' response, and the uniformity of the wire planes' measurement of drifted ionization charge.

\subsubsection{Cold electronics noise}

The goal of the TPC readout system is to measure the \unit{fC}-scale charge of clouds of ionization electrons when they arrive at the wires planes. It is therefore critical that the noise be significantly lower than that in order to achieve a high signal-to-noise ratio.
There are two main sources of noise: intrinsic noise, which is a property of the circuit formed by the wire and the cold electronics readout; and excess noise, which is caused by the circuit picking up spurious signals from other components. 
If not minimized or properly characterized, this noise introduces uncertainty into future analysis.

To quantify the noise in a given TPC channel, we calculate the root mean square (RMS) deviation from the pedestal, where the pedestal represents the channel's median value over the entire readout window. Analog-to-Digital Conversion (ADC) units correspond to the charge collected by the wires.
To isolate noise, we exclude events where the ADC signal deviates by more than \qty[mode=text]{20}{ADC} from the pedestal, effectively removing signals associated with ionization electrons collected on the wires.

\begin{figure}
    \centering
    \includegraphics[width=1.0\linewidth]{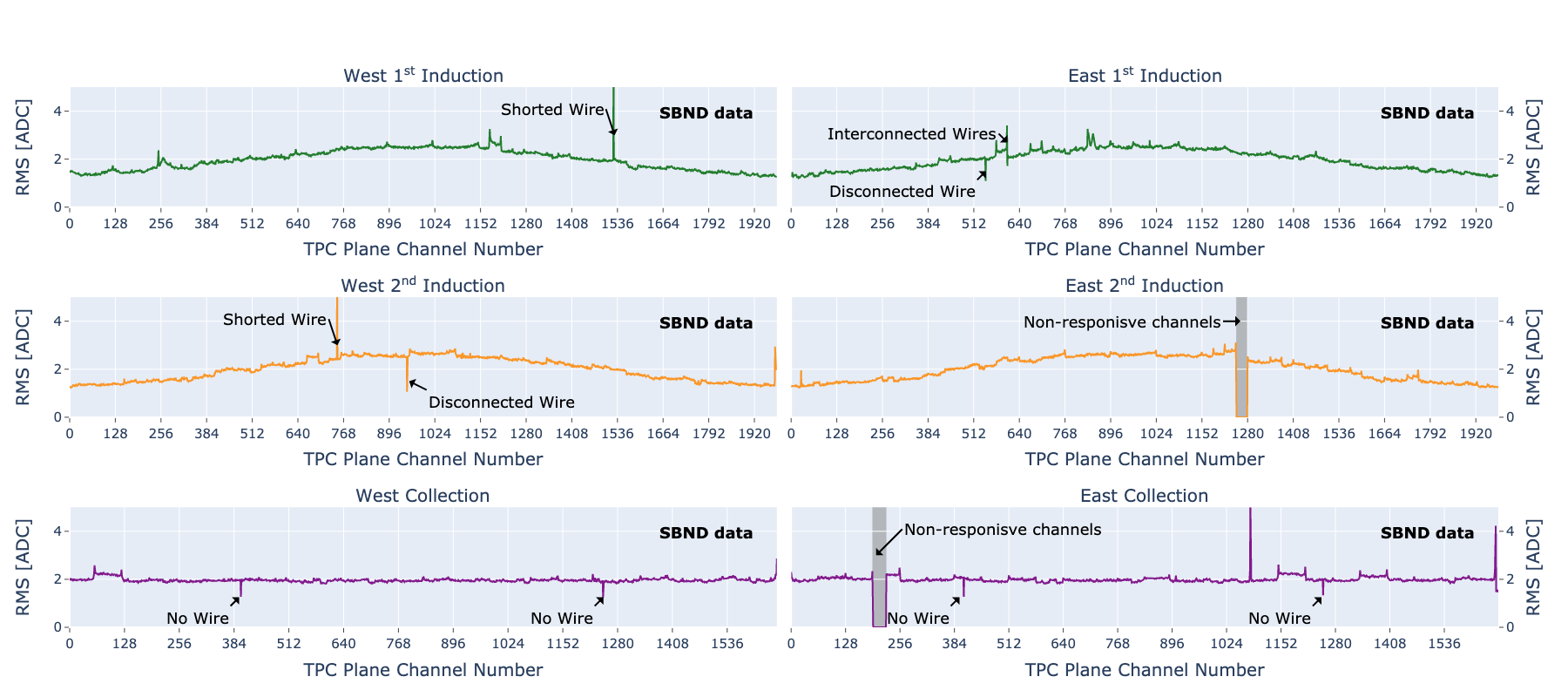}
    \caption{The TPC noise of all the channels in SBND measured by root mean square deviation of the TPC electronics readout from the baseline. 
    Each panels shows channels from a wire plane: green shows the collection (Y) plane, orange the 2nd induction (V) plane, and purple the 1st induction (U) plane, with left and right showing the west and east of the detector respectively.}
    \label{fig:tpc_noise_per_chan}
\end{figure}

To ensure a statistically robust evaluation of noise levels, we analyze the waveforms of each cold electronics channel across hundreds of events.
Figure~\ref{fig:tpc_noise_per_chan} presents the measured noise across all channels, separated by wire plane. Overall, noise variations are minimal, and the detector exhibits consistently low noise levels.
However, certain features stand out, including channels without wires, disconnected channels, and regions where an electronics board is unresponsive. 

There are a total of 80 cold electronics channels with non-standard behavior, which is about 0.7\% of the total number of channels.
These cold electronics channels are described below.
Four channels, one in the middle of the collection plane wires on each half-APA, have no wire attached by design. 
One wire was identified as being disconnected across the APA jumper boundary during construction. The disconnected wire segment was soldered to an adjacent channel, resulting in that channel having elevated noise. 
Sixty-four channels (found in groups of 32 channels on the east second induction plane and east collection plane) were disabled and are non-responsive because of a short on the FEMB that contains these channels identified after the TPC was installed in the cryostat but before cryogenic cool-down. 
Two additional channels were identified as having low noise consistent with being disconnected from their respective wires prior to the cryogenic cool-down. 
One channel has exhibited excessive noise since cool-down. 
Three channels developed excessive noise during the initial ramp-up of the TPC wire bias voltages after the LAr fill had been completed. Two of these (on the west first and second induction planes) are understood to be electrically connected to each other. 
Four additional channels appear to have become disconnected in the first nine months of operations. No additional channels have experienced the same in the following seven months. 

\begin{figure}
    \centering
    \includegraphics[width=1.0\linewidth]{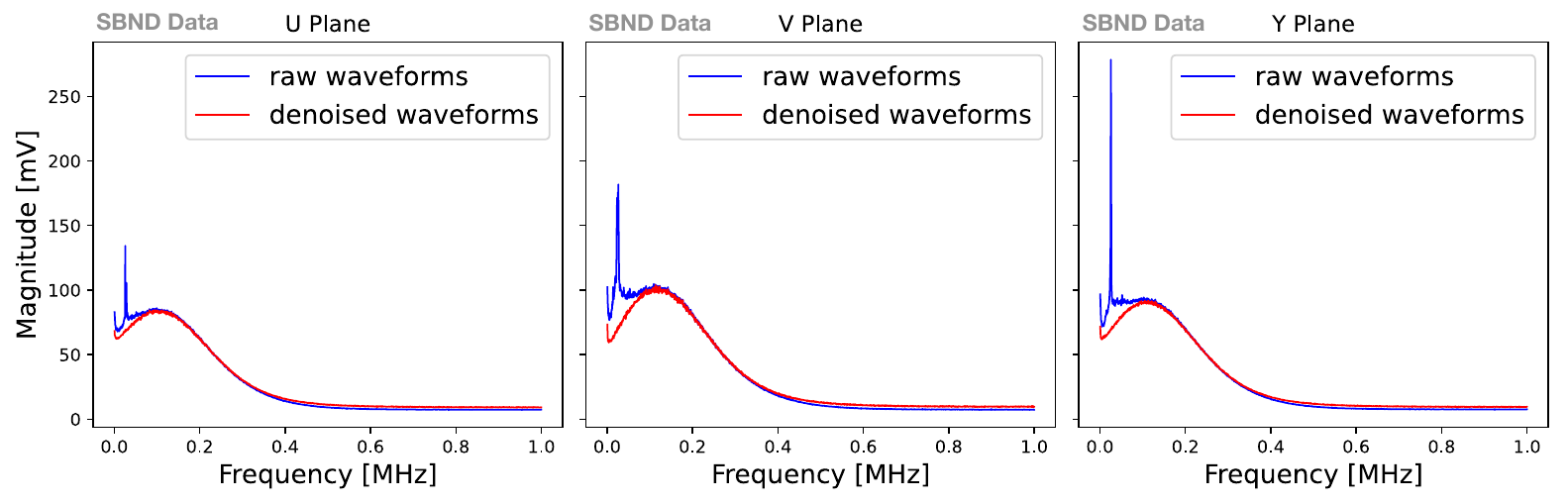}
    \caption{Average noise frequency spectra from a representative group of 256 wires on each of the three wire plane, before (blue) and after (red) correlated noise removal. }
    \label{fig:tpc_fft_cnr}
\end{figure}

Excess noise comes from the TPC readout circuit picking up spurious signals from other components, and it is often coherent or correlated across multiple nearby channels.
In frequency space, excess noise generally appears as distinct spikes overlaid on a smooth Rayleigh spectrum, where the frequency of the spike depends on the source.
As seen in the raw waveforms in figure~\ref{fig:tpc_fft_cnr}, a prominent spike occurs around 25kHz, which we believe is attributable to switching noise from voltage regulators on the FEMBs. However, this noise has minimal impact on data analysis, as it can be effectively identified and removed.

To mitigate correlated excess noise, which often presents across the same FEMBs (either in groups of 32, 64, or 128 channels), we construct a correction waveform across contiguous channels within these groups. This correction waveform is based on the median of each time sample across all wires in the group. Regions that may contain signal are also identified to avoid removing true ionization electrons. The correction waveform is scaled and subtracted from each of the original waveforms in the group.
The FFT of the TPC noise spectrum before and after applying this coherent noise removal procedure is shown in figure~\ref{fig:tpc_fft_cnr}. Correlated noise removal may or may not be incorporated in future data processing, as the procedure has the potential to introduce biases through signal removal, even with signal protection thresholds in place. A careful evaluation of the trade-off between noise mitigation and signal protection is necessary to determine the implementation of correlated excess noise removal.

\begin{figure}
    \centering
    \includegraphics[width=1.0\linewidth]{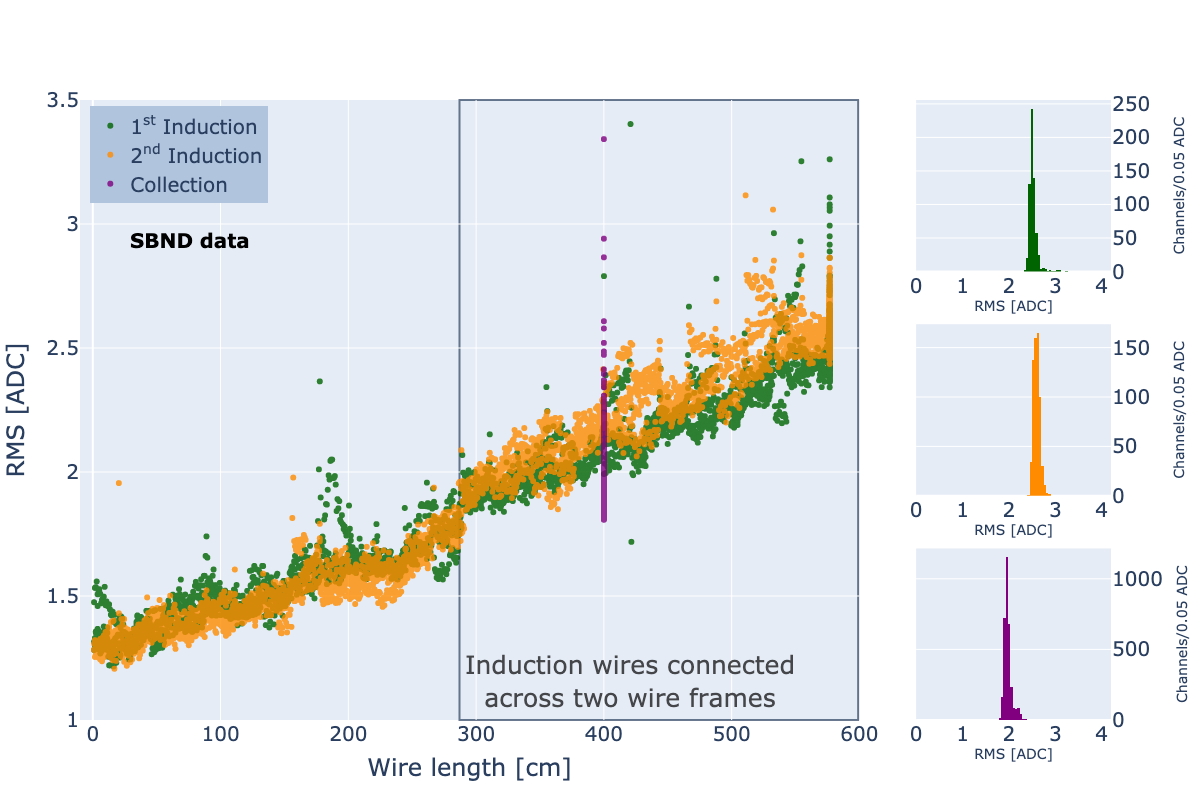}
    \caption{Left: Dependence of the TPC noise on the corresponding active wire length. The gray box indicating the wires that cross two APA frames.
    Right: Histograms of the noise for the longest wires for each wire plane. The wire planes have been separated by color: 1st induction (U) plane (green), 2nd induction (V) plane (orange), and collection (Y) plane (purple).}
    \label{fig:tpc_noise_vs_length}
\end{figure}

Intrinsic noise, on the other hand, can be represented as the Rayleigh distribution in frequency space. Besides the coherent noise spikes, this distribution is evident in figure~\ref{fig:tpc_fft_cnr}. A key characteristic of intrinsic noise is its dependence on wire length due to capacitance. The noise is higher on longer wires. This is apparent in figure~\ref{fig:tpc_noise_per_chan}, where noise  on the induction planes increases for central channels, while edge channels exhibit lower noise levels. This behavior arises because the induction plane wires are angled \qty{\pm 60}{\degree} relative to the vertical collection wires, leading to a broad range of wire lengths. Figure~\ref{fig:tpc_noise_vs_length} further illustrates the relationship between noise and active wire length in SBND, showing a discontinuous jump due to wires being jumpered across APA planes, as is visually shown in figure~\ref{fig:APAphotos}.

The average RMS noise for collection plane wires in the SBND TPC readout is \qty{1.98}{ADC}, which translates to an  ENC of $388\,e^-$, before any offline noise removal.
The expected irreducible intrinsic noise given the design of the circuit and the \qty{4}{m} length for the collection plane wires is \qty{1.94}{ADC}, so the observed noise level is only about 2\% higher.
The observed noise on the induction planes is similarly low.
This performance exceeds the experimental requirement and enables an excellent signal-to-noise ratio in SBND's TPC readout: 14.5, 23.5, 32.5 in the U, V and Y planes, respectively.

\subsubsection{Cold electronics response uniformity}
Another important metric is the uniformity and stability of the TPC cold electronics' response to charge.
To measure the charge response of the cold electronics, a six-bit voltage digital-to-analog converter (DAC) injects a known amount of charge across an input capacitor in short-duration pulses. The injection capacitance has been measured using a calibrated external capacitor. The measurements show that the calibration capacitance is extremely stable, changing from \qty{184}{fF} at room temperature to \qty{183}{fF} at \qty{77}{K}. The goal of the charge injection system is to measure the gain and peaking time of each individual channel to correct for non-uniformities in the signal response prior to offline signal processing~\cite{uboone_sp1,uboone_sp2}.

\begin{figure}[htb]
    \centering
    \includegraphics[width=1.0\linewidth]{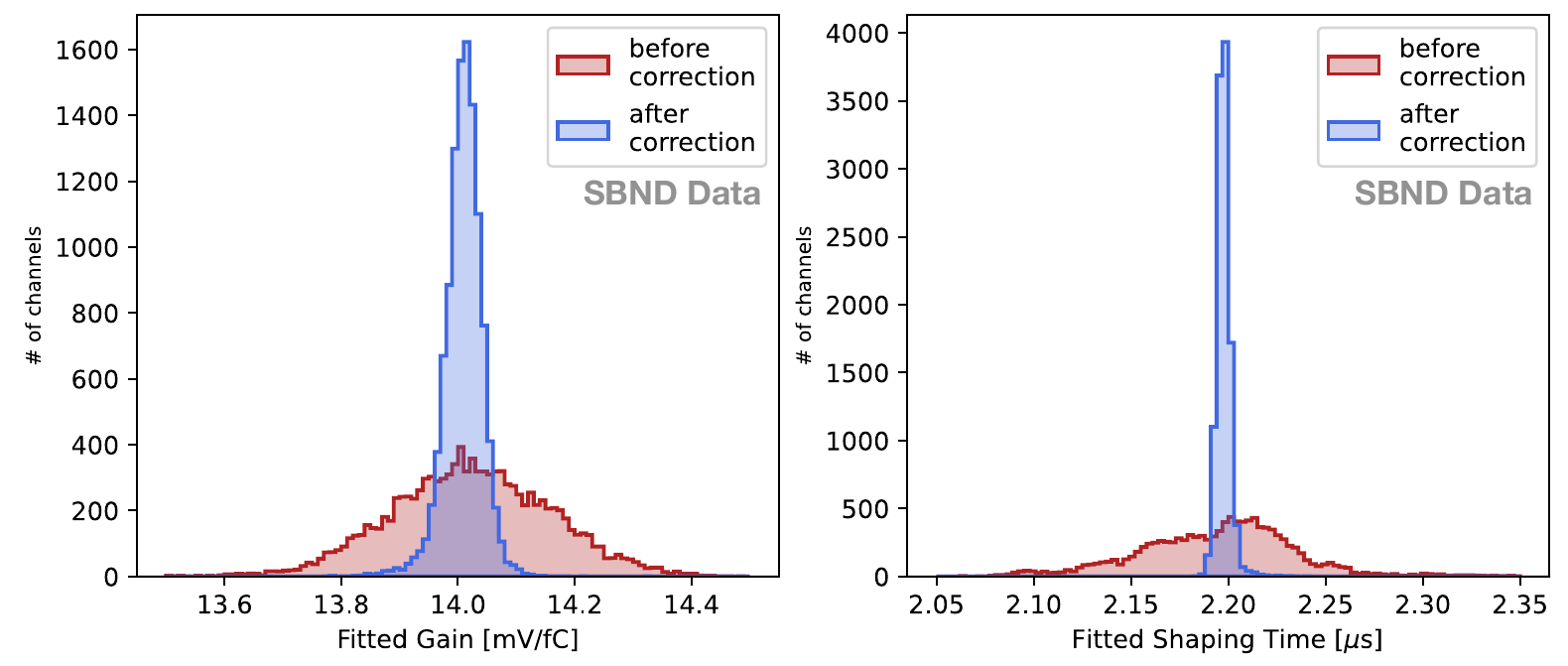}
    \caption{The fitted gain (left) and shaping time (right) of all TPC wires before (red) and after (blue) applying calibration corrections. The nominal gain is \qty{14}{mV/fC} and the nominal shaping time is \qty{2.2}{\micro\second}.}
    \label{fig:ce_gain_and_shaping}
\end{figure}

This measurement is performed by injecting a series of known amounts of charge into the front end of the readout electronics, then fitting each channel's response for the gain and shaping time.
Charge injection up to \qty{3.41}{fC}, within the linear response region of the electronics, is used for the calibration. We obtain an average electronics response for each channel by taking the median waveform of all pulses for each DAC value. Given the ideal electronics response, described by a fifth-order Gaussian anti-aliasing filter~\cite{uboone_sp2}, we perform a fit for each channel's response and obtain a measurement of the amplitude and shaping time. We obtain the gain of each channel by performing a linear fit between the fitted amplitudes vs.\ DAC values in the linear response region. The shaping time obtained for each channel is the weighted average of fitted shaping times for all DAC values. The fitted gain and shaping time before and after calibration corrections are applied is shown in figure~\ref{fig:ce_gain_and_shaping}.
After this correction, the standard deviations of the gain and shaping time distributions decrease from \qty{0.137}{mV/fC} to \qty{0.033}{mV/fC} and from \qty{0.040}{\micro\second} to \qty{0.004}{\micro\second}, respectively.

\subsubsection{Wire plane uniformity}
\label{transparency}

The wire bias voltages ensure that the electric field behind each of the two induction planes is strong enough relative to the field in front of the wire plane to force the drifting electrons to pass between the wires and continue toward the next wire plane.  The following inequality is used as a guideline to ensure that no electrons are collected on the induction planes
\begin{equation}
    \frac{E_B}{E_A} \geq\frac{1+\rho}{1-\rho}~ ,
\end{equation}
\noindent where $E_A$ and $E_B$ are the electric fields before and after each wire plane, $\rho = 2\pi\frac{r}{d}$, $d$ is the wire spacing, and $r$ is the wire radius~\cite{Intransparency3}. The electric fields $E_B$ and $E_A$ depend directly on the wire bias voltages and the wire spacing. When this inequality is violated, the induction planes become partially intransparent to charge~\cite{Intransparency1, Intransparency2, Intransparency3}.

In its nominal configuration, the SBND TPC geometry satisfies this inequality. However, within the tolerance of the wire plane spacing ($3\pm0.5\,$\si{mm}~\cite{SBNDWirePlanes}), it can be violated. During commissioning an exploratory scan of wire bias voltages was carried out to understand if the initially proposed collection plane bias voltage of \SI{420}{V} was optimal for the as-built detector. The spatial uniformity of the calorimetric response of the wire planes to drifting electrons was used to explore the transparency of the induction planes (as was observed in ICARUS, e.g.~\cite{ICARUS:2024hmk}). 

In this study, the charge scale is measured across the detector on each wire plane using the distributions of charge deposition per unit length ($dQ/dx$) from through-going cosmic muons. The reconstruction of cosmic muons is performed by running a hit-finding algorithm tuned to operate on ionization waveforms (as shown in figure \ref{fig:avgpulses}), with those hits fed into Pandora reconstruction~\cite{Pandora, DUNEPandora} as input. Pandora clusters charge depositions on each wire (``hits'') into tracks and reconstructs a value of charge-per-length ($dQ/dx$) per-hit. The charge is reconstructed as the area of the negative lobe of the bi-polar pulse on the induction planes, and as the area of the positive peak of the unipolar pulse on the collection plane. Reconstructed $dQ/dx$ values on each plane are selected by cuts on the local track angle to ensure the stability of the charge reconstruction, and then binned into pixels 

covering the readout plane. These distributions are fit to a Landau distribution convolved with a Gaussian distribution to extract the most-probable-value (MPV) $dQ/dx$, which is interpreted as the charge scale. Performing this analysis directly on raw ionization waveforms enabled quick feedback on the TPC performance before precise signal reconstruction techniques such as deconvolution were available.

\begin{figure}[p]
    \centering  
    \includegraphics[width=0.325\textwidth]{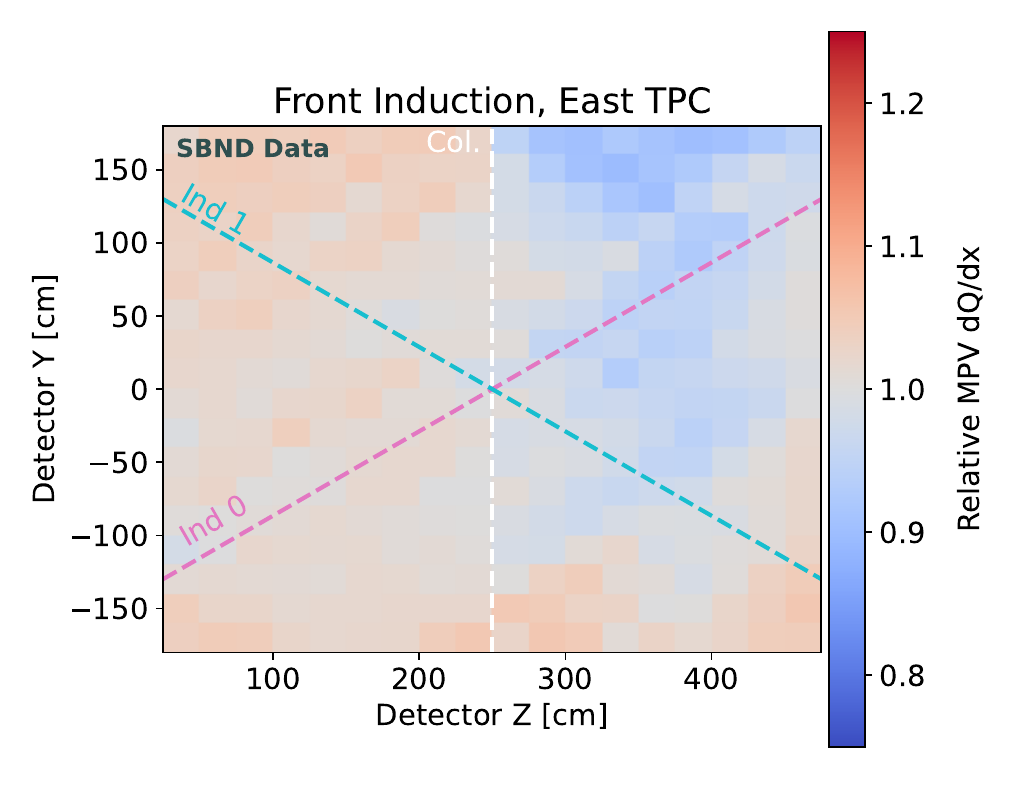}
    \includegraphics[width=0.325\textwidth]{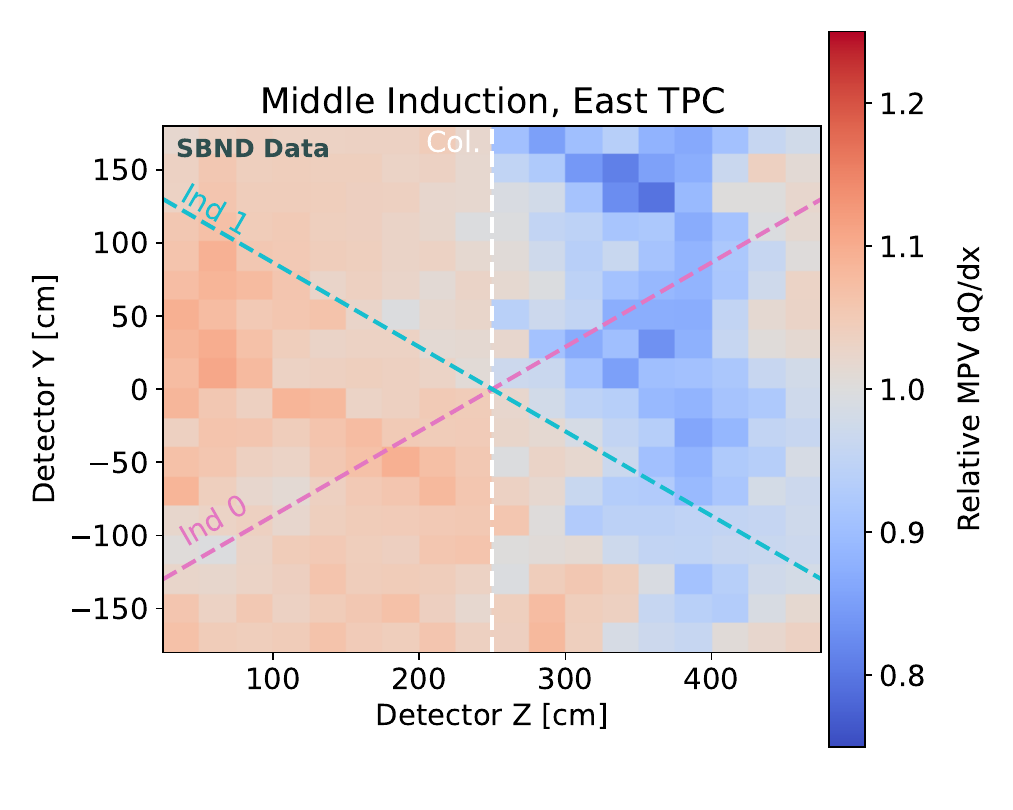}
    \includegraphics[width=0.325\textwidth]{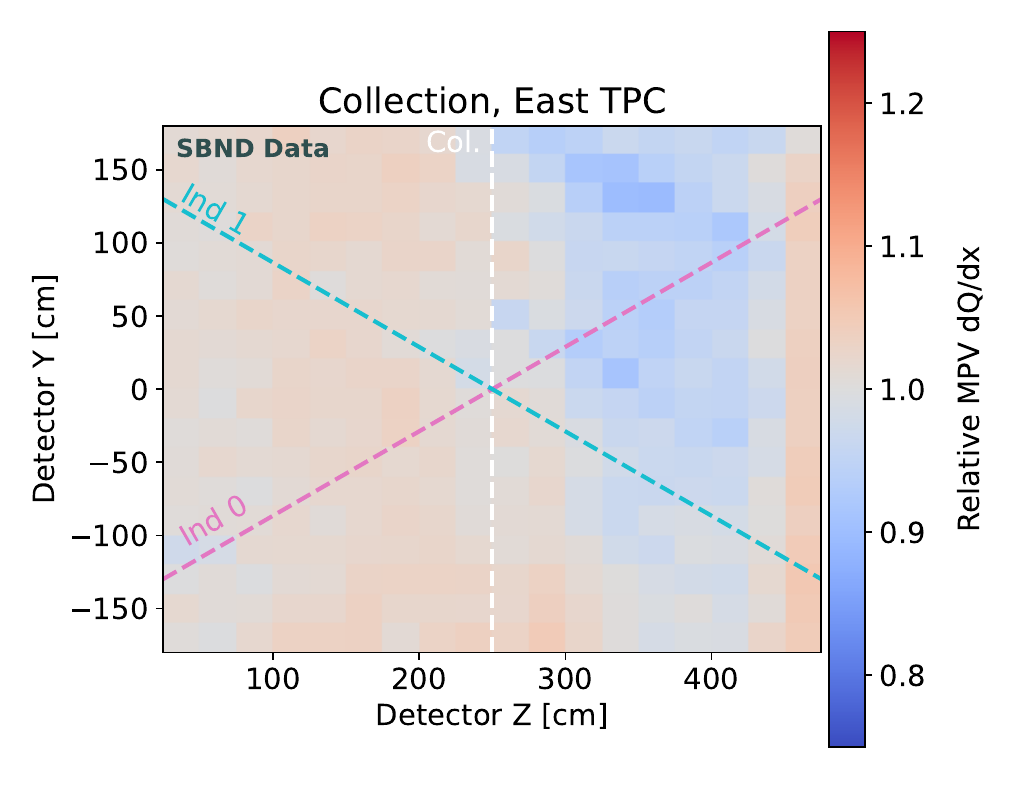}
    \includegraphics[width=0.325\textwidth]{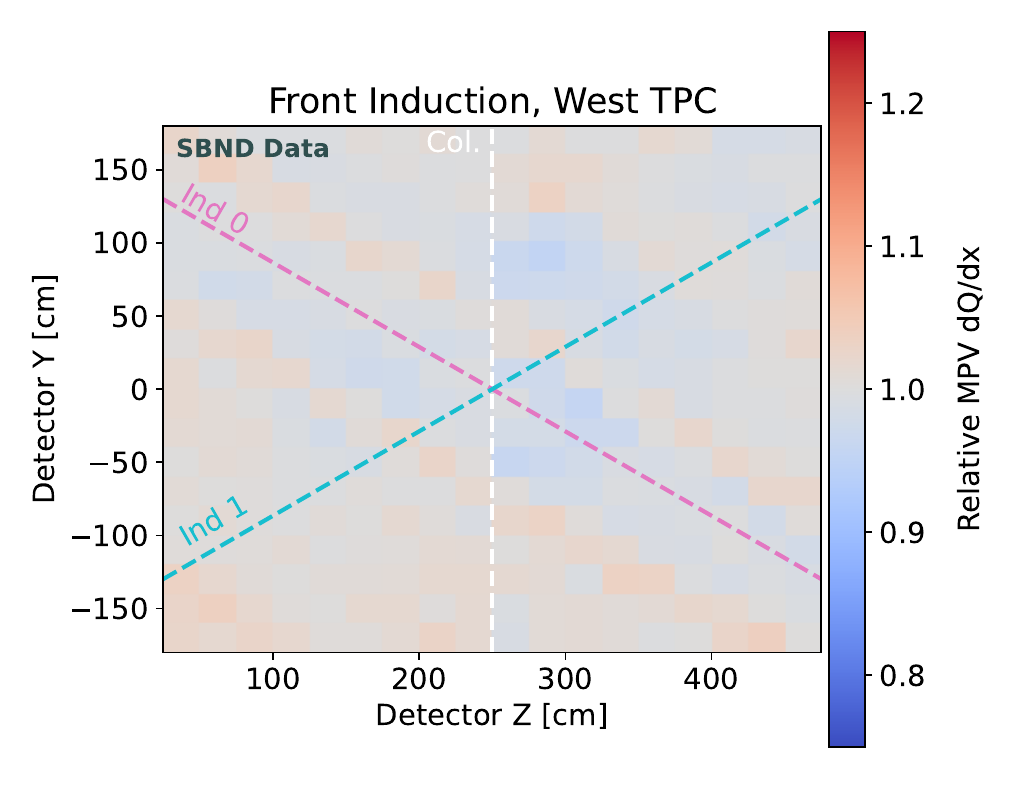}
    \includegraphics[width=0.325\textwidth]{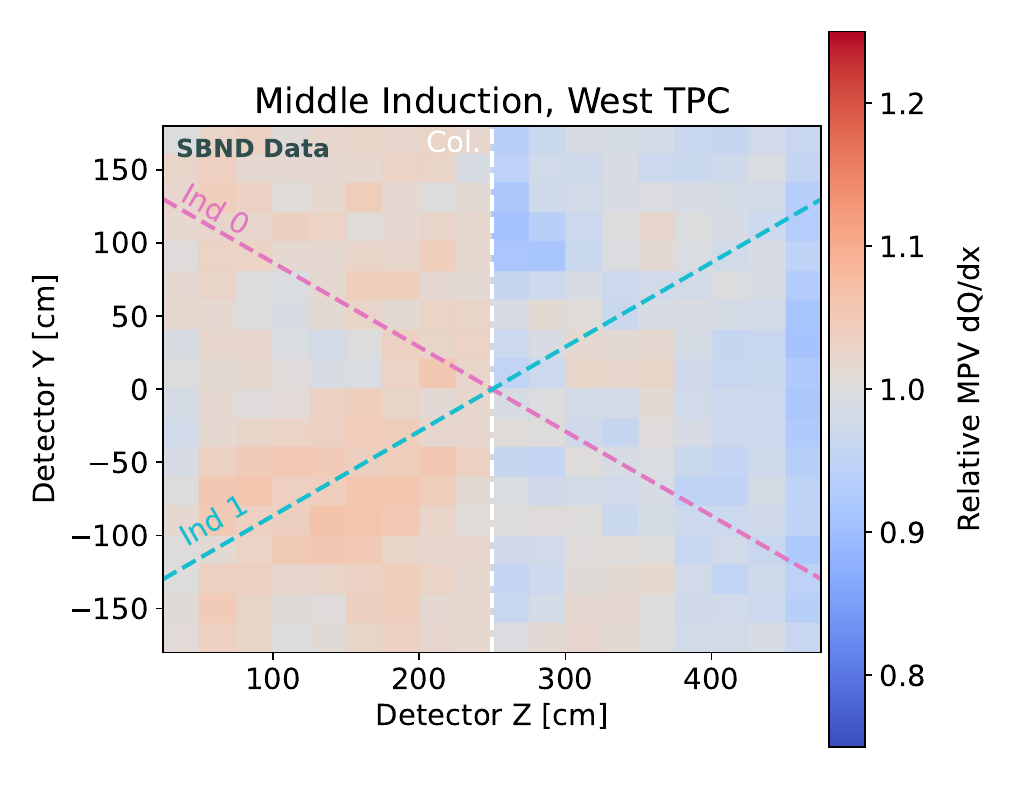}
    \includegraphics[width=0.325\textwidth]{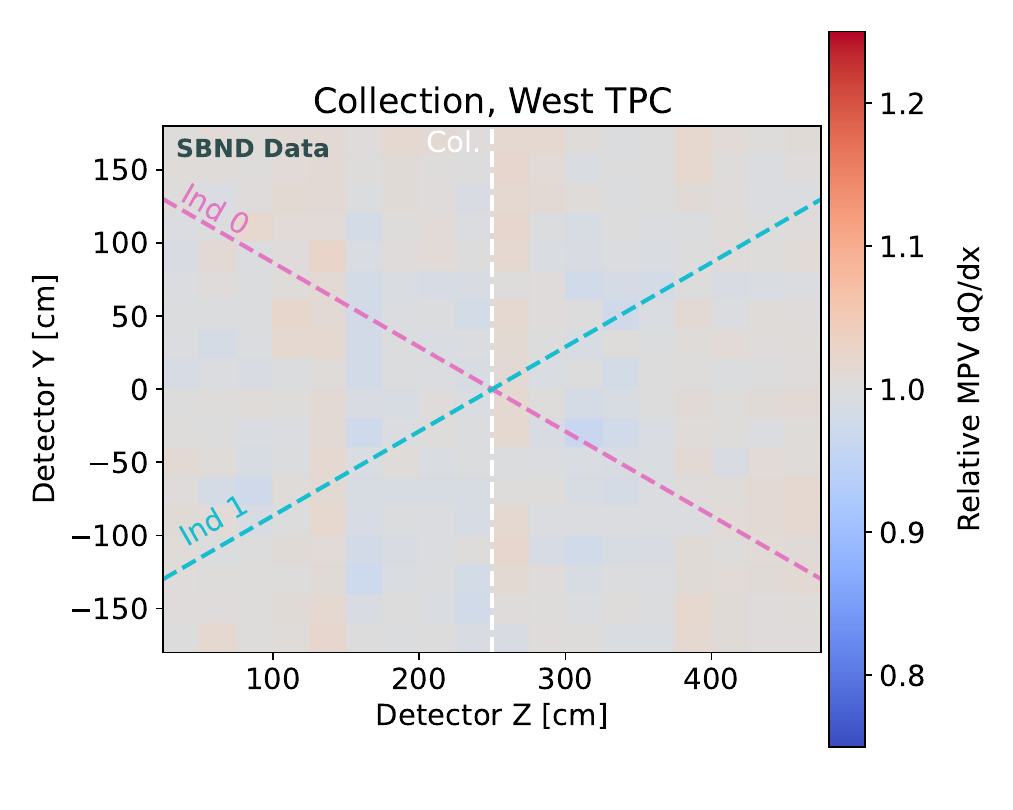}    
    \caption{Relative charge scale on each wire plane at the initial wire bias in the East (top) and West (bottom) TPCs. The relative scale is defined as the Landau-Gaussian MPV in that pixel, divided by the average across the wire plane.}
    \label{fig:tpc-charge-scale-nom}
\end{figure}

Figure~\ref{fig:tpc-charge-scale-nom} shows the relative charge scale across the TPC in its initial wire bias configuration, with the first induction plane biased at \qty{-200}{\volt} and the collection plane biased at \qty[print-implicit-plus]{420}{\volt}~\cite{SBNDWirePlanes}. 

The response of the TPC to charge is quite uniform across the wire planes (\%-level), with the exception of the northern half of the east TPC (TPC NE, henceforth). In this wire bias configuration, there is a decrease in the charge scale in TPC NE, correlated across the three planes, by up to ${\sim}15\%$. 

\begin{figure}[p]
    \centering  
    \includegraphics[width=0.325\textwidth]{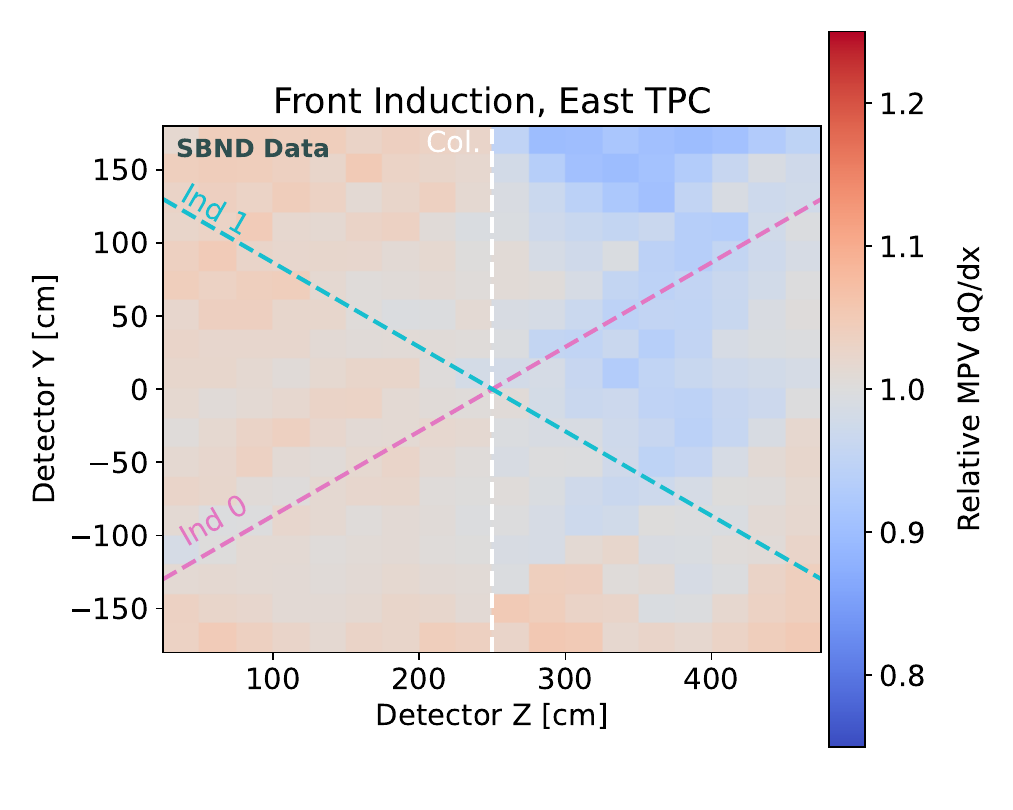}
    \includegraphics[width=0.325\textwidth]{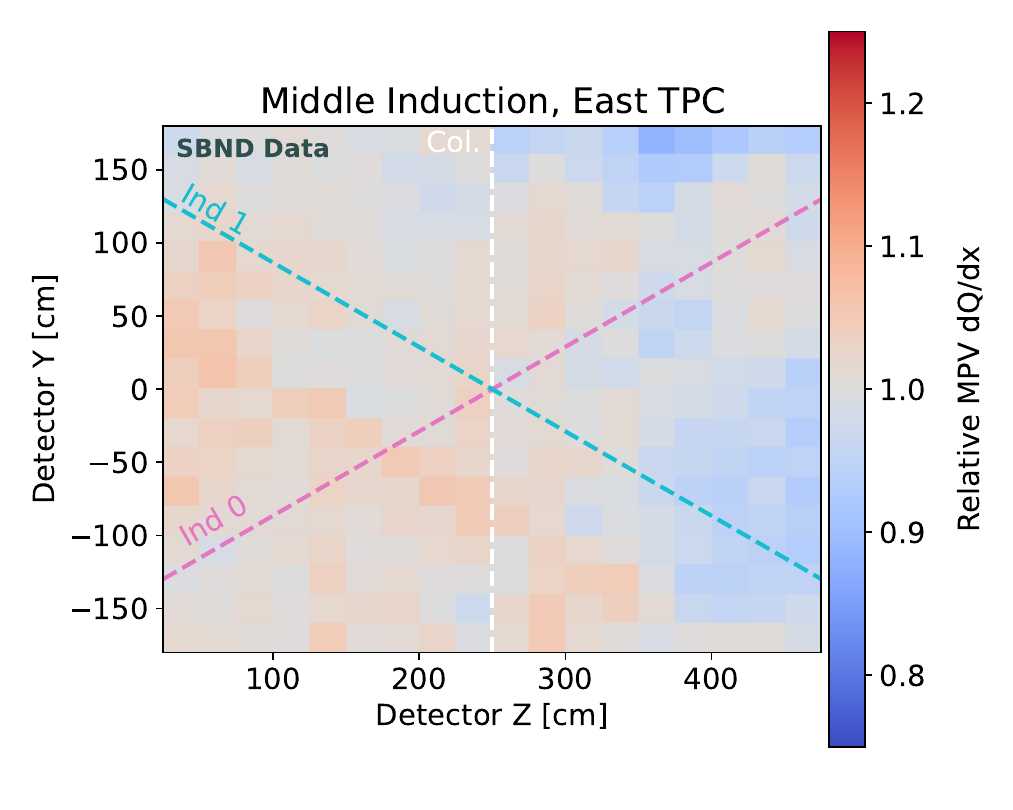}
    \includegraphics[width=0.325\textwidth]{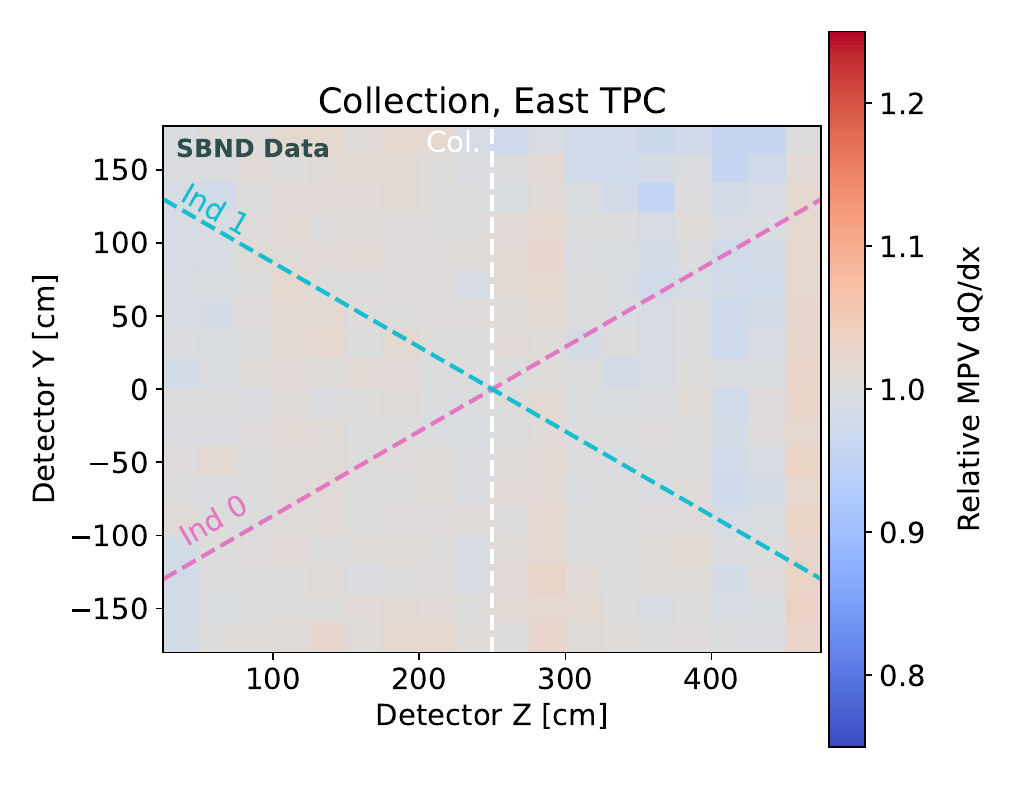}
    \includegraphics[width=0.325\textwidth]{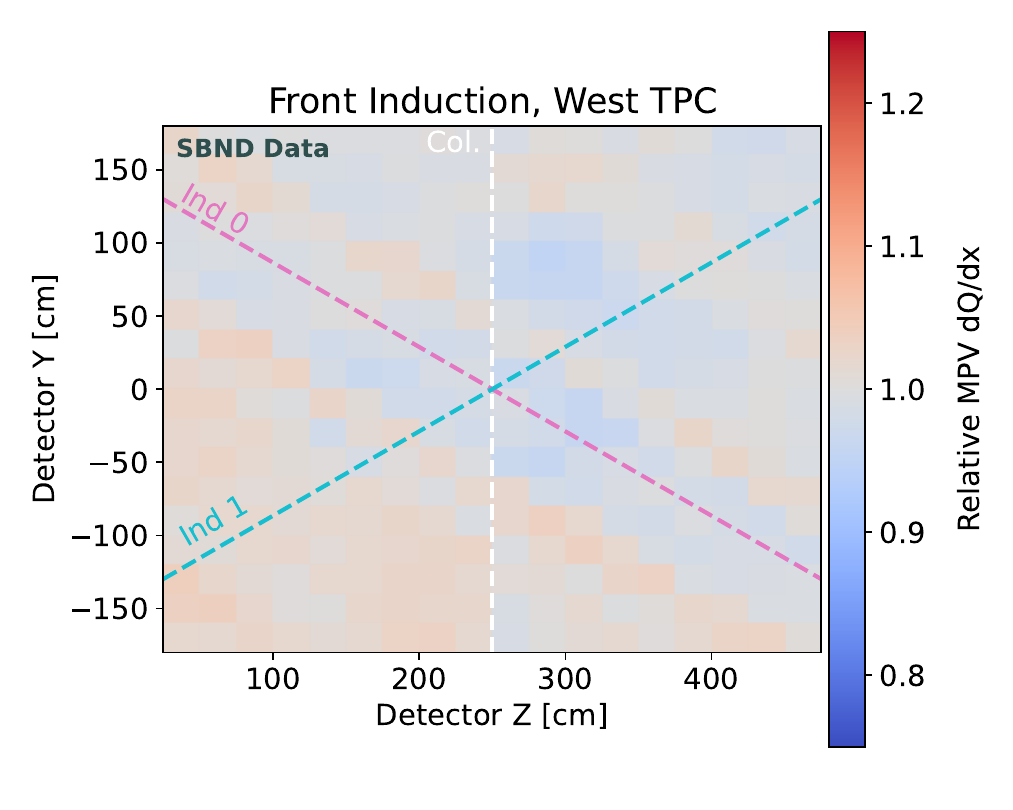}
    \includegraphics[width=0.325\textwidth]{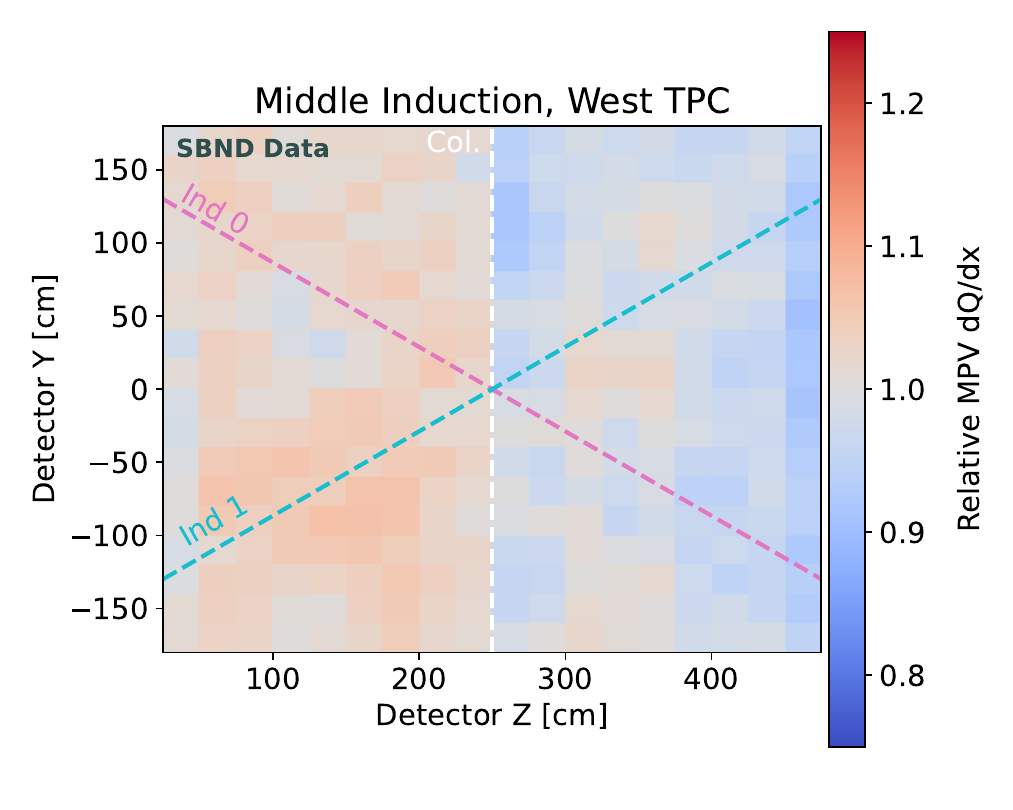}
    \includegraphics[width=0.325\textwidth]{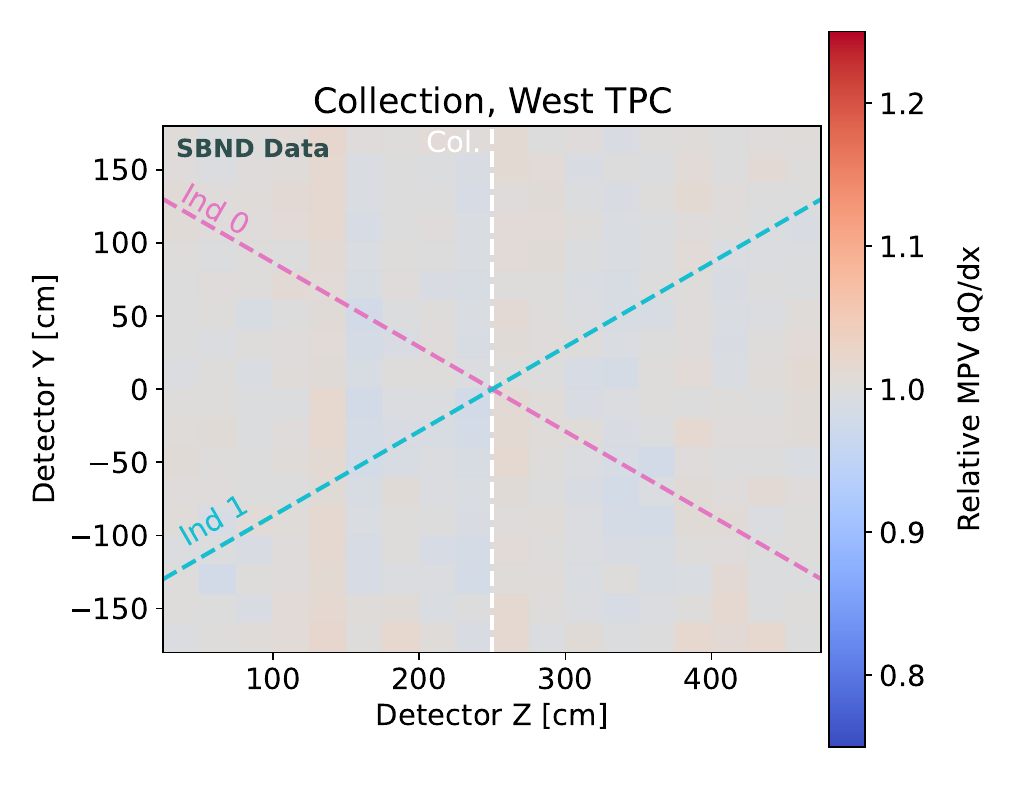}    
    \caption{Relative charge scale at the operational wire bias on each wire plane in the East (top) and West (bottom) TPCs. The relative scale is defined as the Landau-Gaussian MPV in that pixel, divided by the average across the wire plane.}
    \label{fig:tpc-charge-scale-opt}
\end{figure}

To mitigate this feature, we increased the collection (Y) plane voltage to \qty[print-implicit-plus]{600}{\volt} for operations. The relative charge scale across the detector in this configuration is shown in figure~\ref{fig:tpc-charge-scale-opt}. At this increased bias voltage, the non-uniformity in TPC NE on the second induction (V) and collection planes decreases. However, it is not resolved for the first induction (U) plane. Furthermore, through additional tests we found that the residual feature on the first induction plane was not sensitive to any change in the bias voltage on the front induction or collection plane, nor on the magnitude of the drift electric field. We thus concluded that the non-uniformity on the middle induction and collection planes was due to charge intransparency. However, since the charge scale decrease on the front induction plane is not sensitive to the bias voltage, and does not correlate with a residual effect on the collection plane, it cannot be charge intransparency alone.

The only configuration of the geometry that is both within construction specifications and consistent with all of our observations from data taken at different wire plane bias voltages is a shift of roughly 0.5$\,$mm of the  second induction towards the first induction plane in TPC NE.  
In this scenario the field between the first and second planes is stronger than the minimum required for transparency, due to the proximity of the planes. The field lines from the second induction plane cause a ``herding'' effect close to the center of the wire pitch in the first plane, away from the sense wires. This charge herding decreases the amplitude of the induced signal on the first plane, while not impacting the other two planes. Since this effect would be largely geometric, and not strongly sensitive to the bias voltage, it is consistent with the residual non-uniformity observed.

\begin{figure}[b]
    \centering
    \includegraphics[width=0.49\textwidth]{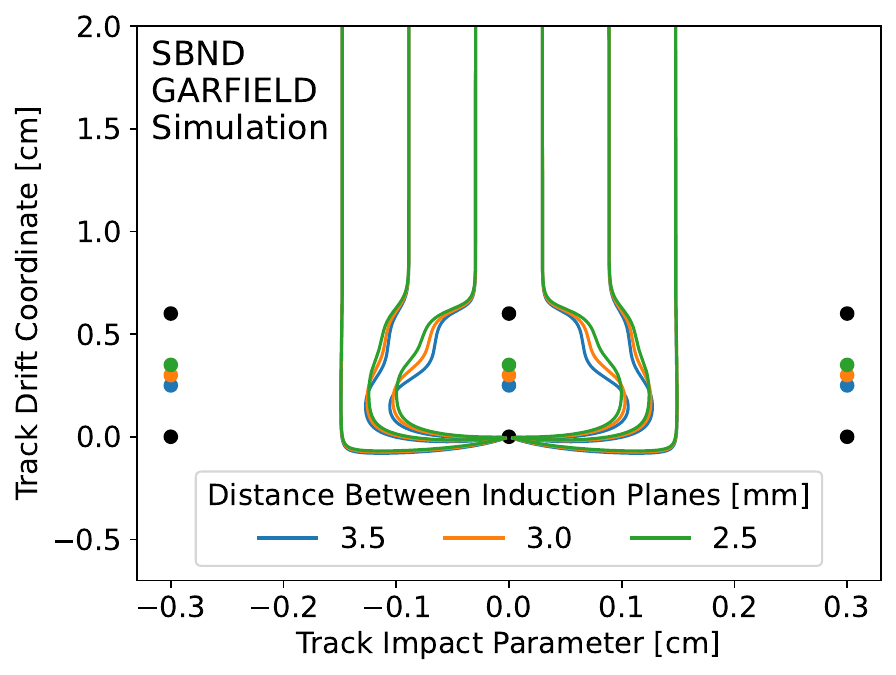}    \includegraphics[width=0.49\textwidth]{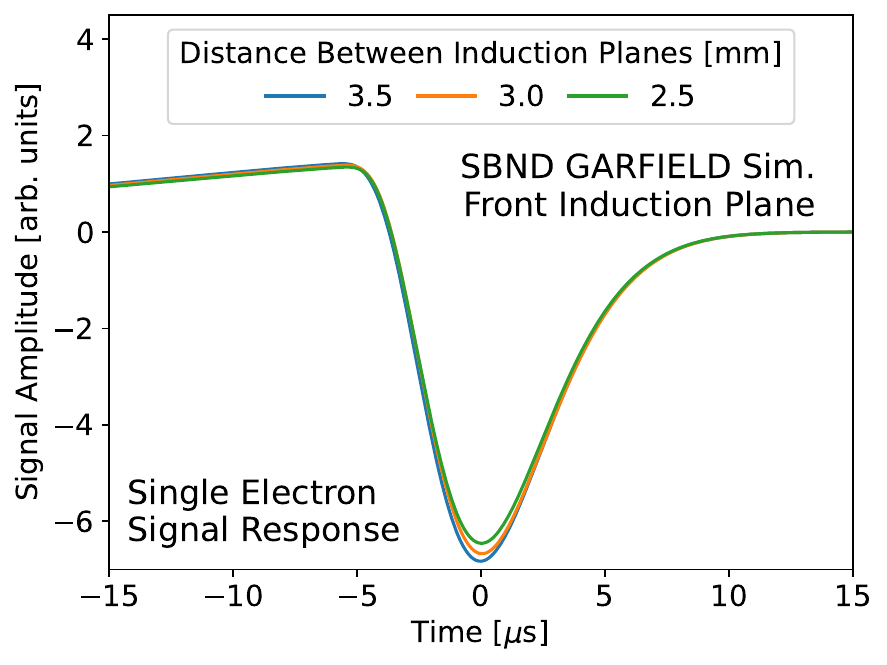}
    \caption{Left: Trajectories of sampled ionization charge paths in the vicinity of the wire planes as simulated by GARFIELD~\cite{Garfield}. Shown for the nominal wire plane spacing (\qty{3}{\milli\meter}, orange), as well as two variations within the specification (\qty{\pm 0.5}{\milli\meter}, blue and green). Solid curves show the trajectories of ionization electrons. Points show the positions of the first induction and collection plane wires (black) as well as the second induction plane wires (colors match corresponding trajectories). Right: Simulated ionization signal shape on the first induction plane. Shown for different spacings between the induction wire planes.}
    \label{fig:garfield}
\end{figure}

To confirm this hypothesis and gain additional insight, GARFIELD~\cite{Garfield} simulations of the SBND wire planes were performed, to calculate the current induced by trajectories of ionization electrons as they traverse the wire planes. The GARFIELD single electron field response (averaged over the ionization pitch) was convolved with the SBND electronics response to obtain the single electron signal response. Figure~\ref{fig:garfield} shows the dependence of the signal response, as well as the individual ionization trajectories, on the spacing between the induction planes. The simulation supports the hypothesis that the spacing of the TPC NE second induction plane is closer to the first induction plane than the collection plane. The GARFIELD simulation indicates that the impact on the front induction plane signal is insensitive to the wire bias voltage, as observed in the detector.  Thus, the non-uniformity from charge herding is a purely geometric effect of the wire planes that will be removed through calibrations of the detector.

While the magnitude of the effect shown in figure~\ref{fig:garfield} is smaller than the charge scale deviation observed in SBND data in figure~\ref{fig:tpc-charge-scale-opt}, the scenarios considered in the study here are only a subset of the possibilities for ways that the wire plane geometry could be different from nominal. This study demonstrates that because the voltage applied to the wire plane is uniform, variations in the spacing of the wire plane result in significant variation in the electric field lines, which in turn impact the induction signal amplitude.
This explanation is qualitatively consistent with our observations of the feature in the NE TPC.

\section{Photon detection system}
\label{sec:PDS}

The SBND photon detection system (PDS) collects vacuum ultra-violet (VUV) scintillation light produced by charged particles interacting with the argon using two active elements, photomultiplier tubes (PMTs) and X-ARAPUCA modules~\cite{Machado:2018rfb} located behind the wires of the TPC anode planes. 
The PMT system is used to trigger on interactions in coincidence with the neutrino beam as described in section~\ref{sec:Trigger}. The X-ARAPUCA system is a SiPM-based light trap device being operated for the first time in a detector placed in a neutrino beam. 
SBND employs two variants of X-ARAPUCA %
providing an R\&D opportunity for the light detection system of DUNE, which will also use this technology~\cite{Paulucci:2019omb}. 
At the TPC cathode, a wavelength-shifting and reflecting passive element is used to increase the uniformity of light collection throughout the detector by reflecting the isotropically emitted scintillation light towards the light collection devices behind the anode planes.

\begin{figure}[hbt]
    \centering
    \includegraphics[width=0.5\linewidth]{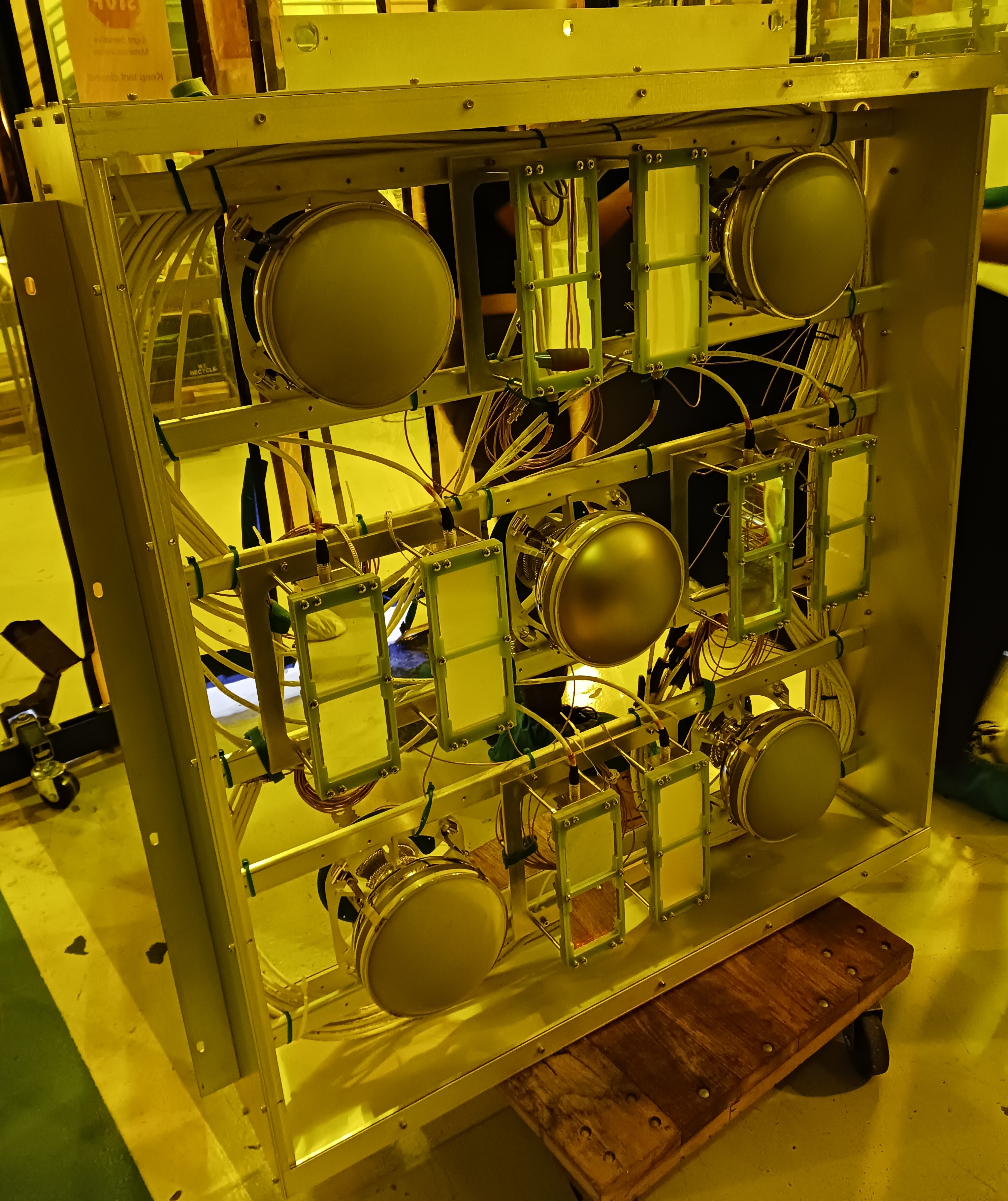}
    \caption{Front view of PDS mounting box. The TPB-coated PMTs are in the top and bottom rows; the single uncoated PMT is located in the center. The VUV sensitive rectangular-shaped X-ARAPUCAs are on the right of each pair while the visible sensitive ones are on the left.}
    \label{fig:PDSmodule}
\end{figure}

The active elements of the PDS are located behind the TPC wire planes in 24 PDS boxes, shown in figure~\ref{fig:PDSmodule}, mounted to the backsides of the APA frames. Each PDS box includes five PMTs in an X shape and eight X-ARAPUCAs that are placed between the PMTs. In total, 24 PDS boxes were installed across the two TPCs (12 per TPC). Each TPC contains in total 60 PMTs and 96 X-ARAPUCA modules integrated within these boxes. The X-ARAPUCAs are not used in the trigger system due to the complexity of different X-ARAPUCA types and the later delivery of the X-ARAPUCA readout systems. 

 The passive PDS element consists of highly reflective plates coated in wavelength-shifter installed on both faces of the CPA. To first order this configuration optically isolates the two TPC volumes from one another. The VUV light that is emitted toward the cathode plane is wavelength shifted and reflected back to the PDS detectors, increasing the overall light yield~\cite{SBND:2024vgn}. As such, both the PMTs and X-ARAPUCAs have been made to be sensitive to the direct VUV light from charged particle interaction with the argon and/or to the reflected visible light.

This is the first time that a photon detection system with dual sensitivity to VUV and visible light has been deployed in a LArTPC in a neutrino beam.

\begin{table}[ht]
    \centering
     \caption{Summary of the main components comprising the SBND photon detection system highlighting important operational characteristics and the three different sensors used for the two different X-ARAPUCA readouts. The channel counts for the X-ARAPUCA systems reflect a 8:1 (32:1) ganging of SiPMs within the APSAIA (ARARA) readout system. The operational voltages for the ARARA have not been determined at the time of writing due to ongoing commissioning.}
    \label{tab:pds_overview}
    \begin{tabular}{c  c  c  c  c}
        \hline \hline
         & \multirow{2}{*}{\textbf{PMT}} & \multicolumn{3}{c}{\textbf{X-ARAPUCA}} \\ \cline{3-5}
         &                      & \textbf{ARARA} & \textbf{APSAIA-HS} & \textbf{APSAIA-VE} \\ \hline \hline
        \makecell{Manufacturer\\Sensor name} & \makecell{Hamamatsu\\R5912-mod2}  & \makecell{SensL\\ MICROFC-30050} & \makecell{Hamamatsu\\S14160-6050-HS} & \makecell{Hamamatsu\\S13360-6050-VE} \\ \hline
        \makecell{Number of\\ channels (sensors)} & 120 (120)& 176 (5632) &8 (64)& 24 (192)  \\ \hline
        %Sensors in series & 1 & 4 ($\times$2) & 4&4  \\ \hline
        \makecell{Operational\\voltage} & 1350-1800\,V & 47-52\,V &46.5-49.7\,V&36.1-41.1\,V  \\ \hline
        \makecell{Coated vs\\uncoated}  & 80\%/20\% & 50\%/50\% &50\%/50\%&50\%/50\% \\  \hline \hline
    \end{tabular}

\end{table}

\subsection{Photomultiplier tube system}

The PMT system consists of 120 Hamamatsu R5912-mod2 PMTs, each with an 8-in hemispherical photocathode and a peak quantum efficiency of about 25\% in the 400–450\,nm range~\cite{sbnd_pmt_spec_sheet}. To enable sensitivity to liquid argon scintillation light, which peaks at 128\,nm, 96 of the PMTs were coated with the wavelength-shifter tetraphenyl butadiene (TPB). The coating was applied by a chemical vapor deposition process in which TPB powder is evaporated and condensed onto the sandblasted glass domes of the PMTs. TPB converts 128\,nm scintillation photons to a wavelength of approximately 425\,nm, aligning with the PMTs’ region of highest efficiency. The remaining 24 PMTs, which were left uncoated, are therefore not directly sensitive to  argon scintillation light but are sensitive to wavelength shifted light from the reflector on the cathode or to Cherenkov light. In  each mounting box, the PMTs in the top and bottom rows are TPB-coated, while the single PMT in the middle row remains uncoated as shown in figure~\ref{fig:PDSmodule}. The AC-coupled PMT base design (shown in figure~\ref{fig:pmt_base} in appendix~\ref{appendixPMT}) is optimized for very linear response up to about 50 photoelectrons (PEs) while maintaining nanosecond-level timing precision. The high level of timing precision will be used to resolve the neutrino beam timing structure as shown in~\cite{SBND:2024vgn}. Based on simulations, the light yield expected at a distance of 2\,m (the TPC cathode) is roughly 12\,PE/MeV for scintillation light and 0.5\,PE/MeV for Cherenkov light. Finally, the PMT domes are rated for a pressure of 700\,kPa; it is expected the deepest PMTs will experience 90\,kPa, well below the safety threshold.

\begin{figure}[h] 
        \centering 
        \includegraphics[width=1\linewidth]{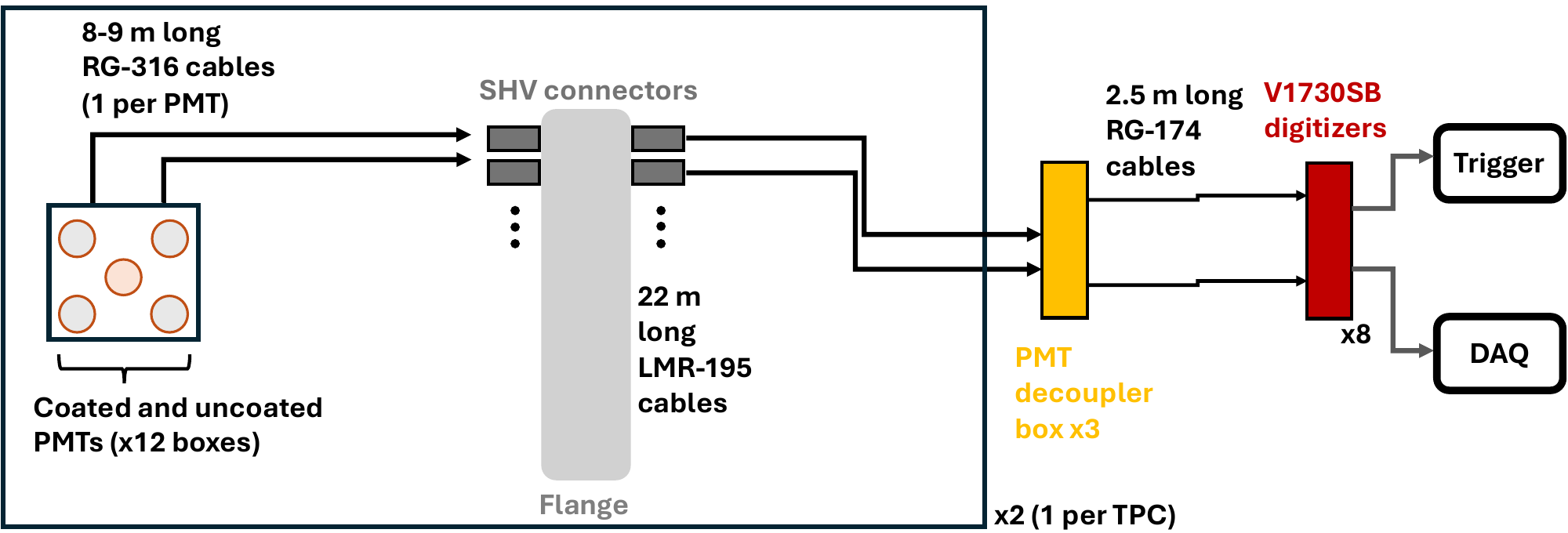} 
        \caption{Signal path diagram of the PMT system.} 
        \label{fig:schematic_pmt_signal} 
\end{figure}

The PMTs are powered by a CAEN SY5527 Power Supply system~\cite{caensy5527} with three CAEN A7030P HV power supplies~\cite{caena7030} containing 48 channels per board. Each power supply has a built-in interlock system to simultaneously power-down all channels in case of emergency. The PMT interlock system is connected to the detector monitoring system from section~\ref{sec:interlocks}. The signal and the bias voltage share the same co-axial cable for each individual PMT and are decoupled using a custom-made decoupling box (circuit shown in figure~\ref{fig:pmt_breakoutbox} in appendix~\ref{appendixPMT}) mounted near the power supply. Approximately 8--9\,m of RG-316 cables run from the PMTs inside the cryostat to the inside of the cryostat flange; approximately 22\,m of LMR-195 cables run from the outside of the flange to the decoupler box for signal processing. The cables were designed to have similar length to facilitate nanosecond timing capabilities, within tolerance. The excess of cable for the upper region of the detector, closer to the flange, is stored inside the PDS boxes. The PMT signals are read out by eight CAEN V1730SB~\cite{caen1730} 16 channel digitizer boards with 14-bits ADC resolution at a 500\,MS/s sampling rate. The CAEN digitizer sends an analog signal to the trigger system (see section~\ref{sec:DAQnTrigger}) that corresponds to how many channel pairs (0-8) have at least one of the two channels in the pair above threshold. The two PMTs at the same Y$-$Z position, one from each TPC, are paired together to send signals to the trigger logic. The full signal path of the PMTs is shown in figure~\ref{fig:schematic_pmt_signal}. All components of the PMT system were successfully tested in the CCM experiment~\cite{PhysRevD.106.012001} before being installed in SBND.

\subsection{X-ARAPUCA system}

\begin{figure}[h] 
    \centering 
    \includegraphics[width=0.55\linewidth]{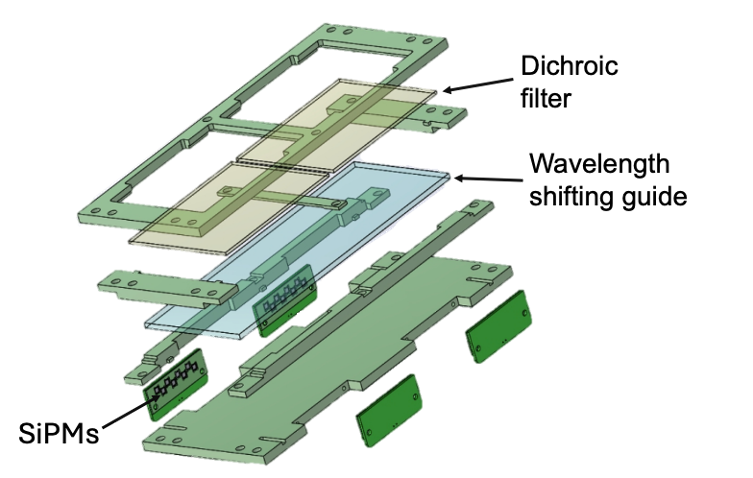}
    \caption{Exploded view of an X-ARAPUCA module highlighting the dichroic filter and wavelength-shifting guide that converts and traps argon scintillation light until they are detected by one of the 4 SiPM readouts in dark green.} 
    \label{fig:sbnd_xarapuca_design} 
\end{figure}

The X-ARAPUCA modules each combine several SiPM sensors with wavelength-shifters, dichroic filters, Vikuiti reflective foil and wavelength-shifting guide bars to trap the incoming scintillation light into the module before being detected by the SiPMs. This results in the effective active area increasing from the few square millimeters of the SiPMs to a total of 145\,cm$^2$\ at the front face of the module. The system is made up of two similar light-trapping modules but utilizing different models of photo-sensor and different amplification systems. Similarly to the PMTs, half of the X-ARAPUCA modules (VUV X-ARAPUCA) are made sensitive to the VUV light using a coating of the wavelength-shifter p-terphenyl (pTP) on the outer layer of the dichroic filter. The choice of pTP was made to match the wavelength cutoff of the filters and the guide bars. The pTP film was evaporated using the same chemical vapor evaporation technique as the PMTs but replacing the substrate and evaporation materials. The other half of the modules (VIS X-ARAPUCA) are made sensitive only to visible light by using filters and wavelength-shifting bars with different adjusted transmittance and reflectance. The VUV X-ARAPUCAs also have some sensitivity to visible light. An exploded view diagram of an X-ARAPUCA module is shown in figure~\ref{fig:sbnd_xarapuca_design}.

\begin{figure}[h] 
    \centering 
    \includegraphics[width=0.5\linewidth]{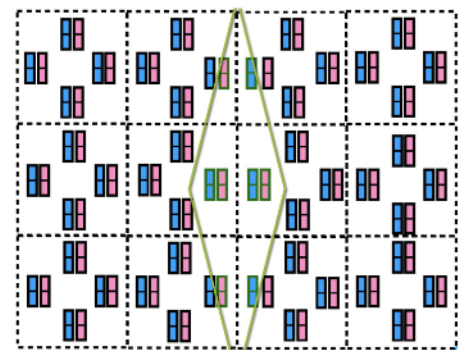} 
    \caption{X-ARAPUCAs placed on a single APA with VUV (pink) and visible (blue) sensitive modules installed in pairs. Dashed lines indicate the boundary of each PDS box. Each APA hosts a total of 96 X-ARAPUCAs, of which 8 are APSAIA-connected X-ARAPUCAs, highlighted within the green diamond.} 
    \label{fig:apa_arapucas} 
\end{figure}

The components of the different X-ARAPUCA module types are detailed in table~\ref{tab:xarapuca_comparison}. The modules are categorized by the amplifier system, located outside the cryostat to which they are matched. The first 16 modules utilize the Arapuca Power Supply And Input Amplifier (APSAIA) system for two different SiPM models from Hamamatsu~\cite{hammaHS, hammaVE} while the remaining 176 modules employ the ARApuca Readout Amplifier (ARARA) system for SensL SiPMs by onsemi~\cite{onsemi} as described in table~\ref{tab:pds_overview}. The manufacturer of the wavelength-shifting light guide bar is also different for the APSAIA (Glass to Power~\cite{glass2power}) and ARARA (Eljen\,\cite{eljen}). The VUV-sensitive X-ARAPUCAs have a 400\,nm cutoff wavelength dichroic filter manufactured by OPTO~\cite{opto} with a pTP-coated surface and a \textit{blue}-colored wavelength-shifting bar. The X-ARAPUCAs designed for visible light employ a 450\,nm cutoff dichroic filter, also from OPTO, and incorporate the \textit{green}-colored wavelength-shifting bar instead. The X-ARAPUCAs are installed in pairs with one VUV and one visible light sensitive in the spaces between the PMTs in each PDS box as shown in figure~\ref{fig:apa_arapucas}.

\begin{table}[ht]
    \centering
    \caption{Comparison between the different components of the two types of X-ARAPUCA module for both VUV and visible (VIS) flavors.}
    \label{tab:xarapuca_comparison}
    \begin{tabular}{c  c  c  c  c}
        \hline \hline
         & \multicolumn{4}{c}{\textbf{X-ARAPUCA}} \\ \cline{2-5}
          & \multicolumn{2}{c}{\textbf{ARARA}} & \multicolumn{2}{c}{\textbf{APSAIA}} \\ \hline \hline % \cline{1-5}
         %Sensor name & \multicolumn{2}{c}{3x3\,mm$^2$ MICROFC-30050 (x8)} & \multicolumn{2}{c}{\makecell{6x6\,mm$^2$ S14160-6050-HS (x4)\\6x6\,mm$^2$ S13360-6050-VE (x4)}} \\ \hline
         Type & VUV & VIS & VUV & VIS \\ \hline 
         Dichroic filter cut off& OPTO 400\,nm & OPTO 450\,nm & OPTO 400\,nm & OPTO 450\,nm \\\hline
        wavelength-shifter & pTP & none & pTP & none  \\ \hline
        Light guide bar & \makecell{EJ286\\(blue)} & \makecell{EJ280\\(green)} & \makecell{Glass to Power\\(blue)} & \makecell{Glass to Power\\(green)} \\
        Number Installed & 88 & 88 & 8 & 8 \\ \hline \hline
    \end{tabular}

\end{table}

\begin{figure}[h] 
        \centering 
        \includegraphics[width=.90\linewidth]{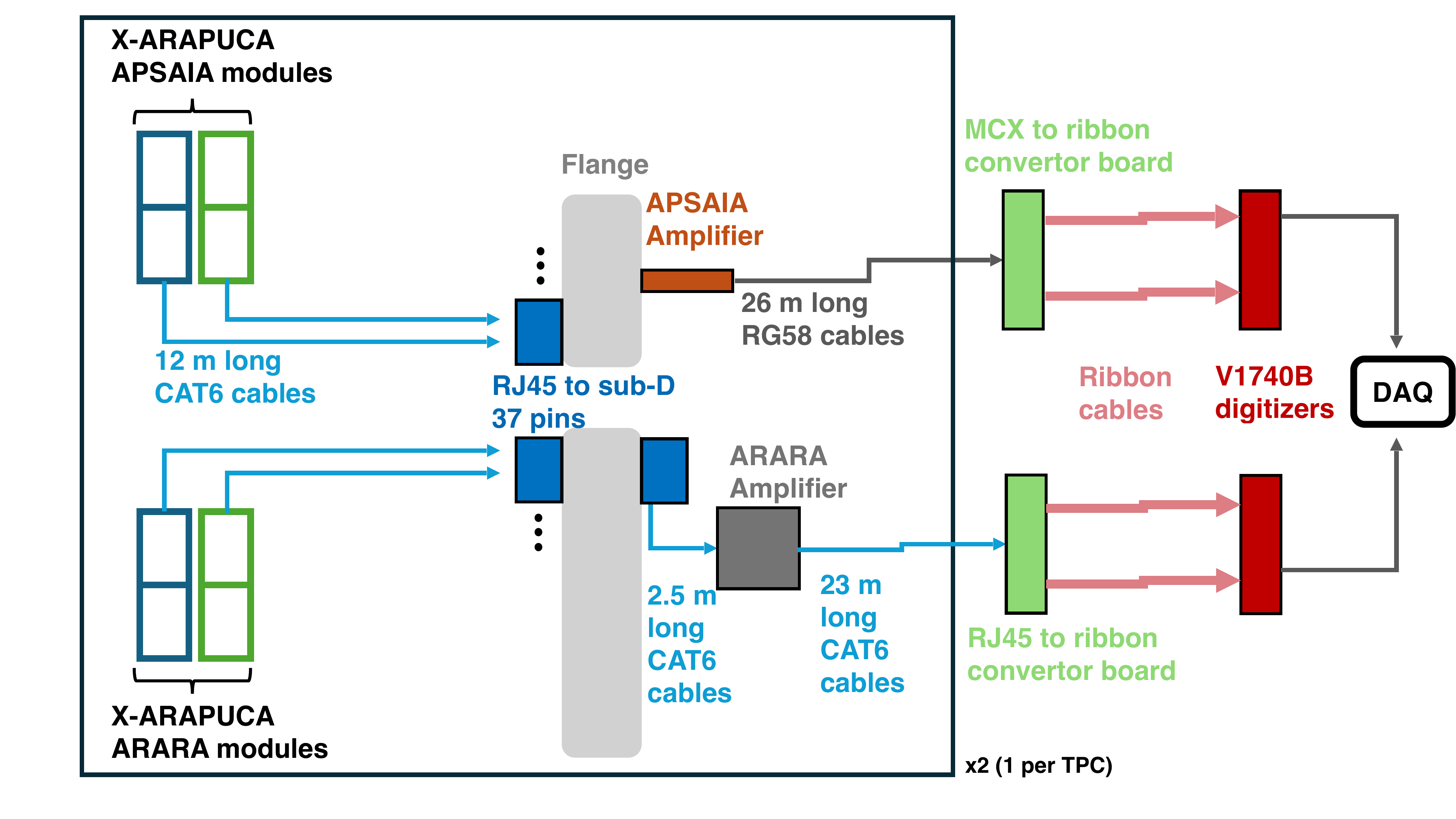} 
        \caption{Signal path diagram of the X-ARAPUCA system. The power cables are not shown but follow the same path to the amplifier from the power source located close by the ribbon convertor boards.} 
        \label{fig:schematic_xarapuca_signal} 
\end{figure}

\begin{figure}[h] 
        \centering 
        \includegraphics[width=0.5\linewidth]{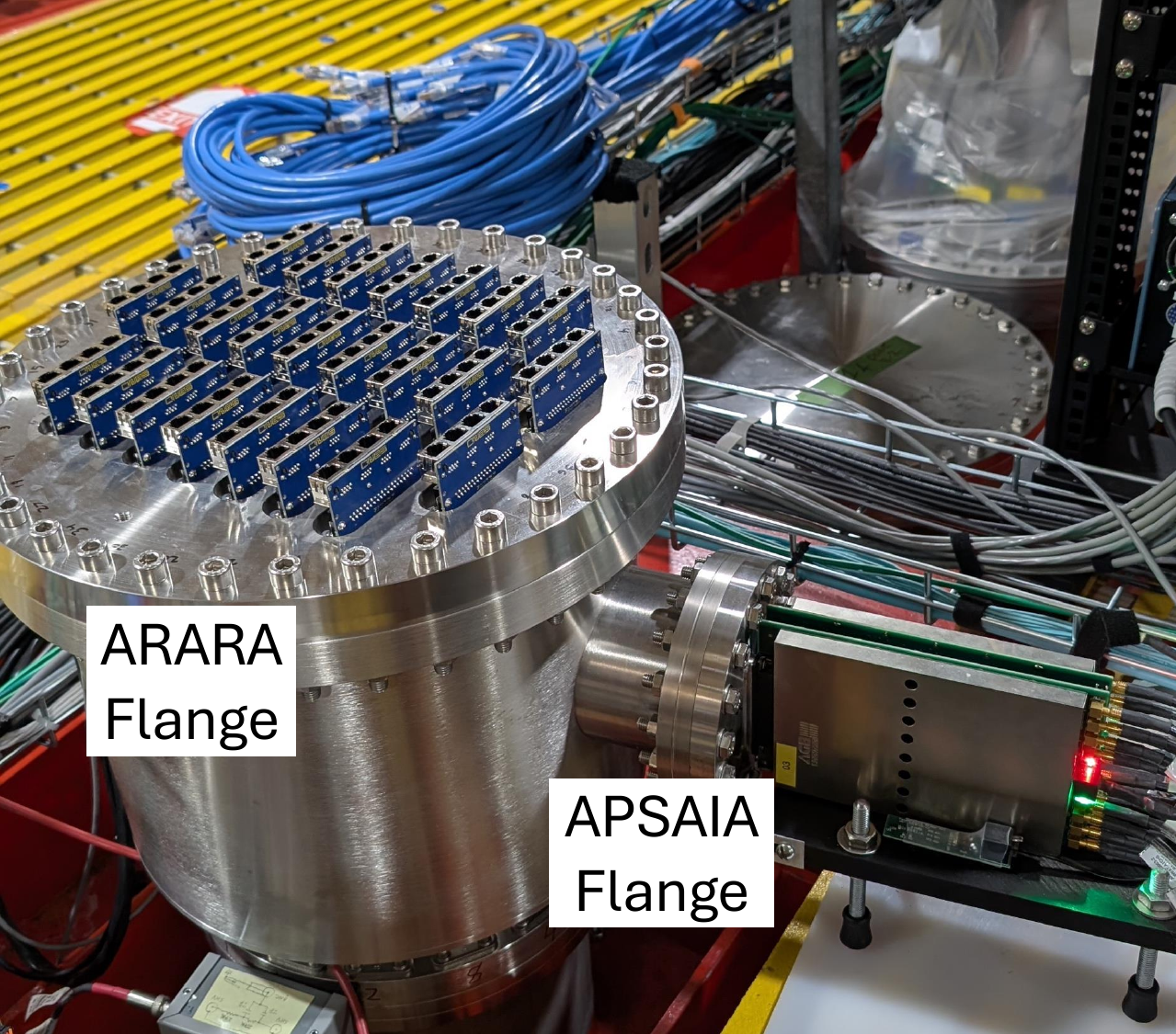} 
        \caption{The west X-ARAPUCA feedthrough during installation. The smaller APSAIA flange is fully populated with amplifiers while the cables from the ARARA flange are not yet populated. Both flanges use sub-D 37-pin connectors on both the cold and warm sides with embedded printed circuit boards ensuring a gas seal.} 
        \label{fig:xarapuca_flange} 
\end{figure}

The readout and powering of APSAIA and ARARA modules use effectively the same design but with somewhat different component choices. 
All X-ARAPUCAs have their SiPMs connected to 12\,m CAT6 cables from the module to the cryostat flange where a sub-D 37 to RJ45 custom connector card receives up to four CAT6 cables, one for each X-ARAPUCA module. As with the PMTs, cables were cut to the same length to equalize time delays. Signals are amplified using two different custom-built amplifiers (one for APSAIA and one for ARARA), mounted on or close to the warm side of the respective cryostat flange as shown in figure~\ref{fig:schematic_xarapuca_signal}. Amplified signals are digitized by CAEN V1740B digitizers\,\cite{caen1740}, featuring 64 input channels with 12-bit ADC resolution and operating at a sampling rate of 62.5\,MS/s. The APSAIA populates a half of a single V1740B (32 out of 64 available channels) while the ARARA populates almost three V1740B (176 out of 192 available channels). All digitizers are read out by the common DAQ after being triggered by the PMT trigger (see section~\ref{sec:Trigger}) which sends a signal to the digitizers to record and save the waveforms. The length of ribbon cables used for input to the digitizers is minimized by using converter boards mounted near the digitizers with higher quality cables used between the amplifier and converters as shown in figure~\ref{fig:schematic_xarapuca_signal}. 
%The full signal path of the X-ARAPUCAs is shown in figure~\ref{fig:schematic_xarapuca_signal}. 
A photograph of the cryostat mounted X-ARAPUCA feedthrough with the APSAIA and ARARA flanges is shown in figure~\ref{fig:xarapuca_flange}.

The two different systems, APSAIA and ARARA, each provide both the amplification of signals from the photodetectors and the voltage required to correctly bias the SiPMs in independent circuitry. In both systems the bias voltage is controllable separately for each amplifier channel using a micro-controller which is interfaced using the RS-232 protocol providing 10\,mV and 100\,mV steps over the required bias voltage range for the ARARA and APSAIA, respectively. The following sections will go over the main hardware differences between the APSAIA and ARARA X-ARAPUCAs.

\subsubsection{APSAIA readout}

\begin{figure}
    \centering
    \includegraphics[width=0.49\linewidth]{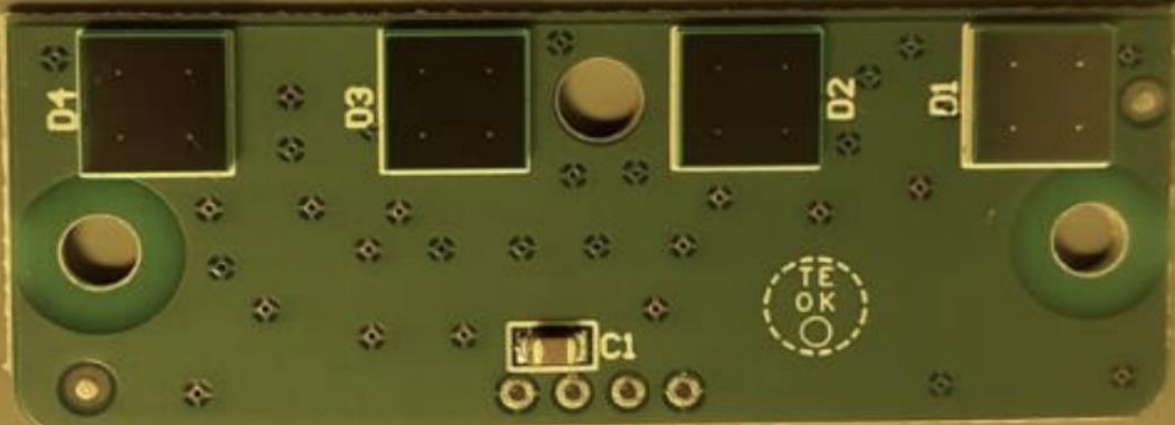}
    \includegraphics[width=0.425\linewidth]{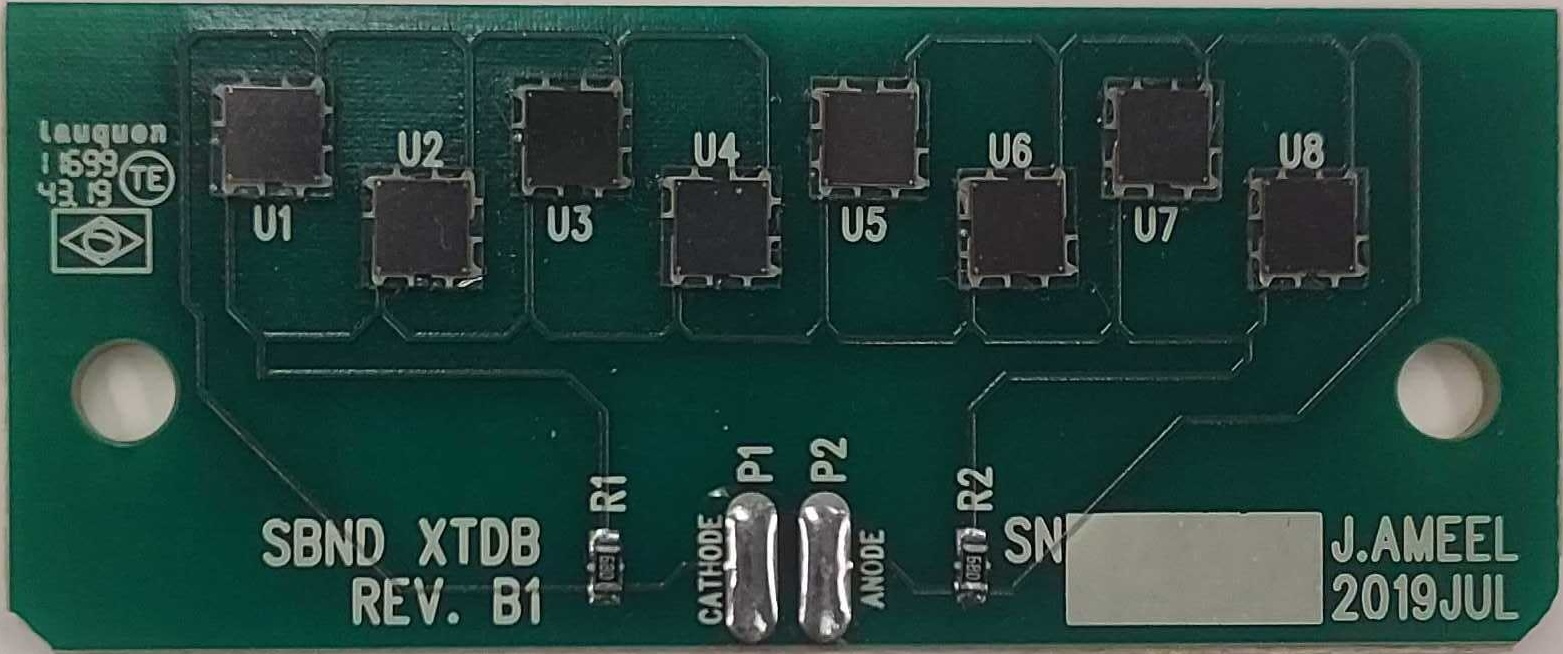}
    \includegraphics[width=0.49\linewidth]{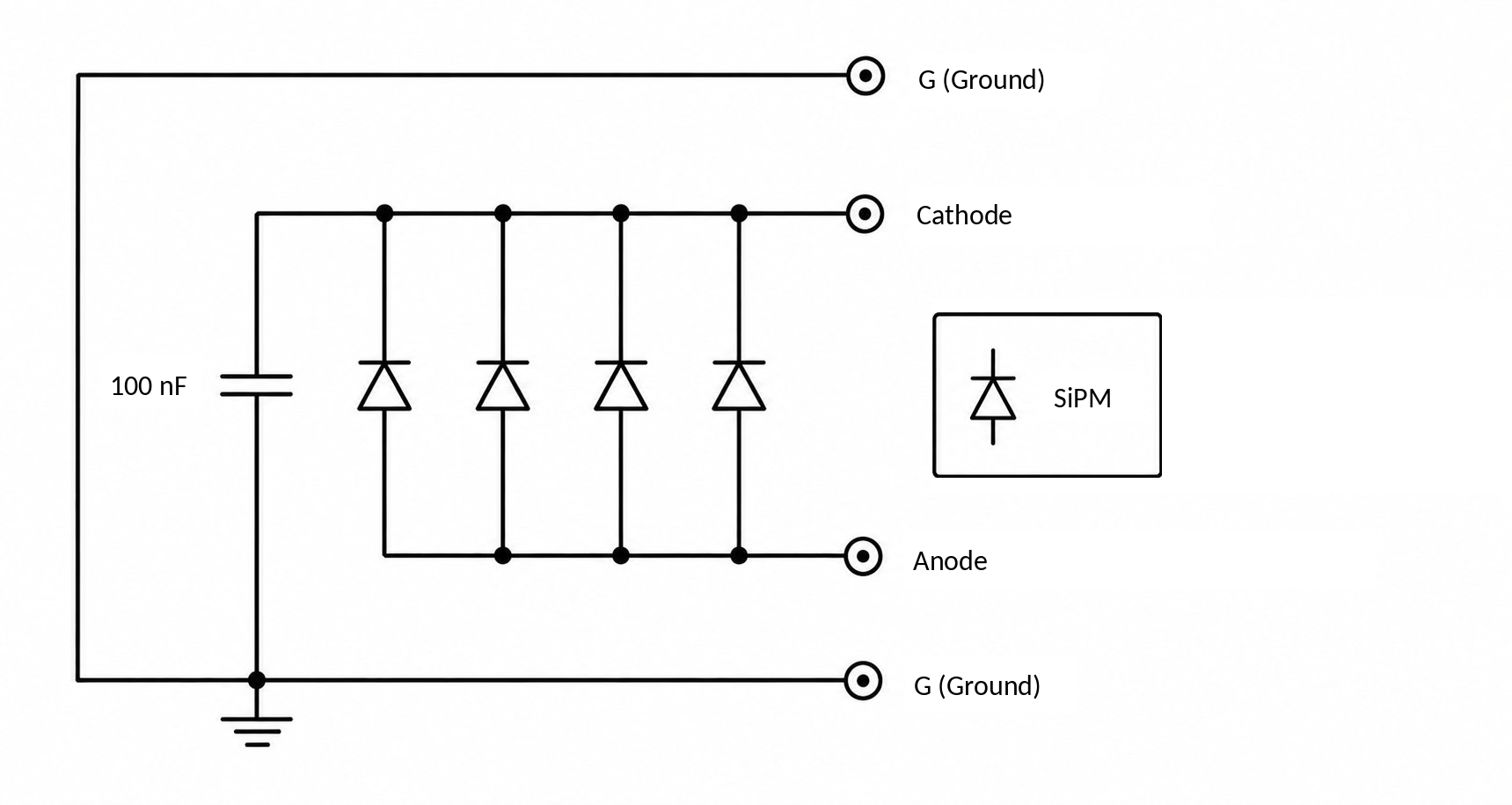}
    \includegraphics[width=0.49\linewidth]{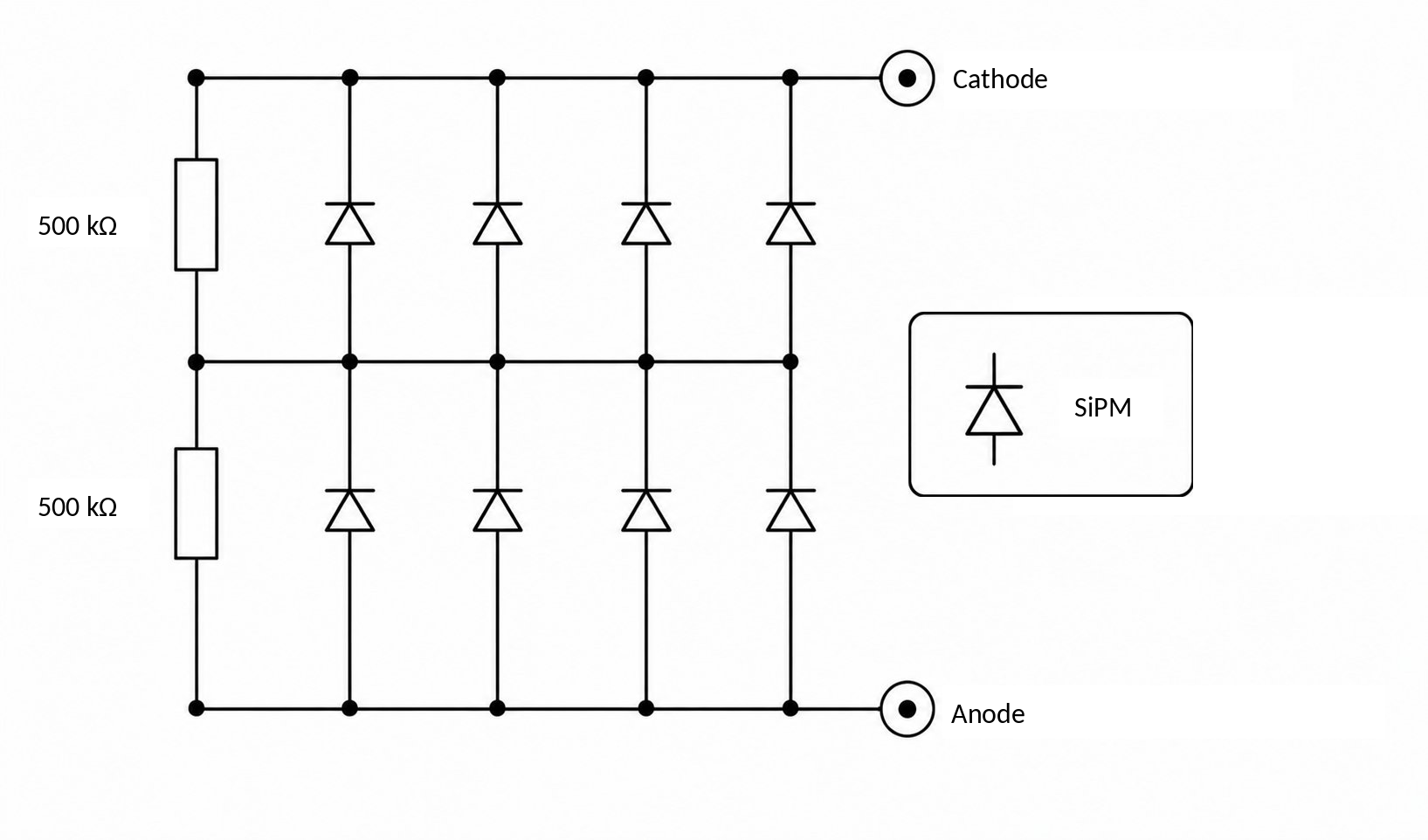}
    \caption{The APSAIA (left) and ARARA (right) SiPM boards and their corresponding circuit diagram illustrating the ganging scheme.}
    \label{fig:xarapuca_sipm_boards}
\end{figure}

The 16 (8 per TPC volume) APSAIA X-ARAPUCA modules are located in the middle of the photon detection plane as schematically shown by the green diamond in figure~\ref{fig:apa_arapucas}. The 8:1 ganging of SiPMs (sensors to channel) described in table~\ref{tab:pds_overview} is implemented within the APSAIA X-ARAPUCA modules themselves. The ganging in the X-ARAPUCA is implemented with
four circuit boards in parallel, with each board connecting two sets of 4 SiPMs in series to form the four outputs per module. One of these SiPM boards is shown in figure~\ref{fig:xarapuca_sipm_boards} (left).

On the external side of the flange, two APSAIA amplifier boards are plugged directly into the sub-D 37-pins as shown in figure~\ref{fig:xarapuca_flange}. Each amplifier board has 8 amplifier inputs, totaling 32 channels for both TPC volumes, which matches the number of signal outputs from the 16 APSAIA-connected X-ARAPUCAs with two outputs per module. This number of signal output matches the number of bias voltages provided to each module resulting in two independent biases per module split into directly facing SiPM boards from figure~\ref{fig:xarapuca_sipm_boards} (left) putting, in total, 8 SiPMs at the same voltage.

The APSAIA amplifier gain can be remotely adjustable via the microcontroller with a low or high gain, which is 20 or 40 (13.01\,dB or 16.02\,dB), respectively. 
The APSAIA is designed for signals with rise times as short as 30\,ns.% It 

\subsubsection{ARARA readout}
The 176 (88 per TPC volume) ARARA X-ARAPUCA modules are read out using the custom-built ARARA amplifiers readout system whose design was just completed at the time of writing. Unlike the APSAIA modules, the ganging of SiPMs for ARARA SiPMs described in table~\ref{tab:pds_overview} is implemented in two stages: an 8:1 ganging within the X-ARAPUCA module and a 4:1 ganging at the input of the ARARA amplifier board. The ganging in the X-ARAPUCA is implemented with four circuit boards each connecting two sets of 4 SiPMs in parallel with the sets then connected in series to form the four outputs per module. One of these SiPM boards showing all 8 sensors is shown in figure~\ref{fig:xarapuca_sipm_boards} (right). For each X-ARAPUCA module, the 4 indiviual outputs are passed through the amplifier where they are ganged into a single output that is then amplified. The ARARA modules have a single amplified channel per module unlike the two per module in APSAIAs. This results in a singular bias voltage per module for all four SiPM boards from figure~\ref{fig:xarapuca_sipm_boards} (right) which puts, in total, 32 SiPMs at the same voltage. This two stage ganging allows for the possibility of a finer granularity readout to be implemented in the future to read out up to four independent channels per X-ARAPUCA module.  
 
On the warm side of the feedthrough flange, short 2.5\,m CAT6 cables transmit the signals from the flange into the ARARA amplifier boards located above the flanges. Each amplifier board has 16 inputs matched to 16 different X-ARAPUCA module cables receiving in total 64 individual channels. The amplifier was designed to process signals with rise times of approximately 50\,ns and provides a gain of 140 through a transimpedance amplification stage.

\subsection{Wavelength-shifting reflector system}
\label{sec:PSDFoilSystem}

\begin{figure}
    \centering
    \includegraphics[width=0.317\linewidth]{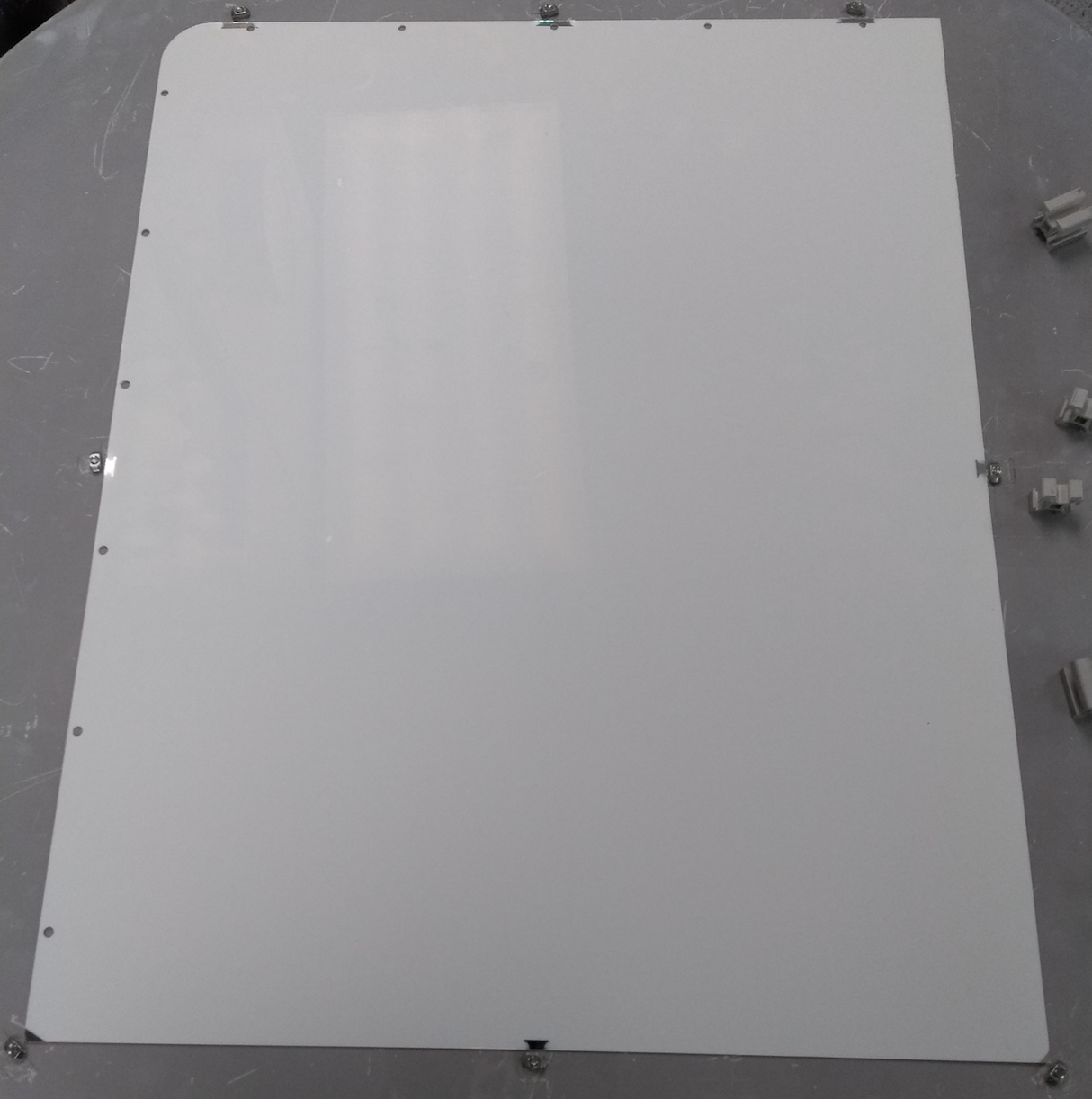}
    \includegraphics[width=0.426\linewidth]{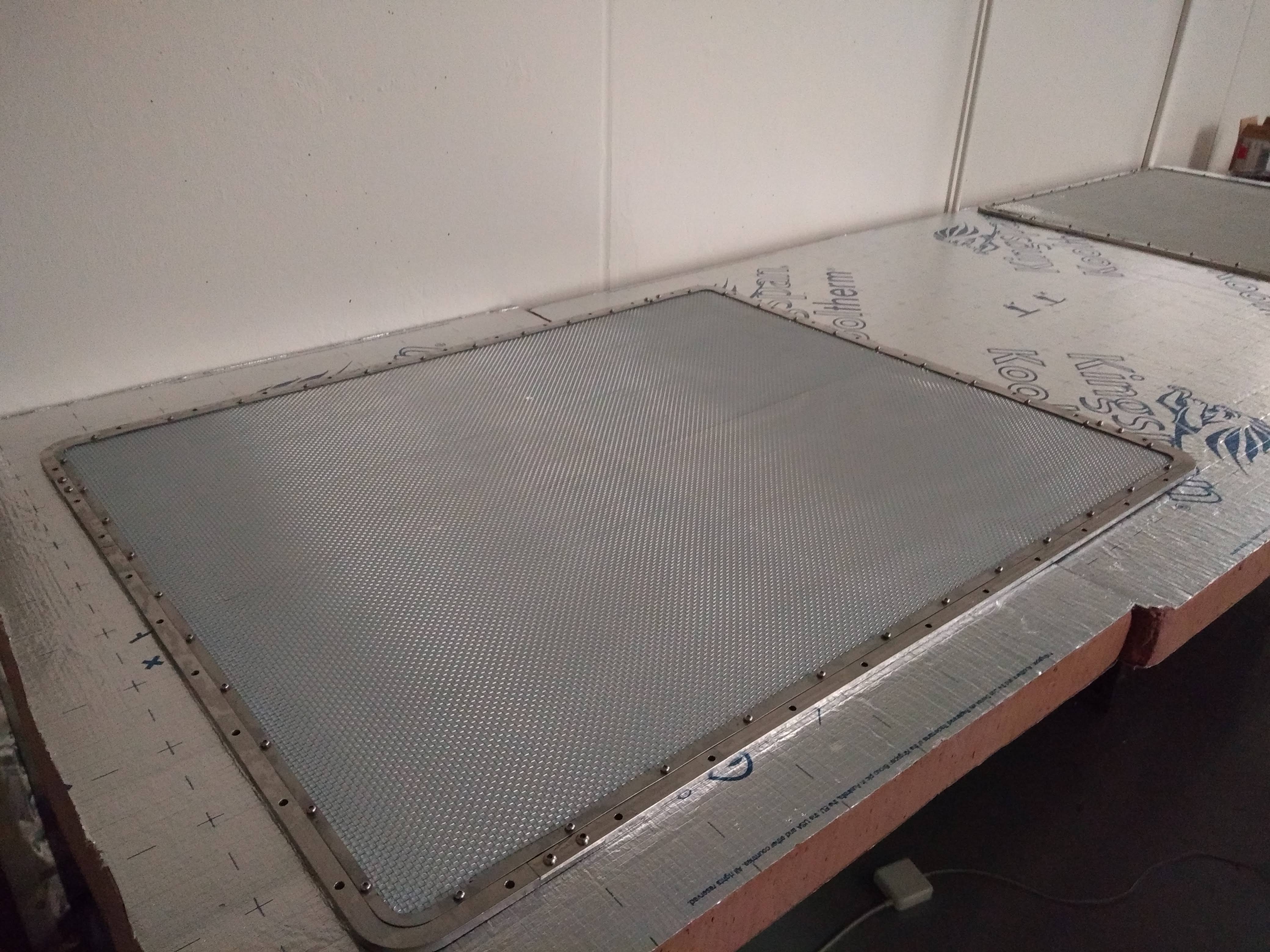}
    \includegraphics[width=0.240\linewidth]{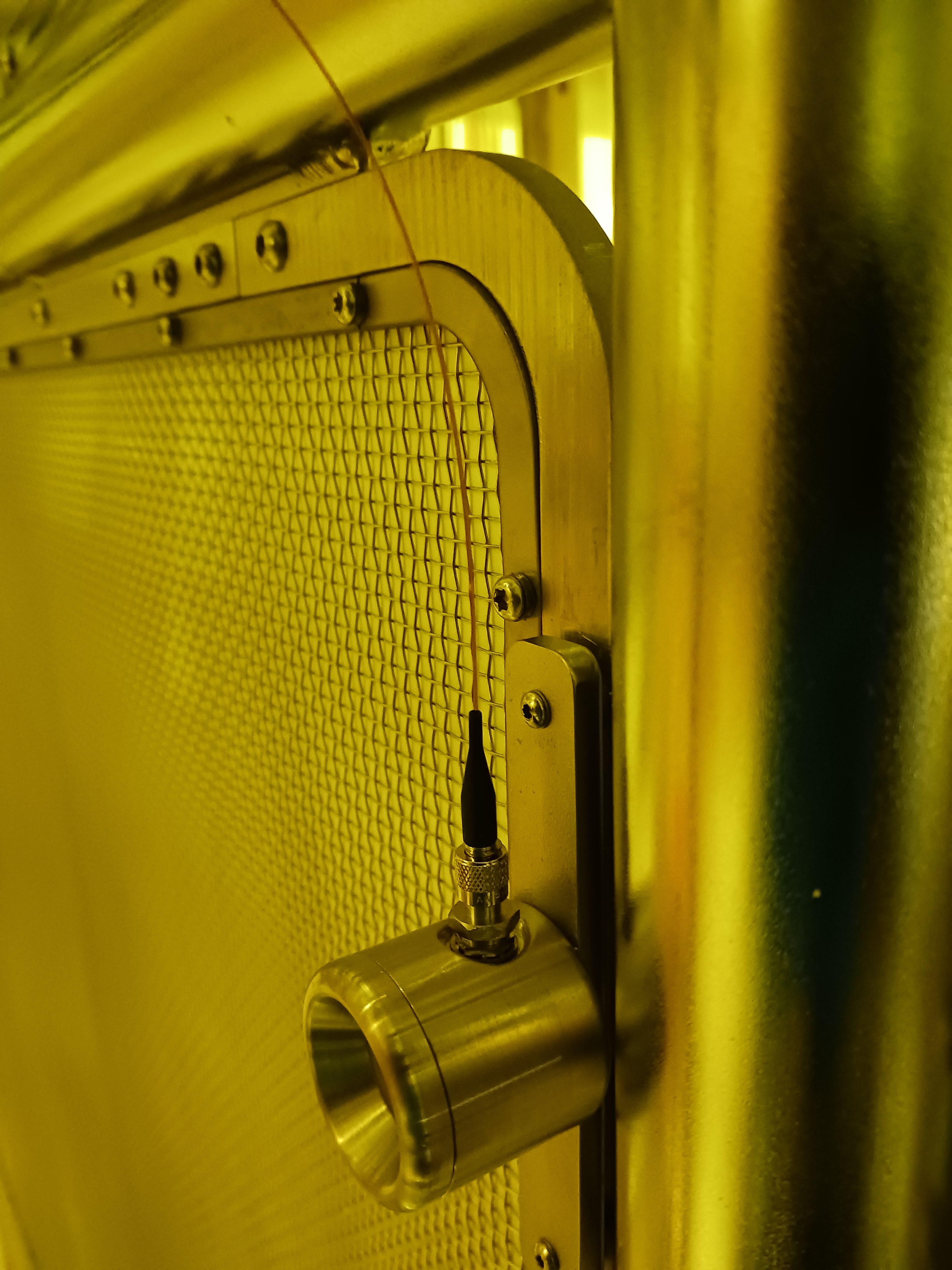}
    \caption{Left: a TPB-coated reflective plate. Center: four plates fully assembled in the CPA mesh assembly. Right: diffuser system after installation on the CPA mounted to the mesh assembly.}
    \label{fig:tpb_plate}
\end{figure}

Double-sided TPB-coated highly reflective FR4 plates laminated with DF2000MA~\cite{df2000} are sandwiched in-between the CPA mesh comprising the passive element of SBND's PDS. Both sides of the plates face the active PDS components (PMTs and X-ARAPUCAs) from 2\,m away. A photograph of a TPB-coated plate is shown in  figure~\ref{fig:tpb_plate}~(left). Each of the 16 CPA windows from figures~\ref{fig:CPADrawing} and \ref{fig:CPADPhoto} is composed of four plates and held down in place by long washer-bars screwed into the mesh assembly frames. A photograph of a fully assembled mesh assembly is shown in figure~\ref{fig:tpb_plate}~(center). A more detailed description of the production, assembly, and installation is available in a dedicated publication~\cite{sbnd_wls_plates}. 

The goal of the coated reflective plates is to recover the scintillation light produced and traveling in the opposite direction of the active elements of the PDS. Light traveling towards the CPA will be absorbed by the TPB layer then re-emitted isotropically and reflected back towards the active elements of the PDS. This leads to an overall enhancement in the amount of light detected throughout the drift region, while also improving the spatial uniformity of the response. A more complete description of the physics advantages of the reflector system in SBND can be found at~\cite{SBND:2024vgn}.

\subsection{Calibration system}
\label{sec:PDSCalibration}
An LED calibration device with a wavelength of 470\,nm, driven by a pulse generator, serves as the light source for PDS timing calibration. A Thorlabs fiber-coupled LED driver (M470F4 at 470\,nm wavelength)~\cite{led_pds} is connected via optical fiber through the PMT flange where fibers were installed in the TPC. These are connected to five UV fused-silica ground glass (220 grit) diffusers~\cite{pds_diffuser} mounted on each side of the CPA to send light to the PDS. The LED driver is stimulated by a function generator which also provides input to the trigger for data acquisition. A photograph of the diffusers components installed on the CPA is shown in figure~\ref{fig:tpb_plate}~(right). A total of 10 diffusers, 5 for each TPC, were installed to maximize the coverage of the whole PDS wall.

The calibration will be performed by utilizing the time of the PDS signal pulse relative to the LED pulse provided by the function generator and the geometry of the detector. A pulse width of 1\,$\mu$s enables the calibration of the signal propagation delay through the full readout path to $\mathcal{O}(\textrm{ns})$. A first in situ test was successfully performed by injecting light into one of the diffuser fiber ports with PMT signals captured on an oscilloscope.

\subsection{PDS performance}
\subsubsection{Initial powering up}

The PMTs were first ramped up after the detector was filled with argon in March 2024. When initially powered, 112 out of the 120 PMTs were able to maintain their nominal voltage without current instability resulting in the PMT power supply tripping. Within a few months of the initial start up, 7 additional PMTs were powered off due to an inability to maintain their operational voltage without tripping their power supply current limit. Based on experience with the same equipment (PMTs, bases, and cryostat flanges) in the CCM experiment~\cite{PhysRevD.106.012001}, we suspect micro-discharges result from humidity originating at the warm and cold connected interfaces of the SHV connectors that leaks through the seal and freezes inside the connector. To prevent further humidity leakage at the connectors, the two PMT flanges where the cold and warm connection meet were enclosed in plastic sheets with a continuous flow of dry air purging the area. Since the addition of the dry air environment in late summer 2024, only one PMT with unstable baseline was powered off in March 2025 resulting in a final tally of 104 PMTs operating stably. 

For the X-ARAPUCA system, three of the APSAIA X-ARAPUCA channels using the Hamamatsu S13360-6050-VE SiPM succumbed to connection failure during the filling of the detector due to component failure in one case and a possible disconnection at the cold-warm interface for the other two. The ARARA X-ARAPUCAs have all passed their QA/QC tests and will be powered in 2026 after the ARARA electronics are installed.  A summary of the nominal number of PDS channels and the number that were in operation at the end of Run\,1 is shown in table~\ref{tab:pds_number_on_off}. The ARARA readout for the remaining X-ARAPUCAs was installed in 2026.

\begin{table}[ht]
    \centering
    \caption{Summary of the number of the PDS channels that are currently powered on compared to the nominal number and the number that were first powered on during commissioning. For the ARARA X-ARAPUCAs, all 704 channels (pre-ganging) have passed the QA/QC but were not powered on during Run\,1.}
    \label{tab:pds_number_on_off} 
    \begin{tabular}{c  c  c  c}
        \hline \hline
         & \multirow{2}{*}{PMT} & \multicolumn{2}{c}{X-ARAPUCA} \\
         &                      & ARARA & APSAIA (HS \& VE) \\ \hline \hline
        Nominal & 120 & 704 & 8 \& 24 \\ \hline
        First power on & 112 & - & 8 \& 21  \\\hline
        Operational & 104 & - & 8 \& 21  \\ \hline \hline
    \end{tabular}

\end{table}

\subsubsection{Photon detection system gains}

For both the PMT and operational X-ARAPUCA systems, gain characterization was performed. The PMT signals in SBND provide essential input to the trigger, and accurate knowledge of the PMT gains is critical for operations.  SBND aims to maintain each PMT at a gain of approximately $5 \times 10^{6}$. 

Conveniently, surface LArTPCs exhibit a high rate of naturally occurring SPE pulses, on the order of $100\,\textrm{kHz})$ per PMT~\cite{microbooneSPE}, which serve as an excellent \textit{in situ} calibration source. These random SPEs enable extraction of the effective gain (i.e. the ADC response per photoelectron) directly from routine experiment data taking without requiring additional instrumentation or dedicated runs. The SPEs can appear either randomly in PMT waveforms without a main beam/cosmic interaction happening or at the tail of a beam/cosmic interaction if they are late enough relative to the main light peak.

\begin{figure}[tbp]
    \centering
    \includegraphics[width=0.585\textwidth]{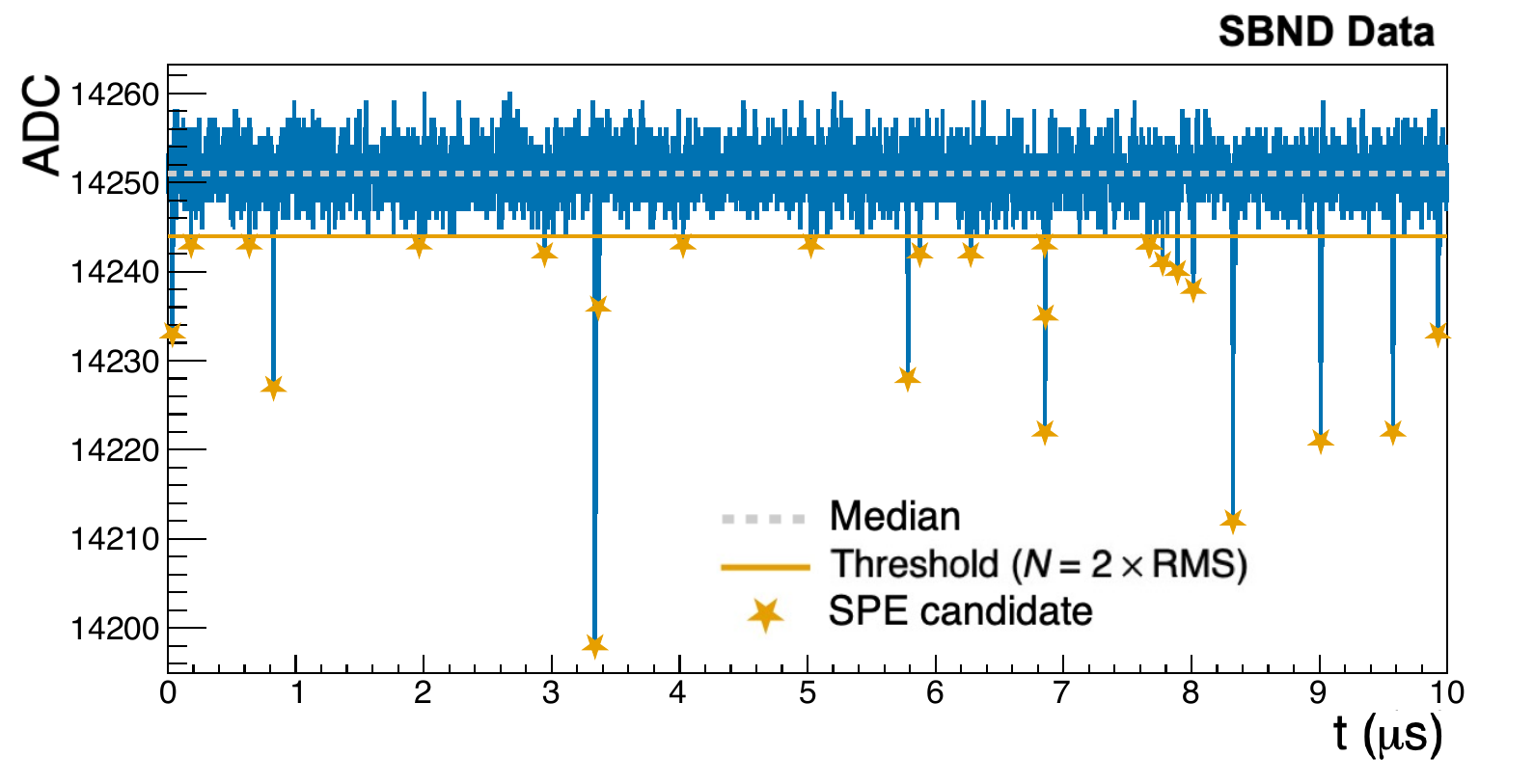}
    \includegraphics[width=0.405\textwidth]{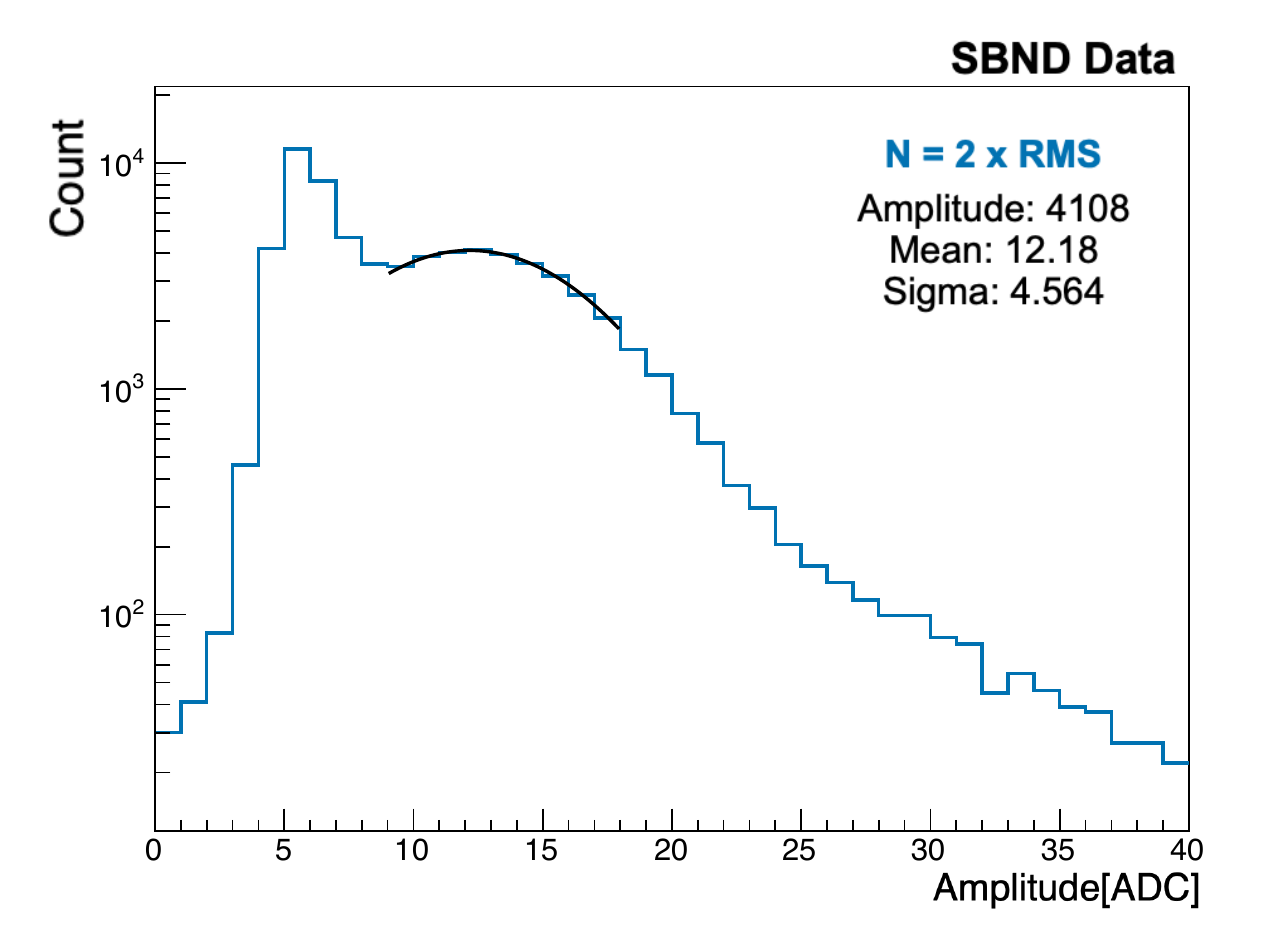}
    \caption{Left: Typical PMT recorded waveform used to find SPE pulses (indicated by yellow stars) passing the set threshold (yellow line). Right: the accumulation of the SPEs to fit the amplitude gain.}
    \label{fig:pmt_spe_plots}
\end{figure}

\begin{figure}[!htbp]
    \centering
    \includegraphics[width=0.70\textwidth]{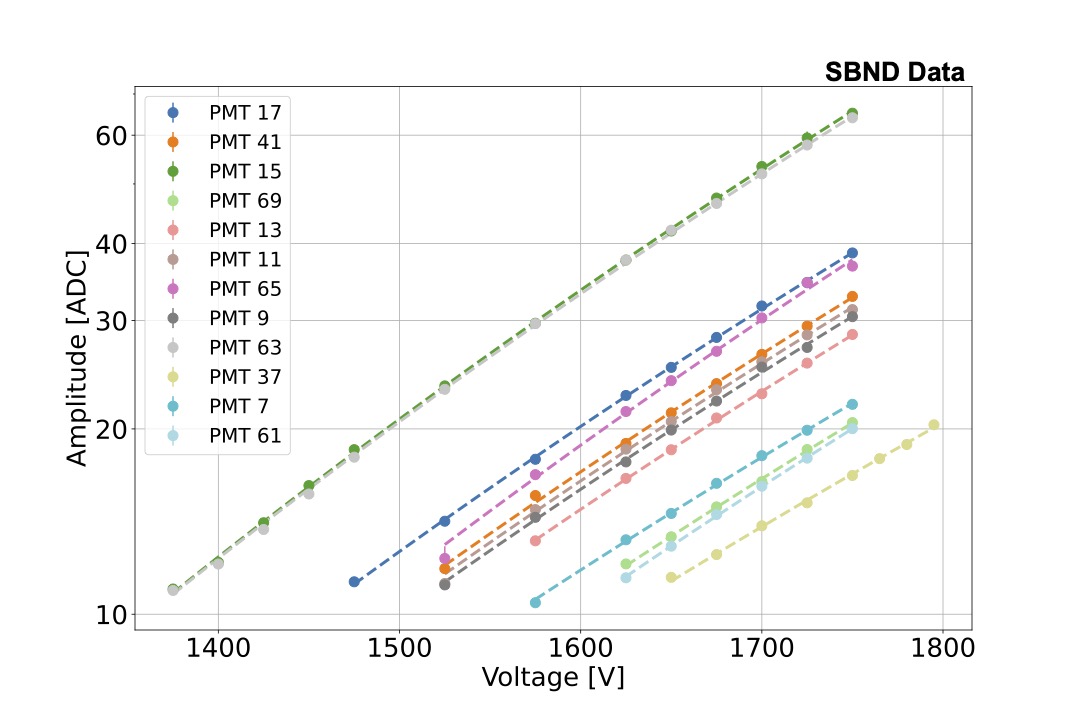}
    \caption{SBND PMT effective gains using the amplitudes of the SPE signals at different high voltage values for a standard set of 3 PMT boxes with 12 operational PMTs.}
    \label{fig:pmt_gain_plots}
\end{figure}

An algorithm was developed to identify these SPEs by applying a threshold to the height of the negative-polarity PMT pulses. The threshold was defined as the baseline minus \textit{N} times the waveform RMS.
This efficient method of using the pulse height to calculate the gain was chosen because the AC-coupling of the PMTs, along with their inherent undershoot following the main pulse, makes calculation of the full pulse integral less reliable without a full understanding of the single electron response which is being worked on at the timing of writing. As shown in figure~\ref{fig:pmt_spe_plots}~(left) all the light signals passing the threshold identified by yellow star markers have their amplitudes saved as SPE candidates. This algorithm was applied to \textgreater 10000 individual waveforms per PMT to produce plots of the SPE amplitudes like that shown in figure~\ref{fig:pmt_spe_plots}~(right). A Gaussian is fitted to the SPE peak which is seen to the right of the pedestal peak centered around 7\,ADC. Since the threshold is applied on a waveform basis, no threshold cutoff is seen as the RMS of some waveforms were lower thus reducing the effective threshold. A choice of $N=2$ was chosen as a middle ground to reduce pulses with ADCs lower than the expected SPE while keeping a visible separation between the pedestal and SPE peaks. This process was repeated for all of the individual operational PMTs. To understand how the gain changes according to the high voltage applied on each PMT, dedicated runs at different high voltages were taken during the commissioning phase. The results of the effective gain measured using the amplitude of the SPEs at different high voltages for the PMTs are shown in figure~\ref{fig:pmt_gain_plots} for a typical three PDS boxes worth of PMTs. While not as accurate as an integral fitting technique, this method allowed for a reproducible gain setting that could be easily applied during the detector commissioning process, enabling commissioning of light based triggers. The same process is used to monitor gain changes over time. In most cases, the PMT gains were obtained for a wide range of high voltage that allows a fit to be made to determine the operating high voltage values at the nominal gain value of $5 \times 10^6$ equivalent to the SPE amplitude of $\approx$12.5\,ADC/PE. During the initial gain setting, the gain could only be measured for a few (higher) voltage settings for five of the PMTs. For these PMTs, the desired nominal gain could not be achieved within the upper limit of the base input voltage (${\sim}1800$\,V) which is driven by the power limitation in the PMT base design of 1\,W. These five PMTs remain powered at the maximum voltage but with below nominal gains. 

\begin{figure}[!htbp]
    \centering
    \includegraphics[width=0.56\textwidth]{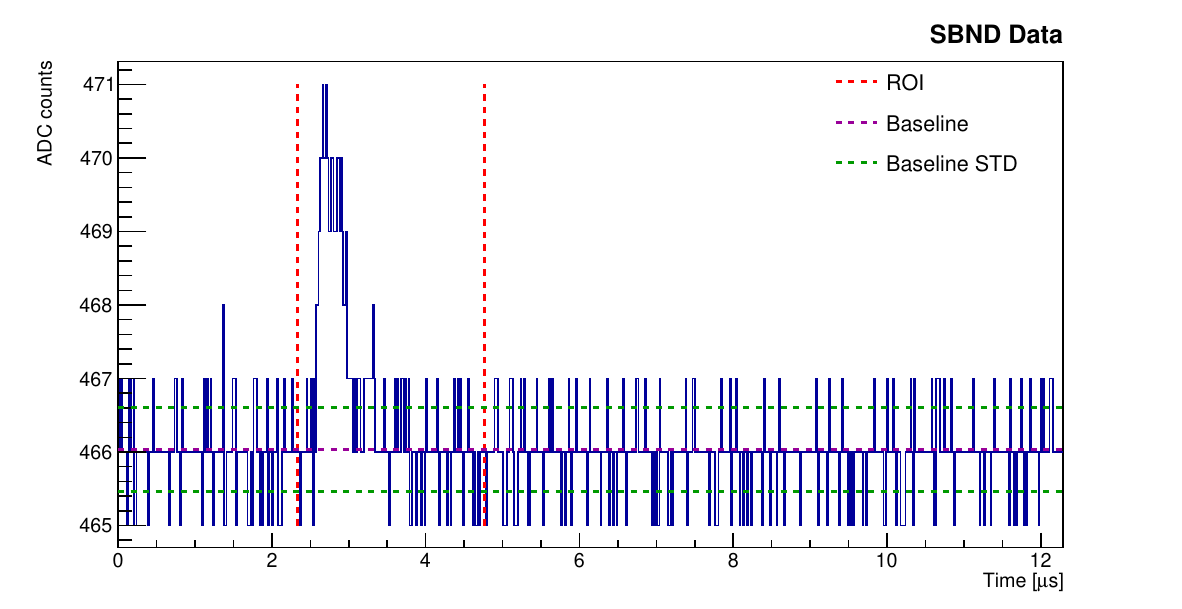}
    \includegraphics[width=0.42\textwidth]{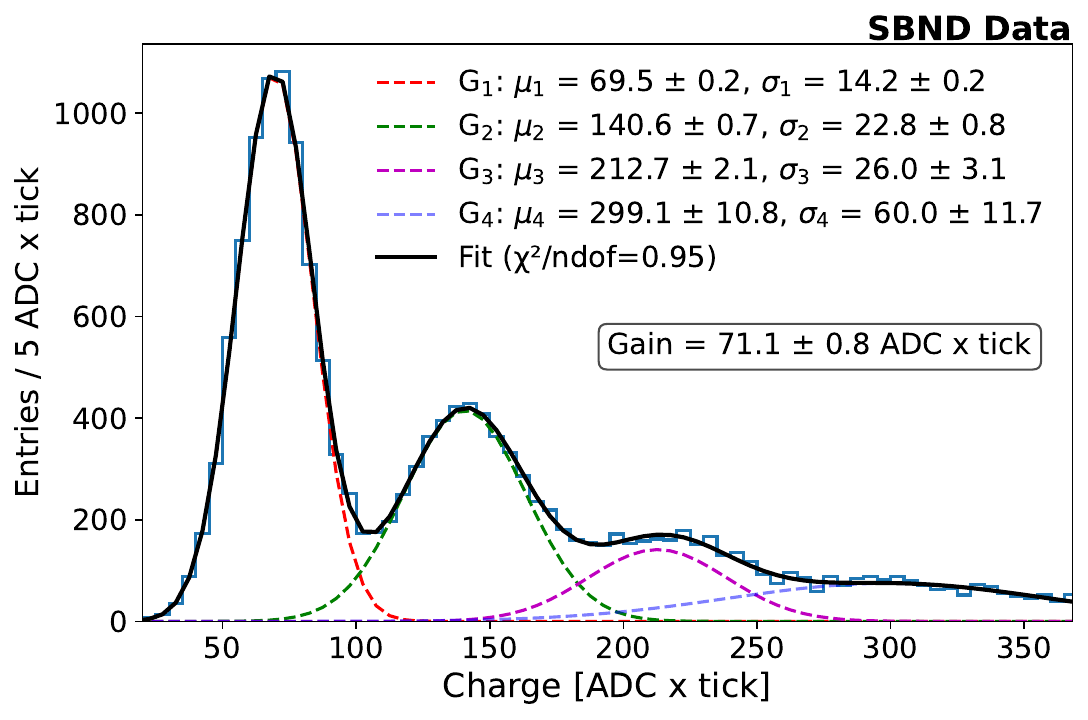}
    \caption{Left: Typical APSAIA-HS X-ARAPUCA recorded waveform with a region of interest (ROI) with a few-PE pulse (indicated by dashed red lines). The estimated baseline (dashed purple) and its standard deviation (dashed green lines) are overlaid. Right: Charge spectrum of an APSAIA-HS channel operated at an over-voltage of 5.5\,V fitted with a sum of Gaussians.}
    \label{fig:xa_spe_plots}
\end{figure}

\begin{figure}[!htbp]
    \centering
    \includegraphics[width=0.70\textwidth]{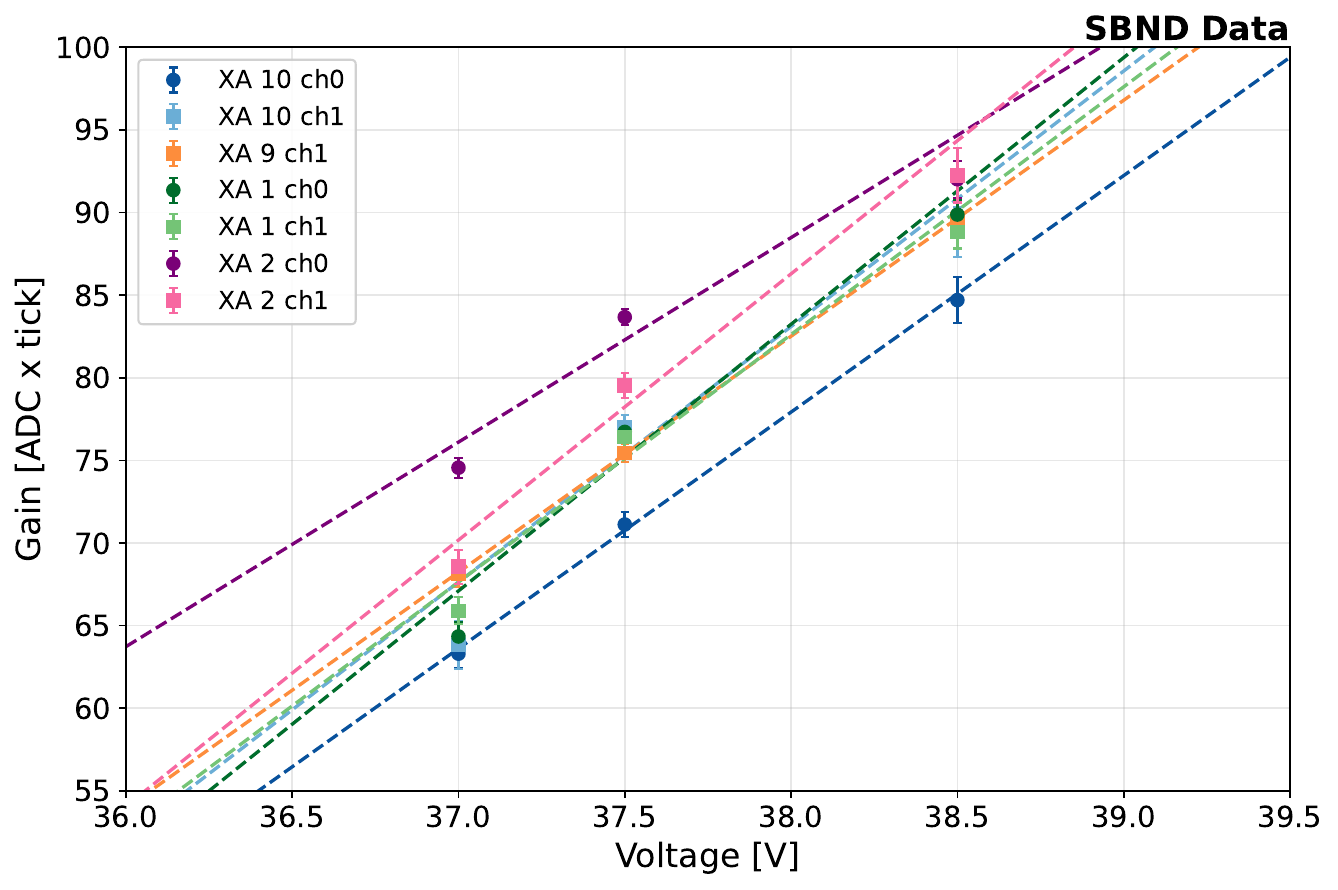}
    \caption{SBND X-ARAPUCA effective gains at different applied voltages for APSAIA-HS channels. A linear fit is performed for every channel. One of the channels is missing since it could not be calibrated during Run 1.}
    \label{fig:xa_gain_plots}
\end{figure}

The gains of the APSAIA X-ARAPUCA channels were calibrated using an algorithm which identified isolated pulses with small amplitude in the readout window as shown in figure~\ref{fig:xa_spe_plots}~(left).
The resulting charge spectrum of integrating the baseline-subtracted pulses is fitted with a sum of Gaussians, shown in figure~\ref{fig:xa_spe_plots}~(right), and the difference between the two most prominent peaks is taken as the gain.
After repeating the procedure for several operation voltages, a relation between the applied voltage and the gain was extracted as shown in figure~\ref{fig:xa_gain_plots}. 
During Run\,1, all APSAIA-HS channels except one were set to an effective gain of $\approx$65\,ADC$\times$Tick/PE, corresponding to a  $3.2 \times 10^6$ factor. APSAIA-VE channels are out of scope for this paper as they were calibrated in later operations. 

\subsubsection{PMT timing}
\label{sec:PMTtiming}

A robust method to evaluate the timing resolution of the SBND PMT system is based on the reconstruction of the well-known temporal structure of the neutrino beam to which the detector is exposed. The neutrino interaction time is reconstructed using PMT signals that register coincident light above a predefined photon threshold. The timing resolution of the PDS is significantly affected by two factors intrinsic to the detector: the spread in time between channels introduced by hardware delays $\mathcal{O}$(10\,ns), and the spread in light propagation between sensors inside the detector $\mathcal{O}$(10\,ns). 

\begin{figure}[!ht]
    \centering
    \captionsetup[subfigure]{margin=2pt}
    {\includegraphics[width=0.49\textwidth]{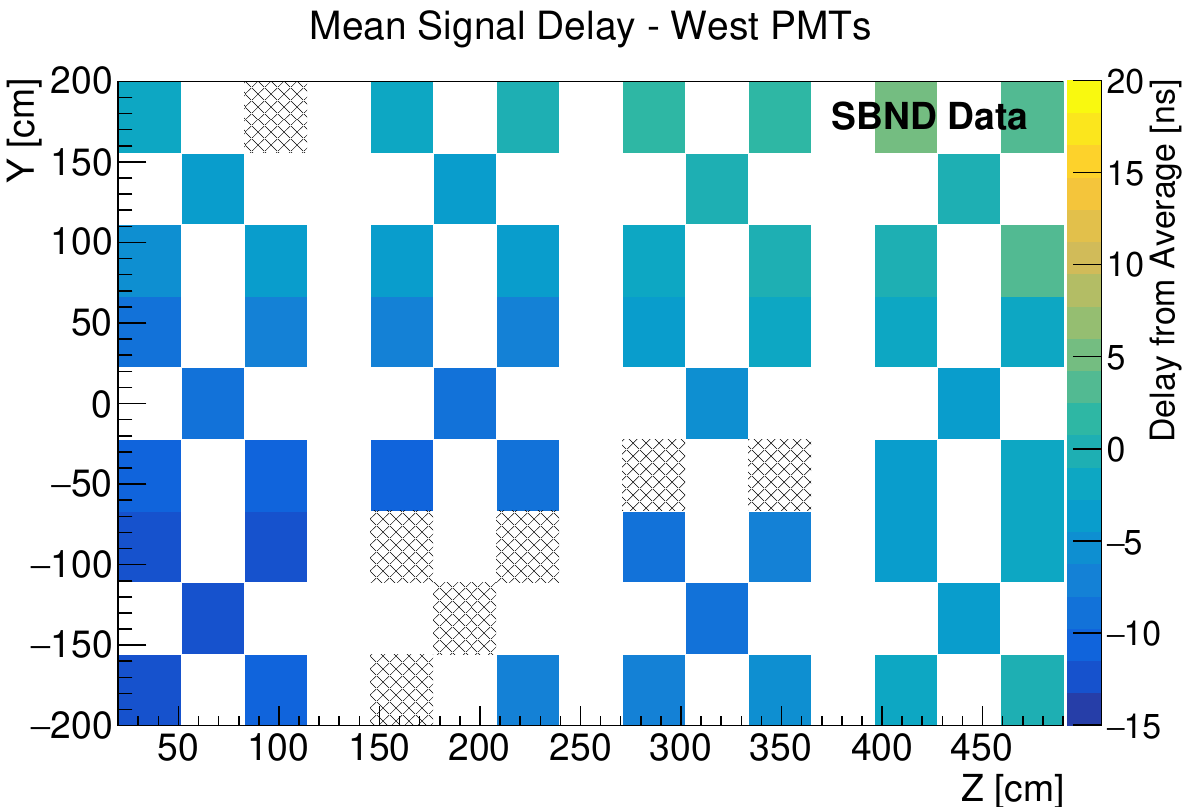}
    \includegraphics[width=0.49\textwidth]{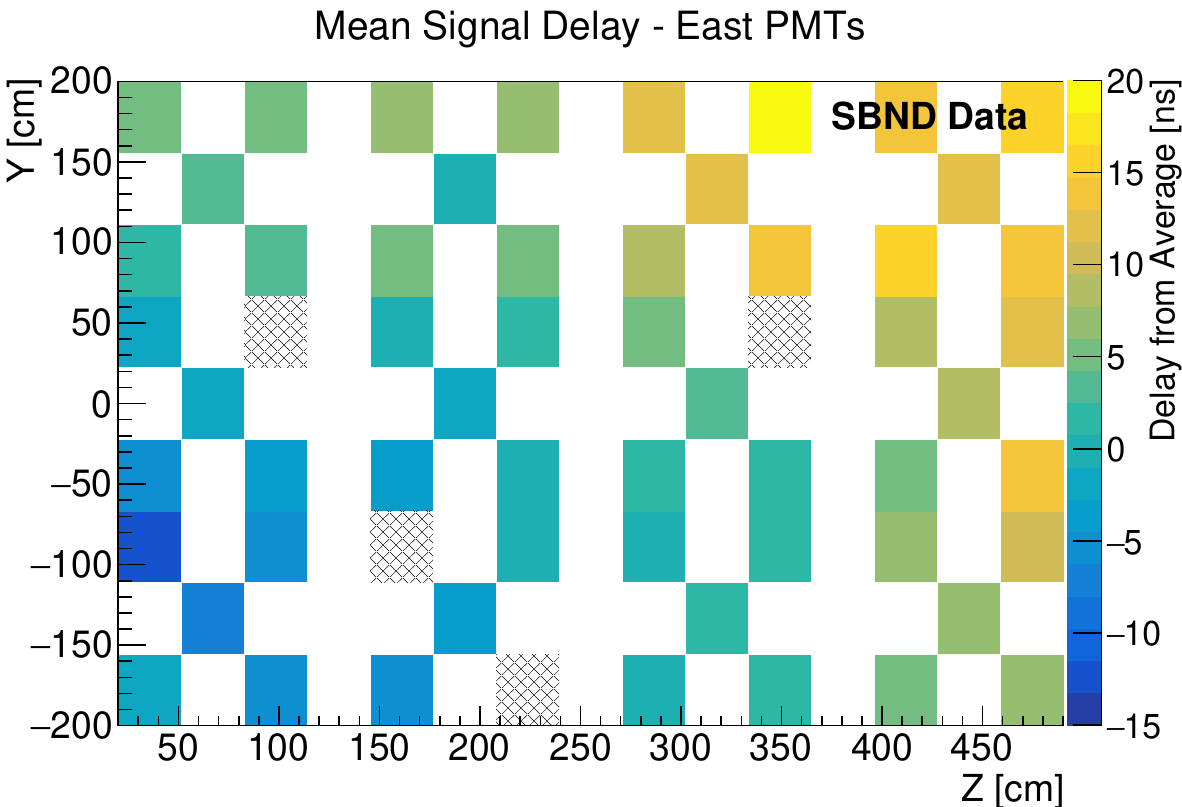}
    }
    {\includegraphics[width=.49\textwidth]{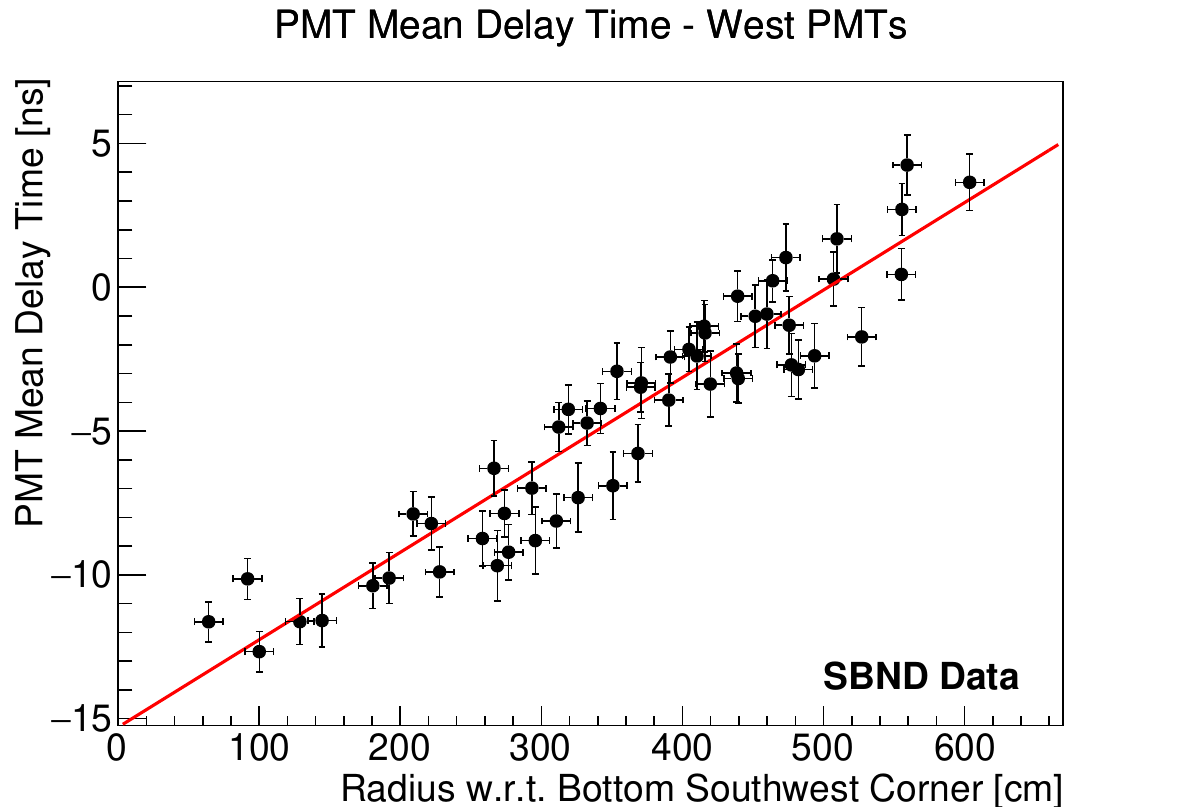}
    \includegraphics[width=.49\textwidth]{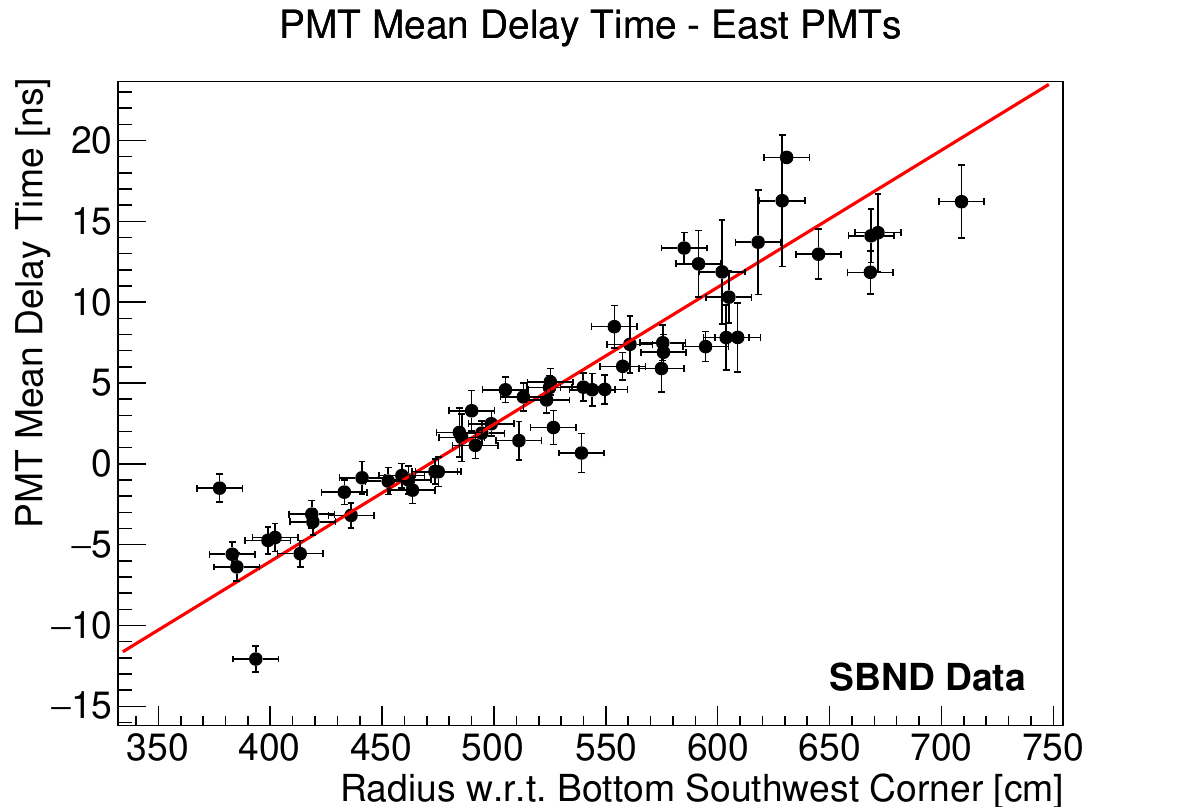}}
    \caption{Top: 2D representation of the mean signal delays measured by the PMTs in the Y$-$Z plane with the non-operational PMTs shown in hashed squares. The change of color from bottom left to top right highlights the propagation of the light source with PMTs within a similar radius having similar delays. Bottom: 1D representation of the mean delay time measured by the PMTs with respect to the bottom south-west corner of the TPC ($y=-200$\,cm and $z=0$\,cm) with a linear fit serving as guide.}
    \label{fig:pmt_timing_plots}
\end{figure}

The PMT timing spread due to hardware delays was measured using a light source located near the bottom south-west corner of the detector around $y=-200$\,cm and $z=0$\,cm. For each event, the start time of each PMT's signal was identified and the difference from the time averaged across all the PMTs was calculated, so to reduce the impact of the timing uncertainty of any one PMT. The time delays for each individual PMT were averaged over all events, as shown in the top of figure~\ref{fig:pmt_timing_plots} in 2D in the Y$-$Z plane, which highlights the physical location of the PMTs. The bottom panel presents the measured time delay as a function of distance from the light source. As expected from simple geometric considerations, the delay grows with increasing separation. This linear dependence is evident in both TPCs and is shown explicitly in figure~\ref{fig:pmt_timing_plots}, where the delay is plotted against the radial distance from the light source. This measurement allowed us to confirm the predicted time delays expected from the measured cable lengths and associated time delays and the channel mapping. This measurement also confirms cable reflection measurements done prior. The larger spread in time delay for the East PMTs is associated with the larger distance uncertainty the photons have to traverse the CPA before being recorded by the PMTs, as expected. More precise measurements will be made using the calibration system described in section~\ref{sec:PDSCalibration}.

In order to recover the intrinsic temporal structure of the beam, it is necessary to correct for delays introduced by the neutrino propagation time within the active detector volume, as well as by the photon time of flight from their point of production to detection. These corrections are purely geometrical and require knowledge of the interaction vertex Z-coordinate and the X-coordinates of the energy depositions associated with the interaction. In addition, channel-dependent timing offsets arising from the different PMT cable lengths are explicitly corrected for in the reconstruction, based on the results shown above.

\begin{figure}[htb]
    \centering
    \includegraphics[width=0.9\linewidth]{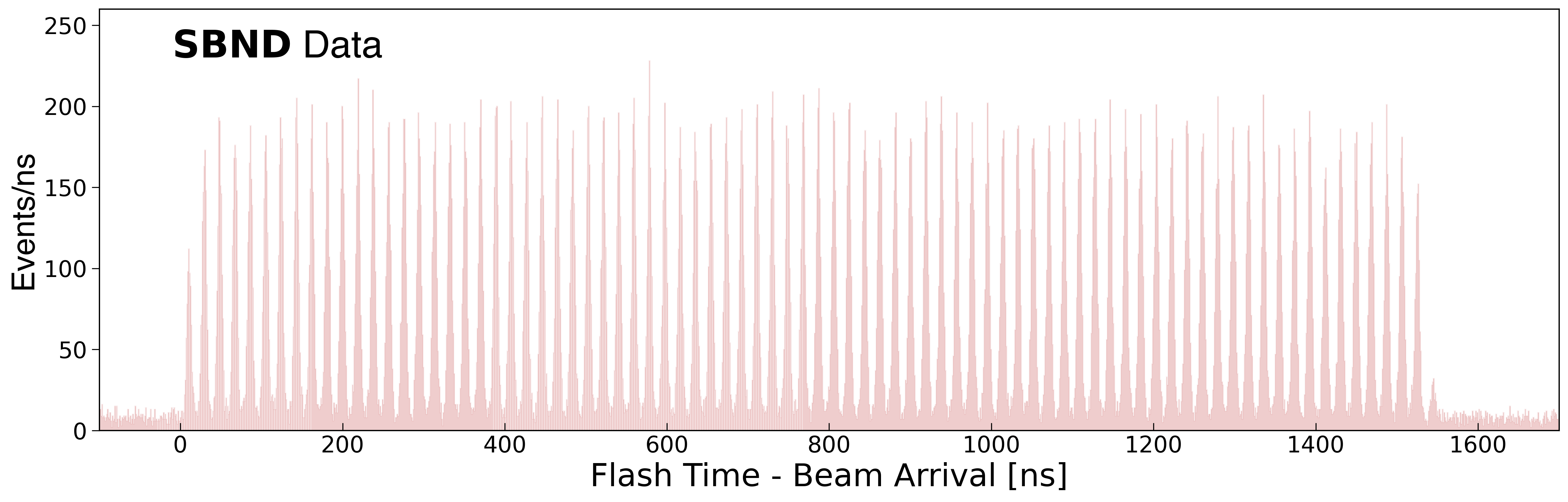}
    \includegraphics[width=0.9\linewidth]{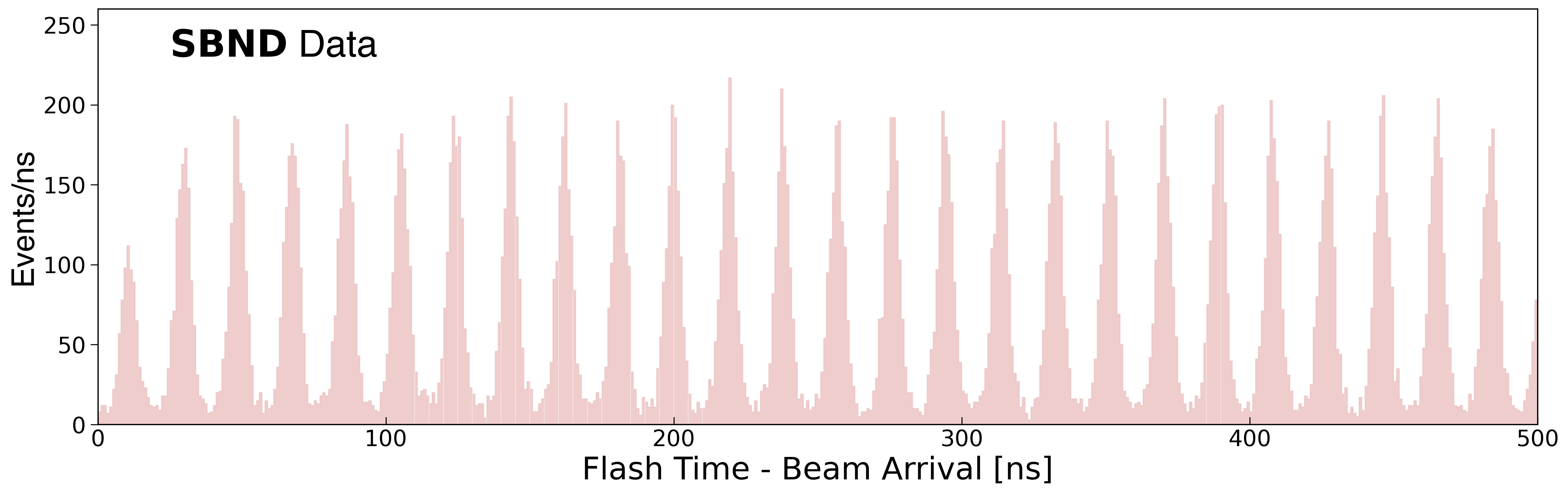}
    \caption{The number of PMT events recorded with respect to the precise beam signal provided by Fermilab Accelerator Division. Top: full beam spill, bottom: zoomed in on a smaller portion for detail}
    \label{fig:pmt_beam_timing}
\end{figure}

As demonstrated in~\cite{SBND:2024vgn}, using only the SBND PMT system it is possible to reconstruct the average interaction coordinates $\langle Z \rangle$ and $\langle X \rangle$, which are sufficient to apply the required propagation-time corrections. The result obtained using this simple, light-only-based approach is shown in figure~\ref{fig:pmt_beam_timing}, where a width of $2.84\,\mathrm{ns}$ is measured for the beam time structure. This value represents an upper bound on the timing resolution achieved by the PMT system, since it includes the intrinsic width of the BNB pulses. It should be noted that this reconstruction does not yet include corrections associated with the spread of individual PMT transit times, nor time-of-flight effects reconstructed using TPC information, including PMT-TPC matching. The inclusion of these effects is expected to lead to a significant improvement in the achieved timing resolution and will be presented in a future publication, where the full calibration and reconstruction potential of the SBND detector will be exploited.

%\clearpage
\section{Cosmic ray tagger}
\label{sec:CRT}

\begin{figure}[ht]
    \centering
    \includegraphics[width=.75\textwidth]{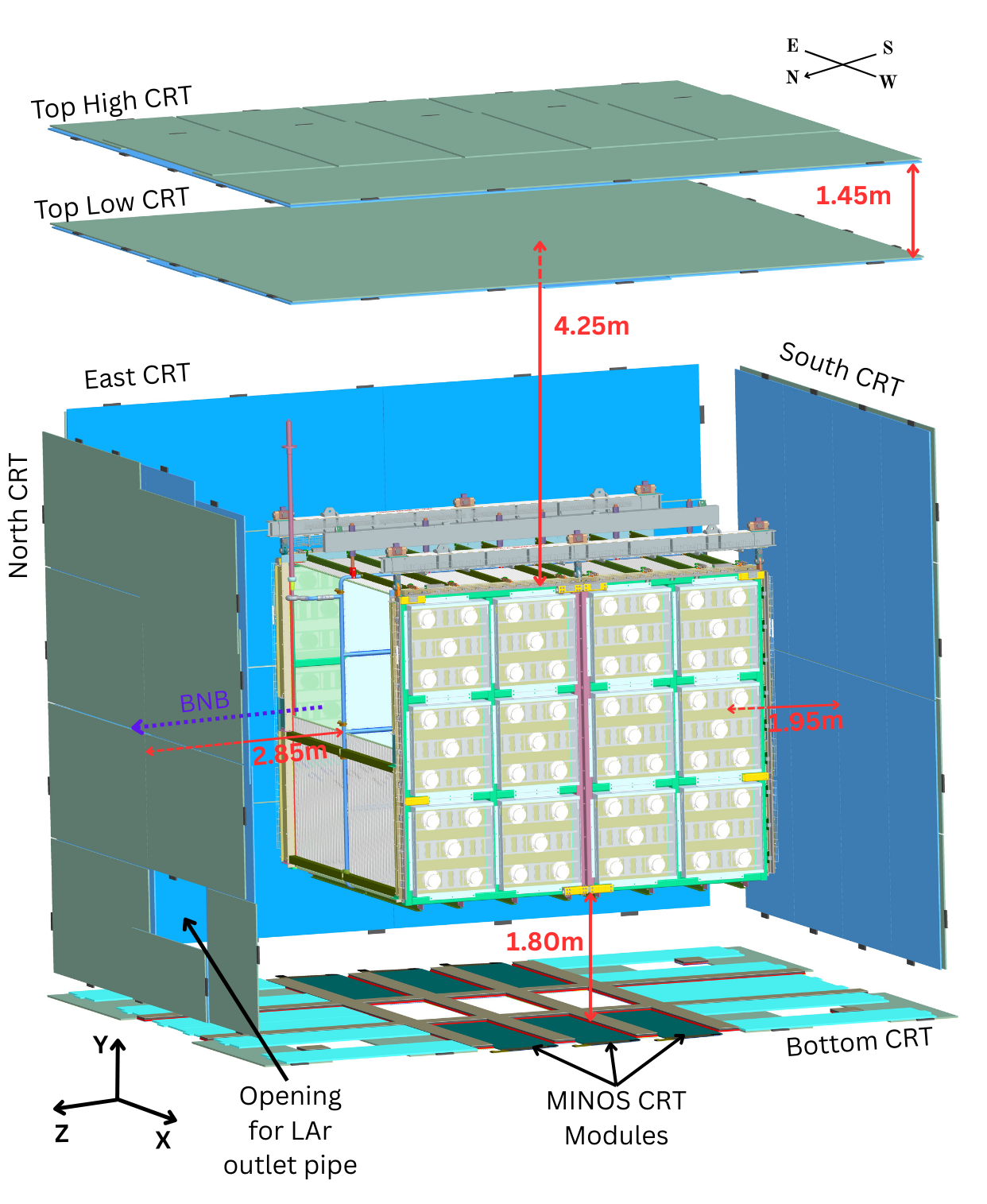}
    \caption{A visualization of six of the seven CRT walls in green and blue positioned around the SBND central detector. Note in particular the telescope arrangement of the two top walls, the perpendicular orientation of modules in the two layers of each wall and the incomplete coverage of the bottom CRT. The west CRT wall and the cryostat have been removed to aid this visualization. The distance between each CRT wall and the TPC volume are shown in red except the east and west walls (2.05\,m).  }
    \label{fig:crtwallgeom}
\end{figure}

The cosmic ray tagger system (CRT) is designed to tag charged particles entering, exiting, or passing through the LAr volume of the SBND detector. The system also provides tagging of beam neutrinos that interact in material upstream or surrounding the SBND detector, and can be utilized by the trigger system to provide high-purity samples of muons that cross the detector in particular directions. These samples were used heavily in the commissioning and calibration of the TPC and PDS systems. The system consists of seven walls of scintillator modules surrounding the outside of the cryostat. Figure~\ref{fig:crtwallgeom} shows the the walls (except one) and their positioning relative to the TPC.

The drift time in the TPC (1.28\,ms) is long enough that charge deposited by several cosmic-ray muons will be collected alongside that deposited by each triggering neutrino interaction or cosmic ray.

The CRT system provides the time and location at which charged particles crossed each wall. This allows for cosmic-ray muons to be removed from consideration in SBND's neutrino analyses either by identifying the particle as being out-of-coincidence with the neutrino beam spill (1.6\,$\mu$s duration) or by identifying its origin as being outside of the TPC. To do this, the CRT information is linked to the corresponding charge deposition in the TPC via an assessment of the geometric consistency.

Matching signals in the CRT and PDS systems can also be used to provide further cosmic rejection such as time-of-flight methods to distinguish exiting and entering muons or the entry of neutral particles induced via cosmic-ray muons that would not be identified through simple track matching.

The CRT timing resolution requirement $\mathcal{O}$(5ns) is dominated by determination of whether the charge was deposited during an individual beam spill bucket and performing time-of-flight assessments alongside the PDS system. Requirements to localize charge deposits in the drift direction and to determine whether the charge was deposited inside or outside of the neutrino beam spill are several orders of magnitude less stringent.    
The few-cm positional resolution is not the limiting factor for the TPC-CRT matching, as uncertainties in the extrapolation of the TPC track direction due to reconstruction edge effects or scattering in the cryostat walls dominate.

\subsection{System description}
\label{sec:crtsysdescr}
The SBND CRT comprises 146 aluminum-encased plastic scintillator modules, 110 of which were manufactured for SBND in Switzerland, 30 of which were previously used in the MicroBooNE experiment, and the other six were adapted from the MINOS experiment~\cite{Michael:2008bc}. The SBND modules are functionally identical to those produced for the MicroBooNE experiment~\cite{Adams:2019bzt}, however, they do vary in size. The different types and sizes of modules are summarized in table~\ref{tab:crtmoduletypes} with their usage across the seven tagger walls summarized in table~\ref{tab:crtmoduleusage} in appendix~\ref{appendixCRT}.

Each of the SBND and MicroBooNE modules comprises 16 optically isolated strips. The strips are created using a polystyrene base and an addition of 1.5\% PTP and 0.04\% POPOP. A wavelength-shifting fiber is embedded in a groove running lengthwise along each side of the strip and is connected at one end to a Hamamatsu S12825-050P silicon photomultiplier (SiPM) to detect the wavelength-shifted scintillation light. This results in 32 SiPM readout channels per module. The far end of the fiber is coated with reflective aluminum to increase the light yield for activity near that end of the strip.

\begin{figure}[ht]
    \centering
    %\captionsetup[subfigure]{margin=10pt}
    %\subcaptionbox{The perpendicular arrangement of modules allows the position of the crossing particle to be constrained.\label{fig:crtperpprinciple}}{
    \includegraphics[width=.42\textwidth]{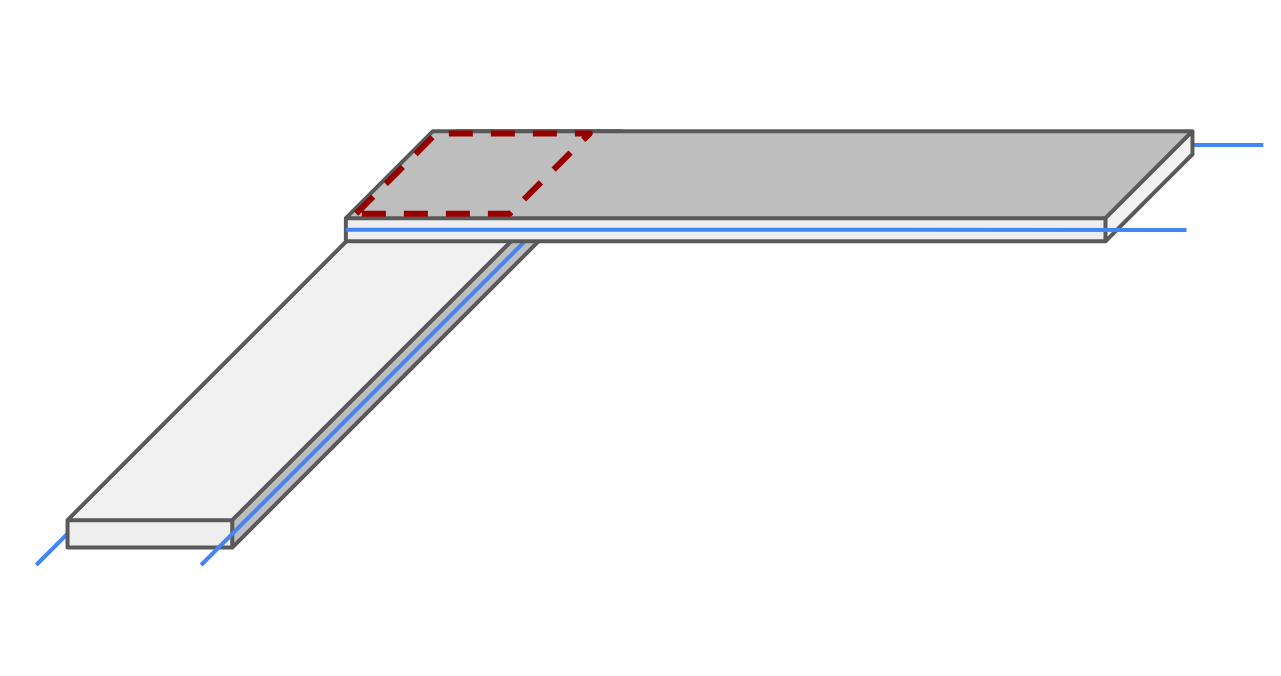}%}
    %\subcaptionbox{A full arrangement of modules in a wall, note the horizontal layer is only visible via the indication of the corresponding readout boards (FEB). The green and purple lines indicate the coincidence cabling loops. \label{fig:crttintoutprinciple}}{
    \includegraphics[width=.48\textwidth]{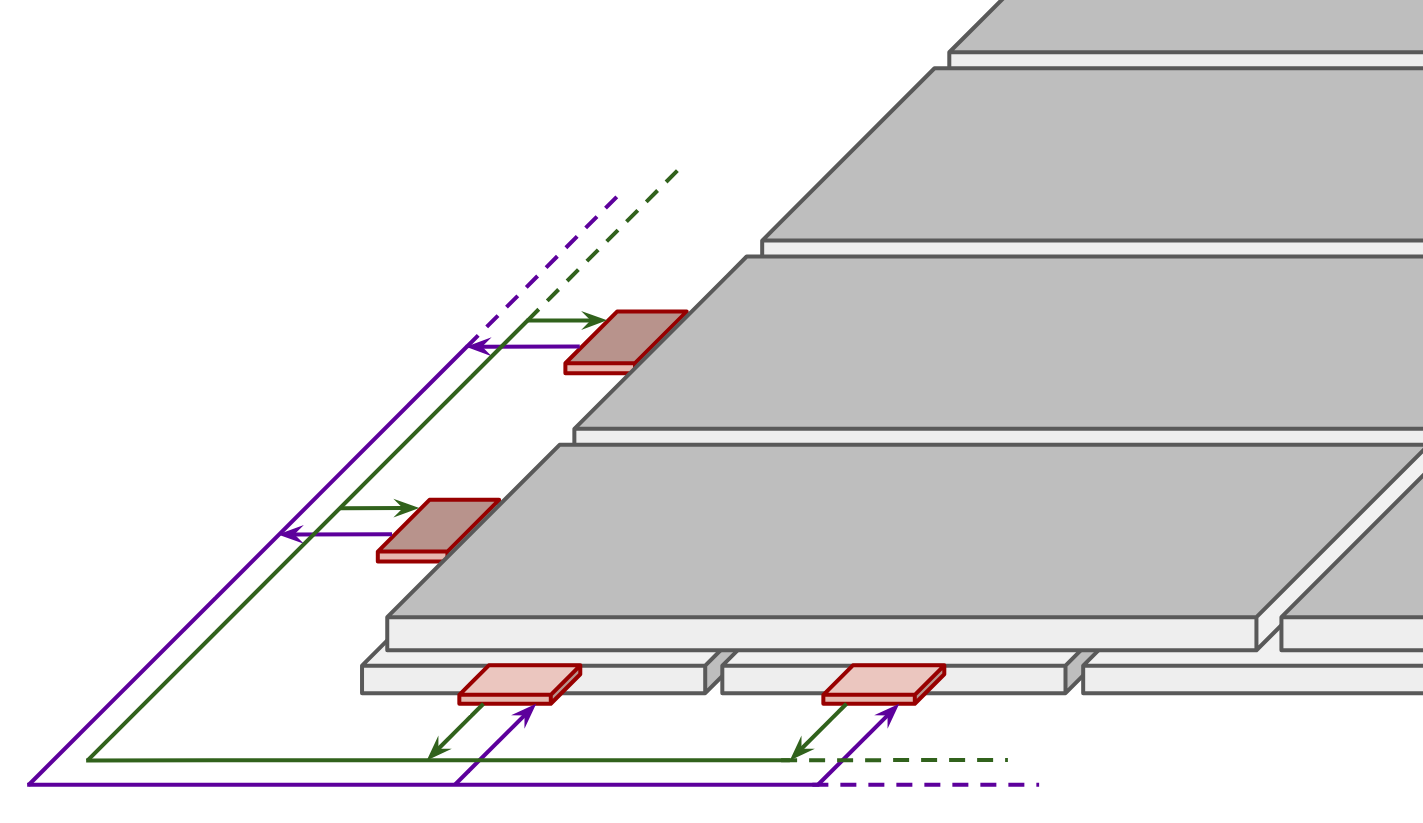}%}
    \caption{Schematics illustrating the arrangement of CRT modules to form a full tagger wall. Left: The perpendicular arrangement of modules allows the position of the crossing particle to be constrained. Right: A full arrangement of modules in a wall, note the horizontal layer is only visible via the indication of the corresponding readout boards (FEB). The green and purple lines indicate the coincidence cabling loops. }
    \label{fig:crtwallprinciple}
\end{figure}

The seven tagger walls are made up of two layers of modules, with the second layer oriented perpendicular to the first. This allows the 2D position at which the particle crosses the wall to be determined, as is indicated by the schematic in figure~\ref{fig:crtwallprinciple}~(left). %figure~\ref{fig:crtperpprinciple}.  

The six modules adapted from the MINOS experiment form part of the bottom CRT, under the cryostat. The adaptation follows the method developed by the ICARUS experiment for their side CRT~\cite{aduszkiewicz2025design}. The MINOS modules contain 20 strips, each 4\,cm in width, with a single wavelength-shifting fiber running down the center. An optical readout module (ORM), external to the module, houses 10 SiPMs each coupled to a pair of adjacent fibers. At MINOS and for most of ICARUS, the modules were read out at both ends of the strip, however, to fit within the spaces under the SBND cryostat, the modules were cut to a shorter length and reflective aluminum placed on the cut end of the fiber.

\begin{table}[ht]
    \centering
    \caption{Dimensions of each module type for the 146 modules in the SBND CRT.}
    \label{tab:crtmoduletypes}
    \begin{tabular}{c c c c c c c c c c}
        \hline \hline
        \multirow{3}{*}{\textbf{Type}} & \multirow{3}{*}{\textbf{Producer}} & \multicolumn{3}{c|}{\textbf{Module}} & \multicolumn{3}{c}{\textbf{Strip}} & \multirow{3}{*}{\textbf{Strips}} & 
        \\ \cline{3-8}
        & & \textbf{Width} & \textbf{Length} & \textbf{Thckn.} & \textbf{Width} & \textbf{Length} & \textbf{Thckn.} & & \textbf{Total} \\
        & & (cm) & (cm) & (cm) & (cm) & (cm) & (cm) & & \textbf{Num.} \\ \hline \hline
        A & SBND & 180 & 360 & 1.8 & 11.2 & 359.2 & 1 & 16 & 50 \\
        B & SBND & 180 & 450 & 1.8 & 11.2 & 449.2 & 1 & 16 & 36 \\
        C & SBND & 180 & 180 & 1.8 & 11.2 & 179.2 & 1 & 16 & 2 \\
        D & MicroBooNE & 180 & 184.5 & 1.8 & 11.2 & 183.7 & 1 & 16 & 5 \\
        E & MicroBooNE & 180 & 364.5 & 1.8 & 11.2 & 363.7 & 1 & 16 & 15 \\
        F & MicroBooNE & 173.6 & 408.3 & 1.8 & 10.8 & 407.5 & 1 & 16 & 5 \\
        G & MicroBooNE & 173.6 & 264.1 & 1.8 & 10.8 & 263.3 & 1 & 16 & 5\\
        H & SBND & 95.8 & 272.1 & 1.8 & 5.95 & 271.5 & 1 & 16 & 22 \\
        I & MINOS & 80.5 & 272 & 1.2 & 4 & 271.9 & 1 & 20 & 6 \\ \hline \hline
    \end{tabular}

\end{table}

\subsubsection{Readout electronics}
\label{sec:crtelectronics}

Each CRT module is coupled to a custom front-end readout board (FEB) designed and produced for the MicroBooNE and SBND experiments then commercialized by CAEN~\cite{caenFEB}. This board supplies the bias voltage (20 - 90\,V), which is separately configurable for each SiPM (at the level of -0.5 - 4.5\,V). It amplifies (up to 600$\times$) and shapes pulses from the SiPMs using a fast (15\,ns) and slow (12.5\,ns to 87.5\,ns) shaping circuit. Hit digitization is initiated based on an adjustable threshold against the output of the fast shaping circuit and a pair of coincidence requirements discussed below. The digitization is performed on the output of the slow shaping circuit for all 32 channels. Their analog pulse height amplitudes are retained in a `track and hold' circuit until the coincidence requirements are met and then digitized with a 12-bit ADC. The hits are timestamped with respect to a pair of reference timing signals, provided on SBND by the White Rabbit timing system (see section~\ref{sec:Timing}), then stored in the board's buffer for collection by the DAQ system.

The first coincidence requirement enforces both channels from a common strip pass the threshold within 30\,ns of each other. This requirement significantly reduces the occurrence of noise hits. A further coincidence requirement can be made between the perpendicular layers of the wall via the `TIN' and `TOUT' connectors on the FEB. The TOUTs of all boards in one orientation are connected in a daisy chain loop to the TINs of all boards in the other orientation. This setup is visualized in figure~\ref{fig:crtwallprinciple}~(right). A signal is sent via the TOUT buffer whenever a strip-pair of SiPMs passes the internal coincidence requirement and this `hit' is only processed if a signal is received by the TIN port within a 160\,ns coincidence window. This external coincidence significantly reduces the dominant radiogenic activity due to their inability to penetrate the $>$1cm of scintillator required to make an above threshold signal in both layers. The signals on each of the coincidence loops are also sent to the SBND trigger system where they can be used to trigger on events in which muons cross particular combinations of walls. Note this external coincidence requirement is not used in the bottom CRT wall where, due to installation restrictions under the cryostat, not all of the active area contains perpendicular overlap (see figure~\ref{fig:crtwallgeom}). However, these modules are connected in a single coincidence loop connected only to TOUT and this signal is also sent to the trigger system.

Each FEB contains a single 250\,MHz oscillator clock synchronized to the delivery of the pulse-per-second (PPS) signal used as the common reference by all SBND systems. The PPS is delivered to the `T0' connector on every FEB via a daisy-chain setup (never more than nine boards per chain). A second timing connection, `T1', utilizes the same setup to deliver a second reference pulse, this time the pulse corresponds to an early (${\sim}1.5$\,ms) warning of the beam (or fake beam). A readout hit is produced whenever a clock reset is received (reset hit) or when the threshold and coincidence conditions are met (physics hit). The readout hit contains two timestamps corresponding to the number of nanoseconds elapsed since the previous reset of that type (T0/T1), an ADC value for every channel (not just the two that fired the readout), a flag to indicate the type of readout (clock reset / physics activity), and counters recording any hits missed due to buffer overflow or deadtime during hit digitization. These fields total 76 bytes per readout hit. More details about the CRT readout is available in section~\ref{sec:CRTDAQ}.

The readout of MINOS modules uses the same electronics as the other modules, with three of the ORMs connected to each FEB. Thus, the six MINOS modules are readout by two FEBs, with a total of 60 channels of interest (two channels per board are left unused). The data format is the same throughout.

\subsubsection{Installation}
\label{sec:crtinstall}
\begin{figure}[ht]
    \centering
    \includegraphics[width=.48\textwidth]{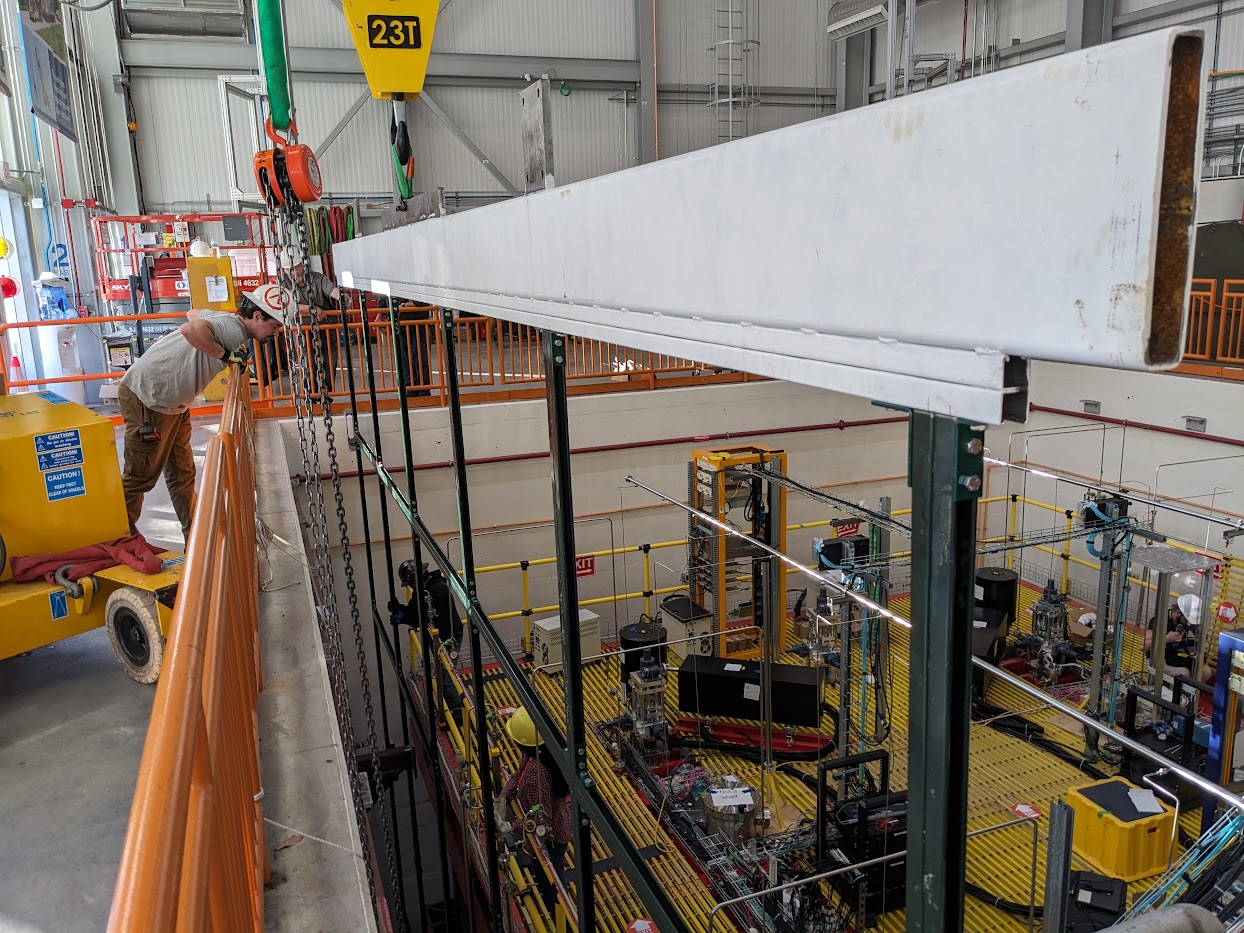}%}%
    \hspace{0.04\textwidth}%
    %\subcaptionbox{Partial north wall installation, May 2023.\label{fig:crtnorthwallinstall}} 
    \includegraphics[width=.48\textwidth]{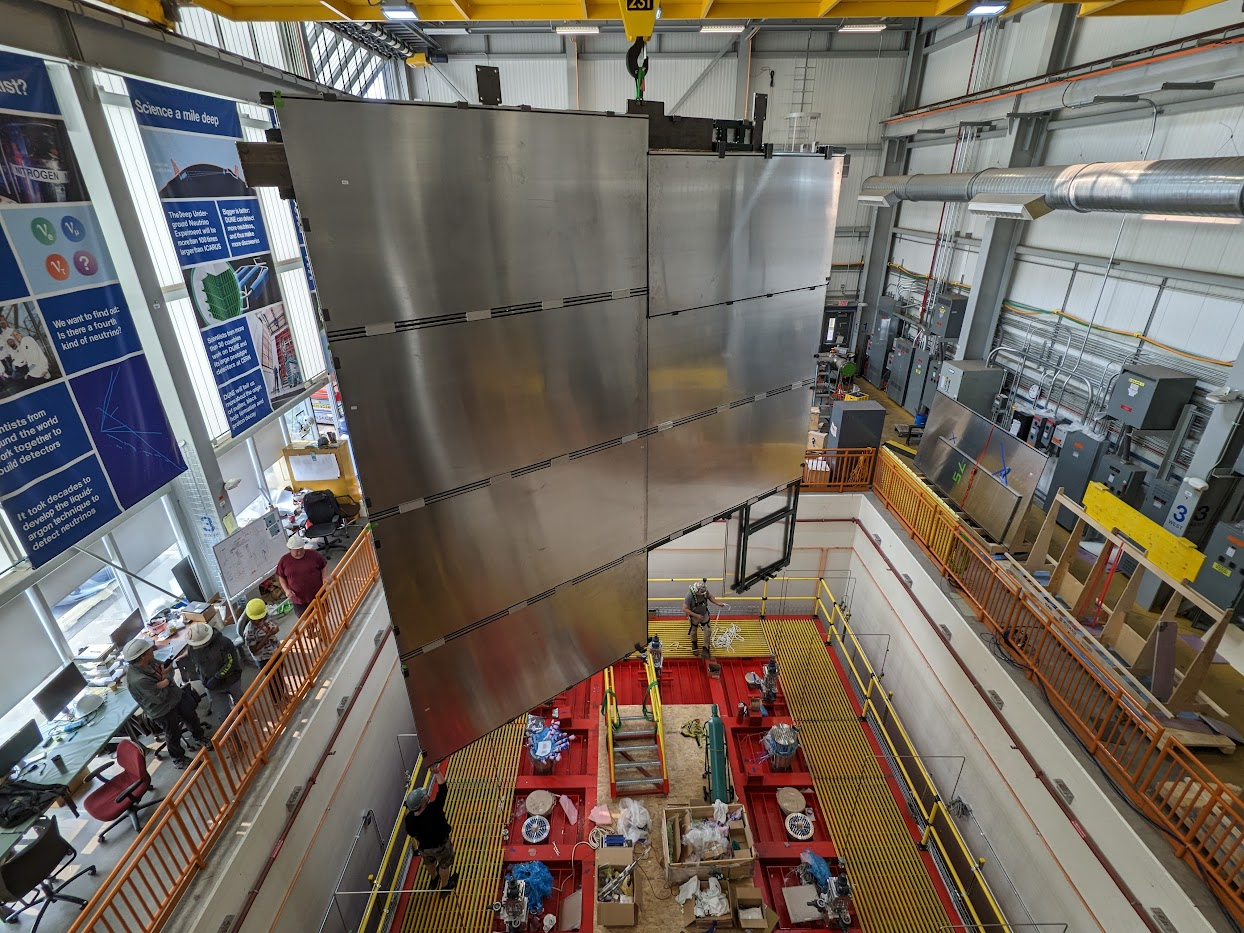}%}
    \caption{Installation of CRT side wall support frames. Left: Test installation of the east wall support frame, December 2023. Right: Partial north wall installation, May 2023.}
    \label{fig:crtsideframeinstall}
\end{figure}

The CRT modules in each side wall are mounted on frames constructed of Unistrut suspended from a steel box beam that is anchored with steel brackets to the pit walls and stabilized with adjustable feet at the floor. Figure~\ref{fig:crtsideframeinstall} shows the test installation of one of these frames without any modules, as well as the installation of the north wall with most modules pre-installed.

\begin{figure}[ht]
    \centering
    %\captionsetup[subfigure]{margin=2pt}
    %\subcaptionbox{The installed support frame for the top low CRT wall, July 2024. \label{fig:crttoplowframeinstall}}{
    \includegraphics[width=.48\textwidth]{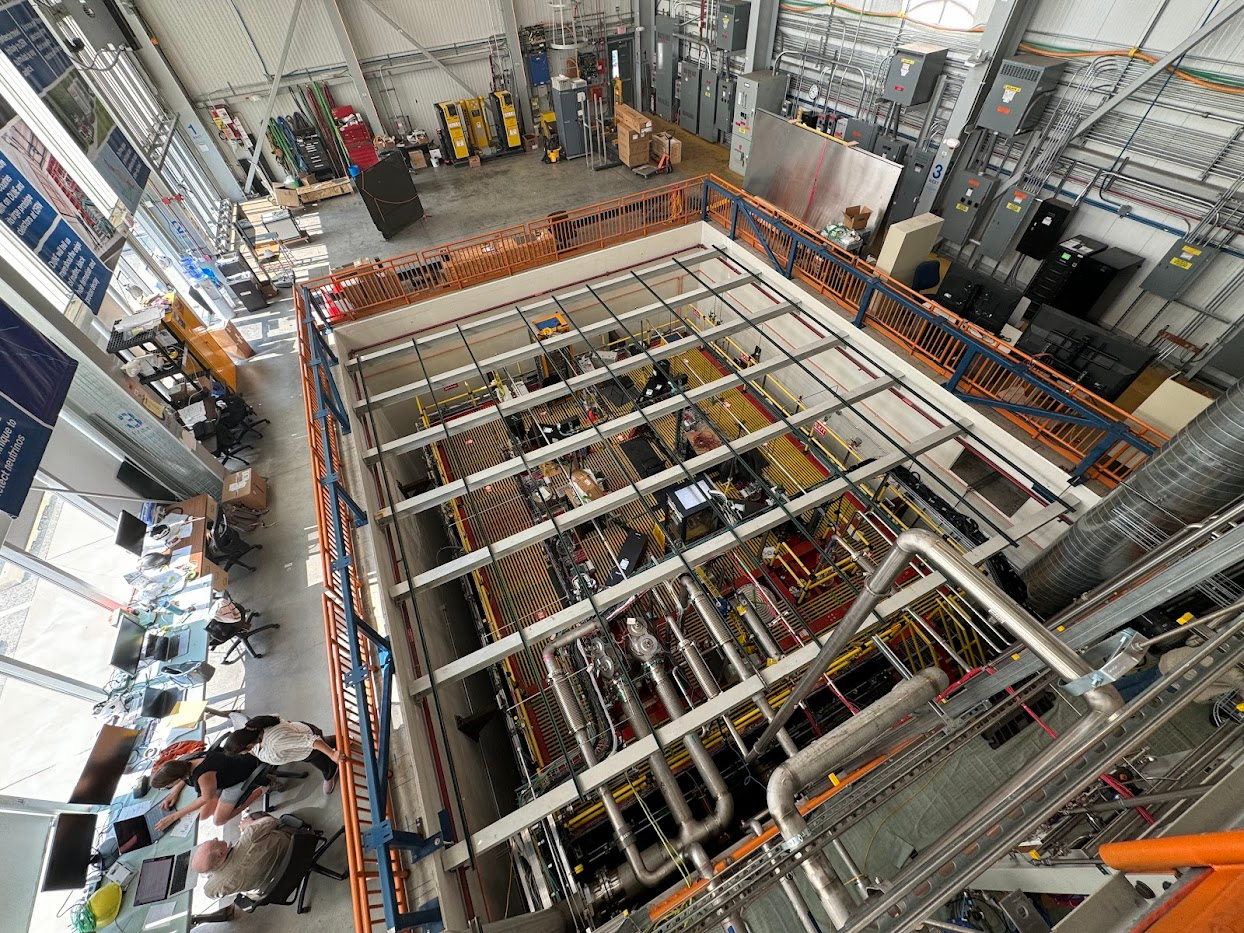}%}%
    \hspace{0.04\textwidth}%
    %\subcaptionbox{Installation of the first modules on the support frame for the top high CRT wall, September 2024. \label{fig:crttophighinstall}}{
    \includegraphics[width=.48\textwidth]{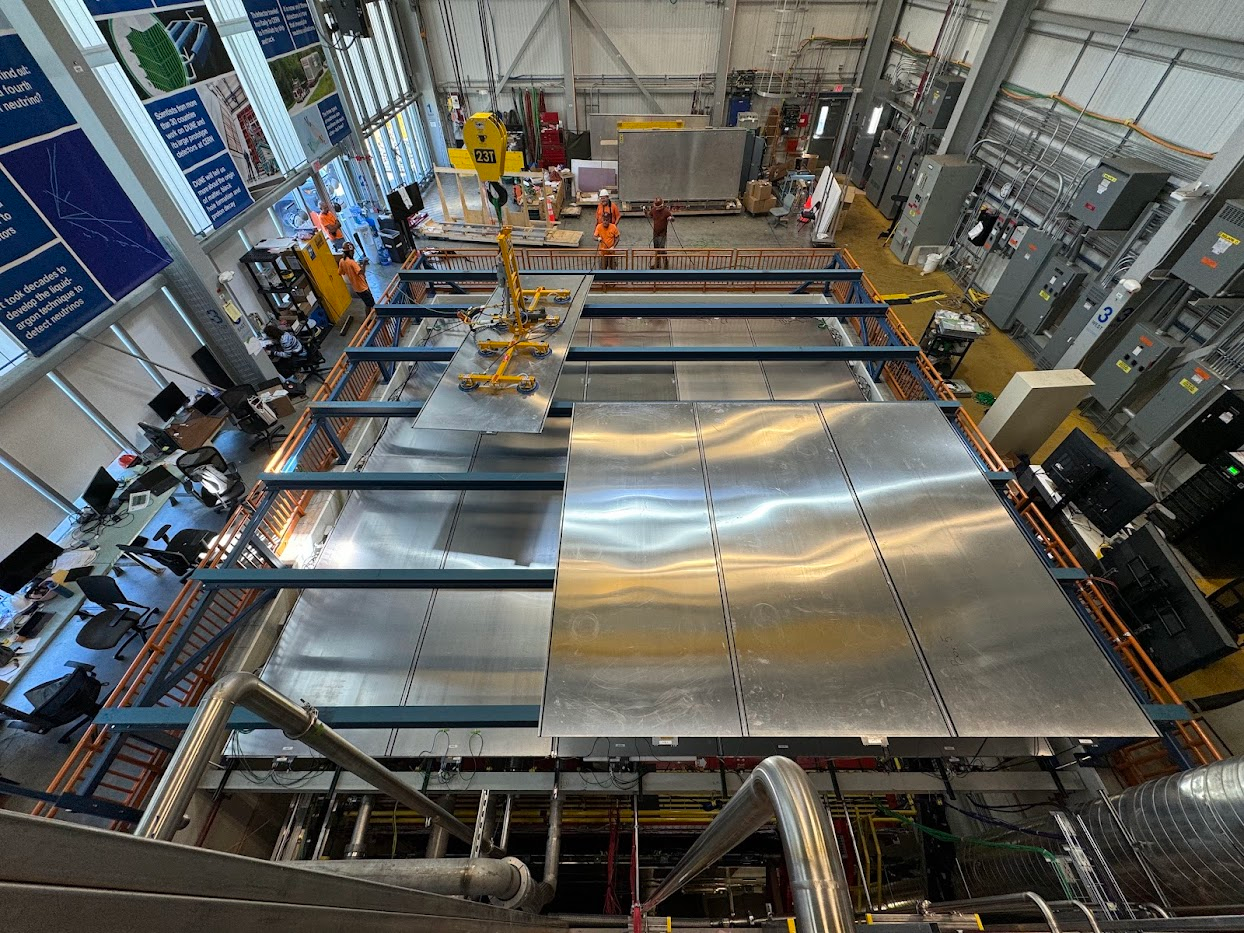}%}
    \caption{Installation of the CRT top wall support frames. Left: The installed support frame for the lower of the two top CRT layers, July 2024. Right: Installation of the first modules on the support frame for the higher of the top CRT layers, September 2024. }
    \label{fig:crttopframeinstall}
\end{figure}

The top CRT uses I-beams to support the modules from below. For the lower of the two top CRT layers these I-beams are bracketed to the pit walls, while a further vertical support structure supported by the surrounding floor holds the higher of the top CRT layers, as shown in figure \ref{fig:crttopframeinstall}. The bottom CRT modules are laid flat along the pit floor under the cryostat. The position of each wall at installation time was measured with tape measures. More precise numbers have been determined via studies of TPC - CRT matching and will be further refined by segmenting the study by the walls' component modules. 
Figure~\ref{fig:crtwallgeom} shows the distances from each CRT wall to the relevant face of the TPC active volume. For supplementary details see table~\ref{tab:crtpositions} in appendix~\ref{appendixCRT}. The asymmetry between the south and north walls occurs because the TPC is located upstream off-center in the cryostat to leave room for cryogenic systems (see section~\ref{sec:cryogenics}). The top walls are located further from the TPC due to equipment and access restrictions on the cryostat top cap. Their separation was chosen following Monte Carlo studies to balance angular resolution with cosmic muon efficiency.

\subsubsection{Grounding and power supply}
The CRT grounding and power distribution was designed to minimize the noise injected into the sensitive TPC and PDS detector elements while also protecting the CRT electronics from any noise impact that could produce artificial hits. The CRT system uses a grounding path (`building ground') independent to that of the TPC and PDS electronics (`detector ground'). The FEBs are isolated from the modules such that their ground returns follow separate paths. The FEBs share a common ground via the daisy chain connections to their T0/T1/TIN/TOUT ports. The module casings are connected together via flat braided wire clamped to each module and then via a copper busbar connected to the building ground.

The CRT power distribution begins with a Wiener PL512 LV low-noise power supply in a ground floor utility rack. Each PL512 output is connected via building ground shielded cable to a custom distribution box, the output of which is a 5-wire flat cable containing two supply and three return channels. This flat cable is used to run along the CRT walls with a number of connection points each supplying the power to two FEBs. No more than 10 FEBs are powered from each distribution box.

\subsection{Performance}
\label{sec:crtperformance}

\subsubsection{Configuration tuning}
\label{sec:crtconftuning}
Each CRT FEB contains a number of configurable parameters that can impact the rate, efficiency and dynamic range. The three used to define the first set of SBND operations parameters were the SiPM bias voltage, the amplifier gain setting and the discriminator threshold. Each of these settings is individually adjustable for every CRT channel and experience from running a previous beam telescope setup with spare CRT modules \cite{CRTLLP} was used to provide informed first-pass settings during installation.

The SiPM bias voltage is critical to the efficiency of muon tagging. During installation every side and top CRT module was tested in a muon telescope setup. This testing arrangement allowed any problematic channels or electronics boards to be identified in advance of full installation. For each muon telescope, a voltage scan was taken in both a standard amplifier setting and again at a high gain amplifier setting. The former allowed for efficiency measurements to be taken by removing a single module from the reconstruction at a time, using the other modules to identify muon tracks that should have passed through the module in question and then looking for the corresponding hit in that module. This efficiency measurement was performed at a range of bias voltage settings and a setting was extracted for each strip which gave $\geq95\,\%$ cosmic muon efficiency. The voltage scan taken at a higher amplifier setting was used in calibration to identify single PE peaks and measure the total gain factor.

Three different threshold settings are used across the entire SBND CRT. One setting is used for the majority of the top and side walls. A higher setting is required for the bottom CRT due to the lack of four-fold coincidence, as described in section~\ref{sec:crtelectronics}, thus a higher rate from radiogenic backgrounds. 
Finally, a lower setting is used for the longest modules (type B from tables~\ref{tab:crtmoduletypes} and~\ref{tab:crtmoduleusage}) in the top and side walls to improve uniformity of response due to the enhanced impact of signal attenuation in these modules. 
These threshold settings were determined by lowering the value until the rate started rising in an accelerated manner due to the presence of noise-induced readouts. The threshold was then set just above this region. 

The amplifier gain setting was tuned following the implementation of the bias and threshold settings in order to standardize the ADC response. The metric used was the ratio between the number of readouts for which the SiPM was saturated to the total number of readouts for that channel. A target of 2\% saturation for horizontally oriented boards and 8\% for vertically oriented boards was used. Typically, saturation occurs due to the passage of cosmic muons at an acute angle with respect to the module, thus extending their path length through the scintillator. Hence, the effect is more pronounced in the vertically oriented boards as a result of the cosmic muon flux decreasing with zenith angle. The absolute level was chosen to minimize saturation while not sacrificing rate, and the relation between horizontal and vertical boards was determined via Monte Carlo studies. The implementation of these amplifier settings also resulted in a more uniform rate being observed across each wall. It is worth noting that the CRT system was designed as a tagging system not a calorimeter, so while the ADC information is used during the positional and timing reconstruction - a detailed energy reconstruction is not intended.

\subsubsection{Observed rates}

\begin{figure}[ht]
    \centering
    %\captionsetup[subfigure]{margin=2pt}
    %\subcaptionbox{\label{fig:crtintrates}}{\vspace{-1em}
    \includegraphics[width=0.5\linewidth]{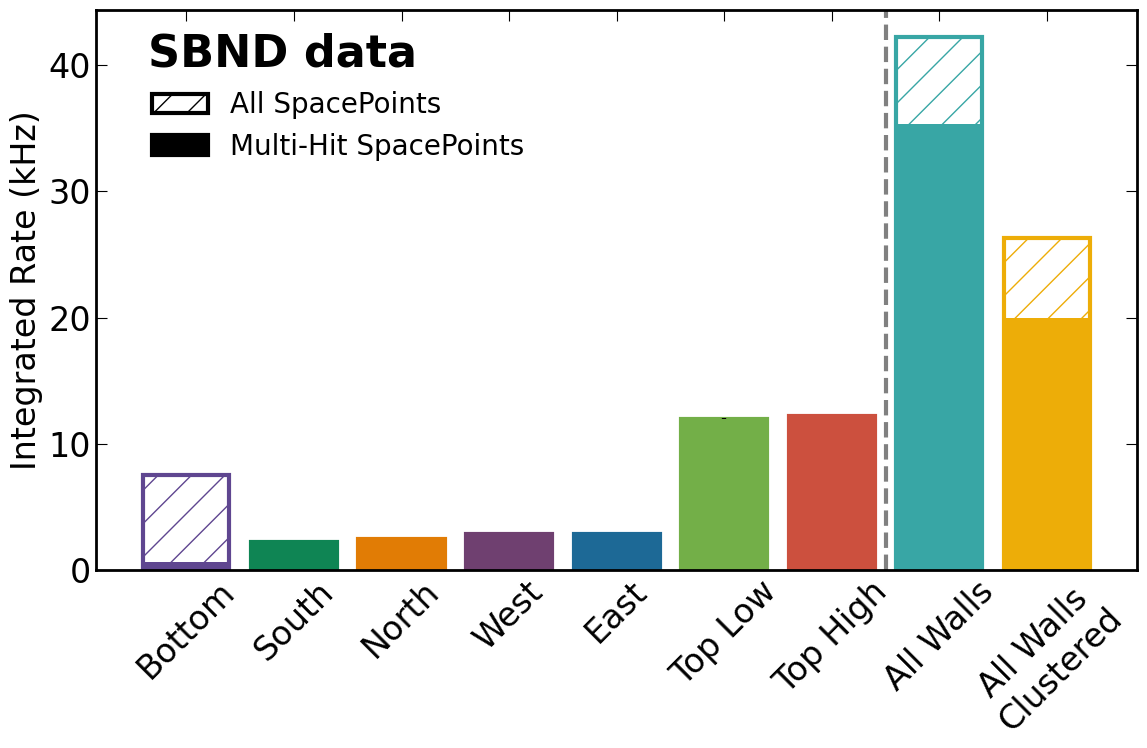}%}%
    %\subcaptionbox{\label{fig:crt3drates}}{
    \includegraphics[width=0.5\linewidth]{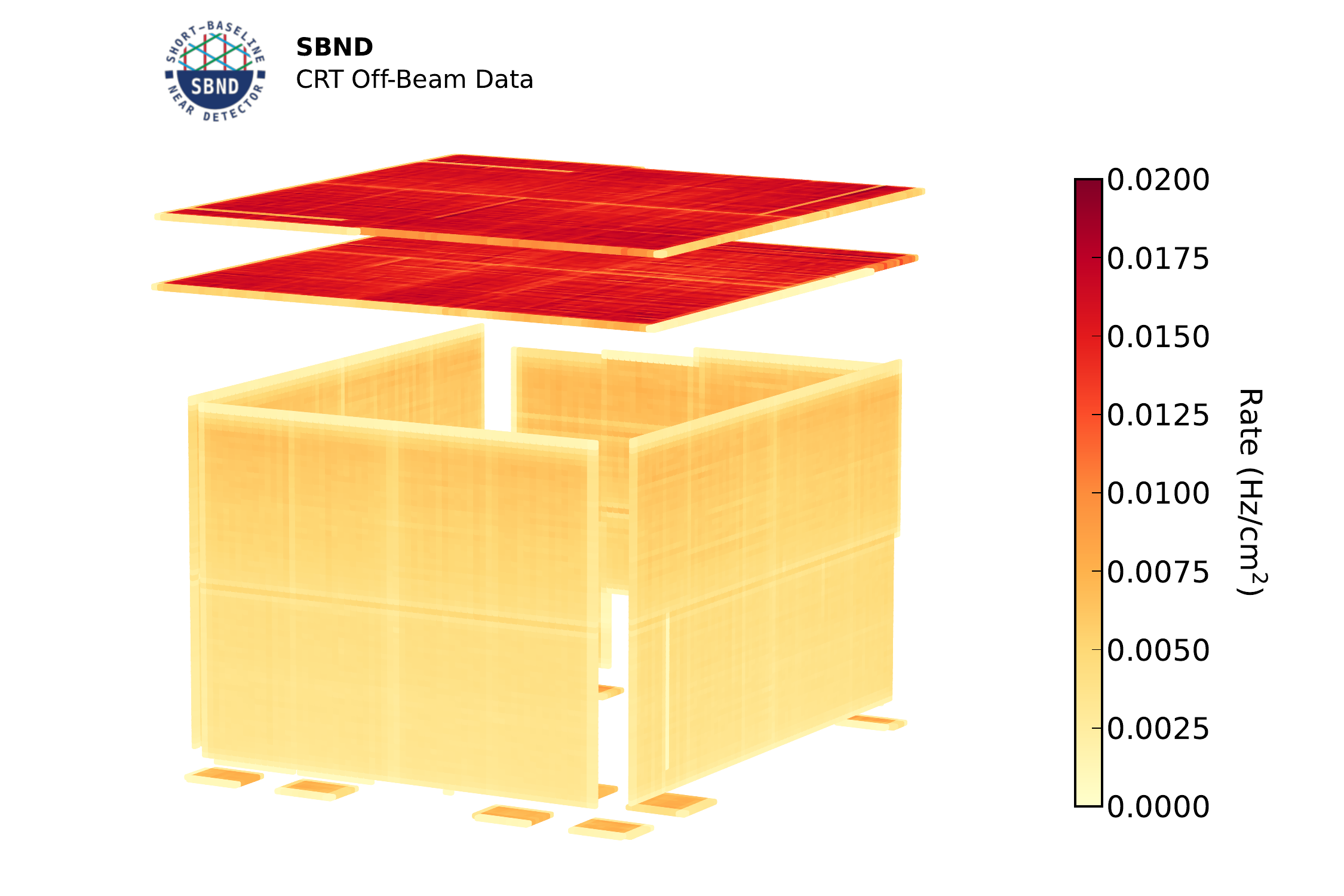}%}
    \caption{Left: The observed integrated rate from each CRT wall during typical operations, %(\ref{fig:crtintrates}). 
    The hatched bars show the rate from any reconstructed space point while the solid bars only consider space points where multiple strip hits were used to constrain the position. The final two bars show the combined rate across all walls, and then the combined rate clustered into 100\,ns coincidences. %\ref{fig:crt3drates} 
    Right: The CRT rates binned across the 2D locations within each wall.}
    \label{fig:crtrates}
\end{figure}

The rates of activity recorded by the seven installed CRT walls are shown in figure~\ref{fig:crtrates}. As expected, the two top walls see the largest rate, as they are the largest walls and are aligned parallel to the ground. The east and west side walls are larger than the south and north side walls and therefore see a slightly higher rate, while the north observes slightly higher than the south due to its reduced shielding by the pit walls. The bottom CRT contains a very small area of overlapping modules, therefore the reconstruction allows for space points to be created from just a single strip hit. The plot in figure~\ref{fig:crtrates}~(left) shows the difference between the total rate for the bottom wall and the rate for space points made using the same overlap criteria as all the other walls. The combined rates are included in figure~\ref{fig:crtrates}~(right). We show both the linear sum combined rate, and also a combined rate in which activity in a 100\,ns coincidence window is clustered to form a single entry. The latter better represents the rate of single-origin activity seen by the CRT. For example, a cosmic muon that crosses three different walls or an air shower that produces particles seen in two different walls would each be a single origin. The observed rates are consistent with expectations from cosmic ray showers. The distribution of these rates across the 2D faces of each wall are shown in figure~\ref{fig:crtrates}~(right) highlighting again the significantly higher rate through the top CRT walls. This representation also demonstrates the uniformity in rate across the top walls and the smoothly decreasing rate due to increased shielding from the pit wall and earth moving down the side walls; an indication of the uniform response achieved by the work in section~\ref{sec:crtconftuning}. Finally, three inactive strips are clearly visible, two in the top high and one in the east wall (right hand side front). A total of just 12 strips, or 0.5\%, are inactive across the entire system.

\subsubsection{Timing performance}

The core goal of the SBND CRT system is to provide tagging of activity that enters, exits or crosses the TPC with few-ns timing resolution. The timing resolution of the CRT system can be assessed using reconstructed space points in which just two strips, of perpendicular alignment, see the same crossing particle (as per figure~\ref{fig:crtwallprinciple}~(left)). The time of flight across the two strips, each of ${\sim}1$\,cm depth, will be negligible, so we would anticipate both strips observing light at the same time. Corrections are applied during reconstruction to account for a number of timing effects: cable delays (systematic shifts per module), light propagation time (hit location dependent) and the time walk effect (hit energy dependent). Following these corrections any residual difference between the time recorded by the two modules is representative of the timing resolution achieved.

\begin{figure}[htp]
    \centering
    \includegraphics[width=.7\textwidth]{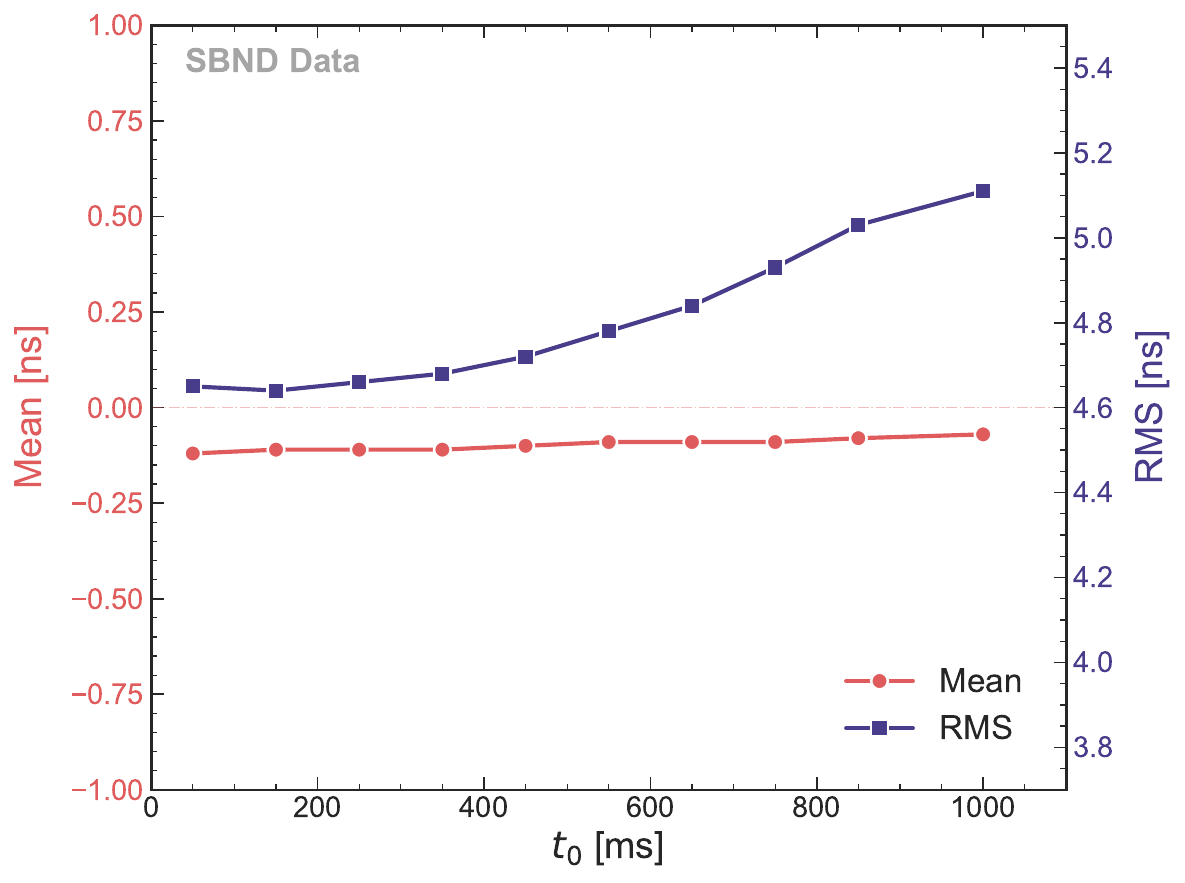}
    \caption{The variation in the mean and RMS of the T0 time difference between two perpendicular strips contributing to the same reconstructed spacepoint, across the full T0 phase space of 1\,s.}
    \label{fig:crttimingresclockdrift}
\end{figure}

The RMS (from a Gaussian fit) of this residual difference is 4.80\,ns and 4.61\,ns for T0 and T1 respectively. This corresponds to single point resolutions of 3.39\,ns and 3.26\,ns respectively. The difference between the T0 and T1 distributions results from the impact of the clock drift effect. As described in section~\ref{sec:crtelectronics}, the T0 and T1 timestamps are recorded with respect to two different reset pulses. The imperfect oscillator clock on each FEB can drift away from the more accurate White Rabbit system (see section~\ref{sec:Timing}) by up to 10\,ns every 1\,s. The scale and direction of this drift can vary randomly on a per-FEB, per-reset basis but each time the clock is reset this drift is eliminated. Given that the T0 clock is reset at a frequency of 1\,Hz and the T1 clock is reset at a higher frequency that averages 10\,Hz, the T1 timestamp at any given time is statistically more likely to have less clock drift than the corresponding T0 timestamp. Figure~\ref{fig:crttimingresclockdrift} shows how the RMS of the $\Delta T_{0}$ variable increases through the 1\,s phase space of the T0 timestamp, as clock drift has a greater impact. The first data point should reflect the behavior of the T1 timestamp RMS, and indeed we see a value of ${\sim}4.6$\,ns.

\begin{figure}[htbp]
    \centering
    %\captionsetup[subfigure]{margin=2pt}
    %\subcaptionbox{\label{fig:crtbucketfit}}{
    \includegraphics[width=\textwidth]
    %{figures/crt/crt_sp_south_wall_beam_buckets.pdf}}
    {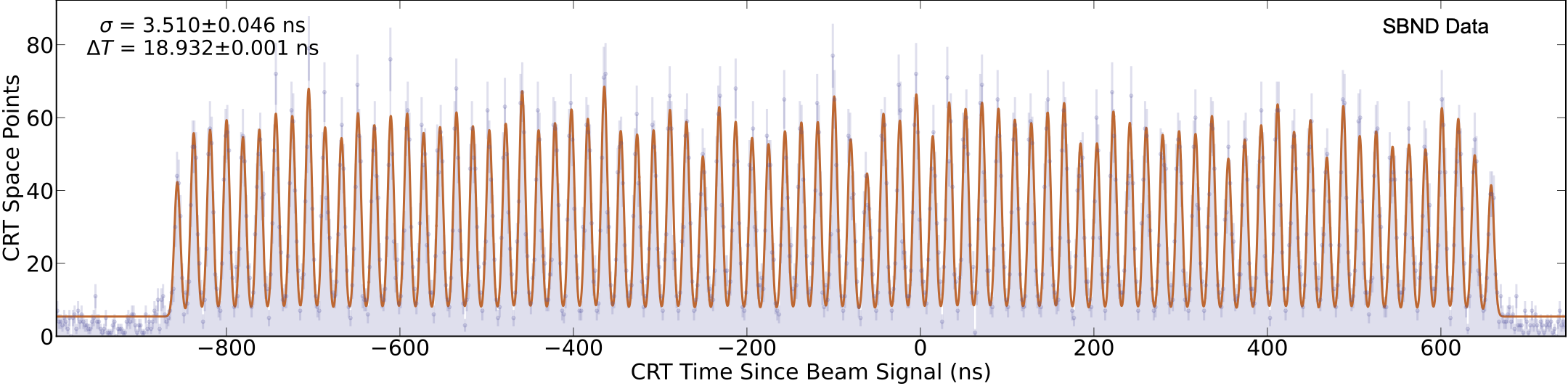}%}
    %\subcaptionbox{\label{fig:crtbucketfitzoom}}{
    
    \includegraphics[width=\textwidth]
    %{figures/crt/crt_sp_south_wall_beam_buckets_zoom.pdf}}
    {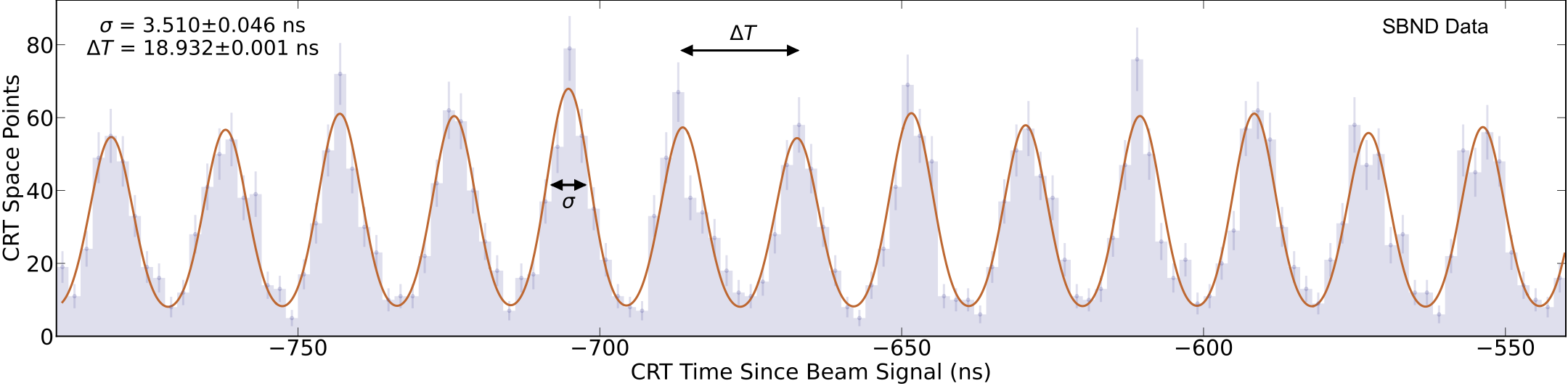}%}
    \caption{The number of CRT space points recorded in the downstream wall with respect to the precise beam signal provided by Fermilab Accelerator Division. Top: %Figure~\ref{fig:crtbucketfit} shows 
    the full beam spill, Bottom: %whilst \ref{fig:crtbucketfitzoom} 
    zooms in on a smaller portion for greater detail.}
    \label{fig:crtbucket}
\end{figure}

This reiterates the role of the T1 reset. It is deliberately positioned ${\sim}1.5$\,ms before each beam and off-beam event trigger (via the early warning beam signals and their corresponding off-beam equivalents). Therefore, any physics that could cause activity in the TPC readout for beam or off-beam events will have a T1 timestamp no greater than 3.5\,ms and the clock drift effect in this region will be negligible. It is clear that with the first iteration of our reconstruction algorithms the CRT is able to provide the few-ns timing resolution necessary to identify cosmic activity as being outside the beam arrival window, or even as being between the BNB buckets which are separated by approximately 19\,ns. This is demonstrated in figure~\ref{fig:crtbucket}, which shows the excess of beam related activity in the CRT downstream wall using the T1 timing, and the resolution is good enough to clearly resolve each of the 81 buckets in the beam substructure.  The combined fit to the peaks, as shown in the zoomed plot of figure~\ref{fig:crtbucket}~(bottom), gives a gaussian width of 3.5\,ns which is an upper limit on the CRT timing resolution since this width is a convolution of the CRT resolution and the width of the accelerator timing signal.

%\clearpage
\section{Data acquisition and trigger}
\label{sec:DAQnTrigger}

\begin{figure}[!htbp]
    \centering
    \includegraphics[width=0.97\textwidth]{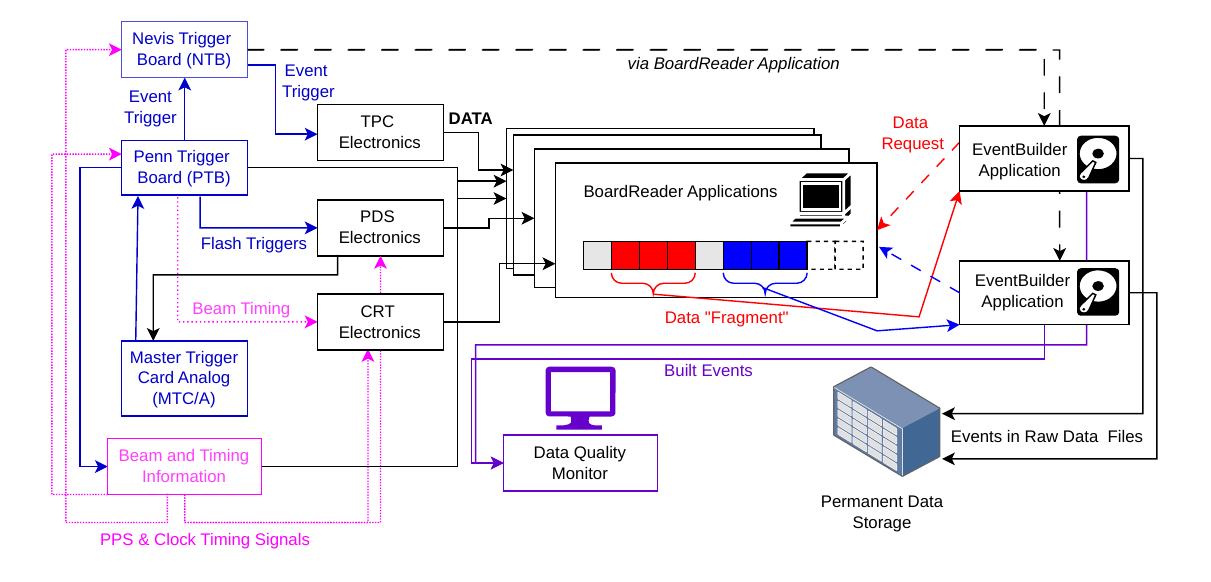}
    \caption{Data flow of the DAQ system. The paths of different types of information are differentiated by color for detector data (black), timing information (magenta), built events (purple) and trigger information (blue).}
    \label{fig:daqarch}
\end{figure}

The data acquisition system (DAQ) uses the \textit{artdaq} software framework~\cite{artdaq2017} to configure detector electronics, collect data from them, build events, and write them in files to local storage. Data from the PDS, CRT, and beam are collected by the trigger system (described in section~\ref{sec:Trigger}), 
which is responsible for making decisions to capture data (a trigger) and communicating those decisions to the readout electronics. The readout electronics for each detector subsystem interface with the DAQ through applications called “BoardReaders” running on DAQ servers. BoardReader applications continuously receive the available data from the readout electronics and store it in circular buffers in application memory. “EventBuilder” applications receive notification of an event trigger from the BoardReader application of the Nevis Trigger Board, and then request data based on event trigger number (for the TPC) or timestamp (all other systems) from all other BoardReader applications. Once all data fragments have been gathered, they are written to a ROOT-formatted file. Each file once closed is then asynchronously transferred to permanent tape storage facilities at FNAL. The flow of various data through the DAQ is shown in figure~\ref{fig:daqarch}. 

\begin{figure}[!htbp]
    \centering
    \includegraphics[width=0.8\textwidth]{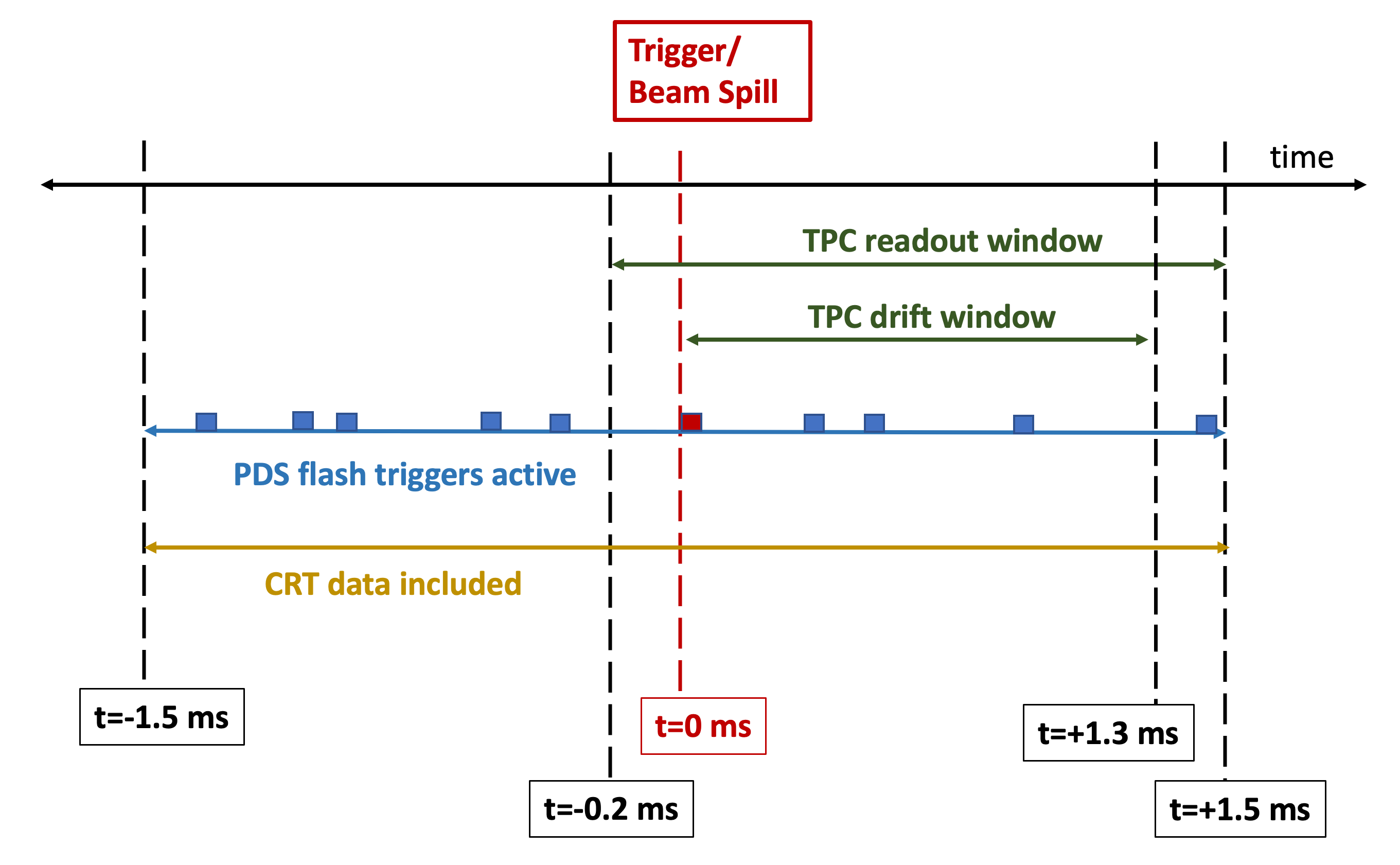}
    \caption{Timing structure of data collection by the SBND DAQ system for each of the detector subsystems. The slow waveforms of TPC data are acquired continuously throughout the TPC readout window. PDS data which consists of much faster signals is acquired in short chunks ($10\,\mu$s), represented by the blue squares, that can be accumulated multiple times per event.  The CRT data is sparsified in the CRT FEBs as described in section~\ref{sec:crtelectronics}.}
    \label{fig:eventtiming}
\end{figure}

Data from all components of the detector (TPC, PDS, CRT, trigger, and timing) and the BNB are assembled into events that represent signals from the same range in time. Figure~\ref{fig:eventtiming} illustrates the timing structure of the DAQ readout for the three detector subsystems. Since the charge collection of the TPC occurs over the full drift time of the TPC, the signal waveforms from the TPC readout are acquired continuously for the full drift time (plus a buffer) starting at the time of beam arrival in the detector. This is represented by a data "Fragment" in the BoardReader application shown in figure~\ref{fig:daqarch}. The PDS and CRT data consist of much faster signals that are digitized at a factor 100 higher rate than the TPC requiring sparsification for practical purposes. Since there can be multiple cosmic rays present in the detector for each beam spill, the PDS and CRT readouts accommodate acquisition of multiple pulses per event in separate fragments. Fragments from each subsystem are assigned timestamps using signals from the Timing system (described in section~\ref{sec:Timing}) and the Network Time Protocol (NTP)~\cite{NTP} timestamp from the DAQ server running the associated board reader process.    

The \textit{Event Building} process starts when the Nevis Trigger Board (NTB), responsible for triggering the TPC readout, receives one of the types of trigger signals which the Penn Trigger Board (PTB)~\cite{ptb} distributes. This type of trigger is called the ``event trigger'' as it results in an event being built. The PTB (described in section~\ref{sec:Trigger}) is responsible for determining when sufficient thresholds have been met to warrant a full detector readout. The PTB also instructs the PDS electronics when to create data fragments through ``flash triggers''.

\begin{figure}[!htbp]
    \centering
    \includegraphics[width=0.8\textwidth]{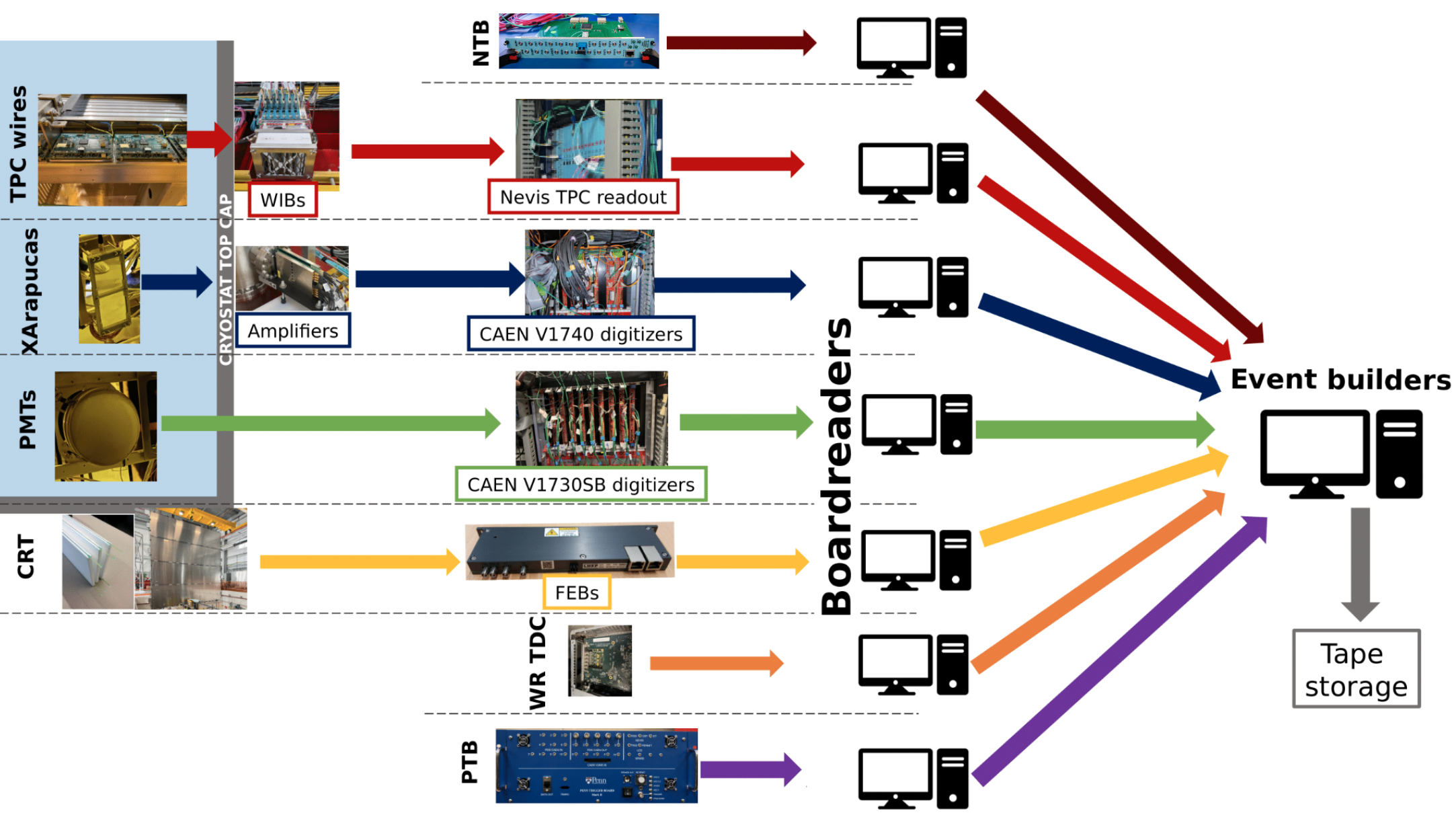}
    \caption{An overview cartoon of the SBND readout hardware. The TPC, X-ARAPUCA, PMT and CRT hardware (left column) are digitized and read out independently. 
    %are read out via digitization methods specific to each subsystem. 
    The TDC and PTB internally produce digitized outputs. From these digitized signals (center column), fragments are created in a set of board readers (right column), each of which run on their own dedicated server(s).  The event building process then combines the fragments from each board reader buffer into complete events, processes these events through the software triggers, and saves them to a file. A more detailed schematic is shown in figure~\ref{fig:DAQDiagram} in appendix~\ref{appendixDAQ}.}
    \label{fig:sbnd_readyfordata}
\end{figure}

The hardware components of the DAQ readout chain for each subsystem are illustrated in figure~\ref{fig:sbnd_readyfordata} and detailed in table~\ref{tab:daqreadouts}. The system uses 26 servers in the data transfer process: 11 for TPC data, 4 for PDS data, 3 for CRT data, 3 for trigger and timing data, and 6 for event building. An additional 9 servers provide the remaining services of the online system. Readout of the TPC, PDS, and CRT are described in sections~\ref{sec:TPCDAQ}, \ref{sec:PDSDAQ}, and \ref{sec:CRTDAQ} respectively. The system performance is summarized in section~\ref{sec:DAQPerformance}. 

\begin{table}[htb!]
\centering
  \caption{Details on the readout hardware used for each subsystem where the DAQ system acquires data. Clocks that are synchronized via phase-locked-loop (PLL) are indicated by *. Trigger data from the PDS subsystems are sampled at 62.5\,MHz.} 
\label{tab:daqreadouts}
\begin{tabular}{c c c c c}
\hline \hline{}
\textbf{Subsystem} & \textbf{Readout Hardware} & \makecell{\textbf{Readout} \\ \textbf{Boards}} & \textbf{Channels} & \makecell{\textbf{Sampling} \\ \textbf{Frequency}} \\ \hline \hline
%TPC & Nevis TPC Readout & \makecell{176 FEMs \\11 crates} & 11264 & \makecell{Data: 2\,MHz \\Trigger: 64\,MHz} \\ 
TPC & \makecell{Nevis TPC Readout FEM \\Nevis Trigger Board (NTB)} & \makecell{176 \\1} & \makecell{11264 \\---} & \makecell{2\,MHz \\16\,MHz} \\ 
\hline
PMT* & CAEN V1730SB & 8 & 120 & 500\,MHz \\ 
\hline
X-ARAPUCA* & CAEN V1740B & 4 & 208 & 62.5\,MHz \\ 
\hline
CRT & FEB & 142 & 4540 & 250\,MHz \\ 
\hline  
Trigger* & \makecell{Penn Trigger Board (PTB)} & 1 & --- & 50\,MHz \\ 
\hline 
Timing* & \makecell{White Rabbit \\Simple PCIe FMC Carrier \\Time to Digital Converter \\(WR SPEC TDC)} & 1 & 5 & 500\,GHz \\
\hline \hline
\end{tabular}

\end{table}

SBND uses a custom version of the \texttt{artdaq} software, which is shared with the ICARUS collaboration~\cite{ICARUS:2023gpo} under the ``SBN DAQ'' framework. 
The \texttt{artdaq} toolkit includes a built-in process for streaming data between event builder machines and online monitoring platforms.
Information from the board readers and event builders is transmitted to these platforms in real time for various monitoring purposes over shared memory or network communications. 
SBND currently employs two online platforms for data quality monitoring (DQM) purposes: (1) Grafana and (2) Minargon. Grafana~\cite{grafana_webpage} provides real time monitoring of the status of DAQ processes, including the ability to display fragment-level information that does not necessarily get recorded in the events. 
Minargon is a monitoring system developed by the SBN experiments that provides higher level monitoring of the quality of data acquired by the DAQ. 
It displays data from a subset of events which have simplified low-level reconstruction applied in order to quantitatively verify the physics characteristics of an event.

\subsection{Time Projection Chamber}
\label{sec:TPCDAQ}

The SBND TPC Back-end Readout System (Nevis TPC Readout) is responsible for receiving digitized TPC data from the SBND TPC Front-end Electronics System (described in section~\ref{sec:TPCreadout}) then pre-processing, packaging, and handing the data to downstream DAQ software processes in the form of two data streams. 

The Neutrino (NU) stream, consisting of losslessly compressed data using Huffman compression with static codes developed by MicroBooNE~\cite{Microboone} to improve speed, is an externally trigger-initiated readout stream that sends out TPC waveform information when a trigger from the PTB is received. 

The system, which is functionally identical to the MicroBooNE TPC Readout System~\cite{uboonedetector}, is composed of 176 Front End Module (FEM) cards, aggregated into a total of 11 custom electronics crates. Each FEM card receives data over optical fibers from 64 TPC wires digitized at 2\,MHz.  This electrically isolates the TPC from the DAQ.
A crate controller card responsible for card configuration and a data transmitter (XMIT) module are also housed in each crate.
The XMIT module distributes data from the FEMs within its crate to one of 11 TPC DAQ servers, via two PCIe cards; a third PCIe card in each server handles I/O communication with the crate controller.

A 16\,MHz clock, specific to the Nevis TPC readout system, is distributed to each crate through a clock distribution card connected to the backplane, as well as being distributed to the WIB to generate the 2\,MHz clock used in the digitizers of the front-end electronics.
One of the crates also houses the NTB, which is responsible for receiving triggers generated by the PTB, distributing them throughout the Nevis TPC Readout System, and informing the event builders of the trigger. The NTB also receives the pulse-per-second (PPS) signal (see section~\ref{sec:Timing}), used to provide an absolute timing reference for matching data to the detector clock used by other readout elements. 

Upon acceptance of a trigger, the NTB generates a signal that is distributed to all 11 crate controllers via a Trigger Fanout module. This also triggers the NTB board reader to make a data fragment, which is pushed to the EventBuilder over the network to initiate the event building process. The TPC board reader programmable waveform length is set to capture the full wire signal waveform over the drift time in the TPC including a buffer before the trigger for baseline correction (see section~\ref{sec:coldelec}). For Run 1, the length was set to 1713.5\,$\mu$s starting 205\,$\mu$s before the trigger. All TPC data (including NTB data) that are read out are timed against the 16\,MHz system clock. 
The TPC readout time is tracked in terms of frames, each 0.57\,ms long, and the sample number subdivision according to the 2\,MHz clock. The trigger and PPS are timestamped with additional precision (64\,MHz), using two phase remainders. The system resets its clock at the beginning of each DAQ run.

In addition to the NU stream, SBND was designed with the capability to operate a supernova (SN) stream which aims to continuously read out compressed waveforms, retrieving temporary, locally stored data only when an external supernova alert is issued~\cite{MicroBooNE:2020mqg} from the Supernova Early Warning System (SNEWS)~\cite{snews}. 
The SN stream uses Huffman compression and a zero-suppression algorithm to reduce the data rates by a target factor of 80. As a result of zero-suppression, only those waveforms with a peak digitized value (amplitude) exceeding a set threshold are saved, and therefore, data in this stream is lossy-compressed data. Unlike the NU stream, the SN stream data is stored locally on disk and does not participate in event building. Commissioning of the SN stream was ongoing, but not completed before the end of Run 1.

\subsection{Photon Detection System}
\label{sec:PDSDAQ}

The SBND PDS readout consists of independent, but similar, readout systems for the PMTs and X-ARAPUCAs.
The 120 PMTs are read out by 8 CAEN V1730SB digitizers~\cite{caen1730}, capable of recording waveforms for 16 channels independently at a sampling frequency of 500\,MHz. 
Each of the eight V1730SB digitizers reads out 15 PMTs in the first 15 readout channels and uses the 16th channel to digitize an attenuated version of the ``flash trigger'' signal sent from the PTB (see section~\ref{sec:Trigger}) to be used in downstream reconstruction to synchronize the timing and remove timing jitter between the digitizers.
The 192 X-ARAPUCAs are digitized by four V1740B digitizers~\cite{caen1740}, that can read out 64 channels at a sampling frequency of 62.5\,MHz.
A 6.25\,MHz clock is distributed to all the V1730SB and V1740B digitizers through a fanout. The digitizers then synchronize to this clock signal via PLL (Phase-Locked-Loop), a process critical to maintaining a $\mathcal{O}$(0.6\,ns) timing resolution.

When the CAEN V1730SB (V1740B) digitizers receive trigger signals (flash triggers, see section~\ref{sec:Trigger}), a 10\,$\mu$s (12.288\,$\mu$s) waveform is recorded in each channel centered such that 20\% of the waveform time duration is from before the trigger signal was received. If additional trigger signals are received during a readout, the recorded waveform is extended to include the full post-trigger waveform for the latest trigger received. The board reader creates fragments for each digitizer and waveform, which are picked up by the event builder if they occur close $\mathcal{O}$(5\,ms) to the TPC readout window of an observed trigger in the NTB.

\subsection{Cosmic Ray Tagger}
\label{sec:CRTDAQ}

The 7 CRT walls are read out using 142 custom-designed Front End Boards (FEB) as described in section~\ref{sec:crtelectronics}. 
The FEBs trigger, construct, store, and output CRT information in the form of ``hits''. There are two kinds of CRT hits: ``reset'' and ``data''. 
When one of two counters (T0 and T1) inside the FEB receives a reset signal via TTL, a ``reset hit'' is triggered, and the internal counter is set back to zero.
These counters are used to construct timestamps for all CRT hits by counting nanoseconds since the last TTL signal was received. The T0 counter is used for the common PPS signal, described in section~\ref{sec:Timing}, and the T1 counter uses a signal from the PTB, derived from beam or off-beam early warning signals, so that the CRT will reset outside the event window and hits can be aligned with the beam spill or equivalent off-beam reference.

A standard data hit occurs when sufficient coincidence conditions are met. A 4-fold coincidence is required for CRT data to be stored. The first layer requires both of the channels in a strip to be above a light threshold, which helps to avoid spurious hits from noisy channels. When this first condition is met, the FEB sends a signal to the modules in the layer orthogonal to it. If the FEB then receives a signal back from one of these orthogonal FEBs within 150\,ns then it records a hit, otherwise the buffer is cleared. This indicates an XY-coincidence in the plane, establishing the second layer of coincidence. A diagram of the coincidence scheme can be seen in figure~\ref{fig:crtwallprinciple}. Further details on CRT layout and daisy chains are found in section~\ref{sec:CRT}.

The BoardReader servers receive CRT hit data via Ethernet connections with the FEBs. Each FEB has one Ethernet port to send data and one to receive it; this allows for multiple FEBs to be read out in a daisy chain. A total of 21 different daisy chains are used for the 142 FEBs, where each chain has its own associated board reader process on the connected server.
Every 10 ms the CRT board reader begins a poll of the FEBs in its daisy chain to collect the previous 80\,ms of CRT hit data, which it then packages into fragments, one per FEB. 
Fragments are given timestamps derived from the median poll time with a correction from the T0 (PPS) time and are added to the boardreaders' buffer for event building.

\subsection{Timing}
\label{sec:Timing}

The timing subsystem is responsible for distributing and recording timing, beam, and trigger signals across all subsystems.
The timing distribution enables readout electronics not only to synchronize with each other but also with the accelerator cycle which extracts protons to the Booster Neutrino Beam (BNB). Meanwhile, recorded timing reference signals provide additional precise handles for downstream data alignment and event reconstruction. The timing subsystem is represented by the magenta connections in figure~\ref{fig:daqarch} with a more detailed view shown in figure~\ref{fig:DAQDiagram} in appendix~\ref{appendixDAQ}. 

The beam interface system is responsible for receiving and distributing signals related to the BNB from Fermilab's Accelerator Division (AD) within the SBND timing system.
The three main signals are the Start of BNB Booster Cycle (\$1D), gated Beam Extraction Signal (BES), and Resistive Wall Monitor (RWM)~\cite{AguilarArevalo:2008yp}. 
The \$1D and BES signals are AD control signals; \$1D signal indicates that the next bunch of protons in the Booster are going to be sent to the BNB, and the BES signal occurs when the Booster extraction is performed.  These are available to SBND through the Multi-Function Timing Unit (MFTU) connected to Fermilab's Accelerator Control Network (ACNET)~\cite{acnet}. 
The RWM signal is a digital signal that indicates that protons have arrived at the resistive monitor immediately upstream of the BNB target, it is the main indicator that a BNB beam spill has occurred. The signal is directly connected from the BNB target hall to the SBND building via an optical fiber, and converted to a TTL signal for distribution. 

\begin{figure}[tbp]
        \centering
          \includegraphics[width=0.85\textwidth]{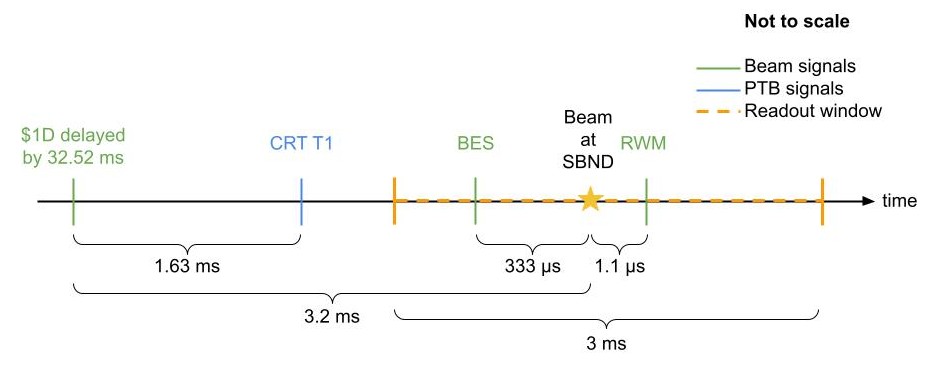}
        \caption{Diagram showing the time intervals across different beam signals arriving at SBND.}
        \label{fig:timing_diagram}
\end{figure}

Both the \$1D and BES signals are utilized by the trigger subsystem to establish the time window of interest for beam activity, as well as to implement a lockout system that guarantees no resets, downtime, or background data collection during the beam spill. 
The \$1D signal comes very early, about 36\,ms before the beam is expected to arrive at the SBND building. The \$1D is delayed by 32.52\,ms in the MFTU so the trigger subsystem sees it 3.2\,ms before the beam arrival. 
The BES comes about 333\,$\mu$s before the beam is expected to arrive at SBND.
Figure~\ref{fig:timing_diagram} summarizes the time intervals across the three beam signals.

The basis for SBND's timing subsystem is CERN's White Rabbit Network (WR)~\cite{whiterabbit}. 
The system starts at the WR master switch, which is synchronized to a GPS-locked, rubidium clock to get an absolute time reference with subnanosecond-level stability. 
The switch distributes this stable time and frequency information to other WR nodes within the network via optical links. 

The first WR node is responsible for sending clock signals that act as referencing frames to readout electronics.
The node contains two Simple VME FMC (FPGA Mezzanine Card) Carrier Fine Delay (SVEC FD) cards~\cite{svec}.
Each card contains four output channels, each of which can be individually programmed to output either a standard PPS signal or a PPS-synced clock signal. 
The PPS signal is distributed to all readout electronics to provide the same reference frame while the electronics' clock signal is a customized frequency for individual readout hardware's internal clocks to latch onto, thereby maintaining their stability. All custom clocks are derived from the WR system. 
SBND uses the SVEC FD outputs to make: (1) PPS signals, (2) a 10\,MHz clock for the trigger system, (3) a 6.25\,MHz clock for the PDS system, and (4) ``fake beam'' signals that mimic the BNB signals described below and are used to collect cosmic background data. 

The second node is responsible for recording subnanosecond precision timestamps for a set of input logic signals, by hosting a Simple PCIe FMC Carrier Time to Digital Converter (SPEC TDC) card~\cite{spectdc}. 
The card interfaces with the DAQ in the same way as other readout electronics, allowing these timestamps to be incorporated into events alongside the TPC, PDS, and CRT data for precise monitoring and alignment of trigger and beam signals. 
The SPEC TDC has five input channels, which are used to record the following signals: the CRT T1 reset signal from the PTB, the gated BES signal and RWM signals for the BNB, and the Event Trigger and Flash Trigger signal from the PTB.

\subsection{Trigger}
\label{sec:Trigger}

The trigger system is responsible for selecting interactions of interest from the background of cosmic rays (on the order of kHz~\cite{uboone_cosmic_rate}), while enabling the capture of a tunable proportion of beam-related interactions, calibration samples, and events for background characterization. 

The main trigger decisions come from the hardware trigger system, which is built around two core components: the Penn Trigger Board (PTB) and the Master Trigger Card Analog (MTC/A). The MTC/A is a multi-channel, multi-threshold analog discriminator board repurposed from the SNO experiment~\cite{SNO}, while the PTB is a versatile and configurable logic trigger board designed for flexible use in particle physics. These two boards operate together to issue triggers based on the inputs from the PDS, the beam signals, and the CRT, all of which are analyzed using programmable logic to reach a final trigger decision. Further refinements of the selected events can be done using software triggers. 

The signature for neutrino interactions is the coincidence of scintillation light detected by the PMTs with a gate opened at the time when neutrinos are expected to arrive at SBND, described in section~\ref{sec:Timing}. Each PMT digitizer is programmed to output a logic signal that is proportional to the number of PMT channel pairs that exceed a configurable threshold. The pairing scheme is described in section~\ref{sec:PDS} and the digitizers are programmed to register a pair over threshold if either of the channels in the pair crosses the threshold. Such an event contributes a fixed-width 80\,ns square pulse to the digitizer’s logic output, with coincident pairs resulting in stacked pulses of higher amplitude. The width of the pulse ensures that light signals from the full extent of the detector will be correctly accounted in the sum. Each of these 8 digitizer-level sums are routed to the MTC/A, which produces a fast analog sum of all these signals, giving a detector-level sum of channel pairs over threshold. The global sum is then compared to 3 programmable thresholds (LO, MED, and HI) and the MTC/A emits a 20\,ns logic pulse to the PTB along the relevant path any time the threshold is exceeded. Finally, the PTB integrates these low level objects, trigger primitives, with the beam signals described in section~\ref{sec:Timing} to decide whether to issue a readout trigger. In Run 1, the thresholds were set to 50\,ADC (roughly 4\,PEs) for each PMT digitizer channel and 3 PMT pairs above threshold for the MTC/A. This combination was chosen to minimize the loss of triggers in time with the expected beam while suppressing triggers on out-of-time light. This results in an event trigger on approximately 10\% of beam spills. The contribution of cosmics rays to the onbeam+light trigger rate is directly measured to be ~25\% with the offbeam+light trigger. 

The PTB issues ``flash triggers'' based on the number of PMT channels above a discriminator threshold applied in the PMT digitizers. Flash triggers are sent to the PDS readout, described in section~\ref{sec:PDSDAQ}, to allow readout of the PDS channels multiple times per event trigger since the PDS readout window is shorter than the TPC readout window as shown in figure~\ref{fig:eventtiming}. 
The flash triggers use the same PMT thresholds and logic in the PMT digitizers as the event triggers. In the MTC/A a separate threshold is available for the flash triggers which was set to 4 PMTs in Run 1. The flash trigger output is sent to all the PDS digitizers to capture additional PDS waveforms as shown by the blue boxes in figure~\ref{fig:eventtiming}. The PDS data is captured at the event trigger time even if the higher flash trigger threshold is not met as shown by the red box in the figure.   

The PTB is a flexible general-purpose board using a MicroZed™ system-on-chip (Xilinx Zynq®-7020) to implement experiment-specific FPGA logic configurable at run time. The board accepts input signals from the PDS, CRT, and beam and timing subsystems using multiple logic standards (NIM, TTL, ECL, LVDS) and standardizes them to 3.3V LVCMOS for internal processing. The firmware architecture is structured around Low-Level Triggers (LLTs) from individual subsystems which are then combined into High-Level Triggers (HLTs). The logic of each LLT and HLT can be configured at the start of each run with a wide range of options (AND, OR, count, etc.), allowing for coincidences, vetoes, and prescaling. Additionally, the PTB supports output and input shaping, masking, and delays, which allows precise coordination with beam spill timing and readout windows, and provides each subsystem with signals of sufficient width. Every LLT and HLT is timestamped, packaged into 128-bit words, and sent to the DAQ, creating a complete record of the trigger decision.

The final level of logic in the PTB firmware, the Trigger Logic Unit, labels each HLT as either an event trigger, flash trigger, or reset. Event triggers are sent to the NTB and hence result in building an event. Since the TPC readout length is much longer than the PDS readout window, multiple flash triggers are sent to the PDS digitizers per event. The flash triggers use an independent MTC/A threshold and can be emitted during a window before each beam spill (determined based on the beam signals) and after each event trigger. The reset HLTs are sent to the CRT as described in section~\ref{sec:CRTDAQ}. 

The PTB allows concurrent operation of multiple trigger types (HLTs). Table~\ref{tab:trigTypes} %\ref{tab:ceprop}
summarizes the event triggers used during the first data taking period in 2025 and their purpose. A ``zero bias'' trigger, which is prescaled during beam data taking, has no requirements on the amount of light present. A trigger with ``+ Light'' in the name requires the 4 PMT pairs above 3\,PE threshold described above.   
The crossing muon trigger requires coincident CRT hits in opposite side walls, providing a data stream of muons for calibrations that is enhanced in track angles not well populated by unbiased cosmic data.  This trigger is inhibited when beam gates are active.

\begin{table}[htb!]
  \caption{SBND Event Triggers. In each case there are also accompanying flash triggers 1.5\,ms after each event trigger emitted to the PDS when there is light in the detectors.  
  Apart from crossing muon triggers, there are also flash triggers 1.5\,ms before the event trigger. Off-beam triggers mimic the same logic as BNB triggers, using ``fake" beam signals from the WR-TDC instead of real \$1D and BES signals.}

\centering
%\begin{tabular}{c c c}
%\hline \hline{}
%\textbf{HLT ID} & \textbf{Trigger Name} & \textbf{Purpose} \\ \hline \hline
%2 & BNB + Light & Neutrino interactions \\ \hline
%1 & BNB Zero Bias & Trigger efficiency \\ \hline
%4 & Off-beam + Light & Cosmic background measurement \\ \hline
%3 & Off-beam Zero Bias & Cosmic background overlays \\ \hline
%14, 15 & Crossing Muon (North-South, East-West)  & Calibration \\ \hline \hline
%\end{tabular}
\begin{tabular}{c c}
\hline \hline{}
\textbf{Trigger Name} & \textbf{Purpose} \\ \hline \hline
BNB + Light & Neutrino interactions \\ \hline
BNB Zero Bias & Trigger efficiency \\ \hline
Off-beam + Light & Cosmic background measurement \\ \hline
Off-beam Zero Bias & Cosmic background modeling \\ \hline
Crossing Muon (North-South, East-West)  & Calibration \\ \hline \hline
\end{tabular}

\label{tab:trigTypes}
\end{table}

    \begin{figure}[tphb]
    \centering
    \begin{subfigure}[b]{0.32\linewidth}
        \includegraphics[width=\linewidth]{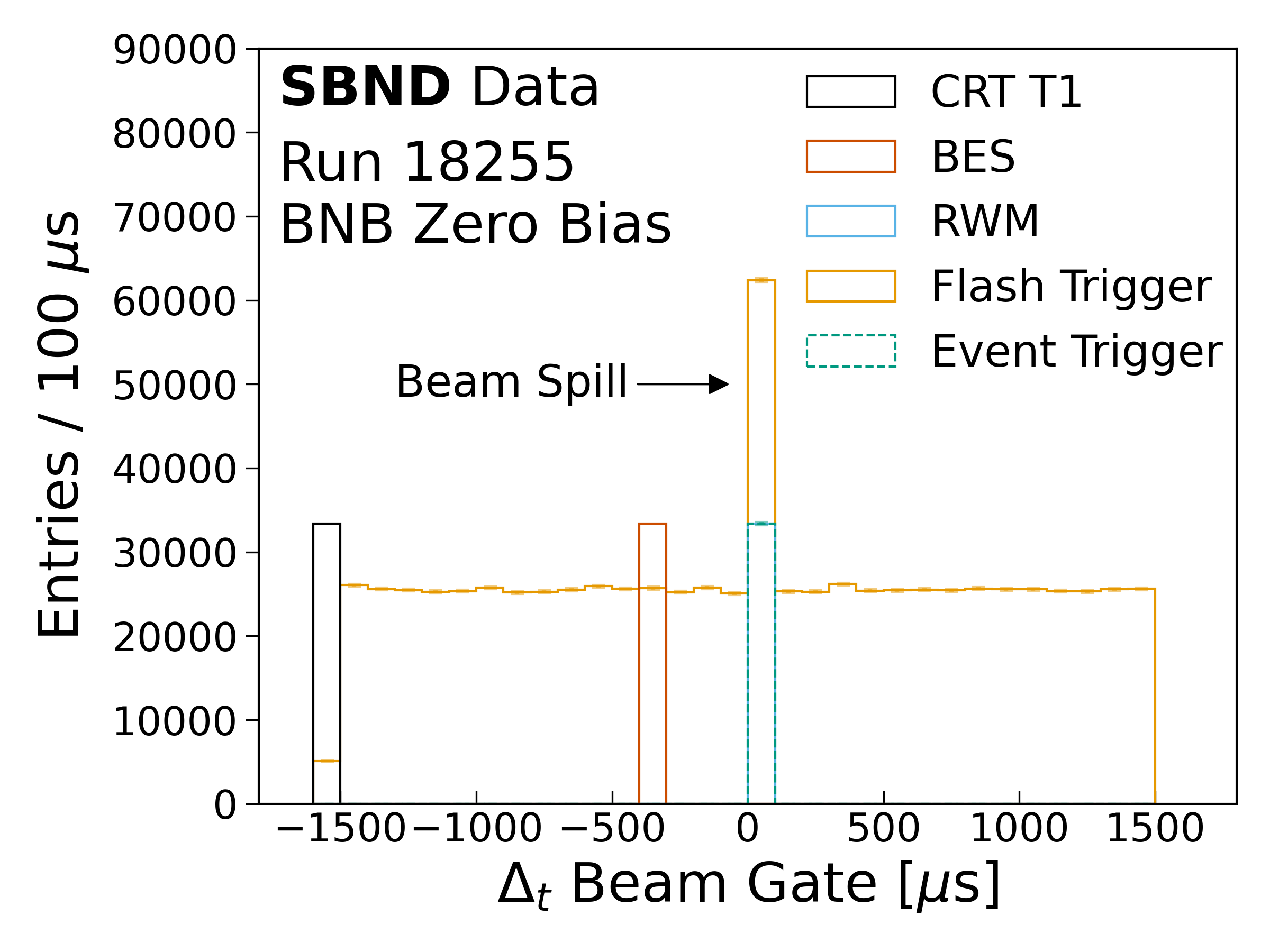}
        %\caption{}
        %\label{fig:ts_beam_spill_zb}
    \end{subfigure}
    \begin{subfigure}[b]{0.32\linewidth}
        \includegraphics[width=\linewidth]{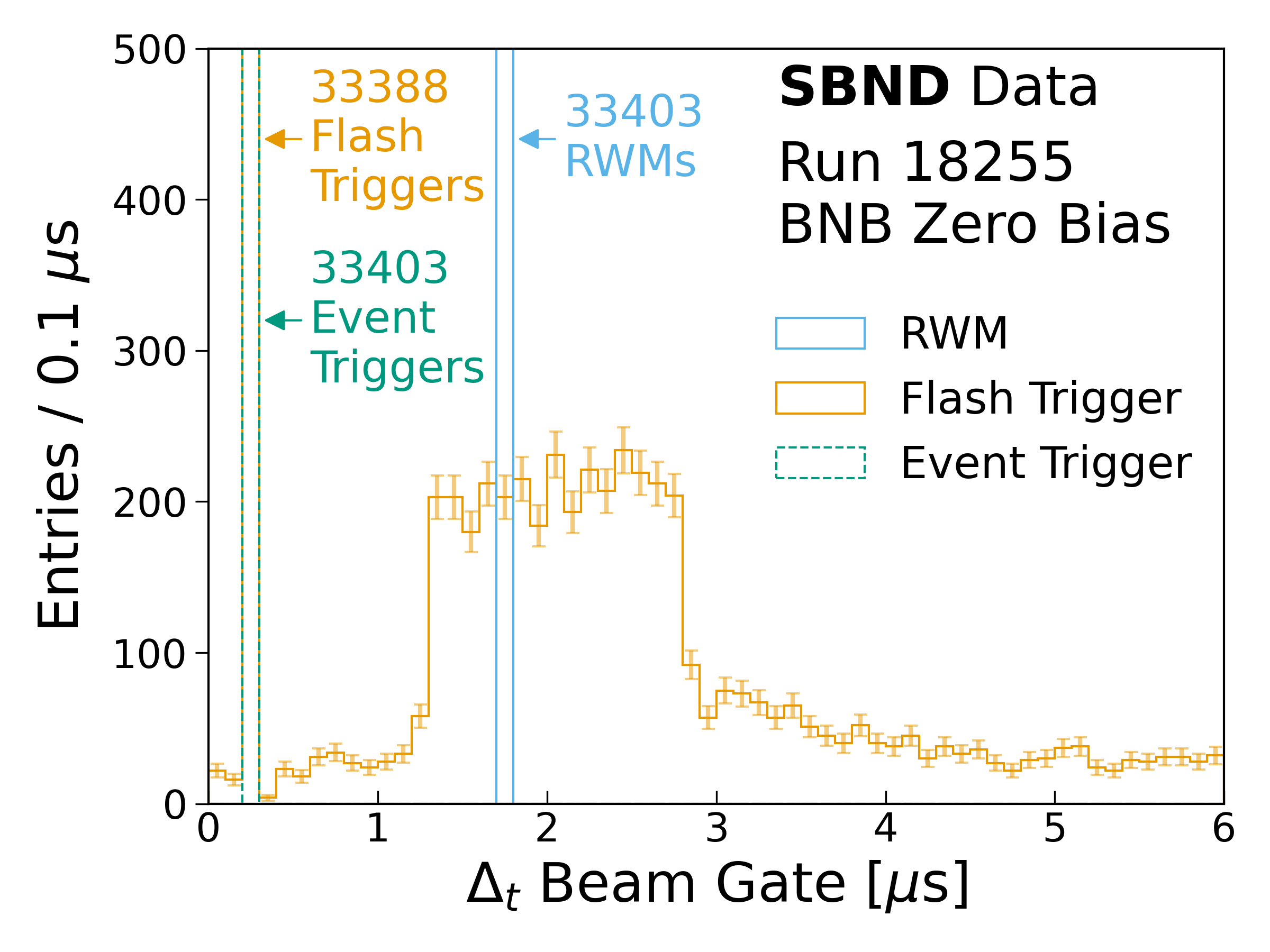}
        %\caption{}
       %\label{fig:ts_beam_spill_zoomed}
    \end{subfigure}
        \begin{subfigure}[b]{0.32\linewidth}
        \includegraphics[width=\linewidth]{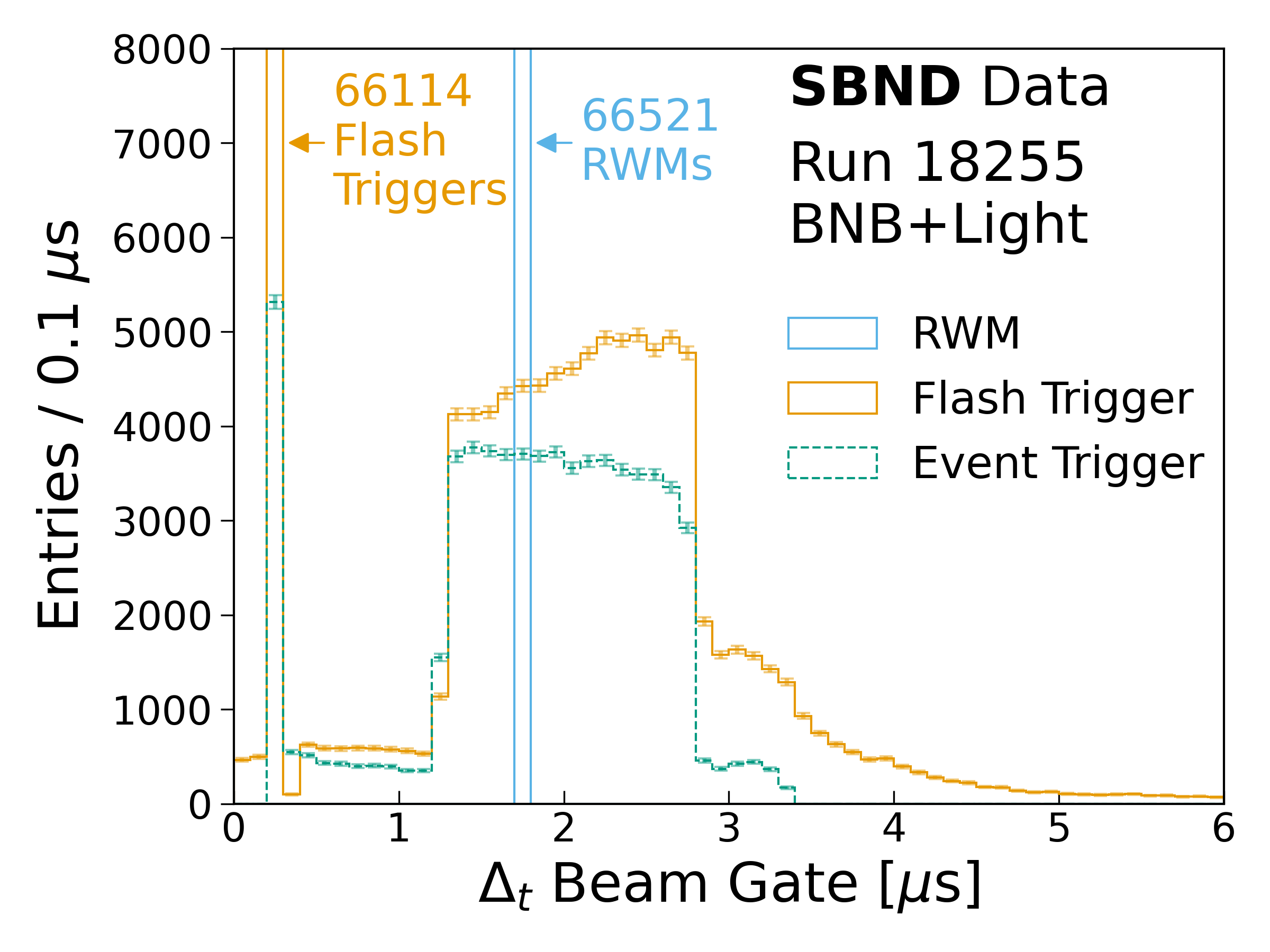}
        %\caption{}
       %\label{fig:ts_beam_spill_zoomed}
    \end{subfigure}
    \caption{All beam and trigger signals for a BNB Zero bias event seen in the SPEC TDC (left). Zoomed in distributions of beam and trigger signals for the beam spill region of [0, 6]\,$\mu$s for BNB Zero Bias triggers (center) where the Event Trigger occurs with the opening of the beam window, and BNB+Light triggers (right) which occur when the PMTs go over the light threshold.}
    \label{fig:ts_beam_spill}
\end{figure}

Figure~\ref{fig:ts_beam_spill} shows the beam and timing signals around the beam gate for a 3\,ms window (left), and a zoomed in 6\,$\mu$s window (center, right) for a run with concurrent BNB Zero Bias (left, center) and BNB+Light (right) trigger streams. %Figure~\ref{fig:ts_beam_spill_zb} 
Figure~\ref{fig:ts_beam_spill}~(left) shows each signal received by the SPEC TDC, where the CRT T1 reset occurs 1.63\,ms after a \$1D is received, the event trigger occurs 333\,$\mu$s after the BES, and the flash trigger (FTRIG) window is open for 1.5\,ms on either side of beam arrival. %In figure~\ref{fig:ts_beam_spill_zoomed} 
The [0, 6]\,$\mu$s window around the beam gate opening is shown for a BNB zero bias trigger in figure~\ref{fig:ts_beam_spill}~(center) and BNB+Light trigger in figure~\ref{fig:ts_beam_spill}~(right). In the zero bias stream, the event trigger is gated to the BES signal such that it is stationary within the window. In the light stream, 
the additional light threshold condition results in the event trigger moving anywhere within the window. 
This demonstrates the clear distinction between the two trigger streams described in table~\ref{tab:trigTypes} that are leveraged for neutrino collection and trigger efficiency studies. 

    The software trigger, a secondary level of triggering after the hardware trigger, uses \texttt{art} modules to form metrics or primitives calculated swiftly from the raw data immediately after the DAQ has assembled it together, but before it is saved in a file as an ``event''. %Since the 
    The software trigger calculates metrics from the assembled data, and uses these metrics to classify the event. 
    Different classifications are written to different output streams or may be dropped if they fail selections. Example metrics include: PTB HLT or LLT trigger words, the total amount of light detected by the PMTs, and the number of CRT hits during the beam spill for each CRT plane. Current output streams correspond to the trigger types outlined in table\,\ref{tab:trigTypes}.

\subsection{DAQ performance}
\label{sec:DAQPerformance}

SBND realized DAQ and trigger performance surpassing the original design specifications of a maximum average data rate of 5\,Hz and 270\,MB/s. During commissioning, stable day-long runs of complex multi-trigger configurations at an 8\,Hz average event rate and an average event size of 50\,MB were achieved. Typical events are comprised of 13\,MB from the TPC, 37\,MB from the PDS and timing, 0.6\,MB from the CRT, and 0.02\,MB from the trigger system. Data compression while creating files for disk storage further reduces the event size by factor of ${\sim}2$\ with largest reduction occurring for the PDS data.     

\begin{figure}[hbt]
  \centering
 % \subfloat{\includegraphics[width=0.60\textwidth]{figures/CryostatMarkIIILayout.png}}
  \subfloat{\includegraphics[width=0.98\textwidth]{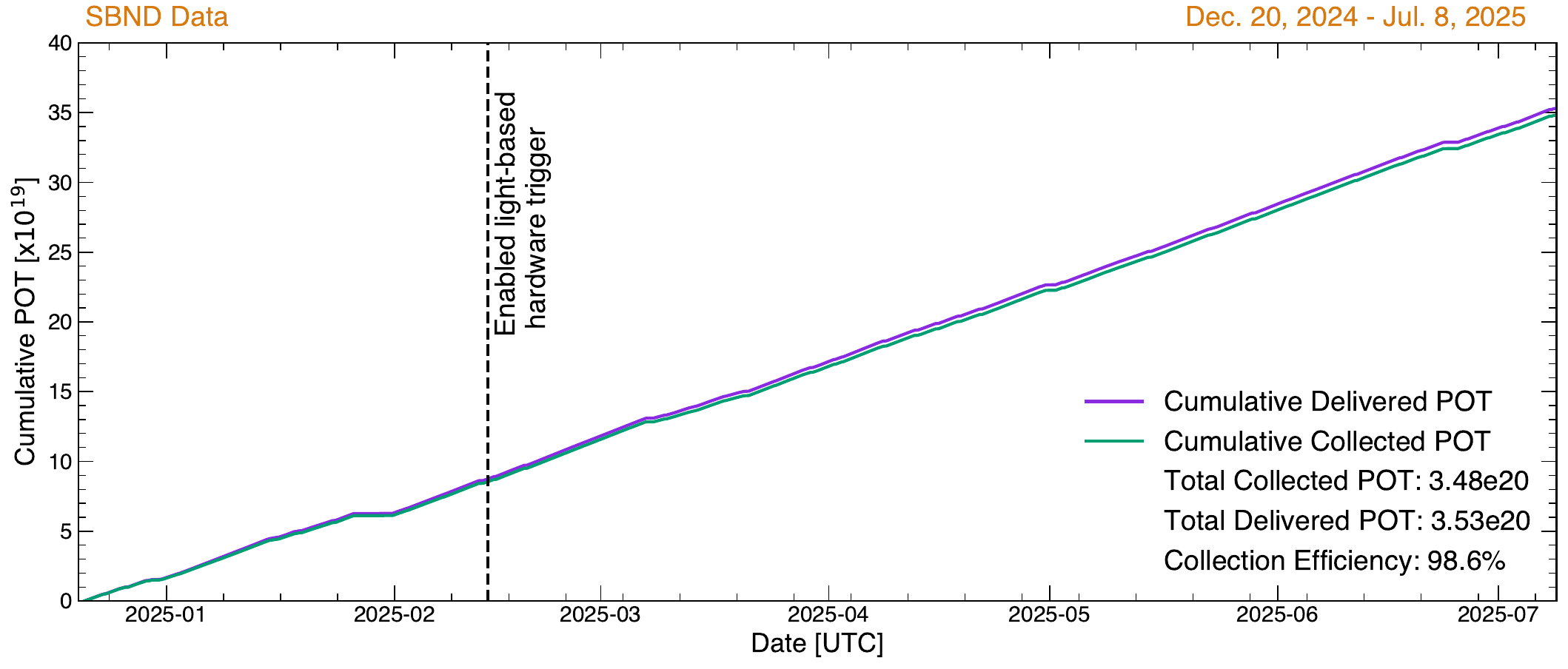}}
\caption{Cumulative protons on target (POT) delivered (purple) to and collected (green) by SBND during Run 1. } 
\label{fig:run1}
\end{figure}

Figure~\ref{fig:run1} shows the beam delivery and accumulation by SBND during Run 1 operations with an overall data collection efficiency of 98.6\%. The data loss of 1.4\% is dominated by three causes of similar weight: DAQ or Trigger system hang or crash (0.65\%); loss while run is stopped due to detector problem, such as power supply trip, HV spike, or noise in TPC (0.31\%), and loss while DAQ run is starting after the return of beam delivery (0.29\%). The remaining 0.14\% of losses occurred due a wide range of different causes. The DAQ system was designed to incur no readout deadtime during normal beam operations, and during Run 1 no significant data loss was observed from deadtime.  However, the configuration during Run 1 had a small error in the length of the trigger inhibit that prevents non-beam triggers in advance of the beam bucket.  This resulted in a loss of about 0.1\% of events and was corrected after Run 1 concluded.  

\begin{figure}[hbtp]
    \centering
    \includegraphics[width=0.7\textwidth]{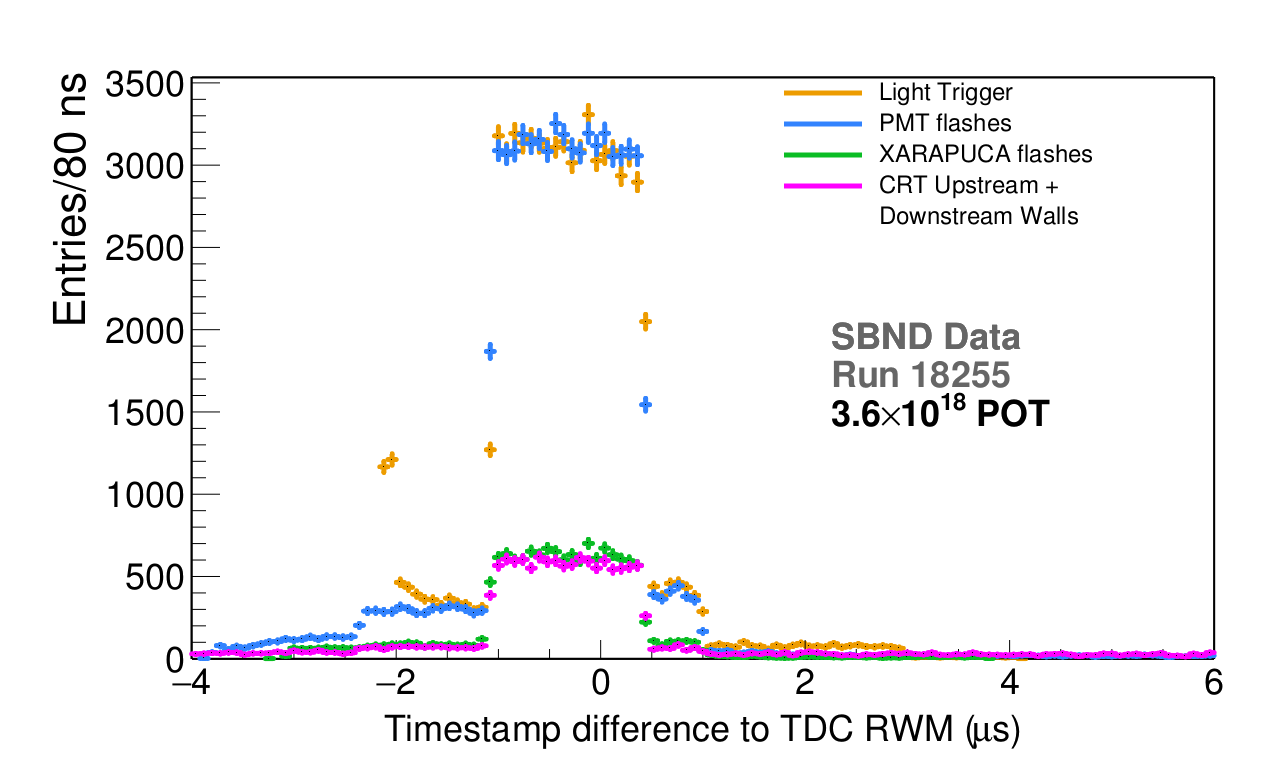}
    \caption{The observed time of neutrino beam arrival signals in various subsystems as referenced to the RWM recorded in the SPEC TDC. 
    Data labeled "Light Trigger" shows the BNB+Light high level triggers from the PTB occurring closest to the RWM. Other signals shown are the instances of flashes from the PMTs and X-ARAPUCAs recorded with CAEN V1730SBs and V1740Bs respectively, and activity from either the upstream or downstream CRT walls.}
      
        \label{fig:daq_combinedTopHat}
\end{figure}

The excellent initial timing resolution was previously shown for the PMT system (figure~\ref{fig:pmt_timing_plots}) and the CRT system (figure~\ref{fig:crtbucket}).  The timing synchronization of subsystems and correct event building is demonstrated through examining the timing of beam dependent signals for each relative to a common beam time reference. If synchronized, all subsystems should synchronously observe the 1.6\,$\mu$s long beam spill. As a common stable reference point, the RWM signal recorded by the SPEC TDC is used, as discussed in section~\ref{sec:Timing}. 
Figure~\ref{fig:daq_combinedTopHat} shows signs of beam arrival above background rates in various subsystems: spacepoints from the upstream and downstream CRT walls, observed flashes in the PMTs and X-ARAPUCAs, and light-based triggers in the PTB; the timing of each is plotted in reference to the SPEC TDC's recorded RWM timestamp. 

The subsystems are clearly synchronized to better than 100\,ns prior to calibration ensuring that data from all subsystems will be correctly recorded by the DAQ for each beam trigger.
The beam window opens at -2\,$\mu s$, resulting in a spike in the number of HLTs from light occurring before the window.

\section{Summary of SBND performance in the first neutrino beam run}
\label{sec:summary}

The preceding sections described the as-built designs of the various subsystems of the SBND neutrino experiment at Fermilab and presented key technical performance results from first operations for each system.
This section summarizes the timeline from detector installation to operation in a neutrino beam and presents the performance of the experiment in the first year of beam operation prior to offline calibration. 
We briefly summarize the major steps of detector commissioning, but a detailed description is beyond the scope of this paper.

The installation of the SBND detector and cryogenics systems was completed at the end of 2023 with the exception of the X-ARAPUCA readout electronics. 
The initial steps of commissioning took place in parallel with installation. Each subsystem was individually tested to ensure basic functionality and proper integration of the detector readout with the DAQ, timing, and trigger systems. 

In January 2024, full commissioning of the cryogenics system commenced which included purging the cryostat with gaseous argon, cooling with argon gas, filling with liquid argon, and finally removing impurities from the argon by circulation through the filtration system as described in section~\ref{sec:cryocooldown}. Throughout the cryogenics commissioning period, the detector subsystems were tested to ensure functionality following each cryogenics commissioning stage. By the end of March 2024, the conditions of LAr purity (see figure~\ref{fig:relativeLifetime}) met the experiment's requirements to start turning on each detector element. 

The fully integrated commissioning of the detector subsystems with neutrinos from the BNB commenced in March 2024. 
With beam available until mid-July 2024, each detector subsystem went through a detailed step-by-step process of power up, testing, and hardware parameter tuning to optimize operations with beam. In parallel with the commissioning of the TPC and PDS systems, installation of the east, south, and west CRT walls was completed. On July 3, 2024 the TPC, PMTs, CRT (except top), DAQ, and trigger were operational with the detectors at nominal voltages and the first beam neutrino events were recorded. From July 3 to July 12 the experiment took data with the neutrino beam collecting initial operational experience.  During four months of the subsequent annual Fermilab accelerator shutdown, each subsystem underwent additional tuning based on this initial data taking to ensure optimal performance for the upcoming run. 
Data from the initial operations in July 2024 were also used to start the process of calibrating each detector subsystem~\cite{calibration} for physics analysis. With stable operation of the cryogenics and central detector established, the top CRT modules were installed and commissioned during the 2024 accelerator shutdown. Installation of the APSAIA readout system for the X-ARAPUCA photon detectors was also completed during this period. By the end of the 2024 accelerator shutdown, commissioning of the TPC, PMT, and CRT systems was complete. 

Data-taking with neutrinos from the BNB resumed for 'SBND Run 1' in December 2024 and continued until July 2025 with a data collection efficiency of 98.6\%.  Figure~\ref{fig:run1} shows the cumulative POT delivered to and collected by SBND. During Run 1, SBND collected a factor of two more $\nu-$Ar interactions than the existing global dataset summed across all previous and current experiments.

As described in the previous sections, all systems performed at or above the design performance. Table~\ref{tab:Performance_summary} describes some key performance parameters during Run 1 compared to the experiment requirements. 
The electron lifetime in the LAr has been continuously monitored 
using tracks that cross from the cathode to the anode. The electron lifetime was greater than 10\,ms throughout Run 1 and greater than 30\,ms for more than 95\% of the run. A few short interruptions in LAr circulation and filtration resulted in two periods with lower lifetime.

\begin{table}[htb!]
\centering
  \caption{SBND detector performance summary during run 1 compared to the experiment requirements.}
\label{tab:Performance_summary}
\begin{tabular}{c c c}
\hline \hline{}
 & \textbf{Requirement} & \textbf{Run 1 Performance}  \\ \hline \hline
LAr Purity (e$^-$\ lifetime) & >3$\,$ms &  >\,30$\,$ms (95\% of run) \\ \hline 
S/N ratio per TPC wire plane (U, V, Y) & >5, >5, >12 & 14.5, 23.5, 32.5 \\ \hline
TPC dead channels & <5\% & 80 (0.7\%) \\ \hline
PMT timing resolution & <5\,ns & <2.84$\,$ns \\ \hline
PMT light yield & >10$\,$pe/MeV  & >10$\,$pe/MeV \\ \hline %(across active volume)\\ \hline
PMT dead channels  & <12 (10\%) & 16 (13\%) \\ \hline
CRT timing resolution & <5\,ns & <3.5$\,$ns \\ \hline
CRT dead strips & <120 (5\%) & 12 (0.5\%) \\ \hline
DAQ bandwidth & >5$\,$Hz & >7$\,$Hz \\ \hline
DAQ deadtime & 0\% & <0.1\% \\ \hline

\hline \hline
\end{tabular}
\end{table}

\begin{figure}[hbt]
  \centering
\includegraphics[width=0.32\textwidth]{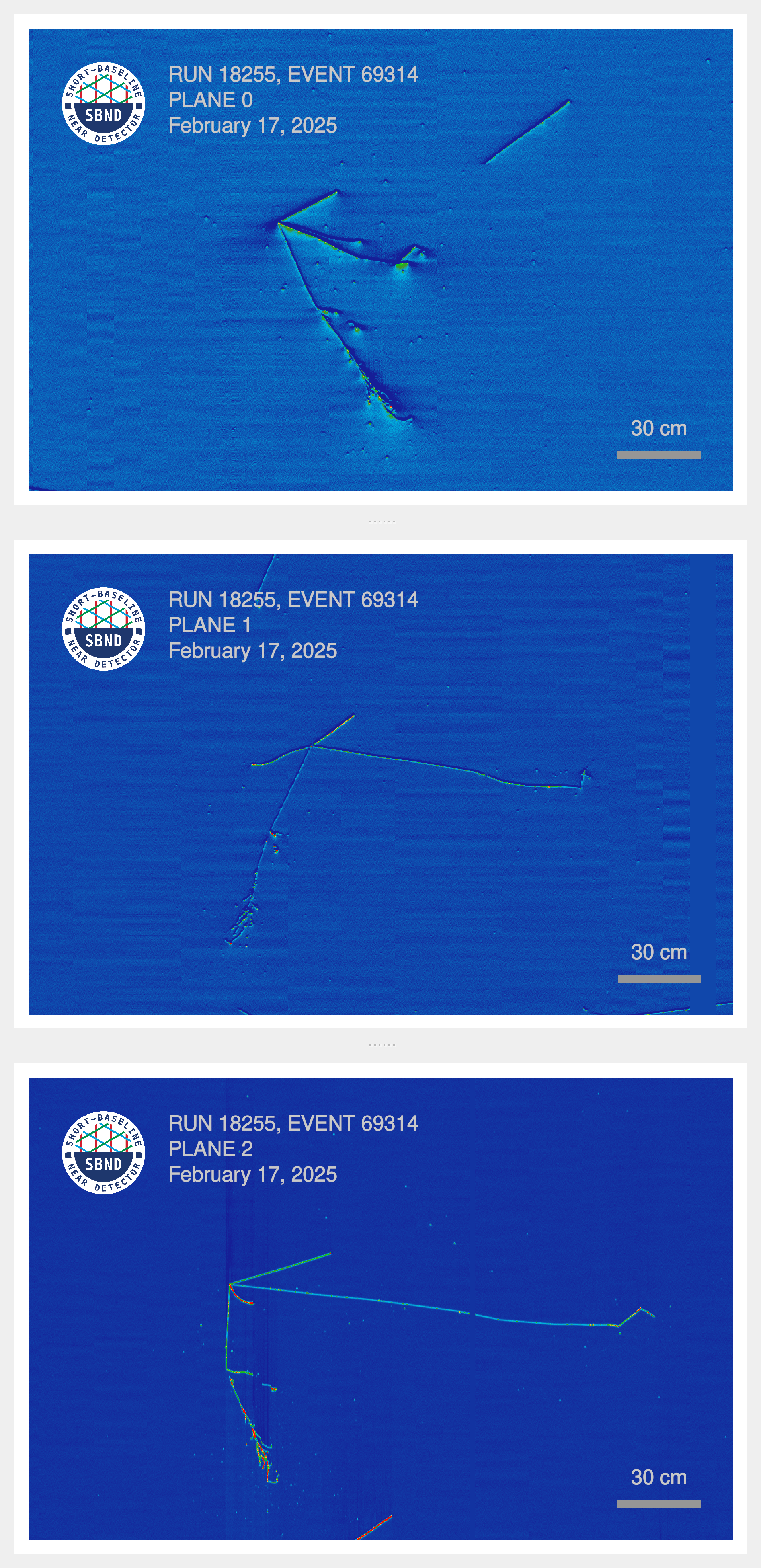}
\includegraphics[width=0.32\textwidth]{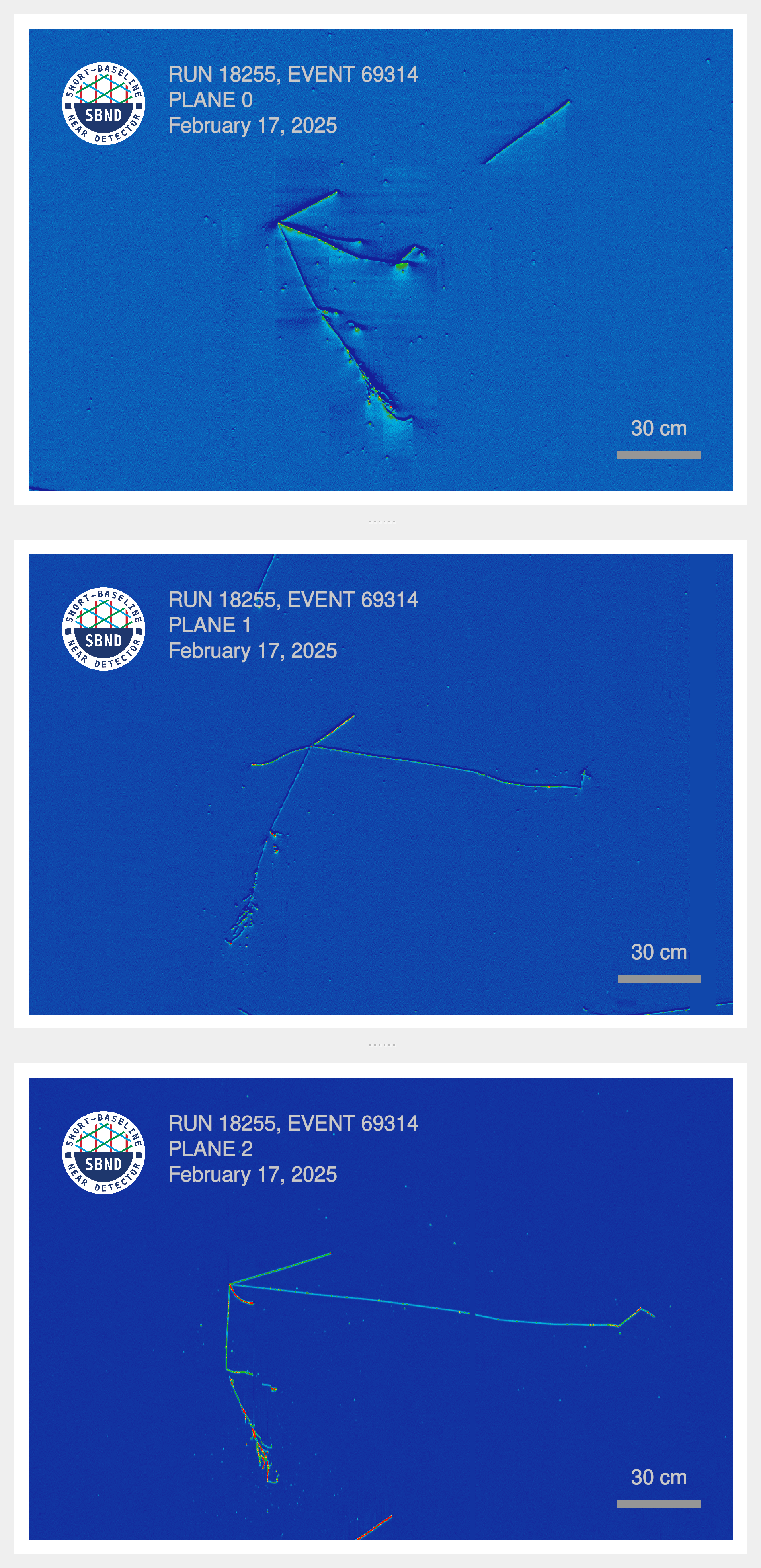}
\includegraphics[width=0.32\textwidth]{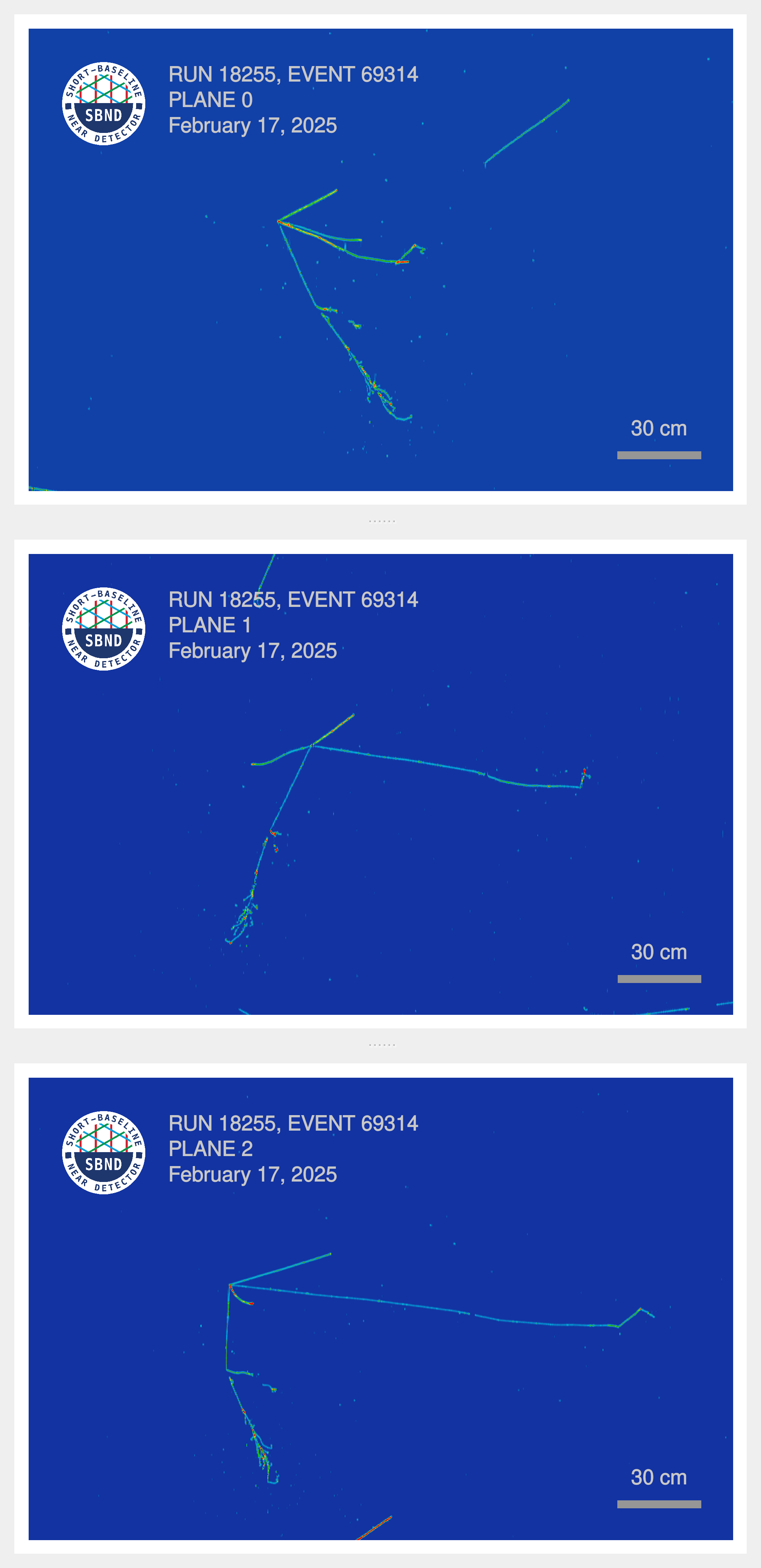}
\caption{Event display of a candidate muon neutrino in the U (top), V (middle), and Y (bottom) TPC planes.  The left column shows the raw data, the center column show the data with coherent noise removed, and the right column shows the data with waveforms deconvolved. 
} 
\label{fig:noiseevent}
\end{figure}

As described in section~\ref{sec:TPCPerformance}, the electronic noise level achieved in the TPC  wire plane readout is essentially at the theoretical floor for the given wire lengths, resulting in signal to noise ratios (SNR) significantly exceeding the experiment requirements.  Combined with the very high LAr purity levels, the high SNRs result in extremely clean events that enhance reconstruction. The low electronic noise levels are illustrated in figure~\ref{fig:noiseevent}, which shows a candidate muon neutrino interaction in the three planes of the TPC. For each plane the event is shown using raw data, data with coherent noise removed, and data after waveform deconvolution. While with raw waveforms the data is very clean, the removal of coherent noise removes many artifacts, particularly on the two induction planes which have lower signal to noise. The TPC dead channels, also described in section~\ref{sec:TPCPerformance}, are dominated by 64 channels of one FEMB that failed prior to the start of cryostat filling with other failures of single wires or pairs spread across the detector. The failure rate during Run 1 operations was negligible. 

The excellent timing resolution of the PMT system was described in section~\ref{sec:PMTtiming}. Even prior to offline calibration, the system can easily distinguish the 19\,ns bunch structure of the BNB beam, providing a handle for cosmic rejection and enabling searches for BSM physics. The light yield of the PMT system is consistently at or above the requirement throughout the volume. The number of PMTs that cannot stably hold voltage without an over-current trip falls short of the design requirements by 3\%, but the rate of failures was significantly reduced after the initial operating period and is currently stable. Since the failed PMTs are spatially distributed, they do not present a significant impact on the PDS system performance. 

Similar to the PMT system, the CRT timing resolution exceeds the requirements prior to offline calibration as shown in section~\ref{sec:crtperformance}. The combination of the PDS and CRT timing resolutions will provide excellent cosmic ray rejection in addition to determining whether tracks traversing the cryostat walls are entering or exiting.  

The DAQ bandwidth and deadtime, described in section~\ref{sec:DAQPerformance}, meet or exceed the experiment requirements. With a bandwidth comfortably larger than the beam delivery maximum of 5\,Hz, it is possible to run the experiment collecting every beam spill rather than just those tagged as containing a neutrino interaction by the trigger.  This provides tremendous flexibility in collecting background data sets and measuring the efficiency of the hardware trigger. 

\begin{figure}[tbp]
  \centering
\includegraphics[width=0.98\textwidth]{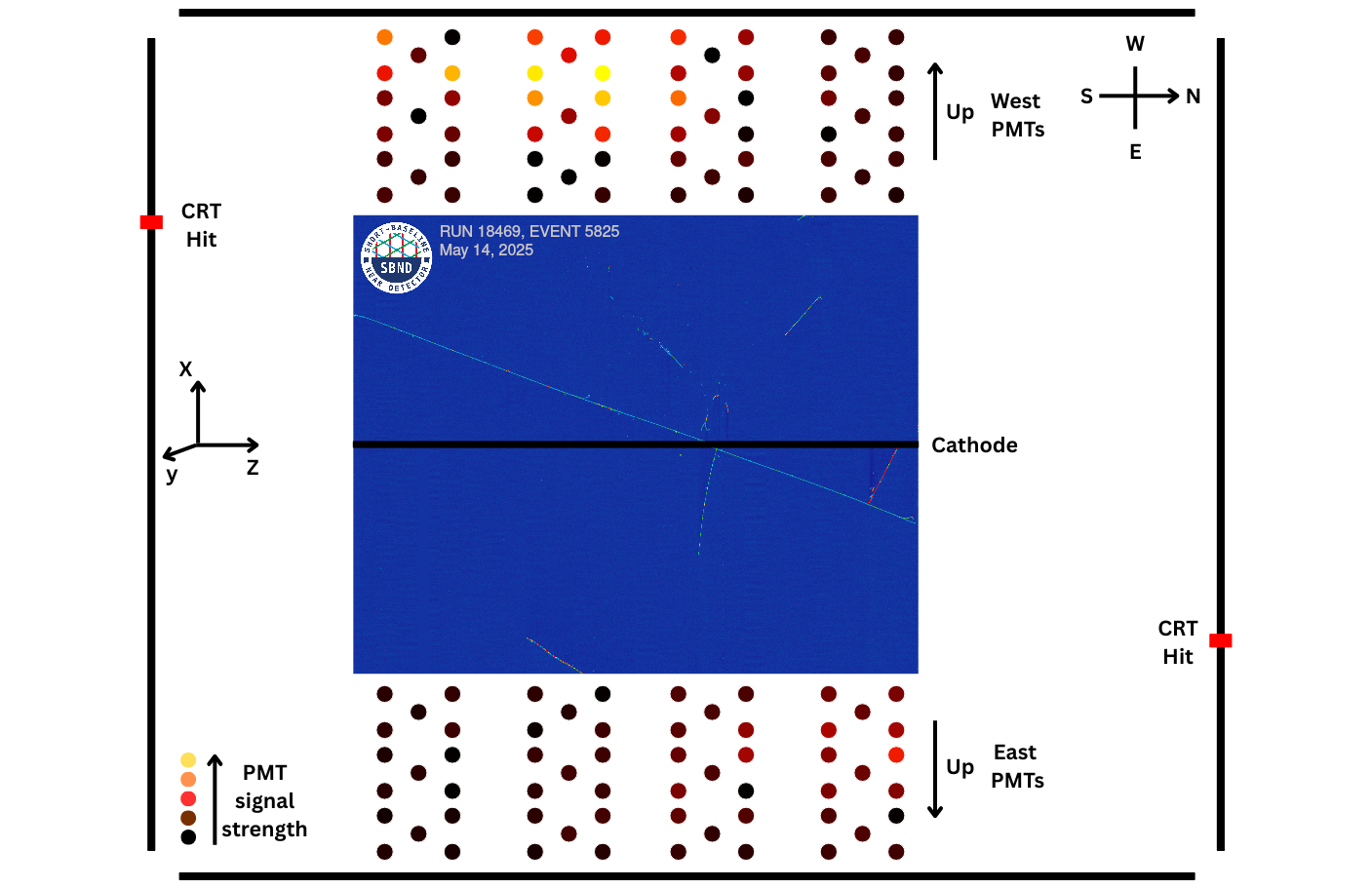}
\caption{Event display of candidate muon track entering from outside the detector pit then passing through the south CRT wall (left), the full TPC volume, and exits through the north CRT wall (right). The view is from the top down with TPC collection (Y) plane shown in full with the cathode in the middle. The PMTs are shown as dots with those closest to the TPC at the bottom with increasing amplitude going from black (no hit) to yellow.  The raw TPC data without coherent noise removal is shown in this display.  
} 
\label{fig:crossing_muon_event}
\end{figure}

Figure~\ref{fig:crossing_muon_event} shows an event display with a candidate muon resulting from a neutrino interacting with the earth or concrete upstream of the detector that passes through the south (upstream) CRT wall, the full 5\,m length of the TPC active volume, both walls of the cryostat, and the north CRT wall. In addition, the muon crosses through the cathode having entered the TPC on the west side and exiting on the east side. Note that the x-coordinate in the TPC is represented by time of arrival of signals on the wires. PMT signals are evident on the upper south end on the west side and lower north end on the east side consistent with the muon crossing the cathode and pointing down as it traverses the detector. These crossing-muon events can be used to calibrate the timing and spatial matching of the TPC, PDS, and CRT detector systems. 

The excellent performance of the SBND detector is further illustrated with displays of three characteristic events recorded in the initial neutrino beam data taking in July 2024 and during early 2025. All of the displays show a zoomed in view of the TPC looking down from the top with the neutrino beam entering from the left. The data displayed is from the collection plane and the raw ADC data has been deconvolved.  

\begin{figure}[tbp]
  \centering
 % \subfloat{\includegraphics[width=0.60\textwidth]{figures/CryostatMarkIIILayout.png}}
 \includegraphics[width=0.85\textwidth]{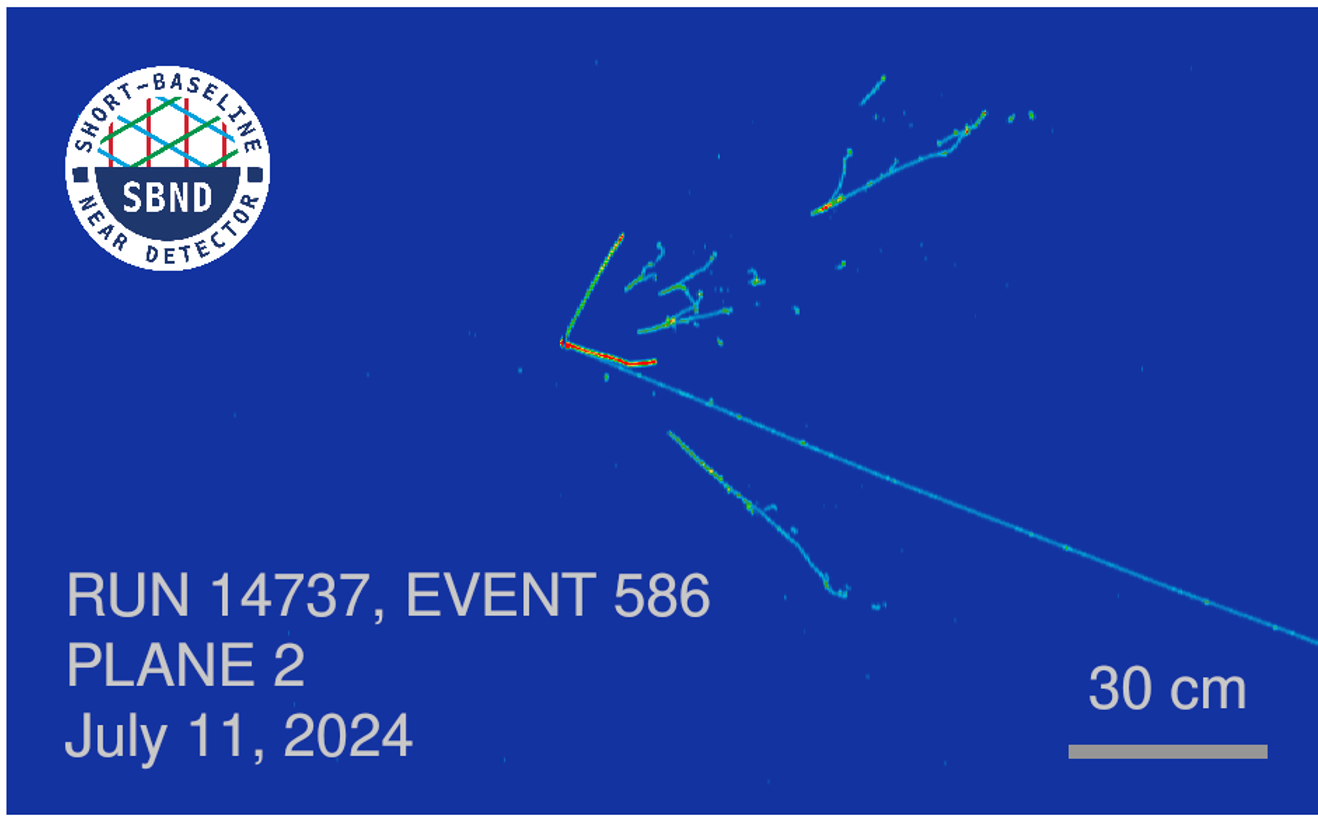}
\caption{Event display of a candidate $\nu_\mu$-argon interaction.  } 
\label{fig:numu_event}
\end{figure}

\begin{figure}[p]
  \centering
 % \subfloat{\includegraphics[width=0.60\textwidth]{figures/CryostatMarkIIILayout.png}}
\includegraphics[width=0.85\textwidth]{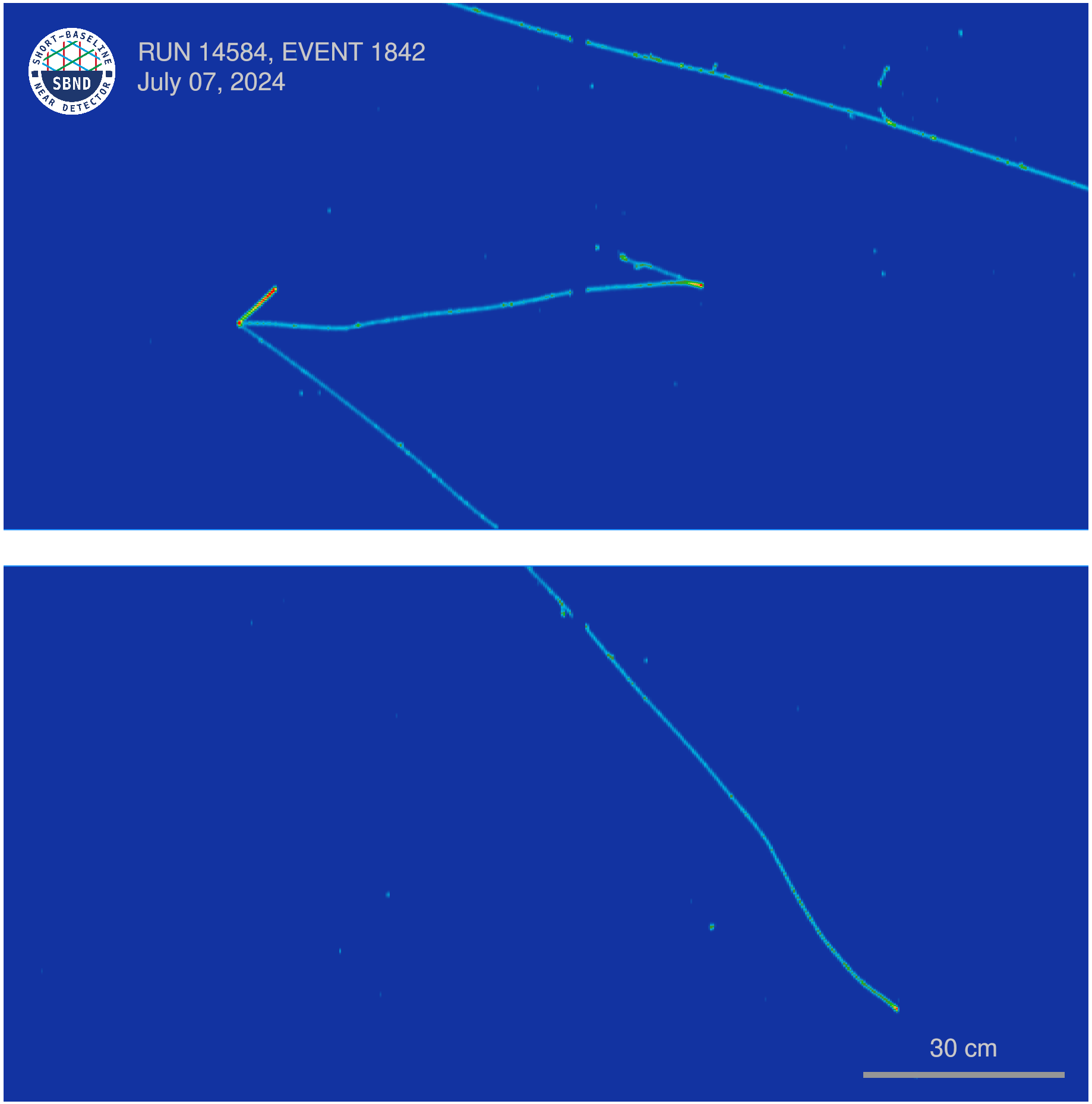}
\caption{Event display of a candidate $\nu_\mu$-argon interaction with the outgoing $\mu$\ track crossing the cathode from one TPC volume to the other.  } 
\label{fig:numu_cathode_crossingevent}
\end{figure}

A $\nu_\mu$\ candidate event from the first week of beam operation is shown in figure~\ref{fig:numu_event}. The resulting muon exits the TPC active volume to the right while a highly ionizing proton leaves a trail of higher intensity hits nearly overlapping the muon track in this projection. Multiple neutral products of the interaction result in showers detached from the primary vertex. Another $\nu_\mu$\ candidate from early operations is shown in figure~\ref{fig:numu_cathode_crossingevent}. In this case the muon crosses the cathode from the west TPC volume to the east TPC volume and is fully contained in the active volume. Such cathode crossing tracks provide an excellent source for cross calibration of the TPCs.  The fact that the track segments on either side of the cathode are roughly in-line demonstrates that the arrival time of the beam signals correctly reflects the actual neutrino arrival time. Also of note is that the hit intensity on the muon track has essentially no change from the primary vertex to the cathode or from the cathode end of the track. This demonstrates the very high argon purity (long electron lifetime) achieved. 

\begin{figure}[tbp]
  \centering
 % \subfloat{\includegraphics[width=0.60\textwidth]{figures/CryostatMarkIIILayout.png}}
  \subfloat{\includegraphics[width=0.9\textwidth]{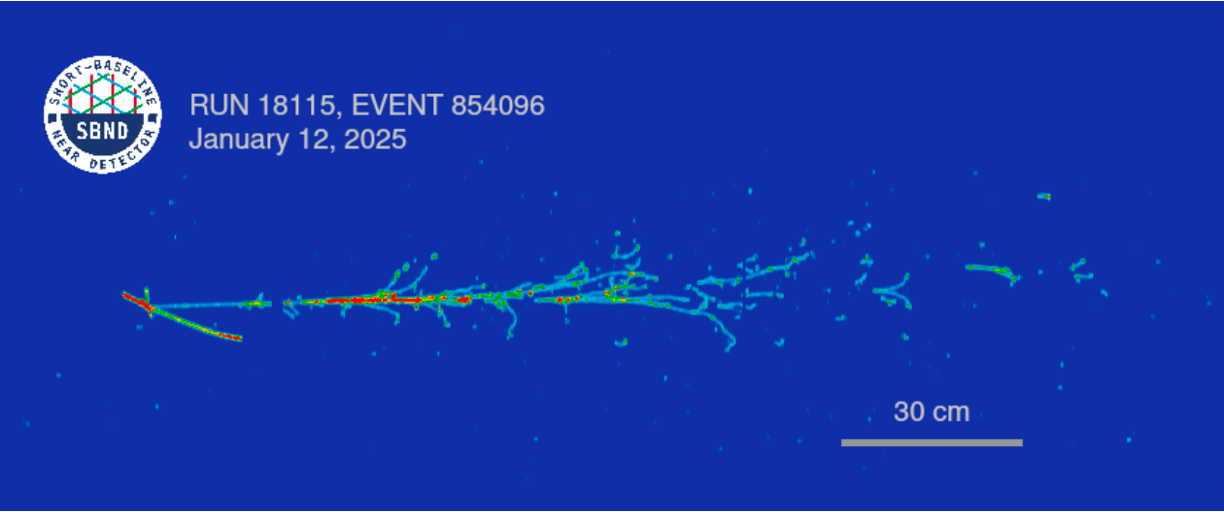}}
\caption{Event display of a candidate $\nu_e$-argon interaction.  } 
\label{fig:nue_event}
\end{figure}

A $\nu_e$\ candidate event from early in Run 1 is shown in figure~\ref{fig:nue_event}. The resulting electron generates an electromagnetic shower while the highly ionizing proton is kicked nearly backward. While much lower frequency than muon neutrinos, there is a non-negligible fraction of the beam that consists of $\nu_e$.  One of the primary functions of the SBND detector, within the SBN program, is to precisely measure the $\nu_e$\ fraction as a control for the $\nu_e$\ appearance oscillation measurement.

% all systems Top Hat  (?) Leave in DAQ, point to it

% crossing muon (calibration, 2 systems) 
% numu event (fab position resolution, has some showers in it)
% numu crossing 2 TPCs (discuss cathode crossing)
% nue (discuss that we do get some!)

% Outlook, vague running plan, total POT, physics prospects
%purity (always greater than X, peak of Y)

\section{Conclusion}
\label{sec:conclusion}

The Short-Baseline Near Detector is the most performant LArTPC built to date. It builds on well-established technology used in MicroBooNE, ICARUS and ProtoDUNE, while paving the way to future large-scale LArTPCs such as DUNE with innovations in detector technology, cryostats, and cryogenics.

SBND boasts several novel achievements, such as the combined use of X-ARAPUCAs and PMTs, the first implementation of wavelength-shifting reflective plates in a large-scale TPC, and the first use of commercial analogue-to-digital converters at cryogenic temperatures.
The excellence in detector performance leverages developments of TPC electronics pioneered by MicroBooNE and ProtoDUNE. 

The SBND detector has been operational since 2024, and is currently taking data. 
Neutrino beamline operations are foreseen until the end of 2027, with a projected total of around $1 \times 10^{21}$\, protons on target of neutrino data, approximately tripling the Run 1 statistics.

Detailed descriptions of the calibration procedures 
and the timing performance of the photon detector system will be presented in forthcoming publications.

\clearpage

%\printglossary[type=\acronymtype]
%\input{acronyms}

\acknowledgments
% Add here all grants and acknowledgements from funding agencies who contributed to detector construction (and ops, etc.)

This document was prepared by the SBND Collaboration using the resources of the Fermi National Accelerator Laboratory (Fermilab), a U.S. Department of Energy, Office of Science, Office of High Energy Physics HEP User Facility. Fermilab is managed by FermiForward Discovery Group, LLC, acting under Contract No. 89243024CSC000002. The SBND Collaboration acknowledges the generous support of the following organizations: the U.S. Department of Energy, Office of Science, Office of High Energy Physics; the U.S. National Science Foundation; Los Alamos National Laboratory for LDRD funding, the Science and Technology Facilities Council (STFC), part of United Kingdom Research and Innovation (UKRI), the UKRI Future Leaders Fellowship (grant number MR/V022407/1), and The Royal Society; the Swiss National Science Foundation; the Spanish Ministerio de Ciencia, Innovacíon y Universidades (MICIU/ AEI/ 10.13039/ 501100011033) under grants No PRE2019-090468, CNS2022-136022, RYC2022-036471-I, PID2023-147949NB-C51 \& C53 and Comunidad
de Madrid (PEJ-2023-AI/COM-28399); the European Union’s Horizon 2020 research and innovation program under GA no 101004761 and the Marie Sklodowska-Curie grant agreements No 822185, 101081478, and 101003460; the São Paulo Research Foundation 1098 (FAPESP), the National Council of Scientific and Technological Development (CNPq) and Ministry of Science, Technology \& Innovations-MCTI of Brazil; the Minas Gerais research foundation (FAPEMIG), grants APQ-00544-23 and APQ-01249-24; the Anusandhan National Research Foundation (ANRF, India) under the Ramanujan Fellowship (Grant No. RJF/2025/000203). An award of computer time was provided by the ASCR Leadership Computing Challenge (ALCC) program. This research used resources of the Argonne Leadership Computing Facility, which is a U.S. Department of Energy Office of Science User Facility operated under contract DE-AC02-06CH11357.

\clearpage
\appendix
\section{Appendix}
\label{appendix}

\subsection{LAr filter regeneration}
\label{appendixCryo}

\begin{figure} [hbt!]
  \centering
  \includegraphics[width=1\textwidth]{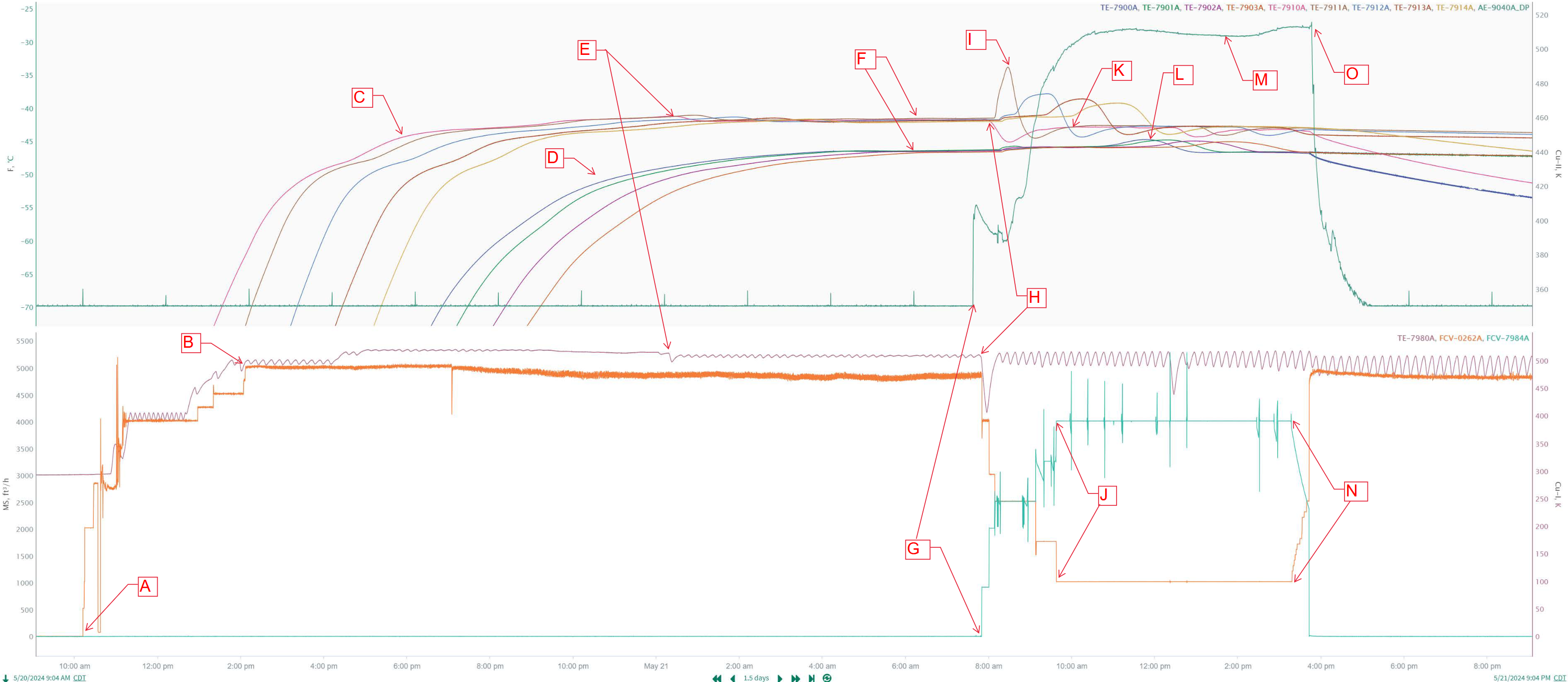}
\caption{Copper media filter regeneration plot: 
\textbf{(A)} Opened flow control valve, FCV-0262A, from argon dewar set to 5000 SCFH and heater, HE-7978, outlet temperature set to 515\,K.
\textbf{(B)} Flow from FCV-0262A reached 5000\,SCFH and heater outlet temperature achieved 515\,K.
\textbf{(C)} Temperatures rising in first Cu filter.
\textbf{(D)} Temperatures rising in second Cu filter.
\textbf{(E)} Heater outlet temperature adjusted from 515\,K to 500\,K due to rise in temperature on first filter media overnight. First Cu filter RTD measurement decreased and settled as a result.
\textbf{(F)} Temperatures on both copper filters equilibrated overnight. First and second Cu filter RTD’s around 460\,K and 440\,K, respectively.
\textbf{(G)} Opened flow control valve, FCV-7984A, from hydrogen/argon tube trailer. Dew point measurement, AE-9040A, rise indicating moisture. Flow from argon dewar and hydrogen/argon tube trailer adjusted simultaneously to maintain consistent overall flow rate, eventually reaching 1:1 ratio.
\textbf{(H)} Heater, HE-7978, tripped because the logic on the heater had mismatch in flow with the monitor vs. the output. As a result, the first RTD on the first Cu filter dropped from 458\,K to 445\,K in approximately 30 minutes before recovering. Exothermic reaction was attenuated as a result.
\textbf{(I)} Exothermic reaction from the hydrogen/argon gas observed in first Cu filter 2nd RTD with an increase of approximately 30\,K.
\textbf{(J)} Increased flow from hydrogen/argon tube trailer to 4000\,SCFH, while simultaneously adjusting argon flow to 1000\,SCFH.
\textbf{(K)} Exothermic reaction observed in first Cu filter 3rd, 4th, and 5th RTD’s resulting in a temperature change of approximately 10\,K.
\textbf{(L)} Exothermic reaction observed in second Cu filter resulting in a temperature change of approximately 5\,K was observed.
\textbf{(M)} Dew Point measurement indicates moisture being generated and released from the filter media.
\textbf{(N)} Stopped flowing hydrogen/argon mixture and increased argon flow rate to begin molecular sieve regeneration.
\textbf{(O)} Dew point reading decreased while flowing through molecular sieve.} 
\label{fig:CuRegen}
\end{figure}

\begin{figure} [hbt!]
  \centering
  \includegraphics[width=1\textwidth]{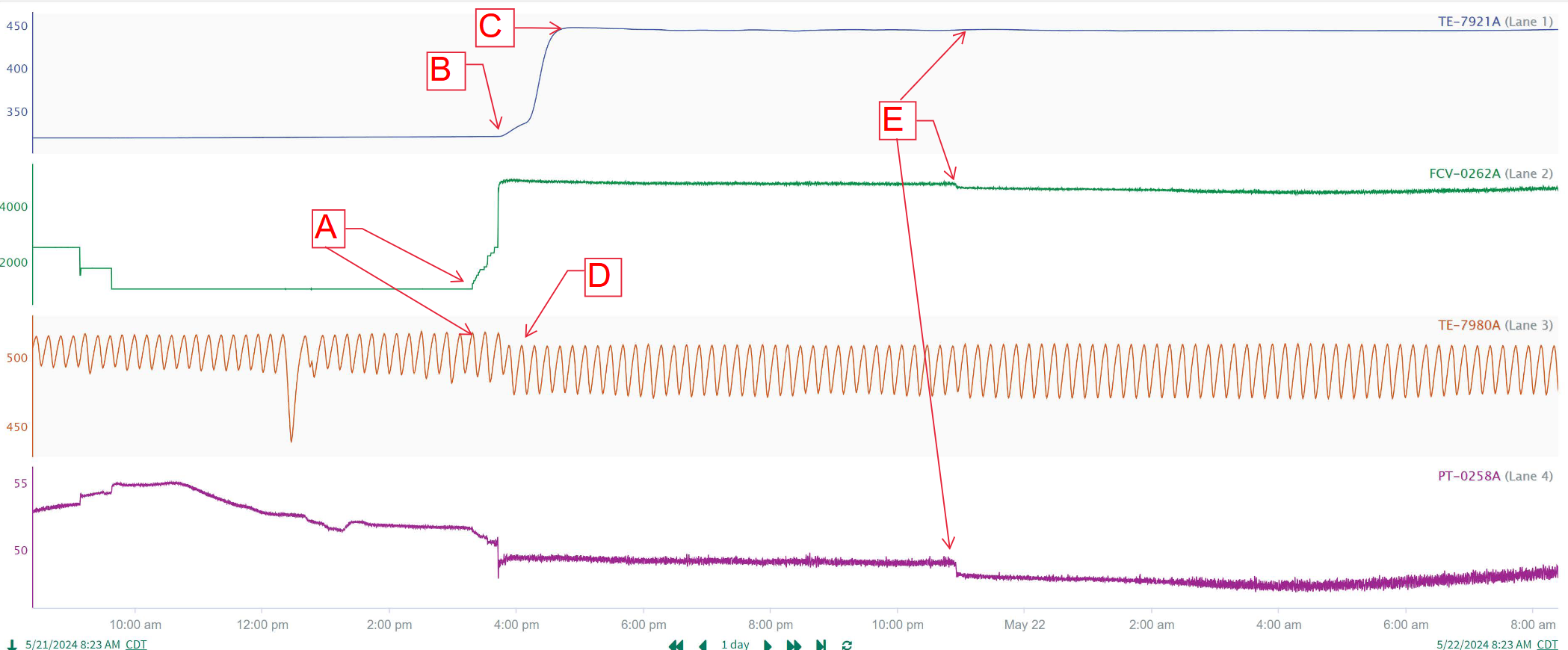}
\caption{Molecular sieve filter regeneration plot: 
\textbf{(A)} Heater, HE-7978, outlet flow still at 500\,K from Cu regeneration; increased flow of argon to 5000\,SCFH via FCV-0262A.
\textbf{(B)} Molecular sieve filter temperature, TE-7921A, increasing.
\textbf{(C)} Molecular sieve filter achieves temperature of 450\,K.
\textbf{(D)} Heater outlet temperature reduced to 490\,K.
\textbf{(E)} Slight flow loss from 4800 to 4650\,SCFH due to the pressure loss from the argon dewar seen.} 
\label{fig:MSRegen}
\end{figure}

\newpage
\subsection{TPC position, non-standard readout channel summary, drift component details}
\label{appendixTPC}

\begin{table}[htb]
    \centering
    \caption{Summary of non-standard TPC readout channels. A total of 80 channels with non-standard behaviour; 0.7\% of the total number of channels.}
    \label{tab:tpcnonstandard}
    \begin{tabular}{c c c c}
    \hline \hline{}
      \makecell{\textbf{Cause of TPC}  \\ \textbf{readout channel problem}} & \textbf{Number of channels} & \textbf{When identified} \\
        \hline \hline
        Missing wires due to support combs & 4 & design\\ \hline
        \makecell{Disconnected wire at jumper connector \\ (electrically connected to adjacent wire)} & 1 (2 affected) & assembly \\ \hline
        Short on FEMB & 64 & installation \\ \hline
        Low noise (consistent with disconnection) & 2 & installation \\ \hline
        High noise & 1 & cooldown \\ \hline
        High noise & 1 & first TPC operation \\ \hline
        Wires touching & 2 & first TPC operation \\ \hline
        Unknown & 4 & first months of operation \\ \hline \hline
    \end{tabular}
\end{table}

\begin{table}[htb]
    \centering
    \caption{Location of the TPC inside of the cryostat.  Distances between TPC component and cryostat walls with a measurement precision of 20mm.}
     \label{tab:tpcposition}
    \begin{tabular}{c c}
    \hline \hline
        TPC component to cryostat wall measurement & Distance (mm) \\ \hline \hline
        West APA to cryostat north wall & 1319 \\ \hline
        East APA to cryostat north wall & 1316 \\ \hline
        CPA to cryostat north wall & 1293 \\ \hline
        West APA to cryostat west wall & 422 \\ \hline
        East APA to cryostat east wall & 417 \\       \hline 
        West APA to cryostat floor & 536 \\ \hline
        East APA to cryostat floor & 538 \\   \hline  
        CPA to cryostat floor & 523 \\ \hline
        CPA beam top to cryostat top cap & 147 \\ \hline
        \hline
    \end{tabular}
    
\end{table}

\begin{table}[htb!]
\centering
  \caption{SBND TPC drift electrical components.}
\label{tab:HVDriftPar-appendix}
\begin{tabular}{c c}
\hline \hline{}
%CPA dimensions (H x L x W) & 4194\,mm x 5213\,mm x 54\,mm \\ \hline
%CPA tubing OD & 54\,mm \\ \hline
%CPA tubing wall thickness & 2\,mm \\ \hline
%CPA mesh aperture & 5\,mm \\ \hline
%CPA mesh wire thickness & 1\,mm  \\ \hline
%CPA mesh transparency & 70 \% \\ \hline
Field Cage resistor model & MFG SM104FE-1000M \\ \hline
Field Cage resistor value & 1\,G$\Omega$ \\ \hline
Field Cage MOV model & ERZ-V14D182 \\ \hline
Field Cage MOV clamping voltage & 1.7\,kV \\ \hline
%N. Field Cage modules per drift volume & 8 \\ \hline
%Field Cage module total resistance & 16.7\,G$\Omega$ \\ \hline
HV Power Supply model & Heinzinger PNChp 150000 -1 neg \\ \hline
HV Power Supply input & 1/N/PE 1\,V \\ \hline
%HV Power Supply output & DC 0–150000V, 0–1mA\\ \hline
%HV Power Supply readback sampling frequency & 20 Hz\\ \hline
HV cable model & Dielectric Sciences 2121 \\ \hline
HV cable gauge & 12\,AWG \\ \hline
HV cable voltage rating & 150\,kV DC \\ \hline
HV Filter resistor model & Vishay Dale DZZ1 \\ \hline
HV Filter resistor value & 1\,M$\Omega$ \\ \hline
Ground break resistor value & 10\,$k\Omega$ \\ \hline
Ground break diode & 1N4005 \\ \hline
\hline 
\end{tabular}

\end{table}

\clearpage

\subsection{PMT circuit diagrams}
\label{appendixPMT}

\begin{figure}[!htbp]
    \centering
    \includegraphics[width=0.95\textwidth]{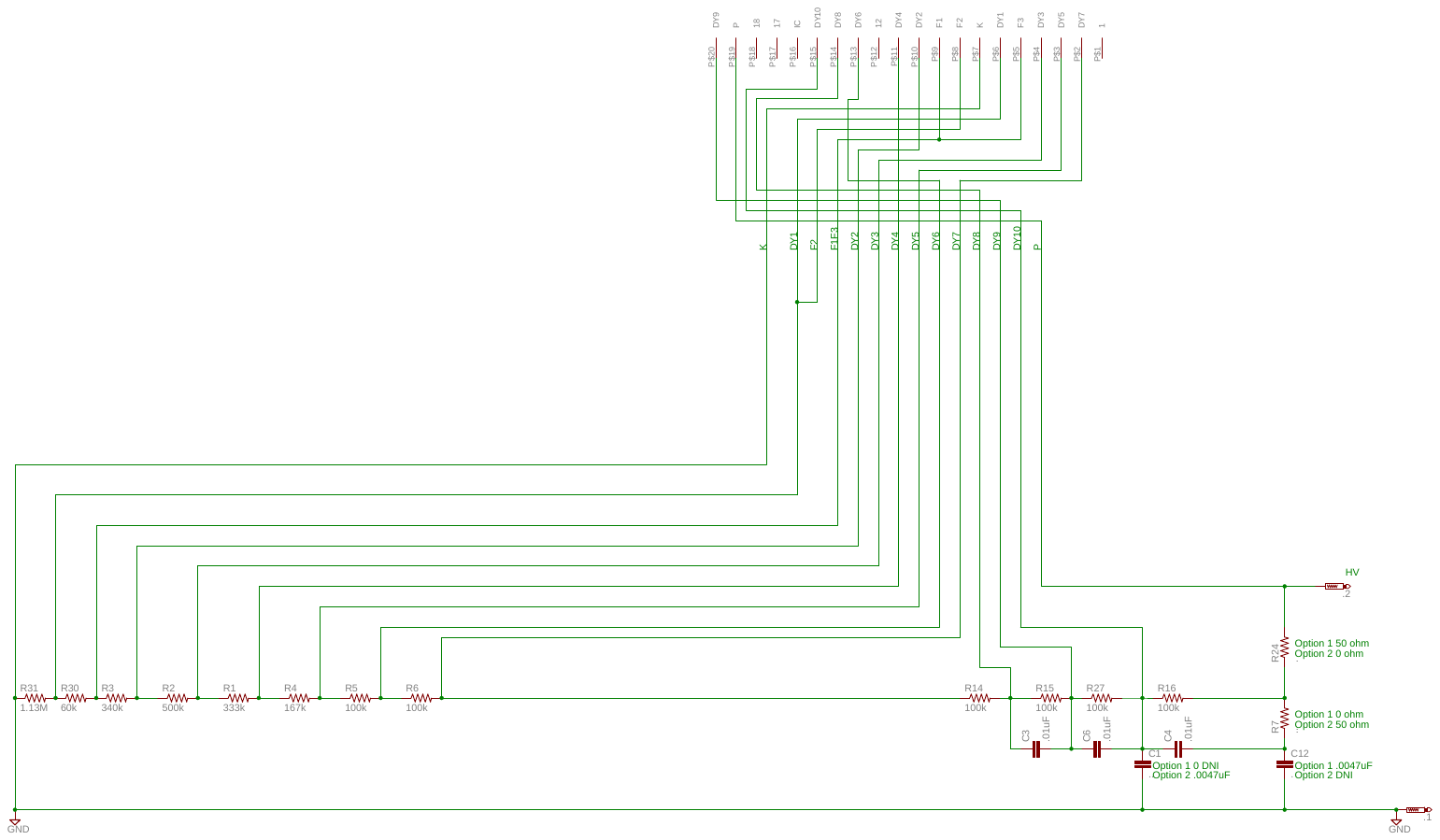}
    \includegraphics[width=0.65\textwidth]{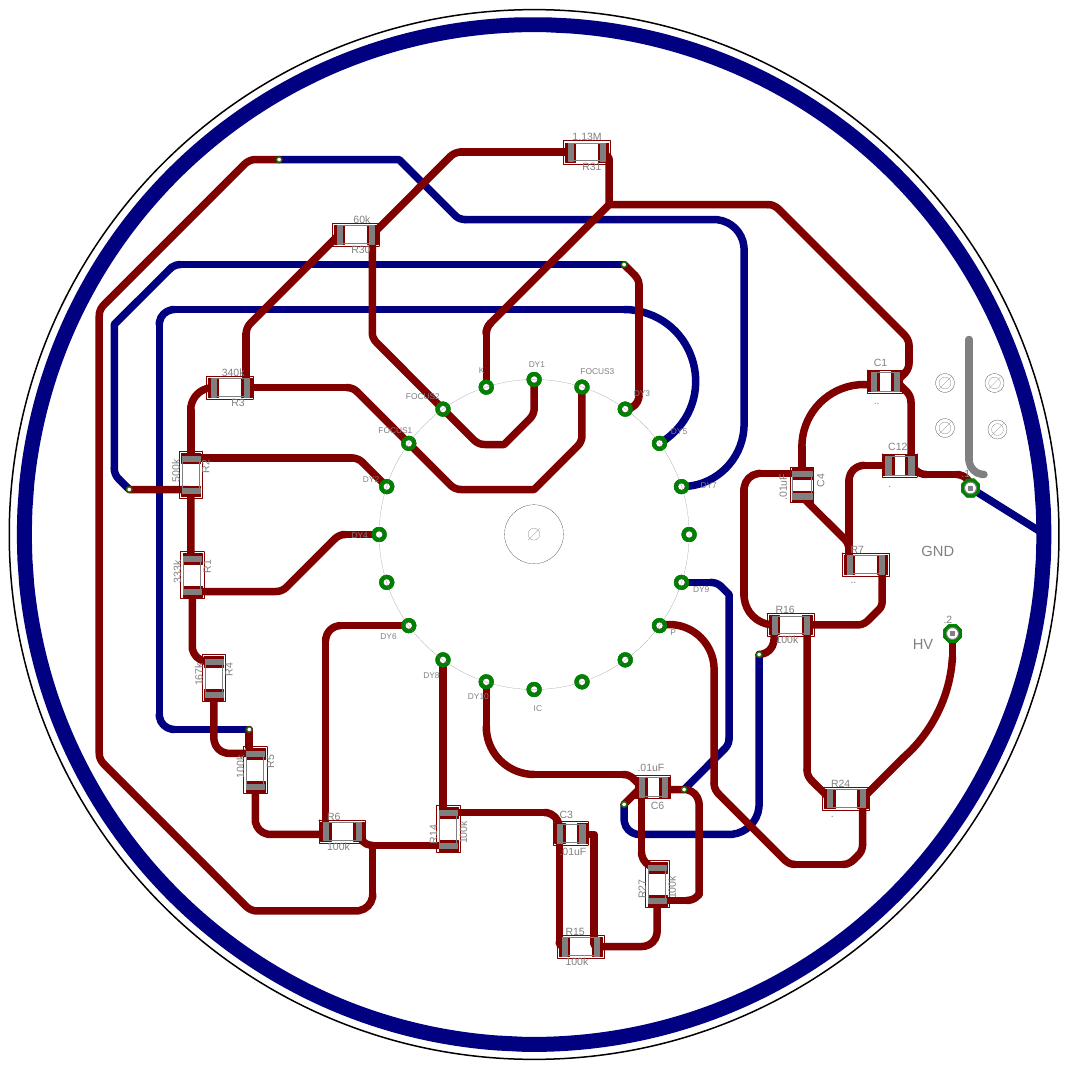}
    \caption{Top: Schematic circuit diagram of the PMT base.  Bottom: layout of the PMT base. }
    \label{fig:pmt_base}
\end{figure}

\begin{figure}[!htbp]
    \centering
    \includegraphics[width=0.98\textwidth]{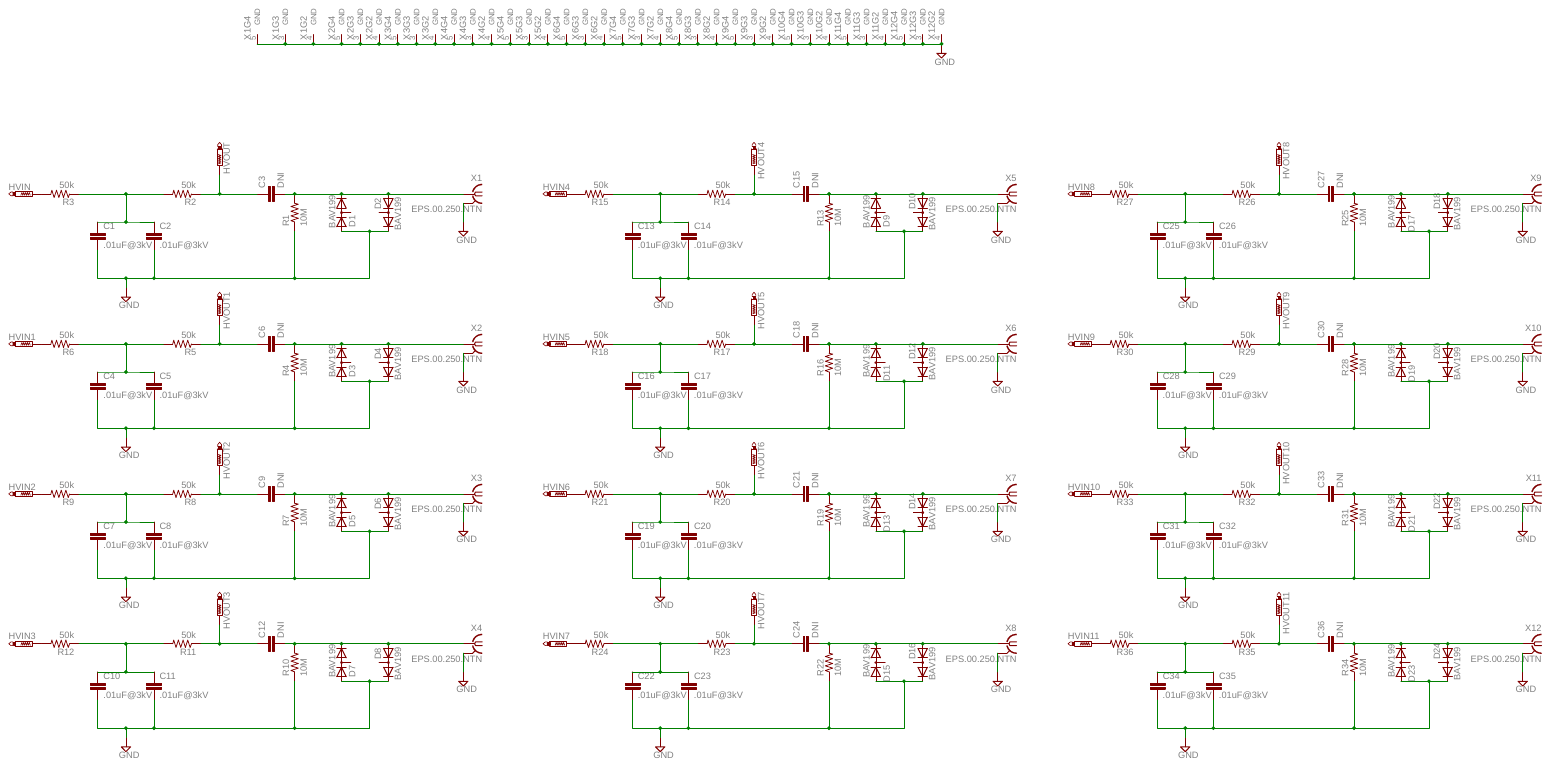}
    \caption{Schematic circuit diagram of the PMT HV breakout box that separates the HV supply from the signal path to the digitizers.}
    \label{fig:pmt_breakoutbox}
\end{figure}

\newpage

\subsection{CRT position and module type details}
\label{appendixCRT}

\begin{table}[ht]
    \centering
     \caption{Distances from the plane of each CRT wall to the plane containing the face it covers on the TPC active volume (AV).}
    \label{tab:crtpositions}
    \begin{tabular}{c c c c c c }
        \hline \hline
        \textbf{Tagger} & \textbf{Distance to TPC AV face (cm)} \\ \hline \hline 
        Bottom & 180 \\ \hline
        South (Upstream) & 195 \\ \hline
        North (Downstream) & 285 \\ \hline
        West & 205 \\ \hline
        East & 205 \\ \hline
        Top Low & 425 \\ \hline
        Top High & 565 \\ \hline \hline
    \end{tabular}

\end{table}

\begin{table}[ht]
    \centering
     \caption{The usage of the different module types in the seven CRT walls.}
    \label{tab:crtmoduleusage}
    \begin{tabular}{c c c c c c c c c}
        \hline \hline
        \multirow{2}{*}{\textbf{Type}} & \multicolumn{7}{|c|}{\textbf{Tagger Wall}} & \multirow{2}{*}{\textbf{Total}} \\ \cline{2-8}
        & Bottom & South & North & West & East & Top Low & Top High & \\ \hline \hline
        A & 0 & 16 & 14 & 10 & 10 & 0 & 0 & 50 \\ \hline
        B & 0 & 0 & 0 & 8 & 8 & 10 & 10 & 36 \\ \hline
        C & 0 & 0 & 2 & 0 & 0 & 0 & 0 & 2 \\ \hline
        D & 0 & 0 & 0 & 0 & 0 & 5 & 0 & 5 \\ \hline
        E & 0 & 0 & 0 & 0 & 0 & 10 & 5 & 15 \\ \hline
        F & 0 & 0 & 0 & 0 & 0 & 0 & 5 & 5 \\ \hline
        G & 0 & 0 & 0 & 0 & 0 & 0 & 5 & 5 \\ \hline
        H & 20 & 0 & 2 & 0 & 0 & 0 & 0 & 22 \\ \hline
        I & 6 & 0 & 0 & 0 & 0 & 0 & 0 & 6 \\ \hline \hline
        Total & 26 & 16 & 18 & 18 & 18 & 25 & 25 & 146 \\ \hline \hline
    \end{tabular}

\end{table}

\newpage

\subsection{DAQ detailed dataflow diagram}
\label{appendixDAQ}

\begin{figure}[!htbp]
    \centering
    \includegraphics[width=0.98\textwidth]{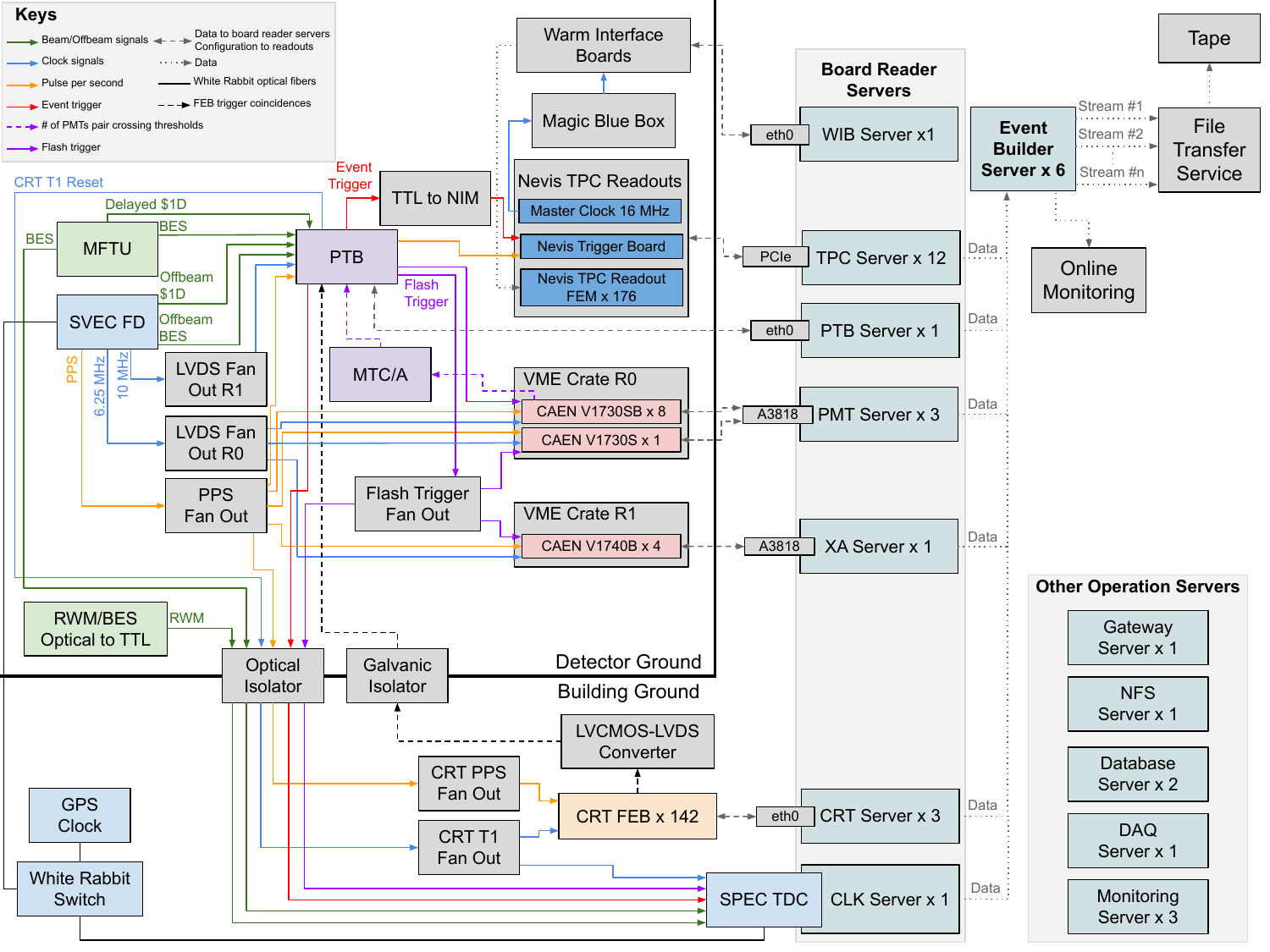}
    \caption{Detailed diagram of the DAQ dataflow. }
    \label{fig:DAQDiagram}
\end{figure}

%\newpage
%\subsection{Appendix B}
%\label{appendix:TPCdaq}
%Changes to TPC readout system between MicroBooNE and SBND to address technical specifications:

%\begin{enumerate}    
%    \item SBND introduced a variable delay in the NTB firmware of up to four 2\,MHz-samples for triggers received near the frame boundary in order to resolve an issue previously seen in MicroBooNE. 
%    This issue was the propagation of the trigger signal from the NTB to the FEM cards failed when the external trigger to the NTB was received on or near the TPC readout frame boundary, resulting in missing event data from some or all of the FEMs. %In SBND, this was resolved by introducing a variable delay in the NTB firmware of up to four 2\,MHz-samples for triggers received near the frame boundary.
%    \item %A related upgrade to XMIT firmware was also made. 
%    SBND also removed a patch to the XMIT firmware which MicroBooNE had required because the NTB firmware fix above was not implemented. 
%    MicroBooNE configured the XMIT to skip data transfers when triggers were received near frame boundaries, by SBND's delay of trigger signals near frame boundaries this intervention is no longer needed. 
    %In MicroBooNE, the XMIT was configured to skip starting a data transfer when a trigger was received at the frame boundary, to avoid trying to send data that was missing. %With the above fix in the NTB firmware, this XMIT firmware patch was no longer needed, and was therefore removed.
%\end{enumerate}

%\paragraph{Note added.} This is also a good position for notes added after the paper has been written.

\clearpage
\bibliographystyle{JHEP} 
\bibliography{biblio}  

\end{document}